\documentclass[a4paper,twoside,12pt,titlepage]{mybook}
\usepackage{a4}

\usepackage{amsmath,amsfonts,amssymb,amsthm}
\usepackage{url}
\usepackage{float}
\usepackage{graphicx}
\usepackage{subfigure}
\usepackage[sectionbib]{bibunits}
\usepackage{algorithmic}
\usepackage{setspace}
\usepackage{fancyhdr}
  \renewcommand{\chaptermark}[1]{\markboth{\chaptername \ \thechapter \ \ #1}{}}
  
\usepackage{adfatitlepage}
\usepackage{rotating}
\usepackage[none]{hyphenat}
\usepackage[intoc]{nomen}
\usepackage{pstricks}
\usepackage{array}
\usepackage{substr}
\usepackage[chapter]{algorithm}
\usepackage{url}
\usepackage{verbatim}
\usepackage{cite}
\usepackage{mathrsfs}
\usepackage{multirow}
\usepackage{bm}
\usepackage{lineno}
\usepackage{array}
\usepackage{color}
\usepackage{latexsym}
\usepackage{amsxtra}
\usepackage{acronym}
\usepackage{longtable}
\usepackage{nomencl}
\usepackage{titletoc}
\usepackage{stfloats}
\usepackage{hhline}
\usepackage{tabularx}
\usepackage{makecell}
\usepackage{enumerate}
\usepackage{epstopdf}
\usepackage{epsfig}

\newtheorem{theo}{Theorem}[chapter]
\newtheorem{lemm}{Lemma}[chapter]
\newtheorem{coro}{Corollary}[chapter]

\newtheorem{rema}{Remark}[chapter]

\newcommand*{\QEDA}{\hfill\ensuremath{\blacksquare}}
\DeclareMathOperator{\sinc}{sinc}

\makenomenclature

\usepackage{geometry}
\begin{document}
\begin{titlepage}
\begin{center}
\vspace*{1cm}
\Huge \hspace{-15mm}\textbf{Millimeter Wave Systems for}\\
\Huge\hspace{-15mm}\textbf{Wireless Cellular Communications}\\
\vspace{1.5cm}
\hspace{-15mm}\normalsize\textbf{Lou Zhao}
\\
\vspace{2cm}
\normalsize
{\hspace{-15mm}A thesis submitted to the Graduate Research School of\\
\hspace{-15mm}The University of New South Wales\\
\hspace{-15mm}in partial fulfillment of the requirements for the degree of\\
\text{ \ }\\
\hspace{-15mm}\textbf{Doctor of Philosophy}}\\
\vspace{2cm}
\hspace{-15mm}\includegraphics[width=0.4\textwidth]{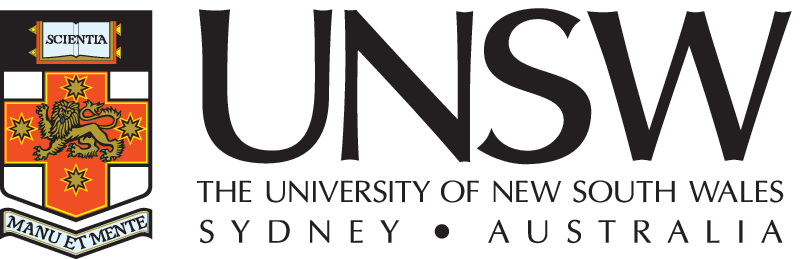}\\
\vspace{2cm}
\textbf{\hspace{-15mm}School of Electrical Engineering and Telecommunications\\
\hspace{-15mm}Faculty of Engineering\\
\hspace{-15mm}The University of New South Wales}\\
\vspace{2cm}
\hspace{-15mm}{August 2018}
\end{center}
\end{titlepage}

\frontmatter
\onehalfspacing
\pagestyle{empty}
\section*{Copyright Statement}

I hereby grant The University of New South Wales or its agents the right
to archive and to make available my thesis or dissertation in whole or part
in the University libraries in all forms of media, now or hereafter known,
subject to the provisions of the Copyright Act 1968. I retain all proprietary
rights, such as patent rights. I also retain the right to use in future works
(such as articles or books) all or part of this thesis or dissertation.

I also authorise University Microfilms to use the abstract of my thesis
in Dissertation Abstract International (this is applicable to doctoral
thesis only).

I have either used no substantial portions of copyright material in my thesis
or I have obtained permission to use copyright material; where permission
has not been granted I have applied/will apply for a partial restriction
of the digital copy of my thesis or dissertation.

\vspace*{1.5cm}
\noindent
\parbox{3in}{Signed \dotfill}

\vspace*{0.5cm}
\noindent
\parbox{3in}{Date  \dotfill}

\vspace{1cm}

\section*{Authenticity Statement}

I certify that the Library deposit digital copy is a direct equivalent of
the final officially approved version of my thesis. No emendation of content
has occurred and if there are any minor variations in formatting, they are
the result of the conversion to digital format.

\vspace*{1.5cm}
\noindent
\parbox{3in}{Signed \dotfill}

\vspace*{0.5cm}
\noindent
\parbox{3in}{Date \dotfill}

\clearpage{\thispagestyle{empty}\cleardoublepage}
\section*{Originality Statement}

I hereby declare that this submission is my own work and to the best
of my knowledge it contains no material previously published or written
by another person, or substantial portions of material which have been
accepted for the award of any other degree or diploma at UNSW or any
other educational institute, except where due acknowledgment is made
in the thesis.  Any contribution made to the research by others, with
whom I have worked at UNSW or elsewhere, is explicitly acknowledged in
the thesis. I also declare that the intellectual content of this thesis
is the product of my own work, except to the extent that assistance from
others in the project's design and conception or in style, presentation
and linguistic expression is acknowledged.

\vspace*{1.5cm}
\noindent
\parbox{3in}{Signed \dotfill}

\vspace*{0.5cm}
\noindent
\parbox{3in}{Date  \dotfill}

\clearpage{\thispagestyle{empty}\cleardoublepage}

\pagenumbering{roman}
\pagestyle{fancy}
\fancyhf{}
\setlength{\headheight}{15pt}
\fancyhead[LE,RO]{\footnotesize \textbf \thepage}
\chapter*{}

\vspace{+5cm}
\begin{center}
    \large \emph{Dedicated to my parents, my wife, and my son.}
\end{center}

\doublespacing
\chapter*{Abstract}
\addcontentsline{toc}{chapter}{\protect\numberline{}{Abstract}}

This thesis considers channel estimation and multiuser (MU) data transmission for time division duplex (TDD) massive multiple-input multiple-output (MIMO) systems with fully digital/hybrid structures in millimeter wave (mmWave) channels. The work reported in this thesis contains novel mmWave transmission schemes and performance analysis, which provides insights into the design of hybrid mmWave networks. It contains three main contributions.

In this thesis, we first propose a tone-based linear search algorithm to facilitate the estimation of angle-of-arrivals (AoAs) of the strongest line-of-sight channel components as well as scattering components of the users at the base station (BS) with fully digital structure. Our results show that the proposed maximum-ratio transmission (MRT) based on the strongest components can achieve a higher data rate than that of the conventional MRT, under the same mean squared errors (MSE) of channel estimation. In addition, we quantify the achievable rate degradation due to phase quantization errors and propose an angular domain user scheduling algorithm for mmWave systems to improve the users' receive signal-to-interference-plus-noise ratio (SINR).

Second, we develop a low-complexity channel estimation and beamformer\\/precoder design scheme for hybrid mmWave systems, which utilizes tone signals and orthogonal pilots for the design of analog beamforming matrices and digital precoding matrices, respectively. In addition, the proposed scheme applies to both non-sparse and sparse mmWave channel environments. We then leverage the proposed scheme to investigate the downlink achievable rate performance. The results show that the proposed scheme obtains a considerable achievable rate of fully digital systems. Taking into account the effect of various types of errors, we investigate the achievable rate performance degradation of the considered scheme.

Third, we extend our proposed scheme to a multi-cell MU mmWave MIMO network. We derive the closed-form approximation of the normalized MSE of channel estimation under pilot contamination over Rician fading channels. Furthermore, we derive a tight closed-form approximation and the scaling law of the average achievable rate. Our results unveil that channel estimation errors, the intra-cell interference, and the inter-cell interference caused by pilot contamination over Rician fading channels can be efficiently mitigated by simply increasing the number of antennas equipped at the desired BS.

\doublespacing
\chapter*{Acknowledgments}

This work would not have been done without the encouragement and the support from some people I have met during my fascinating Ph.D. journey.

First and foremost, I would like to thank my supervisor Professor Jinhong Yuan with my deepest gratitude for offering me a Ph.D. opportunity three and half years ago and having faith in me ever since.
He guides me throughout this work with patience, technical and personal support, and encouragement.
He tolerates my shortcomings and helps me to overcome my weakness.
Besides, he is passionate about new technologies and exciting ideas.
I feel fortunate to pursue my Ph.D. degree under his supervision.

Second, I would like to thank my co-supervisor, Dr. Derrick Wing Kwan Ng, for his guidance, constructive suggestions, and help on my research.
Derrick is talented, extremely self-disciplined, and very hard-working.
I will regard him as the role model of my academic career.

I would also like to thank Dr. Giovanni Geraci and Dr. Tao Yang for the first step of my research.
Many thanks to Dr. Lei Yang, Dr. Yixuan Xie, Dr. Tao Huang, Dr. Mengyu Huang, Dr. Chenxi Liu, Dr. Shuang Tian, Dr. Nan Yang, Dr. Shihan Yan, Dr. Yansha Deng, Dr. Yeqing Hu, and others for having a lot of discussions with me.
Special thanks to my colleagues and friends at the University of New South Wales, you guys make the Ph.D. journey funny and interesting.
I will always hold dear the days and nights spent in the office.

Finally, my deepest appreciation goes to my beloved family for everything they have done for me.
I would like to dedicate this thesis to my parents, my wife, and my son.
To my son, you make my life brighter.

\singlespacing

\chapter*{List of Publications} \label{listofpublications}
\addcontentsline{toc}{chapter}{\protect\numberline{}{List of Publications}}
%

\ifpdf
    \graphicspath{{1_introduction/figures/PNG/}{1_introduction/figures/PDF/}{1_introduction/figures/}}
\else
    \graphicspath{{1_introduction/figures/EPS/}{1_introduction/figures/}}
\fi

\noindent{\large\textbf{Journal Articles:}} \vspace{0.1in}
\begin{enumerate}
\item \textbf{L.~Zhao}, D.~W.~K. Ng, and J.~Yuan, ``Multi-user precoding and channel estimation for hybrid millimeter wave systems,'' {\bf\em IEEE J. Sel. Areas Commun.}, vol.~35, no.~7, pp. 1576--1590, Jul. 2017.
\item \textbf{L.~Zhao}, G.~Geraci, T.~Yang, D.~W.~K. Ng, and J.~Yuan, ``A tone-based {AoA} estimation and multiuser precoding for millimeter wave massive {MIMO},''
   {\bf\em IEEE Trans. Commun.}, vol.~65, no.~12, pp. 5209--5225, Dec. 2017.
\item \textbf{L.~Zhao}, Z.~Wei, D.~W.~K. Ng, J.~Yuan, and M.~C.~Reed, ``Multi-cell hybrid millimeter wave systems: {Pilot} contamination and interference mitigation,'' {\bf\em IEEE Trans. Commun.}, vol. 66, no. 11, pp. 5740-5755, Nov. 2018.
\item Z.~Wei, \textbf{L.~Zhao}, J.~Guo, D.~W.~K. Ng, and J.~Yuan, ``Multi-Beam NOMA for Hybrid mmWave Systems,'' accepted, {\bf\em IEEE Trans. Commun.}, Nov. 2018.
\end{enumerate}
\vspace{0.1in} \noindent{\large\textbf{Conference Articles:}}
\vspace{0.1in}
\begin{enumerate}
\item \textbf{L.~Zhao}, T.~Yang and G.~Geraci and J.~Yuan ``Downlink multiuser massive {MIMO} in {Rician} channels under pilot contamination'' in {\bf\em Proc. IEEE Intern.
  Commun. Conf. (ICC)}, Kuala Lumpur, May 2016, pp. 1--6.
\item \textbf{L.~Zhao}, D.~W.~K. Ng, and J.~Yuan, ``Multiuser precoding and channel estimation for hybrid millimeter wave {MIMO} systems,'' in { \bf\em Proc. IEEE Intern.
  Commun. Conf. (ICC)}, Paris, May 2017, pp. 1--7.
\item \textbf{L.~Zhao}, Z.~Wei, D.~W.~K. Ng, J.~Yuan and M.~C.~Reed, ``Mitigating pilot contamination in multi-cell hybrid millimeter Wave Systems,'' in { \bf\em Proc. IEEE Intern.  Commun. Conf. (ICC)}, Kansas City, May 2018.
\item Z.~Wei, \textbf{L.~Zhao}, J.~Guo, D.~W.~K. Ng, and J.~Yuan, ``A multi-beam NOMA framework for hybrid mmWave systems,'' in { \bf\em Proc. IEEE Intern.  Commun. Conf. (ICC)}, Kansas City, May 2018 (ICC best paper awards).
\end{enumerate}

\chapter*{Abbreviations} \label{abbreviations}
\addcontentsline{toc}{chapter}{\protect\numberline{}{Abbreviations}}
\markboth{ABBREVIATIONS}{}

\begin{longtable}[t]{ll}
\textbf{3-D} \quad\quad&\mbox{Three-dimensional} \vspace{0.1in}\\
\textbf{5G} \quad\quad&\mbox{Fifth-generation} \vspace{0.1in}\\
\textbf{AoA} \quad\quad&\mbox{Angle-of-arrival} \vspace{0.1in}\\
\textbf{AoD} \quad\quad&\mbox{Angle-of-departure} \vspace{0.1in}\\
\textbf{ADC/DAC}  \quad\quad&\mbox{Analog-to-digital converter/Digital-to-analog converter} \vspace{0.1in}\\
\textbf{AWGN} \quad\quad&\mbox{Additive white Gaussian noise} \vspace{0.1in}\\
\textbf{BS} \quad\quad&\mbox{Base station} \vspace{0.1in}\\
\textbf{CSI} \quad\quad&\mbox{Channel state information} \vspace{0.1in}\\
\textbf{CW}  \quad\quad&\mbox{Continuous wave} \vspace{0.1in}\\
\textbf{CRLB}  \quad\quad&\mbox{Cram\'{e}r Rao lower bound} \vspace{0.1in}\\
\textbf{dB} \quad\quad&\mbox{Decibel} \vspace{0.1in}\\
\textbf{DS} \quad\quad&\mbox{Delay spread} \vspace{0.1in}\\
\textbf{D2D} \quad\quad&\mbox{Device-to-device} \vspace{0.1in}\\
\textbf{DPC} \quad\quad&\mbox{Dirty paper coding} \vspace{0.1in}\\
\textbf{FDD} \quad\quad&\mbox{Frequency division duplex} \vspace{0.1in}\\
\textbf{HPBW} \quad\quad&\mbox{Half-power beam width}\vspace{0.1in}\\
\textbf{i.i.d.} \quad\quad&\mbox{Independent and identically distributed} \vspace{0.1in}\\
\textbf{ISI} \quad\quad&\mbox{Inter-symbol interference} \vspace{0.1in}\\
\textbf{I/Q}  \quad\quad&\mbox{In-phase/Quadrature} \vspace{0.1in}\\
\textbf{IoT}  \quad\quad&\mbox{Internet-of-things} \vspace{0.1in}\\
\textbf{LOS} \quad\quad&\mbox{Line-of-sight} \vspace{0.1in}\\
\textbf{LO} \quad\quad&\mbox{Local oscillator } \vspace{0.1in}\\
\textbf{LS} \quad\quad&\mbox{Least squares } \vspace{0.1in}\\
\textbf{MU} \quad\quad&\mbox{Multi-user} \vspace{0.1in}\\
\textbf{mmWave} \quad\quad&\mbox{Millimeter wave} \vspace{0.1in}\\
\textbf{MIMO} \quad\quad&\mbox{Multiple-input multiple-output} \vspace{0.1in}\\
\textbf{MMSE}  \quad\quad&\mbox{Minimum mean square error} \vspace{0.1in}\\
\textbf{MRC} \quad\quad&\mbox{Maximal-ratio combining} \vspace{0.1in}\\
\textbf{MRT} \quad\quad&\mbox{Maximal-ratio transmission} \vspace{0.1in}\\
\textbf{MSE}  \quad\quad&\mbox{Mean squared error} \vspace{0.1in}\\
\textbf{NMSE}  \quad\quad&\mbox{Normalized mean squared error} \vspace{0.1in}\\
\textbf{NLOS} \quad\quad&\mbox{Non line-of-sight} \vspace{0.1in}\\
\textbf{NOMA} \quad\quad&\mbox{Non-orthogonal multiple access} \vspace{0.1in}\\
\textbf{OLB} \quad\quad&\mbox{Open-loop beamforming} \vspace{0.1in}\\
\textbf{PA}  \quad\quad&\mbox{Power amplifier} \vspace{0.1in}\\
\textbf{PDP} \quad\quad&\mbox{Power delay profile} \vspace{0.1in}\\
\textbf{PAC}  \quad\quad&\mbox{Pilot-aided-CSI} \vspace{0.1in}\\
\textbf{PESA}  \quad\quad&\mbox{Passive electronically scanned array} \vspace{0.1in}\\
\textbf{QoS} \quad\quad&\mbox{Quality-of-service} \vspace{0.1in}\\
\textbf{RA} \quad\quad&\mbox{Random angle} \vspace{0.1in}\\
\textbf{RF} \quad\quad&\mbox{Radio frequency} \vspace{0.1in}\\
\textbf{RCS} \quad\quad&\mbox{Radar cross-section} \vspace{0.1in}\\
\textbf{RACH} \quad\quad&\mbox{Random access channel} \vspace{0.1in}\\\textbf{r.m.s.} \quad\quad&\mbox{Root mean square} \vspace{0.1in}\\
\textbf{SS}  \quad\quad&\mbox{Synchronization signals} \vspace{0.1in}\\
\textbf{SLOS}  \quad\quad&\mbox{Strongest LOS} \vspace{0.1in}\\
\textbf{SLPS}  \quad\quad&\mbox{SLOS-plus-scattering components} \vspace{0.1in}\\
\textbf{SNR}  \quad\quad&\mbox{Signal-to-noise ratio} \vspace{0.1in}\\
\textbf{SINR}  \quad\quad&\mbox{Signal-to-interference-plus-noise ratio} \vspace{0.1in}\\
\textbf{SVD}  \quad\quad&\mbox{Singular value decomposition} \vspace{0.1in}\\
\textbf{TDD} \quad\quad&\mbox{Time division duplex} \vspace{0.1in}\\
\textbf{UMi}  \quad\quad&\mbox{Urban micro-cell} \vspace{0.1in}\\
\textbf{UPA}  \quad\quad&\mbox{Uniform panel array} \vspace{0.1in}\\
\textbf{ULA}  \quad\quad&\mbox{Uniform linear array} \vspace{0.1in}\\
\textbf{US}  \quad\quad&\mbox{User scheduling} \vspace{0.1in}\\
\textbf{WLAN}  \quad\quad&\mbox{Wireless local area networks} \vspace{0.1in}\\
\textbf{ZF} \quad\quad&\mbox{Zero-forcing} \vspace{0.1in}\\
\end{longtable}

\chapter*{List of Notations} \label{listofnotations}
\addcontentsline{toc}{chapter}{\protect\numberline{}{List of Notations}}
\markboth{LIST OF NOTATIONS}{}
Boldface upper-case letters denote matrices, boldface lower-case letters denote vectors, and italics denote scalars.
\begin{longtable}[t]{ll}
$\mathbf{X}^T$ \quad\quad&\mbox{Transpose of $\mathbf{X}$} \vspace{0.1in}\\
$\mathbf{X}^{\ast}$ \quad\quad&\mbox{Complex conjugate of $\mathbf{X}$} \vspace{0.1in}\\
$\mathbf{X}^H$ \quad\quad&\mbox{Conjugate transpose of $\mathbf{X}$ } \vspace{0.1in}\\
$\mathbf{X}^{-1}$ \quad\quad&\mbox{Inverse of $\mathbf{X}$ } \vspace{0.1in}\\
$\mathbf{X}\{i,j\}$ \quad\quad&\mbox{The element in the row $i$ and the column $j$ of $\mathbf{X}$ } \vspace{0.1in}\\
$\det\left(\mathbf{X}\right)$ \quad\quad&\mbox{Determinant of $\mathbf{X}$} \vspace{0.1in}\\
$\mbox{tr}\left(\mathbf{X}\right)$ \quad\quad&\mbox{Trace of $\mathbf{X}$} \vspace{0.1in}\\
$|x|$ \quad\quad&\mbox{Absolute value (modulus) of the a complex scalar $x$} \vspace{0.1in}\\
$\mathbb{C}^{M\times N}$ \quad\quad&\mbox{The space of all $M\times N$ matrices with complex
entries} \vspace{0.1in}\\
$\|\cdot\|_{\mathrm{F}}$ \quad\quad&\mbox{Frobenius norm of a vector or a matrix} \vspace{0.1in}\\
$\mathbf{0}$ \quad\quad&\mbox{Zero matrix. A subscript can be used to indicate the dimension} \vspace{0.1in}\\
$\mathbf{I}_{\mathrm{N}}$ \quad\quad&\mbox{$N$ dimension identity matrix} \vspace{0.1in}\\
$j$ \quad\quad&\mbox{Imaginary unit $j = \sqrt{-1}$} \vspace{0.1in}\\
$\mathrm{\mathbb{E}}_{x}\left[\cdot\right]$ \quad\quad&\mbox{Statistical expectation with respect to random
variable $x$} \vspace{0.1in}\\
$\mathcal{CN}$ \quad\quad&\mbox{Complex Gaussian distribution} \vspace{0.1in}\\
$\ln(\cdot)$ \quad\quad&\mbox{Natural logarithm} \vspace{0.1in}\\
$\log_2(\cdot)$ \quad\quad&\mbox{Logarithm in base two} \vspace{0.1in}\\
$\lambda _{i}(\cdot )$ \quad\quad&\mbox{The $i$-th maximum eigenvalue of a matrix} \vspace{0.1in}\\
$\mathrm{diag}\left\{\bm{a}\right\}$ \quad\quad&\mbox{A diagonal matrix with the entries of $\bm{a}$ on its diagonal} \vspace{0.1in}\\
$\lim$ \quad\quad&\mbox{Limit} \vspace{0.1in}\\
$\mathrm{sinc}(x)$ \quad\quad&\mbox{Sinc function with input $x$} \vspace{0.1in}\\
$\max\left\{\cdot\right\}$ \quad\quad&\mbox{Maximization} \vspace{0.1in}\\
$\min\left\{\cdot\right\}$ \quad\quad&\mbox{Minimization} \vspace{0.1in}\\
$\mathrm{s.t.}$ \quad\quad&\mbox{Subject to} \vspace{0.1in}\\
\end{longtable}

\fancyhead[CE]{\footnotesize \leftmark}
\fancyhead[CO]{\footnotesize \rightmark}

\renewcommand{\chaptermark}[1]{%
\markboth{\MakeUppercase{%
\thechapter.%
\ #1}}{}}

\tableofcontents

\listoffigures


\listofalgorithms
\addcontentsline{toc}{chapter}{\protect\numberline{}{List of Algorithms}}



\mainmatter
\doublespacing

\chapter{Introduction}\label{C1:chapter1}

The significantly increased data rate requirement and quality-of-service (QoS) triggered by the proliferation of smartphones and tablets cannot be satisfied by current communication networks.
In particular, the tremendous number of devices and appliances are expected to connect wirelessly to the Internet, e.g. \\device-to-device (D2D) communications and internet-of-things (IoT) applications, the required capacity exceeds the limit of existing networks.
To address the aforementioned requirements, the fifth generation ($5$G), the true revolution of technologies in both the radio access network and the mobile core network, is expected to support $100-1000$ times higher system capacity than current the fourth generation ($4$G) systems and have attracted tremendous interests from both academia and industry, e.g. \cite{Zhao2017,Zhao2017b,Alkhateeb2015,Zhao2018,Kokshoorn2016,Zhao2018a,Zhao2017ss,Zhao2016ss,Wei2018a,Wei2017,Wei2018bb,Xiang2018,He2017ss,Yang2015,Dai2013,Zhu2014,AZhang2015,Ng2017,Akbar2017,Sohrabi2016,Rusek2013,Andrews2014,Zhao2008,HLin2017,CLin2017,Wang2011,Larsson2014a,Marzetta2010,Rappaport2015,Roh2014}.
For example, $5$G should support $10$ Gb/s peak data rate, $50$ Mb/s guaranteed data rate, $10$ Tb/s/km$^2$ mobile data volume, less than $1$ ms end-to-end latency, $1$ million/km$^2$ number of devices, higher than $99.999\%$ reliability, and less than $1$ meter outdoor terminal location accuracy.
As a result, denser network, larger bandwidth, and higher spectral efficiency are necessary to satisfy these demands.
To meet the stringent spectral efficiency requirement, massive multiple-input multiple-output (MIMO) technology is introduced by using hundreds of antennas equipped at the base station (BS) to serve tens of users simultaneously \cite{Marzetta2010,Larsson2014a,Ferrante2016,Xie2017}.
It is proved that, by adopting low computational complexity linear precoding schemes, such as maximal-ratio transmission (MRT) and zero-forcing (ZF), massive MIMO systems can significantly improve the spectral efficiency and fully explore a large amount of available spatial degrees of freedom \cite{Marzetta2010,Yang2015,Bogale2015,Deng2015,Yang2015bb,Dai2013}.
%
Besides, to overcome the bandwidth limitation in licensed band, millimeter wave (mmWave) systems that migrate from sub-$6$ GHz to higher frequencies in unlicensed band can inevitably offer a huge trunk of a bandwidth of the order of gigahertz, e.g. an unlicensed spectrum ranging from $30$ GHz to $300$ GHz, to achieve ultra-high data rate communication \cite{Alkhateeb2015,Dai2015,Bjornson2016,Zhao2017b,Ng2017,Akbar2017,Sohrabi2016,Rusek2013,Bogale2015,Zhang2015a,AZhang2015,Kokshoorn2016,Elkashlan2014}.
In practice, mmWave and massive MIMO can complement each other to overcome some drawbacks in each technology.
For example, due to the short wavelength of mmWave frequencies, mmWave technology can shrink the physical size of antenna arrays of massive MIMO systems.
Therefore, the antenna array of mmWave massive MIMO is small which enables flexible and practical deployment.
On the other hand, massive MIMO, equipping a BS with hundreds of antennas, can provide significant antenna array gains to help mmWave communication systems to compensate the inherent high propagation path loss, low penetration coefficients, and high signal attenuation caused by raindrop absorption \cite{Rahimian2011,Rappaport2015}.
Thus, the combination of mmWave communication systems with massive MIMO systems is considered as one of the promising candidate technologies for 5G communication systems with many potential opportunities for research \cite{Alkhateeb2015,Swindlehurst2014,Bjornson2016,Ng2017,Bogale2015}.

\vspace{-0mm}
\section{Background}

In this section, we will briefly introduce the concepts of massive MIMO and the characteristics of mmWave channels.
\vspace{-0mm}
\subsection{The Concepts of Massive MIMO}
Massive MIMO technology is one of the key technologies for $5$G.
Compared with the conventional multi-user (MU) MIMO systems, massive MIMO is a special form of MIMO systems with hundreds of antennas equipped at the BS to simultaneously serve tens of users in their cells \cite{Ng2012}.
In general, a large amount of experimental and theoretical research shows that several favorable properties can be obtained when the number of antennas equipped at the BS is significantly larger than the number of users.

There are several main advantages when the number of antennas is sufficiently large: (1) the effects of small-scaling fading vanish (also known as channel hardening); (2) simple linear signal processing algorithms can achieve considerable rate performance compared to optimal algorithms, e.g. dirty paper coding (DPC); (3) higher spectral efficiency and higher energy efficiency \cite{Marzetta2010,Hoydis2013aa,Hoydis2013,Muller2014,Yang2013,Mumtaz2017}.
The more antennas the transmitter/receiver are equipped with, the better performance in terms of data rate and link reliability the system can achieve.
However, the cost, the energy consumption, and the complexity of hardware (power amplifier (PA) and analog-to-digital converter/digital-to-analog converter (ADC/DAC)) increase with the increasing number of antennas \cite{Rusek2013,Bjornson2016,Larsson2014a,Yang2015,Bogale2016}.

\subsubsection{Key Aspects of Massive MIMO:}
A lot of efforts have been dedicated to the investigation of massive MIMO from different aspects, e.g. architecture aspects \cite{Ngo2017,Hoydis2013aa,Xiang2014,Zhu2014,Wu2015}, capacity and fundamental aspects \cite{Marzetta2010,Ngo2013,Hoydis2013,Muller2014,Yang2013,Jose2011,Ozgur2013}, channel state information (CSI) acquisition \cite{Ngo2017aa,Fernandes2013,Choi2014,Yin2013,Noh2014,Ma2014,Fan2017,Zhao2017b,Ghavami2017,Yin2016aa}, detection algorithms \cite{Svac2013,Vardhan2008,Wu2014,Wu2014aa,Dai2015}, downlink precoding design \cite{Alrabadi2013,Hong2013,Alkhateeb2017aa,Alkhateeb2015,Sohrabi2016,Bogale2014aa,Liu2014,Choi2015}, channel measurements and modeling \cite{Gao2015,Wu2015,Gao2015b}, resource allocation \cite{Dai2013,Ng2012a,Ng2012,Huang2013,Bjornson2015c,Bjornson2016b,Bjornson2014d}, hardware impairments \cite{Bjornson2014d,Pitarokoilis2012,Zhao2017,Choi2015b,Zhang2016}, beamforming and precoding \cite{Wu2017,Zhu2017,Wu2016,Zhao2017jj,Zarei2017,Saxena2017} and performance analysis \cite{Zhang2017a,Ngo2013b,Huh2011,Yin2014,Chuah2002,Aktas2006,Wagner2012,Yang2013v,Ng2012a,Ng2012}.
These aspects of massive MIMO had been well studied and interested readers may refer to these works for details.

\vspace{+1mm}
\subsection{MmWave Channels Statistics}
The study of mmWave channels can be dated back to decades ago \cite{Yoneyama1981,Pozar1983}.
Due to the high propagation path loss, mmWave technology has not been widely adopted for cellular application in the past.

Nowadays, the spectral resource below $6$ GHz becomes tense while sufficiently large bandwidth is still available at the mmWave band.
Thus, large efforts have been devoted to the research of mmWave communication and there are some practical communication systems utilizing the mmWave band, e.g. mmWave wireless local area networks (WLAN) IEEE $802.11$ ad.
However, these frequencies have not been well-explored for cellular applications.
As mmWave technology is considered as a key feature for $5$G, the features of mmWave channels must be considered carefully for system design.


\vspace{+2mm}
\subsubsection{Path Loss:}

According to the free space propagation path loss model provided by the Friis transmission formula, the received power can be expressed as \cite{Rappa2013}\vspace{-0mm}
\begin{equation}
P_{\mathrm{r}}(d_{\mathrm{r}}) = P_{\mathrm{t}}G_{\mathrm{t}}G_{\mathrm{r}} \left( \frac{\lambda}{4\pi}\right)^{2}d_{\mathrm{r}}^{-n}, \label{Eq:11}\vspace{-0mm}
\end{equation}
where $\lambda$ is wavelength, $d$ is the transmission distance between the desired BS and the user, $n$ is the path loss exponent, and $P_{\mathrm{t}}$ is the transmit power.
In addition, $G_{\mathrm{t}}$ and $G_{\mathrm{r}}$ are the antenna gains of the transmitter and receiver, respectively.

We can rewrite Equation (\ref{Eq:11}) in the form of path loss.
The commonly adopted empirical propagation path loss model is considered as a function of distance and carrier frequency, which is given by\vspace{-0mm}
\begin{equation}
\varpi\text{\ }\mathrm{[dB]}={{10\alpha \log_{10}{d_{\mathrm{r}}}+\xi \log_{10}\left( 4\pi\frac{1}{\lambda}\right)}},\label{Eq_PL}\vspace{-0mm}
\end{equation}
where in a log-log plot of the path loss curve, slope $\alpha $ and intercept $\xi $ are estimated by least squares (LS) linear regression over the measured data \cite{Akdeniz2014,Hur2016}.
For different scenarios, $\alpha $ and $\xi $ may have different values, e.g. line-of-sight (LOS) environment and non line-of-sight (NLOS) environment.
In addition, the propagation path loss model should also take into account the impacts of rain attenuation, atmospheric absorption, and shadowing \cite{Akdeniz2014,Rappaport2015,Hur2016,Xiao2017,Ng2017}.

Generally speaking, the severe path loss of mmWave channels is one of the key technical challenges in preventing mmWave technology to be implemented in conventional cellular communication systems \cite{Akdeniz2014,Rappaport2015,Hur2016,Haneda2016}.
Nowadays, there are two methods to overcome such a challenge, e.g. increasing antenna array gain and decreasing the propagation distance.
In fact, it is expected that small cell will serve as a core structure of future cellular systems \cite{Ngo2017,Bjornson2013as}.
Hence, applying mmWave technology to small-cell urban mobile networks becomes a new trend \cite{MacCartney2017}.
In particular, it can significantly reduce the requirement of transmit power of mmWave systems.
In addition, the high antenna array gain can be enabled by adopting the massive numbers of antennas.


\vspace{+1mm}
\subsubsection{Power Delay Profile and Delay Spread:}
The power delay profile gives the intensity of a signal received through a multi-path channel as a function of time delay.
The shape of power delay profile (PDP) in mmWave channel measurements is a superposition of multiple exponentially decaying spectrums.
The delay spread (DS) is usually quantified through root mean square (r.m.s.) and considered as a collection of the multi-path containing significant energy which spread over the PDP.
Besides, the DS of the channel is an important metric to understand the required system overhead to facilitate communication \cite{Raghavan2017}.
To provide sufficient channel characteristics information for the system design (concerning the length of cyclic prefix for a multi-carrier design), it is necessary to proceed the PDP estimation in mmWave channel measurements, which is called channel sounding.

Channel sounding for the estimation of PDP and calculating DS can be performed with both omni-directional antenna as well as directional antenna array/horn antennas.
The basic procedure of channel sounding with directional horn antennas is listed as follows \cite{Samimi2016,Raghavan2017,Ko2017}:
\begin{itemize}
\item Perform an azimuthal scan and produce $\dfrac{360}{T}$ number of slices (the beamwidth of main lobe of directional horn antenna can cover $T$ degrees)
\item Mark the absolute propagation times for different sounding slices
\item Synthesize the equivalent omni-directional PDP over an absolute propagation time axis
\end{itemize}

The PDP sounding and DS calculation for the indoor environment can be conducted via omni-directional antennas due to the short propagation distance.
The results illustrate that for most indoor scenarios the DS estimated via directional antennas as well as omni-directional antennas are small, e.g. $30-50$ ns indoor office and $50-90$ ns indoor shopping mall \cite{Samimi2016,Raghavan2017,Ko2017}.
However, the outdoor PDP environment sounding should be conducted via directional horn antennas with large antenna-gain to compensate the severe propagation path loss.
Then, by synthesizing the estimated PDP via directional antennas over an absolute propagation time axis, we can obtain the equivalent omni-directional PDP.
For the equivalent omni-directional PDP, there are different outdoor scenarios where a significantly large DS have been observed, e.g. $150-300$ ns for street canyon settings and even up to $800-1000$ ns for open square settings \cite{Samimi2016,Raghavan2017,Ko2017}.
In addition, the simulation results conducted by adopting ray tracing demonstrate that scattering and reflections from small objects affect the mmWave propagation channel \cite{Nguyen2014,Hur2016,Samimi2016}.
These simulation and experiment results for outdoor scenarios can be explained with radar cross-section (RCS) effect, where small objects that do not participate in electromagnetic propagation at low frequencies show up at higher frequencies \cite{Samimi2016,Raghavan2017,Hur2016}.
According to the simulation and experiment results as mentioned earlier, the sparse scattering assumption of mmWave channels holds for most indoor scenarios.
{\color{black}The omni-directional delay spread is comparable with after-beamforming delay spread}.
However, for non-sparse mmWave channels, the omni-directional delay spread is significantly larger than that of after-beamforming delay spread.
Thus, how to support these extremes without incurring a high amount of signaling overhead is an important unsolved problem.
Fortunately, field test results showed that the after-beamforming delay spread could be significantly smaller than the omni-directional delay spread \cite{Samimi2016,Raghavan2017,Ko2017}.
Furthermore, the blocking capability" improves with the increasing number of antennas equipped at the BS/horn antenna gains \cite{Trees2002,Zhao2017}.
Specifically, the beamwidth of the main lobe becomes narrower and the magnitude of sidelobes becomes lower when the number of antennas becomes larger \cite{Trees2002,Hur2016,Zhao2017}.
Thus, the non-sparse scattering environment and the analog beamforming technology should be taken into account in the design of future mmWave systems, e.g. channel estimation and frequency selectivity \cite{Samimi2016,Raghavan2017,Ko2017,Akdeniz2014,Rappaport2015,Hur2016}.

\vspace{-0mm}

\section{Challenges and Motivations}
There are plenty of implementation challenges for mmWave massive MIMO communication systems, such as efficient hardware structures, low-cost hardware elements, hybrid precoding strategies, and channel estimation algorithms.
In this thesis, channel estimation for mmWave hybrid systems as well as the impact of hardware impairments on rate performance are the primary focus of this thesis.
\vspace{-0mm}
\subsection{Hardware Constraint: Fully Digital vs Hybrid Structure}
From the literature, it is certain that conventional fully digital MIMO systems, in which each antenna connects with a dedicated radio frequency (RF) chain, are impractical for mmWave systems due to the prohibitively high cost, e.g. tremendous energy consumption of high-resolution ADC/DAC and PAs \cite{Heath2016a,Andrews2016,MR2015,AZhang2015}.
\begin{figure}[t]
\centering
\includegraphics[width=5.5in]{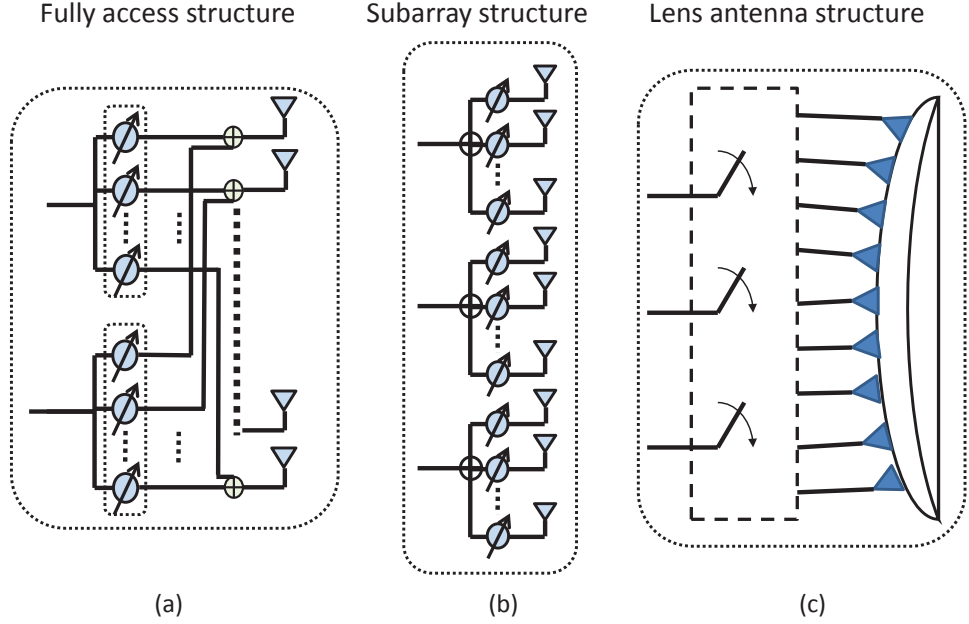}\vspace{-5mm}
\caption
{Illustration of different hybrid structures.}\label{fig:DStru}
\end{figure}
As a result, several mmWave hybrid systems were proposed as compromised solutions which strike a balance between hardware complexity and system performance \cite{Alkhateeb2015,Ni2016,Han2015,Ayach2014,Sohrabi2016,Gao2016,Bjornson2016,Ng2017,Bogale2015}, such as fully access hybrid structure, subarray hybrid structure and lens-based hybrid structure \cite{Andrews2016,AZhang2015,Heath2016a,Dai2015}, as shown in Figure \ref{fig:DStru}.
In particular, the trade-offs between system performance, hardware complexity\footnote{The hardware includes PA, ADC/DAC, phase shifters, and antenna array.}, and energy consumption are still unclear \cite{AZhang2015,Heath2016a}.

In the hybrid structure, $N_\mathrm{S}$ data streams are first through the baseband digital precoding, and then the output is through the $N$ RF chains.
After the RF analog beamforming, the RF signals are outputted to the $M$ antennas.
With $M\gg N$, the number of RF chains can be significantly reduced for practical implementation.
Specifically, the use of a large number of antennas, connected with only a small number of independent RF chains at transceivers, is adopted to exploit the large array gain to compensate the inherent high path loss in mmWave channels \cite{Rappaport2015,Hur2016}.
In the fully access structure shown in Figure \ref{fig:DStru}(a), an RF chain is connected to all the antennas through the network of analog phase shifters.
In the subarray structure shown in Figure \ref{fig:DStru}(b), the array is divided into sub-arrays, and each sub-array is fed by its RF chain.
The lens antenna array structure shown in Figure \ref{fig:DStru}(c), which is formed by an electromagnetic lens with energy focusing capability and a matching antenna array with elements located on the focal surface, can simultaneously realize signal-emitting and phase-shifting.
In general, the hybrid system imposes a restriction on the number of RF chains which introduces a paradigm shift in the design of both resource allocation algorithms, transceiver signal processing, and channel estimation algorithms.

\vspace{-0mm}
\subsection{Hybrid Precoding: Basic Concepts}

For a hybrid mmWave system, the utmost important issues should be developing channel estimation algorithms and hybrid precoding algorithms \cite{Ng2017}.

To address the importance in designing hybrid precoding algorithms, we take a MU hybrid mmWave system with the fully access structure as an example, which consists of one $M$-antenna BS with $N$ RF chains and $K$ single-RF-chain users with $P$-antenna.

During the downlink transmission, the signal received at user $k$ is given by\vspace{-0mm}
\begin{equation}
y_{k} = \bm{f}^{H}_{k} \mathbf{H}_{k}\mathbf{F}_\mathrm{BS}\mathbf{W}_\mathrm{BS}\mathbf{s} + \bm{f}_{k}\mathbf{n}_{k},\vspace{-0mm}
\end{equation}
where $\mathbf{H}_{k}\in\mathbb{C}^{P\times M}$ is the downlink mmWave channel between the BS and user $k$, $\mathbf{F}_\mathrm{BS}\in\mathbb{C}^{M\times N}$ is the analog beamformer at the BS, $\mathbf{W}_\mathrm{BS}\in\mathbb{C}^{N\times N}$ is the digital precoder at the BS, $\bm{f}^{H}_{k}\in\mathbb{C}^{1\times P}$ is the analog beamformer at user $k$, $k \in \{1,\cdots,K \} $, $\mathbf{s}=[
\begin{array}{ccc}
s_{1}, \ldots, s_{N}\end{array}] \in\mathbb{C}^{N\times 1}$ is the transmitted symbols, $\mathrm{\mathbb{E}}[\mathbf{s}\mathbf{s}^{H}] = \frac{E_{\mathrm{s}}}{N} \mathbf{I}_{\mathrm{P}}$, $E_{\mathrm{s}}$ is the total transmit power, $\mathbf{n}_{k}\in\mathbb{C}^{P\times 1}$ is an additive white Gaussian noise (AWGN) vector, $\mathbf{n}_{k}\sim\mathcal{CN}\left( \mathbf{0},{\sigma_{n}^{2}}\mathbf{I}_{\mathrm{P}} \right)$, and ${\sigma}_{n}^{2}$ is the noise variance at each antenna equipped at a user.
The achievable system sum rate is given by \cite{Alkhateeb2015,Ng2017}
\begin{equation}
R_{\mathrm{SUM}}=\overset{K}{\underset{k=1}{\sum }}R_{k}. \label{12345}\end{equation}
In Equation (\ref{12345}), $R_{k}$ is the rate achieved by user $k$ and can be expressed as
\begin{equation}
R_{k}=\log_2\left[1+\frac{\frac{E_{\mathrm{s}}}{N}|\bm{f}^{H}_{k}\mathbf{H}_{k}\mathbf{F}_\mathrm{BS} \bm{w}_{\mathrm{BS},k}|^2}{\frac{E_{\mathrm{s}}}{N}|\bm{f}^{H}_{k}\overset{N}{\underset{j\neq k}{\sum }}\mathbf{H}_{k}\mathbf{F}_\mathrm{BS} \bm{w}_{\mathrm{BS},j} |^2 + \sigma_{n}^{2}} \right],\vspace{-0mm}
\end{equation}
where $\bm{w}_{\mathrm{BS},k}$ is the $k$-th column of $\mathbf{W}_\mathrm{BS}$.

The optimal joint digital precoder and analog beamformer design is to maximize the achievable sum rate.
Thus, the optimal precoders design can be found by solving the following optimization problem \cite{Alkhateeb2015,Sohrabi2016,Dai2015,Mumtaz2017,Ng2017,Xiao2017}:\vspace{-0mm}
\begin{align}
\hspace{-0mm}\left\{\mathbf{W}^{\mathrm{opt}}_\mathrm{BS}, \mathbf{F}^{\mathrm{opt}}_\mathrm{BS}, \{\bm{f}^{\mathrm{opt}}_{k}\}^{K}_{k=1} \right\} &= \arg\max \overset{K}{\underset{k=1}{\sum }}\log_{2}\left[ 1+\frac{\frac{E_{\mathrm{s}}}{N}\left|\bm{f}^{H}_{k}\mathbf{H}_{k}\mathbf{F}_\mathrm{BS} \bm{w}_{\mathrm{BS},k}\right|^2}{\frac{E_{\mathrm{s}}}{N}\left|\bm{f}^{H}_{k}\overset{N}{\underset{j\neq k}{\sum }}\mathbf{H}_{k}\mathbf{F}_\mathrm{BS} \bm{w}_{\mathrm{BS},j} \right|^2 + \sigma_{n}^{2}} \right]\vspace{-3mm}\label{Eq_277}\\
\mathrm{s.t.} & \text{\ \ \ \ }\mathrm{C1:}\text{\ }\left|\bm{f}_{k}\{i\}\right|^2 = 1, \text{\ } \forall k,i,  \notag \\
& \text{\ \ \ \ }\mathrm{C2:}\text{\ }\left|\bm{F}_{\mathrm{BS}}\{i,j\}\right|^2=1, \text{\ } \forall i,j ,  \notag \\
& \text{\ \ \ \ }\mathrm{C3:}\text{\ }\| \mathbf{F}_{\mathrm{BS}}\mathbf{W}_{\mathrm{BS}}\|_{\mathrm{F}}^2 = N \notag, \vspace{-0mm}
\end{align}
where $\mathrm{C1}$ and $\mathrm{C2}$ are the hardware constraint for the design of analog beamformers and the total power constraint is enforced by normalizing $\mathbf{F}_{\mathrm{BS}}$ in $\mathrm{C3}$.

The problem in Equation (\ref{Eq_277}) is a non-convex optimization problem.
In particular, the digital precoder design is under the constraint of the number of RF chains and the designed analog beamformers.
Thus, the optimal solution to the hybrid precoding design is not known in general and only suboptimal iterative solutions exist \cite{Han2015,Alkhateeb2015,Alkhateeb2017aa,Niu2015,Heath2016a}.

In order to avoid the exhaustive search for the optimal analog beamforming and digital precoding design, Ref. \cite{Alkhateeb2015} proposed a two-stage algorithm which designs analog beamformer and digital precoder separately.
In the first-stage, for each user $k$, $ k \in \{1,\cdots,K\}$, the BS and user $k$ select their analog beamformers respectively to solve\vspace{-0mm}
\begin{align}
\left\{  \{\bm{f}^{\mathrm{opt}}_{k}\}_{k=1}^{K}, \{\bm{f}^{\mathrm{opt}}_{\mathrm{BS},k}\}_{k=1}^{K} \right\} &= \arg\max \left\| {\bm{f}^{H}_{k}\mathbf{H}_{k} \bm{f}_{\mathrm{BS},k}}\right\| \label{TAA}\\
\mathrm{s.t.} & \text{\ \ \ \ }\mathrm{C1:}\text{\ }\left|\bm{f}_{k}\{i\}\right|^2 = 1, \text{\ } \forall k,i,  \notag \\
& \text{\ \ \ \ }\mathrm{C2:}\text{\ }\left|\bm{f}_{\mathrm{BS},k}\{j\} \right|^2 = 1,  \forall k,j. \notag\vspace{-0mm}
\end{align}
In fact, the main idea of the first-stage is to jointly design analog beamformers at the user and the BS to maximize the desired signal power of each user.
For the analog beamformer design, Equation (\ref{TAA}) is widely adopted in many works \cite{Heath2016a,Park2015}.
It is true that the power of the received signal before analog beamforming is typically low due to high mmWave path losses, additional blockage, and penetration losses.
Thus, it is reasonable that the proposed analog beamformer design tries to maximize the received power of the desired signal.
However, it neglects the resulting interference among different users \cite{Mumtaz2017,Raghavan2017,Zhao2017}.
Actually, the MU inter-user interference due to the analog beamformers may cause severe impacts and should be taken into account.
Thus, we should consider the trade offs between the received power of the desired signal and the MU interference for the joint analog beamformer design \cite{MacCartney2017,Raghavan2017,Bas2017,Ko2017,Samimi2016,Gao2016}.
Then, a novel low-complexity practical algorithm for the design of analog beamformers is highly desirable.

In the second-stage of the hybrid precoding design, all the users feed back the downlink equivalent CSI to the BS based on the jointly designed analog beamformers:\vspace{+0mm}
\begin{equation}
\mathbf{H}_{\mathrm{eq}} = \left[
\begin{array}{ccc}
\mathbf{h}_{\mathrm{eq},1}, \ldots, \mathbf{h}_{\mathrm{eq},K}
\end{array}\right],\vspace{+0mm}
\end{equation}%
where $\mathbf{h}_{\mathrm{eq},k} = \left(\bm{f}^{\mathrm{opt}}_{k}\right)^{H}\mathbf{H}_{k}\bm{f}^{\mathrm{opt}}_{\mathrm{BS},k}$,  $\forall k \in \{1,\cdots,K \}$.
The digital precoder of the BS based on the feedback of equivalent channel to manage the MU interference\vspace{+0mm}
\begin{equation}
\mathbf{W}_{\mathrm{BS}} =  \mathbf{H}_{\mathrm{eq}}^{H}\left[  \mathbf{H}_{\mathrm{eq}} \mathbf{H}_{\mathrm{eq}}^{H}\right]^{-1}.
\vspace{+0mm}
\end{equation}%
However, explicit CSI feedback from users is still required for estimating these channel.
In practice, CSI feedback may cause high complexity and extra signallings overhead.
In addition, there will be a system rate performance degradation due to the limited amount of the feedback and the limited resolution of CSI quantization.

Therefore, a low computational complexity mmWave channel estimation algorithm, which does not require explicit CSI feedback, is necessary to unlock the potential of hybrid mmWave systems.

\vspace{-0mm}

\subsection{Combination of MmWave Systems and A Small-cell Scenario}

{\color{black} To further improve the network spectral efficiency, the small-cell scenario has been proposed to cooperate with mmWave systems \cite{Andrews2014,Bjornson2016,Niu2015}.
In fact, the combination of hybrid mmWave systems and a small-cell scenario is still unexplored and possesses many exciting research opportunities \cite{Bjornson2016}.
For example, cell shrinking/small-cell scenario can bring numerous benefits for hybrid mmWave systems.
In particular, it facilitates the reuse of the same piece of spectrum across a large geographic area for achieving a high network spectral efficiency \cite{Miao2014} and reducing of the severe large-scale propagation path loss by shortening the distances between transceivers \cite{Andrews2014,Niu2015}.}
{\color{black}
For the multi-small-cell mmWave network, the interference received at the desired user originates from two sources, as illustrated in Figure $3$ of \cite{Niu2015}: interference among different BSs and interference within the desired cell.
It is mentioned in \cite{Andrews2014} that, mmWave beams are highly directional, which completely changes the interference behavior as well as the sensitivity to beams misalignment.
In particular, the interference adopts an on/off behavior where strong interference occurs intermittently \cite{Andrews2014}.
With a shrinking cell radius, distances from neighbouring BSs to the desired user decrease, which may lead to severe inter-cell interference to the desired user.
Recently, it is also mentioned in work \cite{Petrov2017} that the received interference significantly increases when directional transmission is simultaneously adopted in both transceivers for the same amount of total emitted energy.
Besides, due to the impact of imperfect CSI on the design of downlink precoder, the desired BS will cause severe intra-cell interference on the desired user.
Thus, for the multi-small-cell mmWave downlink transmission with imperfect CSI and a large number of antennas, the desired user may suffer severe intra-cell and inter-cell interference.
Hence, it is necessary to study the performance in the multi-small-cell hybrid mmWave network.
}

\vspace{+2mm}
\subsection{MmWave Channel Estimation}
Several channel estimation algorithms are widely adopted in different works, such as multi-resolution hierarchical codebook algorithms, compress sensing algorithms, open-loop beamforming algorithms, and pilot-aided algorithms, cf. \cite{HLin2017,Liu2017,Xiao2016,Shen2017,Noh2014,Noh2017,Wang2011,Wang2016,Biguesh2006,He2017,Gao2017,Gao2016aa,Wang2009,Adhikary2013,Alkhat2014,Heath2016a,Kokshoorn2016,Ma2014,Choi2014,Bjornson2016,Raviteja2017}.
Some of the algorithms mentioned above are based on the assumption of sparse mmWave channels, which may not hold true for some scenarios as shown recent field measurements.
The field test results show that both strong LOS components and non-negligible scattering components may exist in mmWave propagation channels, especially in urban areas, e.g. building valley environment \cite{Samimi2016,Raghavan2017,Ko2017,Akdeniz2014,Rappaport2015,Hur2016}.
Therefore, we introduce two widely adopted algorithms which do not rely on the assumption of sparse mmWave channels.
\vspace{+3mm}
\subsubsection{Pilot-aided Channel Estimation and Pilot Contamination:}
Pilot-aided channel estimation, which relies on orthogonal pilot symbols, is widely adopted by MIMO systems and is suitable for sparse or non-sparse environments.
The majority of contributions in the literature considered that massive MIMO systems rely on pilot sequences to estimate CSI \cite{Rusek2013,Jose2011,Hoydis2013,Marzetta2010,Wagner2012}.
Due to the limited orthogonal pilot resources, they must be reused across different cells.
Thus, the reuse of orthogonal pilot sequences causes interference during channel estimation, which is known as pilot contamination \cite{Jose2011,Marzetta2010}.
As a result, the downlink transmission based on the CSI obtained via contaminated pilots causes severe intra-cell and inter-cell interference in the desired cell.
In fact, pilot contamination is considered as a fundamental performance bottleneck of the conventional multi-cell MU massive MIMO systems, since the resulting channel estimation errors do not vanish even if the number of antennas is sufficiently large, cf. \cite{Marzetta2010,Wu2016,Jose2011,Yang2015cc,Yang2015bb,Yang2013,Sanguinetti2017,Zhao2016}.

{\color{black}
Recently, various algorithms \cite{Jose2011,Ma2014,Akbar2016,Mahyiddin2015,Bogale2015aa,Farhang2014,Yin2013,Yin2016aa,Bjornson2017a} have been proposed to alleviate the impact of pilot contamination, e.g. data-aided iterative channel estimation algorithms, pilot design algorithms, multi-cell {\color{black}minimum mean square error (MMSE)} based precoding algorithm, and so forth.
However, some algorithms, e.g. multi-cell MMSE algorithm, are mostly based on the assumption that the desired BS can have perfect knowledge of covariance matrices of pilot-sharing users in neighbouring cells, which is overly optimistic.
Besides, the condition that the desired BS has the perfect knowledge of covariance matrices is a necessary but not sufficient condition for pilot contamination mitigation, cf. \cite{Yin2013,Yin2016aa,Bjornson2017a}.
The algorithm proposed in \cite{Yin2013} requires that covariance matrices of pilot-sharing users in neighbouring cells are orthogonal, which is unlikely in practice \cite{Bjornson2017a}.
Also, the requirement of \cite{Yin2016aa} for completely eliminating pilot contamination is that the number of antennas equipped at the BS and the size of a coherence time block jointly go to infinity.}
{\color{black}In the literature, most of existing multi-cell massive MU-MIMO works for pilot contamination \cite{Marzetta2010,Jose2011,Andrews2014,Ma2014,Wu2016} assumed that cross-cell channels from pilot-sharing users in neighbouring cells to the desired BS are Rayleigh fading channels with zero means.
However, as discussed in Section IV of \cite{Andrews2012}, it is probably not accurate for modeling the inter-cell interference as a Gaussian random variable with a small-cell setting.
In fact, recent field measurements have confirmed that the strongest angle-of-arrival/angle-of-departure (AoA/AoD) components always exist in the inter-cell mmWave channels in small-cell systems \cite{Akdeniz2014,Rappaport2015,Hur2016,Hur2014}.
Besides, the mean values of cross-cell channels are not zero and different from each other.
In other words, the distribution of cross-cell mmWave channels is different from that of the sub-$6$ GHz channels.
Thus, the results obtained in works mentioned above for pilot contamination mitigation and performance analysis, e.g. \cite{Marzetta2010,Jose2011,Andrews2014,Ma2014,Bjornson2017a,Bjornson2016}, cannot be applied directly.
Furthermore, a thorough study on the impact of pilot contamination in such a practical network system with small cell radius has not been reported yet.}
\vspace{-0mm}
\subsubsection{Open-Loop Beamforming Channel Estimation:}
Channel estimation of practical mmWave systems may rely on AoA estimation in mmWave channels \cite{Bjornson2016,Bogale2015,Xie2017}.
{\color{black}As a result, for mmWave channel estimation, the open-loop beamforming (OLB) channel estimation has been widely adopted.
{{\color{black}Generally, the OLB algorithm is widely used in analog beamforming and hybrid precoding. In particular, the BS transmits $L$ beamforming vectors to its users.
Then, each user reports the index of a beam with the largest gain in the $L$ beams via a feedback link with $\log _{2}L$ numbers of bits.
However, due to the limited amount of feedback bits, the OLB algorithm does not allow the BS to accurately estimate the channel responses of a large number of users.
Interestingly, fully digital systems, which can be considered as a subcase of hybrid systems, can also adopt the OLB algorithm for AoAs estimation \cite{Bjornson2016,Richards2014}.
Thus, the fully digital system, which contains more RF chains than analog and hybrid structures, can simultaneously transmit predesigned beams in all estimation directions.}}, e.g. \cite{He2014,Alkhat2014,Araujo2014,MR2015,Rappaport2015}.
For the OLB channel estimation, the required number of beams depends on the required resolutions and is predesigned by the codebook.
The number of beams for AoA estimation does not necessarily scale up with the number of transmit antennas.
However, to serve an area with an increasing user density, it is expected that the BS should increase the spatial resolution.
For example, if the system requires an half-power beam width (HPBW) coverage for the OLB training, that the required number of beams should be $\dfrac{2M}{1.782}$ \cite{Trees2002}, where $M$ is the number of antennas equipped at the BS. In this case, the required number of beams for the OLB training scales with an increasing number of antennas, which may not be suitable for mmWave systems employing massive MIMO technology.
In fact, it is expected that the required coherence time resources of channel estimation for mmWave massive MIMO systems shall neither scale up with the number of users nor with the number of antennas \cite{Bjornson2016,Swindlehurst2014,Bogale2015}.}
Therefore, a simple AoA channel estimation algorithm with low overhead is highly desirable for hybrid mmWave system design.

\vspace{-0mm}
\subsection{TDD or FDD Feedback?}

In practice, the conventional fully digital systems and the hybrid systems are designed based on different hardware architectures.
Hence, intuitively, the channel estimation algorithms for hybrid mmWave systems are different from that for fully digital systems.
Currently, the majority of contributions in the literature focus on the development of CSI feedback based channel estimation methods for frequency division duplex (FDD) hybrid mmWave systems, e.g. \cite{Alkhateeb2015,Xie2017,Shen2015,Wagner2012,Gao2016,Wei2017}.
This is motivated by the assumption of the sparsity of mmWave channels that the numbers of resolvable AoA/AoD paths are finite and limited.
Thus, the CSI acquisition via feedbacks only leads to a small amount of signaling overhead compared to non-sparse CSI acquisition.
Generally, due to the high propagation path loss, the sparsity may only exist in outdoor long distance propagation mmWave channels \cite{Hur2016}.
In some scenarios, the assumption of the sparsity of mmWave channel may not hold anymore.
For example, for practical urban micro-cell (UMi) scenarios, such as the city center, the number of scattering clusters increases significantly and the channels are expected to be non-sparse.
In \cite{Akdeniz2014,Rappaport2015,Hur2016}, recent field test results, as well as ray-tracing simulation results, have shown that reflections from street signs, lamp posts, parked vehicles, passing people, etc., could reach a receiver from all possible directions in urban micro-cell (UMi) scenarios. In other words, the AoAs of scattering components between the users and the BS are uniformly distributed between $[0, \text{\ }\pi]$.
In \cite{Flordelis2017}, the authors revealed that the presence of significant amount of scatterers in propagation environment has a non-negligible impact on the achievable rate performance of FDD systems when the conventional CSI feedback algorithms are adopted.
Besides, the required amount of feedback signalling overhead increases tremendously with the number of scattering components, consuming a significant portion of system resources \cite{Alkhateeb2015}.
Moreover, recent field test results have also verified that, for both fully digital and hybrid massive MIMO systems in different channel environments, time division duplex (TDD) based beamforming exploiting channel reciprocity for channel estimation can outperform the FDD based beamforming scheme utilizing CSI feedback in terms of achievable sum-rate \cite{Flordelis2017,Raghavan2017}.
To overcome the aforementioned common drawbacks of conventional CSI feedback based FDD mmWave channel estimation algorithms, a TDD-based beamforming channel estimation algorithm for mmWave channels is strongly welcomed.
However, time synchronization and calibration issues between the forward and reverse links need to be taken into account \cite{Ng2017}.
\vspace{-3mm}
\subsection{Hardware Impairments}
In practice, considering the high cost and the high complexity of hardware for high frequency bands, it is likely to build mmWave systems with low-cost components that are prone to hardware imperfection.
Thus, constraints due to non-ideal hardware should be taken account in the designs of mmWave MIMO systems.
{In practice, hardware components may have various types of impairments that may degrade the achievable rate performance.
For example, practical transceiver hardware is impaired by phase noise, limited phase shifters resolution, non-linear power amplifiers, in-phase/quadrature (I/Q) imbalance, and limited ADC/DAC resolution \cite{Bjornson2015bb,Zhang2016,Ying2015,Mo2015,Raviteja2017,MR2015,Zarei2017}.

Generally, three different hardware impairments are considered in the literatures, e.g. additive distortions, incorporate multiplicative phase-drifts, and quantization loss caused by low-resolution ADC/DAC.
{{The authors of \cite{Bjornson2015bb} proved that for a fully digital massive MIMO system, the additive distortion caused by hardware impairments creates finite ceilings on the channel estimation accuracy and on the uplink/downlink capacity.
In other words, the system performance cannot be improved by increasing the number of antennas equipped at the BS as well as the signal-to-noise ratio (SNR).
In addition, the work in \cite{Ying2015} concluded that the impact of phase error on hybrid beamforming has a further reduction on the potential gain brought by mmWave systems.
In fact, the achievable rate degradation caused by phase errors can be compensated by simply employing more transmit antennas}}, e.g. phase errors in phase shifters induced by thermal noise, transceiver RF beamforming errors caused by AoA estimation errors, and channel estimation errors affected by independent additive distortion noises \cite{Bjornson2015bb,Zhang2016,Ying2015}.}
However, the rate performance degradation caused by joint hardware imperfections,  i.e., random phase errors, transceiver analog beamforming errors, and channel estimation errors, has not been thoroughly discussed.

\section{Thesis Outline and Contributions}

The main contributions and the outline of the thesis are summarized as follows.

In Chapter $2$, we investigate channel estimation and MU downlink transmission of a TDD massive MIMO system in mmWave channels.
We propose a tone-based linear search algorithm to facilitate the estimation of AoAs of the strongest LOS (SLOS) channel component as well as  scattering components of the users at the BS.
Based on the estimated AoAs, we reconstruct the SLOS component and scattering components of the users for downlink transmission.
We then derive the achievable rates of MRT and ZF precoding based on the SLOS component and the SLOS-plus-scattering components (SLPS), respectively.
Taking into account the impact of pilot contamination, our analysis and simulation results show that the SLOS-based MRT can achieve a higher data rate than that of the traditional Pilot-aided-CSI based (PAC-based) MRT, under the same mean squared error (MSE) of channel estimation.
As for ZF precoding, the achievable rates of the SLPS-based and the PAC-based are identical.
Furthermore, we quantify the achievable rate degradation of the SLOS-based MRT precoding caused by phase quantization errors in the large number of antennas regime.
We show that the impact of phase quantization errors on the considered systems cannot be mitigated by increasing the number of antennas and therefore the resolutions of RF phase shifters is critical for the design of efficient mmWave massive MIMO systems.

The results in Chapter 2 have been presented in the following publications:
\begin{itemize}
\item \textbf{L.~Zhao}, G.~Geraci, T.~Yang, D.~W.~K. Ng, and J.~Yuan, ``A tone-based {AoA} estimation and multiuser precoding for millimeter wave massive {MIMO},''
   {\em IEEE Trans. Commun.}, vol.~65, no.~12, pp. 5209--5225, Dec. 2017.

\item \textbf{L.~Zhao}, T.~Yang and G.~Geraci and J.~Yuan ``Downlink multiuser massive {MIMO} in {Rician} channels under pilot contamination'' in {\em Proc. IEEE Intern.
  Commun. Conf. (ICC)}, Kuala Lumpur, May 2016, pp. 1--6.
\end{itemize}

In Chapter 3, we develop a low-complexity channel estimation for hybrid mmWave systems, where the number of RF chains is much lower than the number of antennas equipped at each transceiver.
The proposed mmWave channel estimation algorithm first exploits multiple frequency tones to estimate the strongest AoAs at both BS and user sides for the design of analog beamforming matrices.
Then all the users transmit orthogonal pilot symbols to the BS along the directions of the estimated strongest AoAs in order to estimate the channel.
The estimated channel will be adopted to design the digital ZF precoder at the BS for the multi-user downlink transmission.
The proposed channel estimation algorithm is applicable to both non-sparse and sparse mmWave channel environments.
Furthermore, we derive a tight achievable rate upper bound of the digital ZF precoding with the proposed channel estimation algorithm scheme.
Our analytical and simulation results show that the proposed scheme obtains a considerable achievable rate of fully digital systems, where the number of RF chains equipped at each transceiver is equal to the number of antennas.
Besides, by taking into account the effect of various types of errors, i.e., random phase errors, transceiver analog beamforming errors, and equivalent channel estimation errors, we derive a closed-form approximation for the achievable rate of the considered scheme.
We illustrate the robustness of the proposed channel estimation and multi-user downlink precoding scheme against the system imperfection.

The results in Chapter 3 have been presented in the following publications:
\begin{itemize}
\item \textbf{L.~Zhao}, D.~W.~K. Ng, and J.~Yuan, ``Multi-user precoding and channel estimation for hybrid millimeter wave systems,'' {\em IEEE J. Sel. Areas Commun.}, vol.~35, no.~7, pp. 1576--1590, Jul. 2017.

\item \textbf{L.~Zhao}, D.~W.~K. Ng, and J.~Yuan, ``Multiuser precoding and channel estimation for hybrid millimeter wave {MIMO} systems,'' in {\em Proc. IEEE Intern.
  Commun. Conf. (ICC)}, Paris, May 2017, pp. 1--7.
\end{itemize}

In Chapter 4, we investigate the system performance of a multi-cell MU hybrid mmWave MIMO network. Due to the simultaneous reuse of pilot symbols among different cells, the performance of channel estimation is expected to be degraded by pilot contamination, which is considered as a fundamental performance bottleneck of conventional multi-cell MU massive MIMO networks.
To analyze the impact of pilot contamination to the system performance, we first derive the closed-form approximation expression of the normalized MSE of the channel estimation algorithm proposed in \cite{Zhao2017} over Rician fading channels.
Our analytical and simulation results show that the channel estimation error incurred by the impact of pilot contamination and noise vanishes asymptotically with an increasing number of antennas equipped at each RF chain at the desired BS.
Furthermore, by adopting ZF precoding in each cell for downlink transmission, we derive a tight closed-form approximation of the average achievable rate per user.
Our results unveil that the intra-cell interference and inter-cell interference caused by pilot contamination over Rician fading channels can be mitigated effectively by simply increasing the number of antennas equipped at the desired BS.

The results in Chapter 4 have been accepted to appear in the following publications:
\begin{itemize}
\item \textbf{L.~Zhao}, Z.~Wei, D.~W.~K. Ng, J.~Yuan, and M.~C.~Reed, ``Multi-cell hybrid millimeter wave systems: {Pilot} contamination and interference mitigation,'' {\em IEEE Trans. Commun.}, vol. 66, no. 11, pp. 5740-5755, Nov. 2018.

\item \textbf{L.~Zhao}, Z.~Wei, D.~W.~K. Ng, J.~Yuan and M.~C.~Reed, ``Mitigating pilot contamination in multi-cell hybrid millimeter Wave Systems,'' in { \em Proc. IEEE Intern.  Commun. Conf. (ICC)}, Kansas City, May 2018.
\end{itemize}

In Chapter 5, future works and conclusions are presented.

\chapter{Fully Digital mmWave MIMO Systems: {A Tone-based Channel Estimation Strategy} 
}\label{C2:chapter2}

%
\section{Introduction}
In this chapter, we consider a mmWave MU massive MIMO system.
First, we propose a novel and simple channel estimation algorithm, which is inspired by the signal processing of radar and sonar systems.
We illustrate the proposed channel estimation for mmWave channels via analysis and simulation results.
In addition, we analyze the rates achieved by utilizing MRT and ZF precoding, which are based on the estimated components of mmWave channels.
Furthermore, we discuss some hardware constraints that may affect the achievable rate performance, e.g. the limited resolutions of digital RF phase shifters.
Our main contributions are summarized as follows:

\begin{itemize}
\item We propose a novel MU channel estimation scheme via tone-based AoA estimation. By introducing multiple frequency tones, the proposed scheme can simultaneously and efficiently estimate all users' AoAs. More importantly the accuracy of the proposed channel estimation algorithm increases with the number of antennas available at the BS. Compared to existing the pilot-aided channel estimation and the OLB channel estimation methods, our proposed AoA estimation over Rician fading channles neither causes pilot contamination nor requires a channel estimation overhead which increases with the numbers of users and antennas equipped at the BS.
\item Based on various components of the estimated CSI, we derive the achievable rates of the MRT and the ZF precodings. In particular, the SLOS is adopted for the SLOS-based MRT. In addition, the combination of multiple AoAs from the users to the BS are utilized for the SLPS-based ZF downlink transmissions. We also derive closed-form approximations for the achievable rates of the MRT and the ZF precoding strategies in mmWave channels when the traditional PAC estimation is adopted. We analytically show that the achievable rates per user for the SLPS-based and the PAC-based ZF precoding strategies are identical, if the MSEs of CSI are the same. Also, it is interesting to note that the achievable rate per user of the SLOS-based MRT is better than that of the PAC-based MRT.
\item To obtain system design insights, we perform an asymptotic analysis on achievable rates of the MRT and ZF precoding strategies.
    The analysis and simulation results verify that the mmWave massive MIMO can achieve a high sum rate performance even with dense users population. On the other hand, we show that the resolutions of RF phase shifters is critical for mmWave massive MIMO design. In fact, the negative impact of limited resolutions of shifters on the system rate performance cannot be mitigated by increasing the number of antennas.
\end{itemize}

The rest of the chapter is organized as follows. Section 2.2 describes the system
model considered in the chapter. In Section 2.3, we detail the proposed channel estimation algorithm. In Section 2.4, we derive the downlink achievable rate performance based on the CSI estimated by the proposed algorithm. In Section 2.5, we compare rate performance between the conventional algorithms and that investigated in Section 2.4. The impact of imperfect hardware are discussed in Section 2.6. Finally, Section 2.7 summarizes the chapter.

\vspace{-0mm}
\section{System Model}
\begin{figure}[t]
\centering
\includegraphics[width=5in]{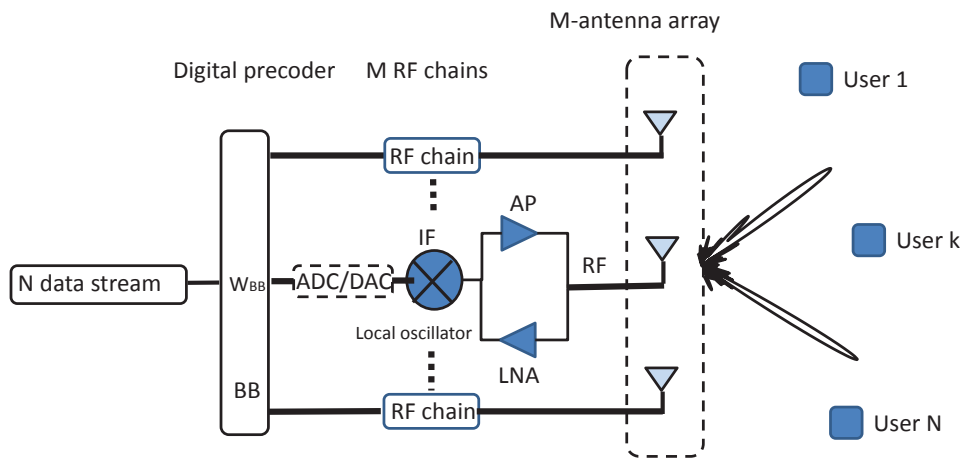}
\caption
{Fully digital massive MIMO architecture.}\label{fig:fully_digi}
\end{figure}
In this chapter, an mmWave massive MU MIMO system is considered, which is shown in Figure \ref{fig:fully_digi}. The system consists of $T$ neighboring cells and there are $N$ users per cell.
The BS in each cell is equipped with $M\geq 1$ antennas to serve the $N$ single-antenna users simultaneously, $M \geqslant N$.
Due to the significant propagation attenuation at mmWave, the system is dedicated to cover a small area, e.g. with a cell radius of approximately $150$ m.
The channels from the BS to the users in the same cell are modeled by Rician fading and with large K-factors ($7$ $\sim $ $17$ dB) \cite{Rappaport2015,Al-Daher2012,Eldeen2010}.
We assume that the users and the BSs in all cells are fully synchronized in time and that the uplink and downlink in each cell adopt TDD \cite{Marzetta2010,Jose2011}.

Let $\mathbf{H}\in\mathbb{C}^{M\times N}$ be the uplink channel matrix between the users and the BS in the desired cell.
{\color{black}{We assume that $\mathbf{H}$ is a narrowband slow time-varying block fading Rician fading channel, i.e., the channel is constant in a time slot but varies slowly from one time slot to another.}}
Each time slot is divided into three phases: channel estimation phase, uplink transmission phase, and downlink transmission phase.
We denote the $k$-th column vector of $\mathbf{H}$ as $\mathbf{h}_{k}\in\mathbb{C}^{M\times 1},$ representing the channel vector between the BS and user $k$ in the desired cell.
According to the field measurement of mmWave channels \cite{Rappaport2015}, a strong LOS component is expected and other propagation paths can be considered as scattering components.
{\color{black}We assume that the large scale path loss can be compensated by using automatic gain control (AGC).}
Thus, the channel vector of user $k$ can be expressed as a combination of a deterministic strongest LOS channel vector $\mathbf{h}_{\mathrm{L},k}\in\mathbb{C}^{M\times 1}$ and a multiple-path scattered channel vector $\mathbf{h}_{\mathrm{S},k}\in\mathbb{C}^{M\times 1}$ , i.e.,\vspace{-1mm}
\begin{equation}
\mathbf{h}_{k}=\left( \sqrt{\dfrac{\upsilon _{k}}{\upsilon _{k}+1}}\right)\mathbf{h}_{\mathrm{L},k}+\left(\sqrt{\dfrac{1}{\upsilon _{k}+1}}\right)\mathbf{h}_{\mathrm{S},k},\label{E_001}\vspace{-1mm}
\end{equation}
where $\upsilon _{k}>0$ is the Rician K-factor of user $k$, which denotes the ratio between the power of the SLOS component and the power of the scattering components.
\vspace{-1mm}
{\color{black}
In general, we can re-express Equation (\ref{E_001}) as\vspace{-1mm}
\begin{equation}
\mathbf{H}= \mathbf{H}_{\mathrm{L}}\mathbf{G}_{\mathrm{L}}+\mathbf{H}_{\mathrm{S}}\mathbf{G}_{\mathrm{S}},\label{E_002}\vspace{-1mm}
\end{equation}%
where $\mathbf{G}_{\mathrm{L}}\in\mathbb{C}^{N\times N}=\mathrm{diag}\left\{\overline{\bm{\upsilon}} _{\mathrm{L}}\right\}$, $\mathbf{G}_{\mathrm{S}}\in\mathbb{C}^{N\times N}=\mathrm{diag}\left\{\overline{\bm{\upsilon}} _{\mathrm{S}}\right\}$.
In addition, we have $\overline{\mathbf{\upsilon}} _{\mathrm{L}} = \left[
\begin{array}{ccc}
\sqrt{\frac{\upsilon _{1}}{\upsilon _{1}+1}}, \ldots, \sqrt{\frac{\upsilon _{N}}{\upsilon _{N}+1}}%
\end{array}%
\right]^{T}$, and $\overline{\mathbf{\upsilon}} _{\mathrm{S}} = \left[
\begin{array}{ccc}
\sqrt{\frac{1}{\upsilon _{1}+1}} \ldots \sqrt{\frac{1}{\upsilon _{N}+1}}%
\end{array}%
\right]^{T}$}.
We adopt uniform linear array (ULA) as in \cite{Alkhat2014,Alkhat2015b}.
Here, $\mathbf{h}_{\mathrm{L,}k}$ $\in\mathbb{C}^{M\times 1}$ is the $k$-th column vector of the SLOS matrix $\mathbf{H}_{\mathrm{L}}$ and it can be expressed as\vspace{-0mm}
\begin{equation}
\mathbf{h}_{\mathrm{L,}k}=\left[
\begin{array}{ccc}
1, & \cdots
, & \text{ }e^{-j2\pi \left( M-1\right) \tfrac{d}{\lambda }\cos \left(
\theta _{k}\right) }
\end{array}%
\right] ^{T},\vspace{-0mm}
\end{equation}%
where $d$ is the distance between the neighboring antennas at the BS, $\lambda $ is the wavelength of the carrier frequency, and $\theta _{k}\in \left[ 0,+\pi\right]$ is the AoA of the SLOS component from user $k$ to the BS in the desired cell.
For convenience, we set $d=\dfrac{\lambda }{2}$ for the rest of the chapter which is an assumption commonly adopted in the literature \cite{Vieira2014,Yue2015}.
The scattering component $\mathbf{h}_{\mathrm{S},k} $ is the $k$-th column vector of the scattering matrix $\mathbf{H}_{\mathrm{S}}$ and can be expressed as\vspace{-0mm}
\begin{equation}
\mathbf{h}_{\mathrm{S},k} = \sqrt{\frac{1}{L}}\overset{L}{\underset{l=1}{\sum }}{\alpha _{k,l}}\mathbf{a}_{k,l}, \text{ } k = 1,\cdots,N,\vspace{-0mm}
\end{equation}%
where $L$ denotes the number of propagation paths, ${\alpha _{k,l}}\sim \mathcal{CN} \left( 0,1\right)$ is the complex path gain and $\theta _{k,l}\in \left[ 0,+\pi\right]$ is the AoA associated to the $(k,l)$-th propagation path \cite{Ayach2014}, and $\mathbf{a}_{k,l}\in\mathbb{C}^{M\times 1}$ is the ${l}$-th propagation path of user $k$ given by\vspace{-1mm}
\begin{equation}
\text{ }\mathbf{a}_{k,l}=\left[
\begin{array}{cccc}
1, & \cdots
, & \text{ }e^{-j2\pi \left( M-1\right) \tfrac{d}{\lambda }\cos \left(
\theta _{k,l}\right) }%
\end{array}%
\right] ^{T}.\vspace{-1mm}
\end{equation}%
We assume that all the users have various AoAs which are uniformly distributed over $\left[ 0,+\pi \right] $ and they are separated by at least
hundreds of wavelengths\cite{Marzetta2010,Nguyen2015}.
To facilitate the investigation of channel estimation and downlink transmission, we assume that perfect long-term power control is performed to compensate for path loss and shadowing between the desired BS and the desired users and that equal power allocation is used among different data streams of the users \cite{Alkhateeb2015,Ni2016,Yang2013}.

\vspace{-0mm}
\section{Channel Estimation and Downlink Transmission}
In this section, we first propose a novel MU mmWave channel estimation algorithm based on various single carrier frequency tones, which is inspired by signal processing of radar and sonar systems \cite{Richards2014,Trees1994}.
Based on the estimated SLOS component and scattering components, we propose the SLOS-based MRT and the SLPS-based ZF precoders for MU downlink precoding.
In addition, the achievable rate performance of these two precoders are derived and used for comparison in the following section.

\subsection{MU MmWave Channel Estimation}

We note that the channel estimation of mmWave channels is equivalent to the AoA estimation of the SLOS and scattering components.
In this section, we propose to use the frequency tone resources to simultaneously estimate the AoAs of the users without causing collision among different cells.
The details are discussed in the following paragraphs.

At the beginning of a coherence time block, all the users register at the desired BS and transmit unique tones separated in frequency.
This initialization can be done during the time synchronization in the handshaking between the desired BS and the users.
The frequency tone of user $k$ for the AoA estimation is a single carrier continuous wave (CW) signal, $s_{k}=\sin (2\pi f_{k}t)$, where $f_{k}$ is the carrier frequency for user $k$ and $t$ is the time variable of the tone signal.
We also denote that $f_\mathrm{c}$ is the system carrier frequency, $BW$ is the total available bandwidth for all tone frequencies, and $f_\mathrm{c}$ is at the very center of the total bandwidth.

{\color{black}
For the channel estimation, we choose to transmit tone signals from the users to the BS via one of the omni-directional antennas equipped at the users.

The reasons are three:
\begin{itemize}
                 \item The utilizing of only one omni-directional antenna at the user to transmit the tone signal is to sound mmWave channels in every direction;
                 \item In practice, the number of antennas equipped at the users is much smaller than the number of antennas equipped at the BS, $M$. As long as a sufficiently large number of antennas, $M$, is equipped at the BS, the power gain loss can be compensated;
                 \item For the AoA estimation, tone signal is robust to the background noise. It is known that the thermal noise power is determined by the signal bandwidth and the noise spectral density level. Due to the extreme narrow bandwidth of the tone signal, the receive thermal noise power is low. Thus, it is easy to detect the tone signal even with a slightly low receive signal power \cite{Richards2014}.
\end{itemize}

\begin{figure}[t]
\centering
\includegraphics[width=4.5in]{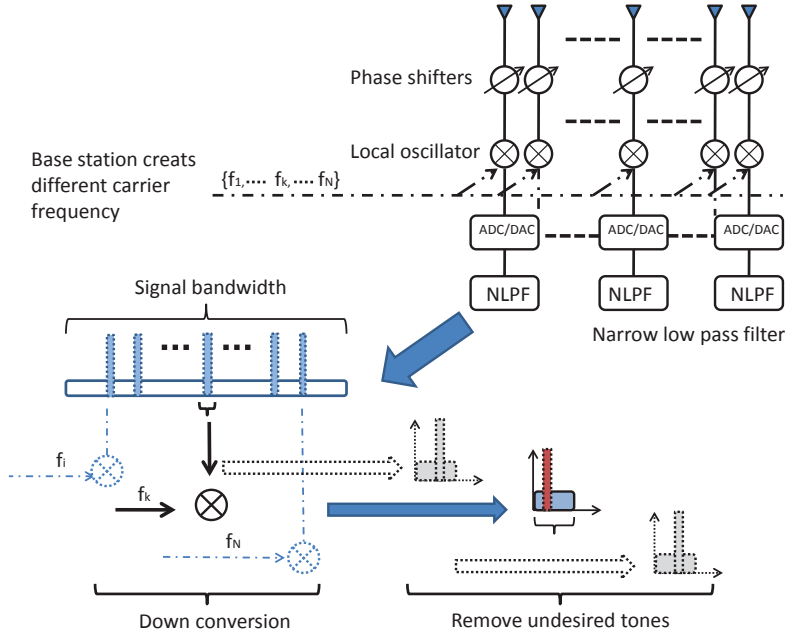}
\caption
{Tone-based AoA channel estimation.}\label{fig:tone_based}
\end{figure}

For any two different frequency tones $f_{i}$ and $f_{j}$, $\forall i\neq j$, in the $BW$, if the frequency tones satisfy the following condition,\vspace{-1mm}
\begin{equation}
\frac{f_{i}-f_{j}}{f_\mathrm{c}} < 2\times10^{-4},\vspace{-1mm}
\end{equation}%
the AoA estimation performance by using ULA with any tones in $BW$ is identical\footnote{The difference of different carrier wavelengths is less than $2$ micrometers for a carrier frequency of $30$ GHz.}.
The condition also can be re-expressed as $\frac{BW}{f_\mathrm{c}}<2\times10^{-4}$\cite{Trees2002}. For example, if the system operates at a carrier frequency of $30$ GHz, then $BW$ is $6$
MHz. Any frequency tones in the frequency range of $\left[ 30\text{ }\text{GHz}-3\text{ }\text{MHz}, \text{ }30\text{ }\text{GHz}+3\text{ }\text{MHz}  \right]$ have an identical performance in the AoA estimation.
However, the frequency tones will be affected due to users' doppler frequency shift $f_{d}$.

In addition, to overcome the collisions caused by users' doppler frequency shift $f_{d}$ between neighboring frequency tones, a frequency protectional gap $B_{k} = |f_{k}-f_{k+1}|$ need to satisfy\vspace{-1mm}
\begin{equation}
B_{k}>2f_{d}.\vspace{-1mm}
\end{equation}%
For example, if the system operates at $30$ GHz and the
maximum speed of users is $72$ km/h, then the minimum frequency protection gap is $B_{k} = 4$ kHz.
In practice, a single carrier frequency CW signal can be easily generated by a commercial $30$ GHz signal generator\cite{8257DPSG}, which has a narrow bandwidth of $100$ $\sim $ $200$ Hz.
Therefore, for a bandwidth of $6$ MHz we can generate $1400$ single carrier CW signals, which can estimate $1400$ users' AoAs simultaneously.
For mmWave massive MIMO systems, according to the 3GPP medium user density standard \cite{NSN2010}, the number of users in each cell ranges from $40$ to $400$.
Therefore, the frequency resources can support at least three cells without causing any collision in the AoA estimation among different cells.

{\color{black} The basic idea of utilizing frequency tone signal for the AoA estimation is borrowed from single-carrier monopulse radar systems \cite{Richards2014}.
The key idea is that the narrow-band single-carrier frequency CW signal can be performed for the AoA estimation. Therefore, orthogonal pilot symbols are no longer required for the AoA estimation of mmWave channels. In addition, different users can be distinguished at the BS via different carrier frequencies.
At the beginning of a coherence time block, all the users register at the desired BS. At the same time, the desired BS instructs all the users' unique tone carrier frequencies. At the user's side, according to the allocated tone carrier frequencies, a single-carrier frequency CW pass-band signal is created in the user's RF chain and broadcasted via one of its antennas.
For example, the frequency tone of user $k$ for the AoA estimation is $s_{k}=\sin (2\pi f_{k}t)$, where $f_{k}$ is the carrier frequency for user $k$ and $t$ is the time variable of the tone signal.
The pass-band received signal of user $k$ at the BS is given by\vspace{-2mm}%
\begin{equation}
\bm{\varphi }_{k}=\left( \sqrt{\dfrac{\upsilon _{k}}{\upsilon _{k}+1}}%
\mathbf{h}_{\mathrm{L},k}+\sqrt{\dfrac{1}{\upsilon _{k}+1}}\mathbf{h}_{%
\mathrm{S},k}\right) s_{k}+\widehat{\mathbf{z}},\vspace{-2mm}
\end{equation}%
where $\widehat{\mathbf{z}}\in\mathbb{C}^{M\times 1}$ is the AWGN vector at the BS antennas, whose entries follow independent and identically distributed (i.i.d.) $\mathcal{CN}\left( 0,\sigma ^{2}\right)$ and $\sigma ^{2}$ is the noise variance at each antenna of the BS.
At the BS, different frequency tones will be generated as the reference signals for the down-conversion of received signals, as shown in Figure \ref{fig:tone_based}. Take user $k$ for example, the reference signal is $s_{k}=\sin (2\pi f_{k}t)$ at the BS and the received signal after down-conversion is given by
\begin{align}
\mathbf{\varphi }_{k}\sin (2\pi f_{k}t)=&\left(\sqrt{\dfrac{\upsilon_{k}}{\upsilon_{k}+1}}\mathbf{h}_{\mathrm{L},k}+\sqrt{\dfrac{1}{\upsilon _{k}+1}}\mathbf{h}_{\mathrm{S},k}\right) s_{k} \sin (2\pi f_{k}t)+\widehat{\mathbf{z}}\sin (2\pi f_{k}t) \notag \\
=& \left( \sqrt{\dfrac{\upsilon _{k}}{\upsilon _{k}+1}}%
\mathbf{h}_{\mathrm{L},k} +\sqrt{\dfrac{1}{\upsilon _{k}+1}}\mathbf{h}_{%
\mathrm{S},k}\right)\frac{1}{2}\left[ 1- \cos(4\pi f_{k}t)\right]+\widehat{\mathbf{z}}s_{k}.
\end{align}%
Then the mixed signal will pass through different narrow low pass filters with an appropriate filter bandwidth to remove high frequency components \cite{Richards2014}.
After the high frequency components are filtered and removed, the received base-band signal which contains the AoA information can be expressed as\vspace{-0mm}%
\begin{equation}
\mathbf{r }_{k}=\left( \sqrt{\dfrac{\upsilon _{k}}{\upsilon _{k}+1}}%
\mathbf{h}_{\mathrm{L},k}+\sqrt{\dfrac{1}{\upsilon _{k}+1}}\mathbf{h}_{%
\mathrm{S},k}\right) + {{\mathbf{z}}}.\vspace{-0mm}
\end{equation}}%
To facilitate the estimation of AoAs, we perform a linear search in the angular domain ranged from $0^{\circ}$ to $180^{\circ}$ with an angle search step size of $\frac{180}{J}$, where $J$ is the maximum number of search steps.
\vspace{0mm}
{\color{black}The AoA detection matrix for all users is $\bm{\Gamma}\in\mathbb{C}^{M\times J}=\left[
\begin{array}{ccccc}
\overline{\bm{\gamma}}_{1} \ldots \overline{\bm{\gamma}}_{i} \ldots \overline{\bm{\gamma}}_{J}%
\end{array}%
\right]$, where $\overline{\bm{\gamma}}_{i}\in\mathbb{C}^{M\times 1}$ is a column vector of matrix $\bm{\Gamma}$ given by\vspace*{-0mm}
\begin{equation}
\overline{\bm{\gamma}}_{i}=\dfrac{1}{\sqrt{M}}\left[
\begin{array}{cccc}
1,
& \cdots , & \text{ }e^{j2\pi \left( M-1\right) \tfrac{d}{\lambda }\cos
\left( \widehat{\theta }_{i}\right) }%
\end{array}%
\right]^{T},
\end{equation}%
where $\widehat{\theta }_{i}=(i-1)\frac{180}{J}$, $i \in\{1,\cdots,J\}$, stands for a potential AoA of user at the BS.}
{\color{black}We note here, the coefficient $\frac{1}{\sqrt{M}}$ appears in Equations (2.11) is the normalization factor.}
The size of matrix $\bm{\gamma}$ represents the computational complexity of AoA estimation.
We then utilize different AoA detection vectors, e.g. $\overline{\bm{\gamma}}_{i}^{T},$ to obtain detection output of user $k$.
Specifically, for $i\in \{1,\cdots,J\}$,\vspace*{-2mm}
\begin{align}
&\bm{\omega} _{k,i}=\overline{\bm{\gamma}}_{i}^{T}\mathbf{r}_{k}=\sqrt{%
\dfrac{\upsilon _{k}}{\upsilon _{k}+1}}\overline{\bm{\gamma}}_{i}^{T}\mathbf{%
h}_{\mathrm{L},k}+\sqrt{\dfrac{1}{\upsilon _{k}+1}}\overline{\bm{\gamma}}%
_{i}^{T}\mathbf{h}_{\mathrm{S},k}+\overline{\bm{\gamma}}_{i}^{T}{%
\mathbf{z}}_{k}.\vspace*{-2mm}
\end{align}%
Now we obtain the detection outputs $[\bm{\omega} _{k,1}\ldots\bm{\omega}_{k,i}\ldots\bm{\omega} _{k,J}]$.
The potential AoA of the SLOS, which leads to the maximum value among the $J$ observation directions, i.e.,\vspace{-0mm}%
\begin{equation}
\widehat{\mathbf{h}}_{\mathrm{L},k}=\widetilde{\bm{\gamma}}_{k}^{\ast}=\underset{\forall \overline{\gamma}_{i}, \text{\ }i\in \{1,\cdots,J\}}{\arg \max \left\vert \bm{\omega} _{k,i}\right\vert },  \label{RFBE}
\end{equation}%
is considered as the SLOS component of user $k$.}
If the number of search steps $J$ is sufficiently large, we accurately approximate the channel of user $k$, $\mathbf{h}_{k}$, by using AoAs from the $D$ strongest detection outputs.
We first sort the detection outputs $[\bm{\omega} _{k,1}\ldots\bm{\omega}_{k,i}\ldots\bm{\bm{\omega}} _{k,J}]$ in descending order and obtain the ordered detection output vector as $[\bm{\bm{\omega}} _{k,(1)}\ldots\bm{\bm{\omega}}_{k,(i)}\ldots\bm{\bm{\omega}} _{k,(J)}]$, where $|\bm{\bm{\omega}} _{k,(i)}| \geqslant |\bm{\bm{\omega}} _{k,(i+1)}|$.
{\color{black}
Then, we can obtain the estimated channel $\widehat{\mathbf{h}}_{k}$ as\vspace{-0mm}
\begin{equation}
\widehat{\mathbf{h}}_{k}= \sqrt{\frac{\widehat{\upsilon}_{k}}{\widehat{\upsilon}_{k}+1}}\widetilde{\bm{\gamma}}_{k}^{\ast} + \sqrt{\frac{1}{\widehat{\upsilon}_{k}+1}}\overset{D-1}{\underset{p=1}{\sum }}\frac{\bm{\bm{\omega}}_{k,(p)}}{\bm{\bm{\omega}}_{k,(1)}}\widetilde{\bm{\gamma}}_{k,p}^{\ast},\vspace{-0mm}
\end{equation}%
where $\widetilde{\bm{\gamma}}_{k,p}^{\ast}$ is the $p$-th strong AoA given by\vspace{-1mm}
\begin{equation}
\widetilde{\bm{\gamma}}_{k,p}^{\ast}=\dfrac{1}{\sqrt{M}}\left[
\begin{array}{cccc}
1, & \cdots , & \text{ }e^{j2\pi \left( M-1\right) \tfrac{d}{\lambda }\cos
\left( \widehat{\theta }_{k,p}\right) }%
\end{array}%
\right]^{T},\vspace{-0mm}
\end{equation}%
and the Rician K-factor is given by\vspace{-0mm}
\begin{equation}
\dfrac{1}{\widehat{\upsilon}_{k}} = {{\overset{D-1}{\underset{p=1}{\sum }}{\left|\frac{\bm{\bm{\omega}}_{k,(p)}}{\bm{\bm{\omega}}_{k,(1)}}\right|}^{2}}}.\vspace{-0mm}
\end{equation}}\vspace{-0mm}%

We note that the exact number of combined AoAs $D$ is hard to obtain in practice.
As a result, the number of combined AoAs is set as a reasonable value for practical implementation. In the sequel, we set as $D=15$ for simplicity \cite{Hur2016}.
In fact, the tone-based AoA channel estimation error is mainly caused by the uncertainty of $D$.

Utilizing the estimated SLOS component as the estimated channel, the MSE of the proposed algorithm in the high SNR regime is given by\vspace{-0mm}
\begin{align}
\mathrm{MSE}_{\mathrm{L},k}\overset{\mathrm{SNR\rightarrow \infty }}{=}&\frac{1}{M}\mathrm{\mathbb{E}}\left[
\left( \mathbf{h}_{k}-\widehat{\mathbf{h}}_{\mathrm{L},k}\right) ^{H}\left(
\mathbf{h}_{k}-\widehat{\mathbf{h}}_{\mathrm{L},k}\right) \right]
=2\left( 1-%
\sqrt{\dfrac{\upsilon _{k}}{\upsilon _{k}+1}}\right).  \label{Eq_2}\vspace{-0mm}
\end{align}%

\begin{figure}[H]
\centering
\subfigure[Illustration of AoAs estimation.]
{\label{Fig003_0}
\includegraphics[width=2.75in,]{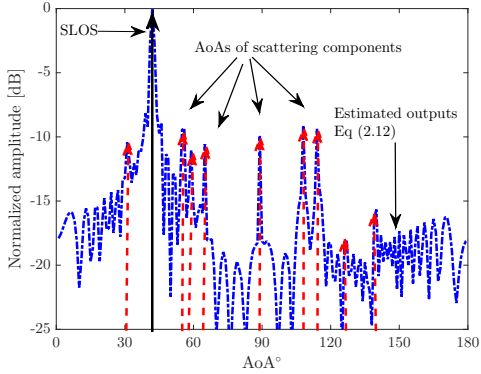}}
\subfigure[Illustration of AoAs estimation.]
{\label{Fig003_1}
\includegraphics[width=2.75in,]{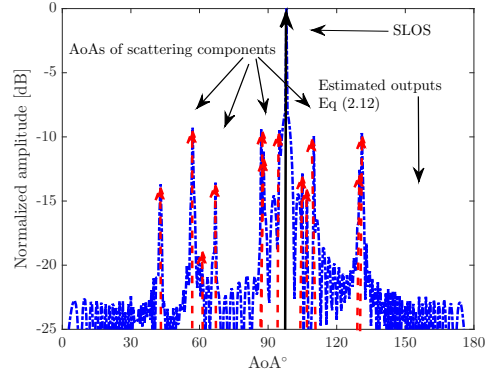}}\\
\subfigure[Successful rate of SLOS estimation.]
{\label{Fig003_2}
\includegraphics[width=2.75in,]{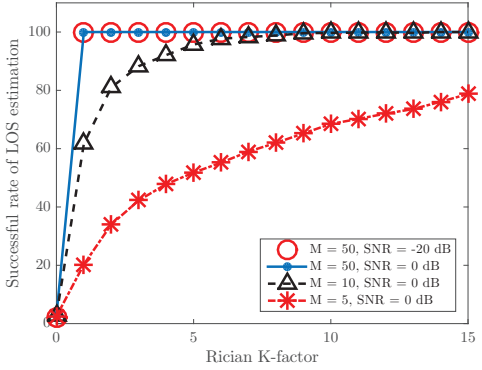}}%
\subfigure[Normalized MSE versus Rician K-factor.]
{\label{Fig003_3}
\includegraphics[width=2.75in,]{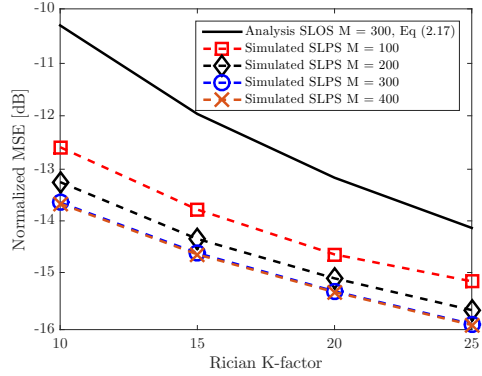}}%
\caption{(a) Illustration of AoA estimation for the SLOS component and scattering components for $M=100$ and $\upsilon_{k}=10$; (b) Illustration of AoA estimation for the SLOS component and scattering components for $M=200$ and $\upsilon_{k}=10$; (c) Successful rate of the SLOS estimation versus the Rician K-factor for various number of antennas; (d) Normalized MSEs of channel estimation versus the Rician K-factor for various number of antennas. For (a), (b), (c), and (d), we set the number of combination AoAs as $D=15$ and the number of search steps as $J=360$.}
\end{figure}

We demonstrate and validate the accuracy of the proposed AoA estimation via simulation in Figures \ref{Fig003_0}, \ref{Fig003_1}, \ref{Fig003_2}, and \ref{Fig003_3}.
The accuracy of the AoA estimation is determined by the number of search steps $J$ as well as the number of strongest AoAs $D$.

\vspace{-1mm}
Searching for $\widehat{\theta }_{i}$ can be processed via baseband digital signal processing.
More details on the Cram\'{e}r-Rao lower-bound of the ULA AoA estimation can be found in \cite{Trees1994}.
In practice, larger number of antennas $M$ means higher spatial resolution for AoAs estimation.

In Figures \ref{Fig003_0} and \ref{Fig003_1}, we illustrate the proposed AoA estimation for different $M$. {\color{black}The SNR of tone signal adopted in the simulations for Figures \ref{Fig003_0} and \ref{Fig003_1} is $5$ dB.}

With a large $M$, the impact of noise $\sigma^{2}$ on the AoA estimation is mitigated, as shown in Figures \ref{Fig003_0} and \ref{Fig003_1}.
These figures indicate that the proposed algorithm is robust to the AWGN when the BS is equipped with a sufficiently large number of antennas.
If the norm of the angle difference between the estimated SLOS component and the actual SLOS component is less than $0.1 \%$, we denote the estimation of the SLOS component as successful.

Figure \ref{Fig003_2} shows the successful rate of the AoA estimation achieved by the proposed algorithm versus the Rician K-factor for different number of antennas $M$.
It is clear that the successful rate increases with the number of antennas as well as with the Rician K-factor.
In addition, we also see that the successful rates of the SLOS estimation for $\mathrm{SNR}=0$ dB with $M=50$ are almost the same as those for $\mathrm{SNR}=-20$ dB with $M=50$.
Figure \ref{Fig003_3} shows the normalized MSEs of channel estimation versus the Rician K-factor for different number of antennas $M$.
It shows that the SLPS algorithm has better MSE performance than the SLOS algorithm.
In Figure \ref{Fig003_3}, the correctness of Equation (\ref{Eq_2}) is also verified.
In Figures \ref{Fig003_2} and \ref{Fig003_3}, we also observe that the improvement in MSE is saturated for a large number of antennas.
Compared to conventional orthogonal pilot-aided and the OLB algorithm approaches for channel estimation \cite{Marzetta2010,Jose2011,Bjornson2016}, the proposed tone-based AoA estimation scheme is simpler and the required coherence time resources will neither increase with the number of users nor the number of antennas.
{\color{black}
In addition, the proposed channel estimation relies on the antenna array gain for AoA estimation and the tone-based AoA estimation can be extended to hybrid systems by following a similar approach as in \cite{Zhao2017}.}
Literally, the more antennas used, the more antenna array gain can be obtained.
Therefore, it is expected that the required power of the transmitted tone is low in the considered massive MIMO mmWave system.
More importantly, the proposed tone-based AoA estimation algorithm can avoid the system performance degradation due to pilot contamination.
With a sufficiently large number of antennas, the proposed algorithm can accurately estimate the SLOS component as well as scattering components.

\begin{figure}[t]
\centering
\includegraphics[width=5.5in]{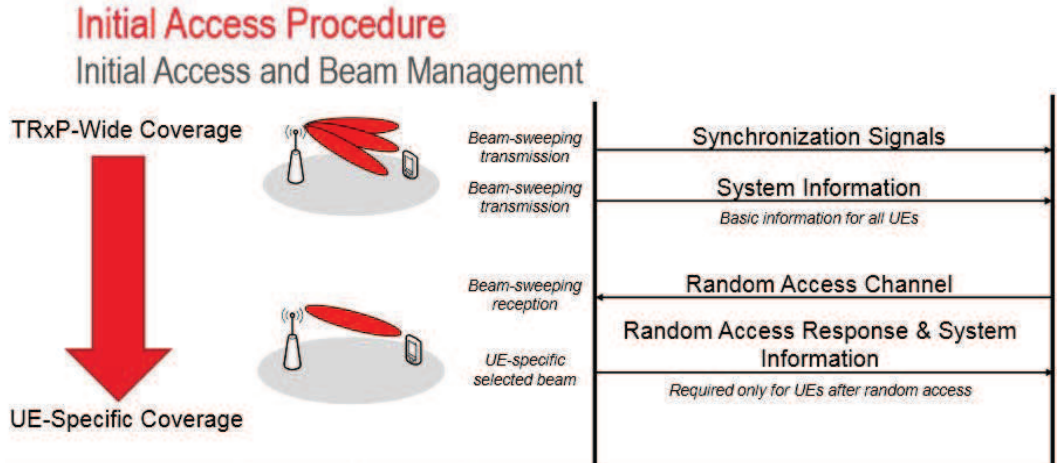}
\caption{Illustration of the practical 5G NR CSI acquisition/initial access procedure.}\label{fig:5G_NR}
\end{figure}

{\color{black}In mmWave systems, channel estimation is important for unlocking the potential of the system performance. Majority of contributions in the literature focus on the development of CSI feedback based channel estimation methods for hybrid mmWave systems, e.g. \cite{Alkhateeb2015,Xie2017,Shen2015,Wagner2012,Gao2016,Wei2017}. In practice, the 5G NR specification includes a set of basic beam-related procedures for the above 6 GHz CSI acquisition \cite{Giordani2018}. The beam management of 5G NR mmWave systems consists of four different operations which are shown in Figure \ref{fig:5G_NR}, i.e., beam sweeping, beam measurement, beam determination, and beam reporting. Taking a single antenna user in 5G NR stand-alone (SA) downlink scheme for example, the BS sequentially transmits synchronization signals (SS) and CSI reference signals (CSI-RSs) to the users by using a predefined codebook of directions. Then, the users search and track their optimal beams by measuring the collected CSI-RSs. At the end of CSI acquisition phase, beam reporting, the users feed back the determined beam information (beam quality and beam decision) and the random access channel (RACH) preamble to the BS.

\begin{figure}[t]
\centering
\includegraphics[width=4.5in]{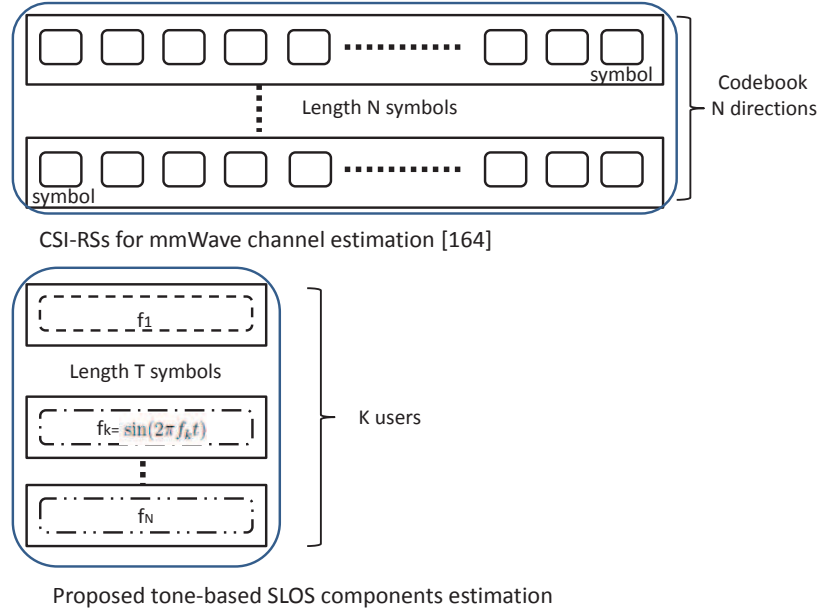}
\caption{Illustration of the overhead signalling difference between the practical 5G NR CSI acquisition procedure and the proposed algorithm.}\label{fig:diff}
\end{figure}
Mainly, the difference between the channel estimation algorithm adopted in practical mmWave systems and the proposed tone-based algorithm is the overhead signalling consumed for the channel estimation phase. Figure \ref{fig:diff} illustrates the overhead signalling difference between our proposed tone-based algorithm and the practical 5G NR mmWave channel estimation protocol. It shows that utilizing tone signals can significantly save the overhead signalling resource.
In addition, the MSE performance of aforementioned algorithms are the same. The reason is simple and straightforward. The antenna array adopted in these two algorithms to facilitate the channel estimation is same and has the same Cram\'{e}r Rao lower bound (CRLB).
}

In the following sections, we analyze the system rate performance of downlink transmission based on the estimated CSI via the proposed tone-based AoA estimation scheme.

\vspace{-0mm}
\section{Performance Analysis}

{\color{black}For the downlink transmission performance analysis, we first discuss the adopted assumptions. In reference \cite{Akdeniz2014}, which is based on recent field test results, the authors suggested that the angles between the users and the BS are uniformly distributed between $(0,\text{\ }\pi)$. In addition, reference \cite{Hur2016} adopted ray-tracing model to characterize the AoAs of users at different locations. The ray-tracing simulation results are verified and supported by field test results (Figure 5 of \cite{Hur2016}). It is shown that reflections from street signs, lamp posts, parked cars, passing people, etc., could reach the receiver from all directions. Based on these field test results, it is clear that the AoAs between the BS and the users are caused by many clusters in the propagation directions. In fact, the clusters around different separated users are different with high probability in the urban areas as shown in Figure 2 of \cite{Hur2016}. Although some users may share part of the clusters \cite{Shen2015}, to simplify the average achievable rate performance analysis, we assume that the set of AoAs of multi-path between different users will be different with high probability \cite{Buzzi2016b} in the large number of antennas regime. In other words, the scattering components of different users are assumed not highly correlated due to the random clusters in the urban scenario in the considered system. Specifically, in the large number of antennas regime, for $\theta _{k,l}\neq \theta _{j,i}$, it is expected that \vspace{-0mm}%
\begin{equation}
\mathbf{a}_{k,l}^{H}\mathbf{a}_{j,i}\overset{M\rightarrow\infty}{\rightarrow} 0,\text{ } k \neq j.\vspace{-0mm}
\end{equation}
We assume that given the random location of the scatters, the set of AoAs of multi-path between different users will be different with high probability \cite{Buzzi2016b}.
Therefore, for $k \neq j$, we have the following result:\vspace{-0mm}
\begin{equation}
\mathbf{h}_{\mathrm{S},k}^{H} \mathbf{h}_{\mathrm{S},j}= \frac{1}{L}\left(\overset{L}{\underset{l=1}{\sum }}{\alpha _{k,l}}\mathbf{a}_{k,l}^{H}\right)\left(\overset{L}{\underset{l=1}{\sum }}{\alpha _{j,l}}\mathbf{a}_{j,l}\right)\overset{M\rightarrow\infty}{\rightarrow} 0 .\vspace{-0mm}
\end{equation}}%
\vspace{-00mm}
\subsection{SLOS-based MRT Precoding}
MRT is considered as the simplest precoding strategy, due to its low computational complexity.
In \cite{Yue2015}, the authors proved that single-user massive MIMO systems are more suitable to exploit strong LOS components of Rician fading channels by comparison with pure Rayleigh fading environments.
In this section, we study the rate performance of the considered MU mmWave systems when MRT downlink transmission is employed and designed based on the estimated SLOS component.
To facilitate the following study, we assume that the perfect SLOS component is available\footnote{It was verified by simulation that, for the proposed SLOS estimation with a sufficiently large number of antennas at the BS, e.g. $50$ antennas, our proposed SLOS estimation algorithm is highly accurate. The simulation results are shown in Figure \ref{Fig003_2}.} and the downlink MRT precoder is given by $\widetilde{\mathbf{W}}_{\mathrm{L}}=\mathbf{H}_{\mathrm{L}}^{\ast }$.
The rate performance achieved by using the MRT precoder $\widetilde{\mathbf{W}}_{\mathrm{L}}$ can be considered as the system performance benchmark.
We then express the received signal of user $k$ as\vspace{-0mm}
\begin{equation}
y_{\mathrm{LMRT}}^{k}=\widetilde{\beta }\mathbf{h}_{k}^{T}%
\mathbf{h}_{\mathrm{L,}{k}}^{\ast }x_{k}+\widetilde{\beta }\mathbf{h%
}_{k}^{T}\overset{N}{\underset{j=1,j\neq k}{\sum }}\mathbf{h}_{%
\mathrm{L,}{j}}^{\ast }x_{j}+z_{k},  \label{E_3}\vspace{-0mm}
\end{equation}%
where $\widetilde{\beta }=\sqrt{\tfrac{1}{\mathrm{tr}(\widetilde{\mathbf{W}}_{\mathrm{L}}\widetilde{\mathbf{W}}_{\mathrm{L}}^{H })}}$ is the
power normalization factor, $x_{k}\in\mathbb{C}^{1\times 1}$ is the transmitted signal from the BS to user $k$ in the desired cell, $\mathrm{\mathbb{E}}(x_{k})=0$, and $\mathrm{\mathbb{E}}\left( x_{k}^{2}\right) =E_{s}$ is the transmit power.
We then present the signal-to-interference-plus-noise ratio (SINR) expression of user $k$ as\vspace{-0mm}
\begin{align}
&\hspace*{-1mm}\mathrm{SINR}_{\mathrm{LMRT}}^{k}
={M^{2}}\left[\dfrac{1}{\upsilon _{k}}%
\overset{N}{\underset{j=1}{\sum }}\mathrm{\mathbb{E}}_{\mathbf{h}_{\mathrm{S}}}\left[ \mathbf{h}_{\mathrm{S,}k}^{H }\mathbf{h}_{\mathrm{L,}j}%
\mathbf{h}_{\mathrm{L,}j}^{H }\mathbf{h}_{\mathrm{S,}k}\right]\right.\notag\\
&\hspace*{+20mm}\left. +\overset{N}{\underset{j=1,j\neq
k}{\sum }}\mathbf{h}_{\mathrm{L,}k}^{H }\mathbf{h}_{\mathrm{L,}j}\mathbf{%
h}_{\mathrm{L,}j}^{H }\mathbf{h}_{\mathrm{L,}k} +%
\dfrac{\upsilon _{k}+1}{\upsilon _{k}}\dfrac{\sigma ^{2}}{E_{s}\widetilde{\beta} ^{2}}\right]^{-1} , \label{E_4}\vspace*{-0mm}
\end{align}%
where $\dfrac{E_{s}}{\sigma ^{2}}$ represents the transmit SNR.
In the large number of antennas and high receive SNR regime, the approximated average achievable rate of the MRT precoding can be expressed as\vspace{-0mm}
\begin{equation}
R_{\mathrm{LMRT}}^{k}\underset{\mathrm{SNR}\rightarrow \infty
}{\overset{M\rightarrow \infty }{\approx }}\log_{2}\Big[ 1+ \Big[ \underset{\mathrm{Part}\text{ }\mathrm{1}}{\underbrace{%
\dfrac{1}{M^{2}}\overset{N}{\underset{j=1,j\neq k}{\sum }}\mathbf{h}_{%
\mathrm{L,}k}^{H }\mathbf{h}_{\mathrm{L,}j}\mathbf{h}_{\mathrm{L,}%
j}^{H }\mathbf{h}_{\mathrm{L,}k}}}+\underset{\mathrm{Part}\text{ }\mathrm{2%
}}{\underbrace{\dfrac{N}{M\upsilon _{k}}}}\Big] ^{-1}\Big],  \label{E_41}\vspace{-0mm}
\end{equation}
where $\mathrm{Part}$ $\mathrm{1}$ in Equation (\ref{E_41}) is considered as the interference to the desired user $k$ caused by SLOS components of different users.
In addition, $\mathrm{Part}$ $\mathrm{1}$ decreases significantly with an increasing number of antennas $M$.
$\mathrm{Part}$ $\mathrm{2}$ in Equation (\ref{E_41}) is considered as the interference caused by the scattering parts, which is determined by the Rician K-factor and the number of users to the number of antennas ratio.
\vspace{-0mm}
\subsection{SLPS-based ZF Precoding}

ZF precoding is widely adopted for MU systems due to its capability in interference suppression.
It was verified by simulation that the estimated SLPS can achieve a better MSE performance in channel estimation than the estimated SLOS, which is shown in Figure \ref{Fig003_3}.
While the estimated SLPS is adopted for the design of ZF precoder, the achievable rate of the SLPS-based ZF can be considered as the rate performance benchmark.
Based on the estimated SLPS-based uplink channel $\widehat{\mathbf{H}}$, the downlink precoder is expressed as\vspace{-0mm}
\begin{equation}
\widehat{\mathbf{W}}=\left( \widehat{\mathbf{H}}\right) ^{\ast }\left[
\left( \widehat{\mathbf{H}}\right) ^{T}\left( \widehat{\mathbf{H}}\right)
^{\ast }\right] ^{-1}.\vspace{-0mm}
\end{equation}%
{\color{black}The received signal at user $k$ in the desired cell is given by\vspace{-0mm}
\begin{equation}
y_{\mathrm{ZF}}^{k}=\widehat{\beta }\mathbf{h}_{k}^{T}\widehat{\mathbf{w}}%
_{k}x_{k}+\widehat{\beta }\mathbf{h}_{k}^{T}\overset{N}{\underset{j=1,j\neq k%
}{\sum }}\widehat{\mathbf{w}}_{j}x_{j}+\widehat{\beta }\overset{T}{\underset{%
p=1}{\sum }}\rho _{p,k}\mathbf{h}_{p,k}^{T}\widehat{\mathbf{w}}%
_{p,k}x_{p}+z_{k},  \label{Eq_5}\vspace{-0mm}
\end{equation}%
where the variable $p$ is the number of neighbouring cells, $\widehat{\mathbf{w}}_{k}\in\mathbb{C}^{M\times 1}$ is the $k$-th column vector of $\widehat{\mathbf{W}}$ and $\widehat{\beta }=\sqrt{\tfrac{1}{\mathrm{tr}\left( \widehat{\mathbf{W}}\widehat{\mathbf{W}}^{H}\right) }}$ is the normalization factor.}
For the ZF precoder under the perfect CSI, the precoder matrix is given by $\mathbf{W=H}^{\ast }\left( \mathbf{H}^{T}\mathbf{H}^{\ast }\right) ^{-1}$. Let $\mathbf{w}_{k}\in\mathbb{C}^{M\times 1}$ denote the $k$-th column vector of $\mathbf{W}, \forall
i\neq k,$ $\mathbf{h}_{i}^{T}\mathbf{w}_{k}=0$, and $\mathbf{h}_{k}^{T}\mathbf{w}_{k}=1$.
We re-express Equation (\ref{Eq_5}) as\vspace{-0mm}
\begin{equation}
y_{\mathrm{ZF}}^{k}=\underset{\mathrm{desired}\text{ }\mathrm{signal}}{\underbrace{\widehat{%
\beta }x_{k}}}+\underset{\mathrm{intra-cell}\text{ }\mathrm{interference}}{%
\underbrace{\widehat{\beta }\mathbf{h}_{k}^{T}\Delta \mathbf{Wx}}}+z_{k},\label{Eq_5_1}\vspace{-0mm}
\end{equation}%
where $\Delta \mathbf{w}_{j}\in\mathbb{C}^{M\times 1}$ denotes the $j$-th column vector of the ZF precoder error matrix $\Delta \mathbf{W}=\widehat{\mathbf{W}}-\mathbf{W}$, and $\mathbf{x}=[x_1,\ldots,x_N]^T$ denotes the transmitted signal vector. We then express the SINR of user $k$ as\vspace{-0mm}
\begin{align}
\mathrm{SINR}_{\mathrm{ZF}}^{k}=\frac{{\widehat{\beta }^{2}}E_{s}}{{\widehat{\beta }^{2}}E_{s}\mathbf{h}%
_{k}^{T}\mathrm{\mathbb{E}}_{\mathrm{\Delta \mathbf{H}}}\left[ \Delta \mathbf{W}\Delta \mathbf{W}%
^{H}\right] \mathbf{h}_{k}^{\ast }+{\sigma ^{2}}}.\label{Eq_151}\vspace{-0mm}
\end{align}%
Now we summarize the achievable rate per user in the large number of antennas and high SNR regime in the following theorem.

\begin{theo}
For a large receive SNR and large number of antennas, the average achievable rate of the ZF precoding based on the estimated CSI can be approximated  by \vspace{-1mm}
\begin{align}
R_{\mathrm{ZF}}^{k}&\underset{\mathrm{SNR}\rightarrow \infty}{\overset{M\rightarrow \infty }{\approx}}\log _{2}\left[ 1+\left[ \left( \sqrt{%
1+\delta ^{2}}-1\right) ^{2} +\frac{\left( \sqrt{1+\delta
^{2}}\right) \left( 2-\sqrt{1+\delta ^{2}}\right) \delta ^{2}\mathrm{tr}%
\left( \mathbf{K}^{-1}\right) }{1+\delta ^{2}\mathrm{tr}\left( \mathbf{K}%
^{-1}\right) }\right. \right.  \notag \\
&\text{\ \ \ \ \ \ \ \ \ }\left. \left.-\frac{2\sqrt{1+\delta ^{2}}\left( \sqrt{%
1+\delta ^{2}}-1\right) \delta ^{2}M\eta _{kk}}{1+\delta ^{2}\mathrm{tr}%
\left( \mathbf{K}^{-1}\right) }\right] ^{-1}\right],\label{Eq_17_0}\vspace{-2mm}
\end{align}
where $\mathbf{K}=(\mathbf{H}^{T}\mathbf{H}^{\ast })$, $\eta _{kk}$ represents the $k$-th diagonal element of $\mathbf{K}%
^{-1}$, and $\delta ^{2}$ is the normalized MSE of proposed channel estimation algorithm.
\end{theo}\vspace{-3mm}
\emph{\quad Proof: } Please refer to Appendix A. \QEDA

Observing Equation (\ref{Eq_17_0}), the approximated average achievable rate of the ZF precoding is bounded by above. The finite achievable rate ceiling is mainly determined by the normalized MSE of the estimated SLPS.

\vspace{-0mm}
\section{Comparison with PAC-based Precoding}
\vspace{-0mm}
In this section, we analyze the achievable rates performance of the MRT and the ZF precoders in multi-cell mmWave systems via pilot-aided estimated CSI.
The adopted CSI for the design of these precoders is under the impact of pilot contamination, which is caused by the reuse of orthogonal pilot symbols among different cells.
The analytical results obtained in this section will be used as a performance reference in the subsequent sections.
In particular, our analysis quantifies the connection between rate performance and the MSE of channel estimation.
\vspace{-0mm}
\subsection{PAC-MRT and PAC-ZF Precoding under Pilot Contamination}
\vspace*{+0mm}
Here we study the performance of the MRT precoding based on pilot-aided CSI in the presence of pilot contamination in mmWave channels.
The pilot contamination leads to two types of interference in the considered system: intra-cell interference and inter-cell interference \cite{Marzetta2010,Jose2011}.
For the uplink channel estimation, we denote the received pilot sequences at the BS in the desired cell as\vspace{-1mm}
\begin{equation}
\mathbf{Y}_{\mathrm{u}}=\mathbf{H}\mathbf{\Psi }+\overset{T}{\underset{p=1}{\sum }}\mathrm{diag}\left(\rho _{p,k}\right)
\mathbf{H}_{p}\mathbf{\Psi }+\mathbf{Z},  \label{Eq_1}\vspace{-1mm}
\end{equation}%
where $\mathbf{\Psi}\in\mathbb{C}^{N\times N}$ stands for the reused orthogonal pilot sequences matrix and $\rho_{p,k}$ is the cross-cell path loss attenuation coefficient.
In Equation (\ref{Eq_1}), $\mathbf{H}_{p}$ is the channel matrix from the users in the $p$-th neighboring cell to the BS in the desired cell.
According to the field test, the probability of existing LOS component decreases significantly with increasing propagation distance \cite{Rappaport2015}. In this case, we model the entries of cross-cell channels $\mathbf{H}_{p}$ as i.i.d. random variable $\mathcal{CN}{\left(0,1\right)}$.

Conventionally, the MMSE channel estimation, which requires the prior channel information, can achieve a better channel estimation performance than LS channel estimation.
However, the prior information of mmWave channels, e.g. the Rician K-factor, is hard to obtain.
Therefore, we adopt the LS algorithm for channel estimation instead of the MMSE algorithm, the estimated channel can be expressed as\vspace*{-0mm}
\begin{align}
\widehat{\mathbf{H}}&=\mathbf{Y}_{\mathrm{u}}\mathbf{\Psi }^{H}\left[ \mathbf{\Psi }\mathbf{\Psi }^{H} \right]^{-1}=\mathbf{H}+\underset{\mathrm{pilot}\text{ }\mathrm{contamination}}{\underbrace{\overset{T}{\underset{p=1}{\sum }}\mathrm{diag}\left(\rho _{p,k}\right)\mathbf{H}_{p}}}+\widehat{\mathbf{Z}}=\mathbf{H}+\Delta \mathbf{H}+\widehat{\mathbf{Z}},\vspace*{-0mm}
\end{align}\vspace*{+0mm}%
where $\widehat{\mathbf{Z}}= \mathbf{Z}\mathbf{\Psi }^{H}\left[ \mathbf{\Psi }\mathbf{\Psi }^{H} \right]^{-1}$ and $\Delta \mathbf{H}\in\mathbb{C}^{M\times N}$ is the equivalent channel estimation error matrix due to the impact of pilot contamination.
Let $\Delta\mathbf{h}_{k}\in\mathbb{C}^{M\times 1}$ be the $k$-th column of the channel estimation error matrix $\Delta \mathbf{H,}$ and the expression of $\Delta \mathbf{h}_{k}$ is given
by\vspace*{-0mm}
\begin{equation}
\Delta \mathbf{h}_{k}=\overset{T}{\underset{p=1}{\sum }}\rho _{p,k}
\mathbf{h}_{p,k},\vspace*{-0mm}
\end{equation}%
where $\mathbf{h}_{p,k}$, with entries follow $\mathcal{CN}\left( 0,1\right)$, is the channel vector from the $p$-th neighboring cell.

{\color{black}
The relationships between $\Delta \mathbf{h}_{k}$, $\mathbf{h}_{p,k}$, and $\delta^{2}_{\mathrm{P}}$ can be expressed as\vspace*{-0mm}
\begin{equation}
\dfrac{1}{M}\mathrm{\mathbb{E}}\left[ \Delta \mathbf{h}_{k}^{H}\Delta \mathbf{h}_{k}\right] =%
\dfrac{1}{M}\overset{T}{\underset{p=1}{\sum }}\rho _{p,k}^{2}\mathrm{\mathbb{E}}\left[
\mathbf{h}_{p,k}^{H}\mathbf{h}_{p,k}\right]=\overset{T}{\underset{p=1}{\sum }}\rho _{p,k}^{2}=\delta ^{2}_{\mathrm{P}}, \label{E_519}\vspace*{-0mm}
\end{equation}
where $\delta ^{2}_{\mathrm{P}}$ is the normalized MSE of channel estimation \cite{Khansefid2015}.

In addition, we have $\mathcal{CN}(\varsigma,1)$ for the desired signal in the desired cell. The reason is that the perfect long-term power control is performed to compensate for path loss and shadowing between the desired BS and the desired users to facilitate the investigation of channel estimation and downlink transmission \cite{Alkhateeb2015,Ni2016,Yang2013}.
Due to the cross-cell propagation path loss, the received energy of desired pilot symbols at the desired BS is significantly larger than the energy of pilot symbols from neighboring cells to the desired BS.

For example, the $200$ meters mmWave path loss is approximately above $140$ dB and the $100$ meters propagation path loss is about $120$ dB \cite{Hur2016,Rappaport2015}. The received energy of desired pilot symbols is about $20$ dB larger than that of the energy of contaminated pilot symbols (Table IV of \cite{Hur2016}).
Thus, the impact of pilot contamination on the channel estimation is assumed as $0<\delta ^{2}_{\mathrm{P}}<1$.}

The received signal at user $k$ with the MRT precoder $\widetilde {\mathbf{W}}_{\mathrm{PAC-MRT}}=\mathbf{H}^{\ast
}+\left( \Delta \mathbf{H}\right) ^{\ast }$ is given by\vspace*{-2mm}
\begin{align}
y_{\mathrm{PAC-MRT}}^{k}&=\underset{\mathrm{desired}\text{ }\mathrm{signal}}{\underbrace{\beta
\mathbf{h}_{k}^{H}\mathbf{h}_{k}x_{k}}}+\underset{\mathrm{%
inter-cell}\text{ }\mathrm{interference}}{\underbrace{\beta \overset{T}{%
\underset{p=1}{\sum }}\rho _{p,k}\mathbf{h}_{p,k}^{H}\widehat{\mathbf{w}}_{p,k}^{\ast }x_{p,k}}}\notag \\
&+\underset{\mathrm{intra-cell}\text{ }%
\mathrm{interference}}{\underbrace{\beta \overset{N}{\underset{j=1,j\neq k}{%
\sum }}\mathbf{h}_{j}^{H}\mathbf{h}_{k}x_{j}+\beta \overset{N}{\underset{j=1}%
{\sum }}\Delta \mathbf{h}_{j}^{H}\mathbf{h}_{k}x_{j}}}+z_{k},  \label{E_521}\vspace*{-2mm}
\end{align}%
where $\beta=\sqrt{\frac{1}{\mathrm{tr}(\widetilde{\mathbf{W}}_{\mathrm{PAC-MRT}}\widetilde{\mathbf{W}}_{\mathrm{PAC-MRT}}^{H})}}$ is the normalization factor, $\widehat{\mathbf{w}}_{p,k}^{\ast }$ is the downlink precoder of user $k$ in $p$-th cell, and$\ z_{k}\sim \mathcal{CN}\left( 0,\sigma^{2}\right) $ accounts for AWGN.
Substituting Equation (\ref{E_001}) into Equation (\ref{E_521}), the received SINR at user $k$ can be expressed as\vspace*{-2mm}
\begin{align}
&\hspace*{-0mm}\mathrm{SINR}_{\mathrm{PAC-MRT}}^{k}= \left| \mathbf{h}_{k}^{H}\mathbf{h}_{k}\right| ^{2}\left[\left( \overset{N}{\underset{j=1,j\neq k}{\sum }}\mathbf{h}%
_{j}^{H}\mathbf{h}_{k}\mathbf{h}_{k}^{H}\mathbf{h}_{j}\right) \right.\notag \\
&\hspace*{+30mm}\left.+\overset{N}{\underset{j=1}{\sum }}\mathrm{\mathbb{E}}_{\Delta\mathbf{h}}\left[ \Delta \mathbf{h}_{j}^{H}%
\mathbf{h}_{k}\mathbf{h}_{k}^{H}\Delta \mathbf{h}_{j}\right] +M\overset{T}{\underset{p=1}{\sum }}\rho _{p,k}^{2}+\dfrac{E_{s}}{\beta^{2}\sigma ^{2}}\right]^{-1}.\label{E_52}\vspace*{-2mm}
\end{align}%
We next present an asymptotic SINR of Equation (\ref{E_52}) in the following corollary.
\begin{coro}\label{Coro_21}
As SNR and the number of antennas approach infinity, the average achievable rate of the MRT precoding under pilot contamination can be approximated by\vspace{-0mm}
\begin{align}
&\mathrm{R}_{\mathrm{PAC-MRT}}^{k}\underset{\mathrm{SNR}\rightarrow \infty}{\overset{M\rightarrow \infty }{\approx }}\notag\\
&\log _{2}\left[1+\left[ \underset{\mathrm{Part}\text{ }\mathrm{1}}{\underbrace{%
\overset{N}{\underset{j=1,j\neq k}{\sum }}\dfrac{\mathbf{h}_{\mathrm{L,}%
j}^{H}\mathbf{h}_{\mathrm{L,}k}\mathbf{h}_{\mathrm{L,}k}^{H}\mathbf{h}_{%
\mathrm{L,}j}}{M^{2}}}}+\underset{\mathrm{Part}\text{ }\mathrm{2}}{%
\underbrace{\dfrac{\left( N-1\right) \left( 2\upsilon _{k}+1\right) }{%
M\left( \upsilon _{k}+1\right) ^{2}}}}+\underset{\mathrm{Part}\text{ }%
\mathrm{3}}{\underbrace{\dfrac{N\delta ^{2}_{\mathrm{P}}}{M}}}+\underset{\mathrm{Part}%
\text{ }\mathrm{4}}{\underbrace{\dfrac{\delta ^{2}_{\mathrm{P}}}{M}}}\right] ^{-1}\right] .
\label{E_513}
\end{align}
\end{coro}\vspace{-3mm}
\emph{\quad Proof: } Please refer to Appendix B. \QEDA

$\mathrm{Part}$ $\mathrm{1}$ and $\mathrm{Part}$ $2$ of Equation (\ref{E_513}) represent the interference caused by the corresponding SLOS components and scattering components of different users, respectively.
In addition, $\mathrm{Part}$ $3$ and $\mathrm{Part}$ $4$ of Equation (\ref{E_513}) denote the intra-cell and inter-cell interference caused only by channel estimation error due to pilot contamination, respectively.
It is interesting to note that, the intra-cell interference in $\mathrm{Part}$ $3$ is much larger than that of the inter-cell interference in $\mathrm{Part}$ $4$.

It is clear that with an increasing number of antennas $M$, the achievable rate increases. Besides, with increasing the Rician K-factor $\upsilon _{k}$, the interference caused by scattering components in $\mathrm{Part}$ $2$ of Equation (\ref{E_513}) vanishes.

{For the downlink ZF precoding, similar to the signal processing in Equations (\ref{Eq_5})$-$(\ref{Eq_5_1}), $\widetilde{\mathbf{w}}_{\mathrm{PAC-ZF}}$ is the downlink ZF precoder and $\overline{\beta}=\sqrt{\frac{1}{\mathrm{tr}(\widetilde{\mathbf{W}}_{\mathrm{PAC-ZF}}\widetilde{\mathbf{W}}_{\mathrm{PAC-ZF}}^{H})}}$ is the normalization factor.

We then express the SINR of user $k$ under the PAC-based ZF precoding as\vspace{-0mm}
\begin{align}\hspace*{-0mm}
&\mathrm{SINR}_{\mathrm{PAC-ZF}}^{k}=\notag\\
&\left[\underset{\mathrm{inter-cell}\text{ }\mathrm{interference}}{\underbrace{\overset{T}{\underset{p=1}{\sum }}\rho _{p,k}^{2}\mathrm{\mathbb{E}}_{ \mathbf{h}_{p,k}}\left[ \mathbf{h}_{p,k}^{T}\widehat{\mathbf{w}}_{p,k}\widehat{\mathbf{w}}_{p,k}^{H} \mathbf{h}_{p,k}^{\ast }\right]}}+\underset{\mathrm{intra-cell}\text{ }\mathrm{interference}}{\underbrace{\mathbf{h}%
_{k}^{T}\mathrm{\mathbb{E}}_{\Delta \mathbf{H}}\left[ \Delta \mathbf{W}\Delta \mathbf{W}%
^{H}\right] \mathbf{h}_{k}^{\ast }}} +\dfrac{\sigma ^{2}}{E_{s}\overline{\beta }^{2}}\right]^{-1}.\vspace{-0mm}
\end{align}}%

Now we summarize the average achievable rate per user in the large number of antennas and high SNR regime in the following corollary.\vspace{+2mm}
\begin{coro}
For a large receive SNR and a large number of antennas, the average achievable rate of the ZF precoding under pilot contamination can be approximated by\vspace{-0mm}
\begin{align}
&R_{\mathrm{PAC-ZF}}^{k}\underset{\mathrm{SNR}\rightarrow \infty}{\overset{M\rightarrow \infty }{\approx }} \log _{2}\left[ 1+\left[ \left( \sqrt{%
1+\delta ^{2}_{\mathrm{P}}}-1\right) ^{2}-\frac{2\sqrt{1+\delta ^{2}_{\mathrm{P}}}\left( \sqrt{1+\delta ^{2}_{\mathrm{P}}}-1\right) \delta ^{2}_{\mathrm{P}}M\eta _{kk}}{1+\delta ^{2}_{\mathrm{P}}\mathrm{tr}\left( \mathbf{K}^{-1}\right) } \right.\right.  \notag \\
&\hspace*{+25mm}\left. \left.+\frac{\delta ^{2}_{\mathrm{P}}}{M}+\frac{\left( \sqrt{1+\delta ^{2}_{\mathrm{P}}}\right) \left( 2-\sqrt{1+\delta ^{2}_{\mathrm{P}}}\right) \delta ^{2}_{\mathrm{P}}\mathrm{tr}\left( \mathbf{K}^{-1}\right) }{1+\delta ^{2}_{\mathrm{P}}\mathrm{tr}\left( \mathbf{K}%
^{-1}\right) }\right] ^{-1}\right],\label{Eq_17_1}\vspace{-0mm}
\end{align}%
where $\mathbf{K}=(\mathbf{H}^{T}\mathbf{H}^{\ast })$, $\eta _{kk}$ represents the $k$-th diagonal element of $\mathbf{K}^{-1}$, and $\delta ^{2}_{\mathrm{P}}$ is the normalized MSE of LS channel estimation given by Equation (\ref{E_519}).
\end{coro}\vspace{-0mm}
\emph{\quad Proof: } Following a similar approach as in Appendix A.\QEDA

Equations (\ref{E_513}) and (\ref{Eq_17_1}) represent the ceiling of the achievable rate of MU mmWave massive MIMO systems when the conventional PAC is adopted.
The performance gap between Equations (\ref{E_513}) and (\ref{Eq_17_1}) will be quantized via simulation in Figure \ref{Fig00434} in the following sections.

\vspace{-0mm}
\subsection{Achievable Rate Comparison}
In this section, we compare the achievable rates of the PAC-based MRT and the PAC-based ZF precoders to the achievable rates of the SLOS-based MRT and the SLPS-based ZF precoders, respectively.
{\color{black}For mmWave cellular systems, the deployment of small cell is necessary to facilitate high data rate communications \cite{Hur2016}.
In our work, the radius of the cell is assumed approximately less than $150$ meters.
Under such a scenario, for a reasonably large transmit power and the use of large number of antennas, high SNR is expected to the system operating regime\footnote{By calculating the link budget \cite{Hur2016}, it can be shown that the considered system operates at high SNR regimes.}.
In fact, the derived downlink achievable rate analysis is based on the high SNR assumption with practical interests for high data rate communication.}
For the comparison between the achievable rates of the PAC-based ZF precoding and the SLPS-based ZF precoding, we observe the mathematical similarity between Equations (\ref{Eq_17_1}) and (\ref{Eq_17_0}). While the MSEs of two different channel estimation are identical, i.e., $\delta^{2}= \delta^{2}_{\mathrm{P}}$, the difference between $R_{\mathrm{ZF}}^{k}$ and $R_{\mathrm{PAC-ZF}}^{k}$ is negligible.

\begin{figure}[t]
\begin{center}
\includegraphics[width=4.5in]{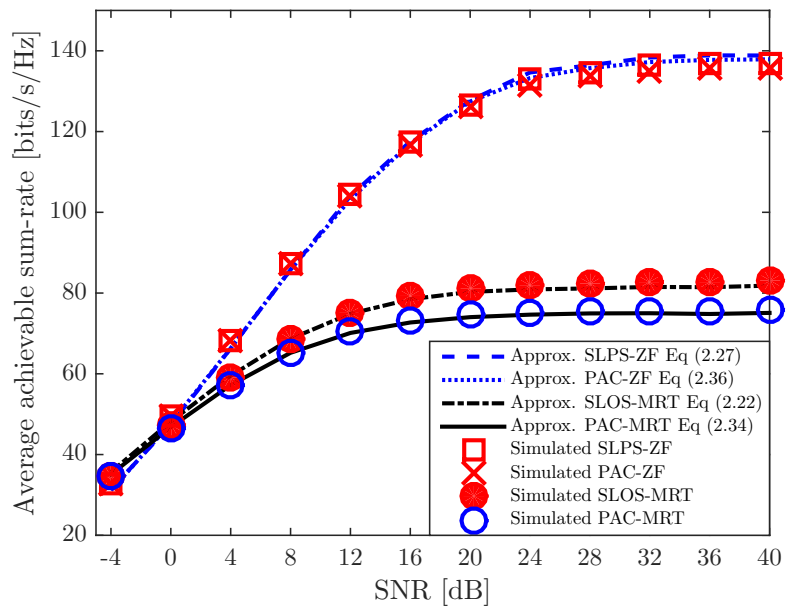}
\caption{Average achievable sum-rate [bits/s/Hz] versus SNR [dB] for
different precoding schemes with $M=100$ and $N=20$. We set the Rician K-factor as $15$ and the normalized $\mathrm{MSE}=0.02$ for the SLPS-based ZF, the PAC-based ZF, and the PAC-based MRT.}
\label{Fig00434}
\end{center}\vspace{-0mm}
\end{figure}

This observation will also be verified via simulation in Figure \ref{Fig00434}.
Then, we compare the achievable rates between the SLOS-based MRT and the PAC-based MRT and summarize the performance comparison in the following theorem.
\begin{theo}
If the Rician factor satisfies $\upsilon _{k}>2$, the rate performance gap between the proposed SLOS-based MRT
and the PAC-based MRT is always larger than $0$, even if the MSE of pilot-aided channel estimation is $\delta_{\mathrm{P}} ^{2} = 0$, i.e.,\vspace{-0mm}
\label{theo2}%
\begin{equation}
\Delta R=\log _{2}\left[ \frac{1+\mathrm{SINR}_{\mathrm{LMRT}}^{k}}{1+\mathrm{SINR}_{\mathrm{PAC-MRT}}^{k}}\right] >0.  \label{E_412}
\end{equation}
\end{theo}\vspace{-0mm}
\emph{\quad Proof: } The result follows by substituting (\ref{E_513}) and (\ref{E_41}) into (\ref{E_412}). \QEDA

Intuitively, the required coherence time resource for the conventional PAC estimation, increases with an increasing number of users $N$.
Interestingly, the time resource consumption of the proposed AoA channel estimation algorithm, $T_\mathrm{CE}$, does not increase with the number of antennas nor the number of users.
In this case, we note that the coherence time saved by adopting the proposed algorithm is $N-T_\mathrm{CE}$ symbols, which can be used for downlink data transmission.
{In Figure \ref{Fig00434}, we present a comparison between the achievable sum-rates of the SLOS-based MRT, the PAC-based MRT, the SLPS-based ZF, and the PAC-based ZF for the same MSEs of channel estimation $\delta ^{2}=\delta ^{2}_{\mathrm{P}}=0.02$.
This result verifies the accuracy of our derived analytical expressions in Equations (\ref{E_41}), (\ref{Eq_17_0}), (\ref{E_513}), and (\ref{Eq_17_1}).
The achievable rates of the SLPS-based and the PAC-based ZF precoders are almost identical.
As expected, the ZF precoding can achieve a higher rate performance than the MRT precoding due to its interference suppression capability, especially in the high SNR regime.
It is interesting that in the high SNR regime, the SLOS-based MRT can achieve a significant performance gain in terms of achievable rate than the PAC-based MRT.
The reasons are two-fold.
First, for the SLOS-based MRT precoding, the signal energy focuses on the strong LOS propagation path leading to a smaller amount of signal energy leakage into the scattering paths.
In fact, the less signal energy leaks into the scattering part, the smaller inter-user interference it will cause.
In contrast, pilot-aided CSI naturally contains scattering components, which leads to a inter-user interference.
Second, the PAC-based channel estimation error may cause extra MU interference.
However, the proposed tone-based AoA channel estimation error, which is due to not taking into account scattering components, does not cause extra MU interference as shown in Equation (\ref{E_41}).
Thus, the SLOS-based MRT can achieve a better SINR performance than the PAC-based MRT.}
\vspace{-2mm}
\subsection{Large-system Analysis}
\begin{figure}[t]
\centering
\includegraphics[width=4.5in]{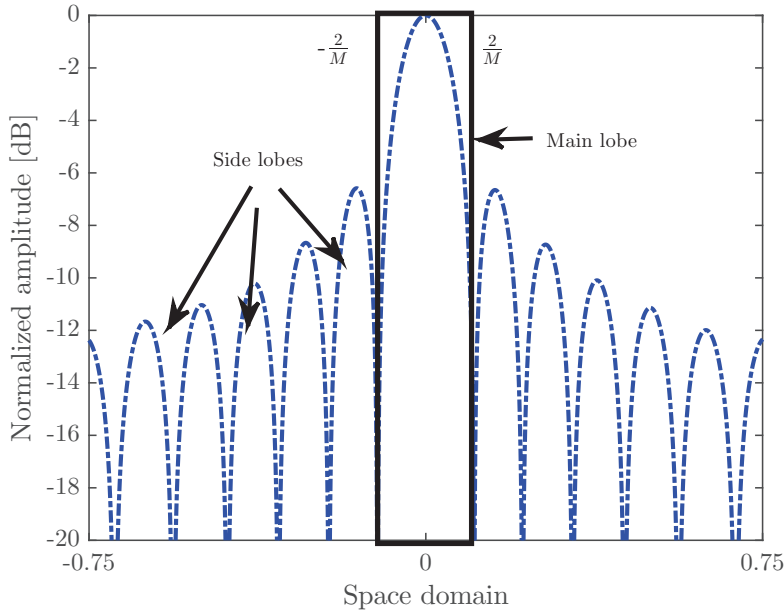}
\caption{Illustration of the main lobe and side lobes of ULA in the space domain with respect to the BS.}\label{fig:illustration}
\end{figure}
In this section, we study the system performance while the numbers of users and antennas are large.
In fact, the achievable rate approximation expressions of the MRT and the ZF in Equations (\ref{E_41}), (\ref{Eq_17_0}), (\ref{E_513}), and (\ref{Eq_17_1}) still rely on instantaneous channel realization of the SLOS components of all the users.
These channel dependent expressions are inconvenient to provide useful guidelines for system design.

Fortunately, similar to conventional massive MIMO systems while the numbers of users and antennas are large, the law of large number can bring us some interesting results \cite{Wagner2012,Hugo2016}.
Now, we focus on a scenario where the number of users and the number of antennas are large while their ratio is bounded above by a finite constant.
To facilitate the analysis, we assume that all the users are uniformly distributed \cite{Dai2016} with certain degrees separation in angular domain.
To illustrate the concepts of angular separation, the main lobe and side lobes of ULA is shown in Figure \ref{fig:illustration}.

The range of the main lobe in the space domain\footnote{The spatial domain constraint is given by $-1 \leqslant\cos(\theta _{k})\leqslant 1$ \cite{Trees2002}.} for user $k$\cite{Yin2013}, $\Theta_{k}$, is given by\vspace{-0mm}
\begin{equation}
\cos (\theta _{k})-\frac{2}{M}<\Theta _{k}<\cos (\theta _{k})+\frac{2}{M},\vspace{-0mm}
\end{equation}%
where $\theta _{k}$ is the AoA of the SLOS of user $k$. If the AoA of the SLOS of user $i$ is not in the main lobe of user $k$, we consider that user $i$ is in the side lobes of user $k$ from the BS point of view.
The minimum angular separation between different users from the BS point of view in a far-field region can be expressed as\vspace{-0mm}
\begin{equation}
\vspace*{-0mm}
\left\vert \cos (\theta _{k})-\cos (\theta _{i})\right\vert
\geqslant \frac{4}{M},\text{ }\forall i\neq k.  \label{Eq_60}\vspace{-0mm}
\end{equation}%
Let us consider the extreme case that all the users are separated one by one and with the minimal angular separation $\left\vert \cos (\theta _{k})-\cos (\theta _{i})\right\vert= \dfrac{4}{M},\text{ }\forall i\neq k$. The maximum number of users $N_{\mathrm{max}}$, which can be supported by the BS equipped with $M$ antennas, should satisfy the following constraint:\vspace{-0mm}
\begin{equation}
\frac{4}{M}\left( N_{\mathrm{max}}-1\right) \leqslant 2.  \label{Eq_601}\vspace{-0mm}
\end{equation}%
Therefore, the maximum number of users $N_{\mathrm{max}}$ is approximately $N_{\mathrm{max}} \approx \dfrac{M}{2}$.
Here, we justify that the density of users due to the angular separation assumption in Equation (\ref{Eq_60}) is reasonable with the following example.

For typical mmWave massive MIMO systems operating at $30$ GHz, the cell radius is roughly $150$ meters.
For a BS equipped with a $200$-antenna ULA array serving $100$ users simultaneously, the user density is $3500$ devices/$($km$)^{2},$ which is the medium user density as stated in 3GPP \cite{NSN2010}.
Besides, the $200$-antenna ULA array, which is deployed for half wavelength separation, is only $0.5$ meter long.
Under such an assumption, we note that the channels between any two users are not orthogonal.
Such non-orthogonality has already been verified via field experiments \cite{Nguyen2015,Gao2015}.

Based on the angular separation assumption in Equation (\ref{Eq_60}),
we can then make a further approximation \cite{Trees2002} for the achievable rates of the MRT and the ZF precoding in Equations (\ref{E_41}), (\ref{Eq_17_0}), and (\ref{E_513}).
$\mathrm{Part}$ $\mathrm{1}$ of Equation (\ref{E_41}) and $\mathrm{Part}$ $1$ of Equation (\ref{E_513}) can be rewritten as\vspace{-0mm}%
\begin{equation}
\overset{N}{\underset{j=1,j\neq k}{\sum }}\mathbf{h}_{\mathrm{L,}k}^{H}%
\mathbf{h}_{\mathrm{L,}j}\mathbf{h}_{\mathrm{L,}j}^{H}\mathbf{h}_{%
\mathrm{L,}k}\overset{M,N\rightarrow \infty }{\approx }\dfrac{N^{2}}{4}.\vspace{-0mm}
\end{equation}%
While the number of antennas $M$ and the number of users $N$ are large with
their ratio $B=\dfrac{N}{M}$ fixed, the asymptotic SINR expressions of the SLOS-based MRT
and the PAC-based MRT in the high SNR regime can be expressed as\vspace{-1mm}
\begin{equation}
\hspace*{-4mm}\mathrm{SINR}_{\mathrm{LMRT}}^{k}\underset{\mathrm{SNR}\rightarrow \infty
}{\overset{M,N\rightarrow \infty }{\approx }}\left[ \frac{1}{4}\left( \dfrac{%
N}{M}\right) ^{2}+\dfrac{1}{\upsilon _{k}}\frac{N}{M}\right] ^{-1},\text{\ and}
\label{E_61}\vspace{-1mm}
\end{equation}%
\begin{equation}
\mathrm{SINR}_{\mathrm{PAC-MRT}}^{k}\overset{M,N\rightarrow \infty }{\underset%
{\mathrm{SNR}\rightarrow \infty }{\approx }}\left[ \dfrac{1}{4}\left( \frac{%
\upsilon _{k}}{\upsilon _{k}+1}\right) ^{2}\left( \dfrac{N}{M}\right) ^{2}+%
\dfrac{N}{M}\dfrac{\left( 2\upsilon _{k}+1\right) }{\left( \upsilon
_{k}+1\right) ^{2}}+\dfrac{N}{M}\delta ^{2}\right] ^{-1},  \label{E_71}\vspace{-1mm}
\end{equation}%
respectively.

Interestingly, if Rician K-factor $\upsilon _{k}$ does not approach infinity, the result shows that Equation (\ref{E_61}) is always larger than Equation (\ref{E_71}) for a fixed ratio $B=\dfrac{N}{M}$.

Now we focus on the asymptotic rate performance of the PAC-based ZF precoding and the SLPS-based ZF precoding.
To this end, we have\vspace{-1mm}
\begin{equation}
\mathrm{\mathbb{E}}[\mathbf{H}^{T}\mathbf{H}^{\ast }]=\left( \mathbf{G}_{\mathrm{L}}\right) ^{2}%
\mathbf{H}_{\mathrm{L}}^{T}\mathbf{H}_{\mathrm{L}}^{\ast }+\left(
\mathbf{G}_{\mathrm{S}}\right) ^{2}\mathbf{H}_{\mathrm{S}}^{T}\mathbf{H}_{%
\mathrm{S}}^{\ast }.\vspace{-1mm}
\end{equation}%
We can obtain several properties of symmetric matrix $\mathbf{H}_{\mathrm{L}}^{T}\mathbf{H}_{\mathrm{L}}^{\ast }$ via singular value decomposition (SVD) as\vspace{-1mm}
\begin{equation}
\mathrm{tr}\left[ \mathbf{H}_{\mathrm{L}}^{T}\mathbf{H}_{\mathrm{L}%
}^{\ast }\right] =\mathrm{tr}\left[ \mathbf{H}_{\mathrm{L}}^{H }\mathbf{H}%
_{\mathrm{L}}\right]=\mathrm{tr}\left[ \mathbf{U\Sigma U}^{H }\right] =%
\overset{N}{\underset{i=1}{\sum }}\lambda _{i}\left( \mathbf{\Sigma }\right)
=MN,\vspace{-1mm}
\end{equation}%
where $\lambda _{i}\left( \mathbf{\Sigma }\right) $ is the non-negative
eigenvalues of $\mathbf{\Sigma}$ and $\mathbf{UU}^{H }=\mathbf{I}_{%
\mathrm{N}}=\mathbf{U}^{H }\mathbf{U}$. We can obtain\vspace{-1mm}
\begin{equation}
\mathrm{tr}\left[ \mathbf{h}_{\mathrm{L,}i}^{H }\mathbf{h}_{\mathrm{L,}i}%
\right] =M=\lambda _{i}(\mathbf{\Sigma }),\text{ }\forall i\in \lbrack 1,N].\vspace{-1mm}
\end{equation}%
where $\mathbf{h}_{\mathrm{L,}i}$ is the $i$-th column vector of $\mathbf{H}_{\mathrm{L}}.$
As we assume that all the users are distinguishable in the angular domain, we have the following expression:\vspace{-2mm}
\begin{equation}
\mathrm{tr}\left\{ \left[ \mathbf{H}_{\mathrm{L}}^{H }\mathbf{H}_{\mathrm{L}}\right] ^{-1}\right\} =\mathrm{tr}\left[ \left( \mathbf{U}^{H }\right)^{-1}\mathbf{\Sigma }^{-1}\mathbf{U}^{-1}\right]=\mathrm{tr}\left( \mathbf{\Sigma }^{-1}\right) =\overset{N}{\underset{i=1}{\sum }}\frac{1}{\lambda_{i}\left( \mathbf{\Sigma }\right) }\approx \frac{N}{M}. \label{Eq_711}\vspace{-2mm}
\end{equation}%

Based on the previous results Equations (\ref{Eq_17_0}) and (\ref{Eq_711}), the SINR expression of user $k$ under the SLPS-based ZF precoding in the high SNR regime is given by\vspace{-2mm}
\begin{align}
\mathrm{SINR}_{\mathrm{ZF}}^{k}&\underset{\mathrm{SNR}\rightarrow \infty
}{\overset{M,N\rightarrow \infty }{\approx }} \left[ \left( \sqrt{1+\delta ^{2}}-1\right) ^{2} -\frac{2\delta ^{2}\sqrt{%
1+\delta ^{2}}\left( \sqrt{1+\delta ^{2}}-1\right) \upsilon _{k}}{\left(
1+\delta ^{2}\dfrac{N}{M}\right) \left( \upsilon _{k}+1\right) }\right.\notag\\
&\hspace*{+10mm}\left.+\frac{%
\left( \sqrt{1+\delta ^{2}}\right) \left( 2-\sqrt{1+\delta ^{2}}\right)
\delta ^{2}}{\left( 1+\delta ^{2}\dfrac{N}{M}\right) }\dfrac{N}{M}\right]
^{-1}.\label{Eq_82}\vspace{-2mm}
\end{align}%
With the fixed system load, the number of users to the number of antennas ratio $B=\dfrac{N}{M}$, we have the observations: a) Equation (\ref{Eq_82}) is determined by the MSE of channel estimation $\delta ^{2}$; b) $\mathrm{SINR}_{\mathrm{ZF}}^{k}$ is saturated with respect to $\upsilon_{k}\rightarrow \infty $.
\begin{figure}[t]
\centering
\includegraphics[width=4.5in,]{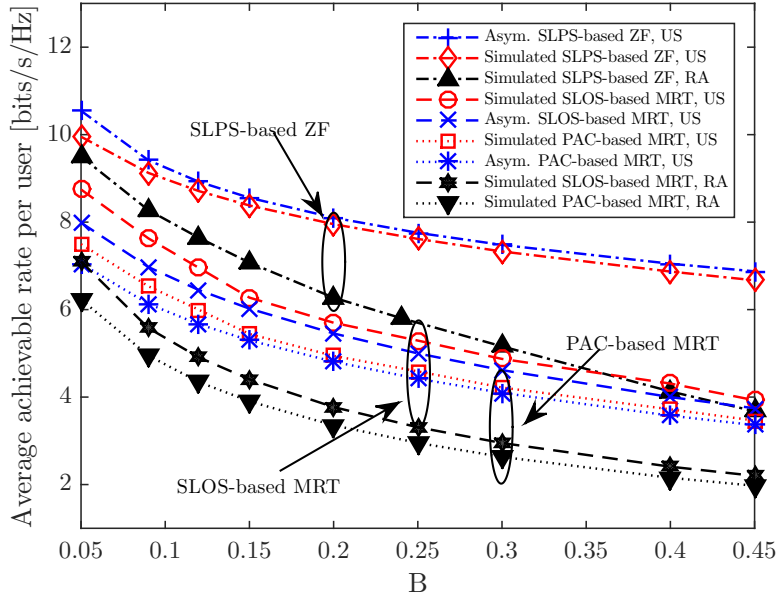}
\caption{Average rate per user [bits/s/Hz] based on deterministic channel versus different $B =\frac{N}{M}$ for $M=200$, the Rician K-factor $15$, $\mathrm{SNR}=40\text{\ }\mathrm{dB}$, and the normalized $\mathrm{MSE}=0.02$.}\label{Fig007}
\end{figure}

\begin{figure}[t]
\centering
\includegraphics[width=4.5in,]{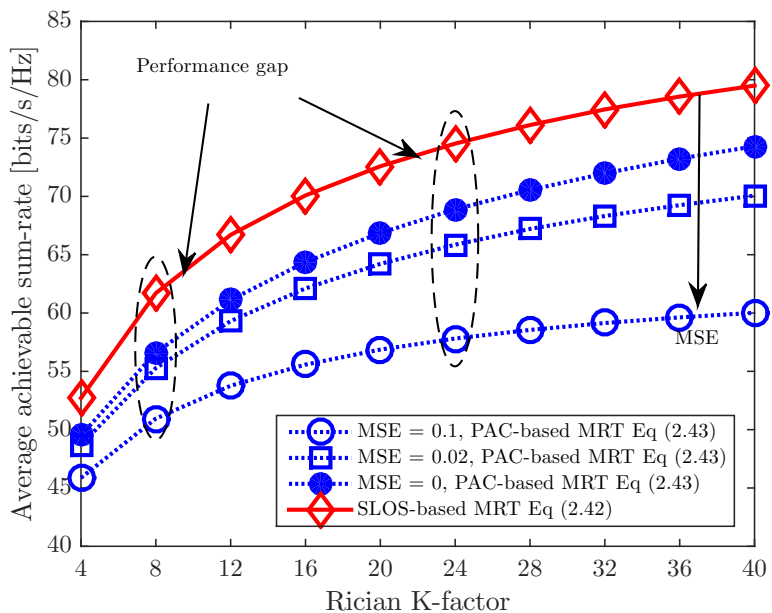}
\caption{Average achievable sum-rate [bits/s/Hz] versus Rician K-factor comparison with different $\mathrm{MSEs}$, $N=10$, and $M=100$.}\label{Fig011}
\end{figure}

In Figures \ref{Fig007} and \ref{Fig011}, we validate Theorem \ref{theo2} and the asymptotic analysis in Equations (\ref{E_61}), (\ref{E_71}), and (\ref{Eq_82}).
In Figure \ref{Fig007}, user scheduling (US) stands for that the AoAs of all the users' SLOS components satisfy the assumption in Equation (\ref{Eq_60}) and random angle (RA) stands for that the AoAs of all the users' SLOS components are uniformly distributed from $0^{\circ}$ to $180^{\circ}$.
Figure \ref{Fig007} shows that the proposed SLOS-based precoder outperforms the conventional imperfect PAC-based precoder for a wide range of scenarios.

In Figure \ref{Fig011}, interestingly, for a wide range of Rician K-factor, the proposed SLOS-based MRT can achieve a significant higher rate performance than that of the PAC-based MRT, even the pilot-aided estimated CSI is perfect.
As the MSE of channel estimation increases, the sum-rate gap between the SLOS-based MRT and the PAC-based MRT increases significantly.
The results shown in Figure \ref{Fig011} indicates that the SLOS-based MRT is a better choice than the conventional PAC-based MRT for mmWave massive MIMO systems.
It is interesting to note that, while the assumption in Equation (\ref{Eq_60}) is satisfied, the achievable rate performance under the assumption of US can significantly outperforms the achievable rate performance under RA assumption.
These simulation results imply that, an appropriate US algorithm selecting a set of users satisfying Equation (\ref{Eq_60}) can improve the achievable rate performance for mmWave massive MIMO communication systems.
Therefore, Equation (\ref{Eq_60}) can be considered as a guideline of US.
\vspace{-0mm}

\section{Impact of Imperfect Hardware: Phase Quantization Error}
Hardware impairments in massive MIMO systems causing performance degradation have been widely discussed \cite{Bjornson2014d,Zhang2015a}.
The hardware structure is widely adopted as a fully digital conventional MIMO scheme that $M$ RF chains with $M$ phase shifters are connected to $M$ antennas.
In particular, most of the works in the literature modeled that the impact of phase and amplitude errors are additive distortions and hardware imperfections are uncorrelated with the desired signals.
Some works, e.g. \cite{Bjornson2014d}, proved that the additive distortion created by hardware impairments at the BS will not vanish even the number of antennas goes large.
In this section, we analyze the impact of phase quantization errors on the average achievable rate performance in the high SNR and large number of antennas regime.
The large-system analysis results provide important design guidelines for mmWave massive MIMO systems.
{\color{black}Amplitude quantization errors are critical important for the performance of MIMO systems. However, amplitude quantization errors, which are partially contributed by the limited resolution of ADC/DAC and hardware imperfection attenuation, are nonlinear and very hard to model. Currently, to our best knowledge, no existing mathematical model is precise enough for capturing the characteristics of amplitude quantization errors. Thus, we leave this amplitude quantization errors problem as a future work.}

Transceiver RF chains need to provide accurate phase information for different antennas at the BS.
Yet, digital/analog RF chains have limited resolutions in practical systems.
Thus, phase quantization error may cause strong impacts on mmWave massive MIMO systems.
For example, if the digital phase shifters of each RF chain can only provide a $2$-bit of resolution for quantization, the quantized phase responses are $\left[0,90,180,270\right]$ degrees.
If the exact phase responses in different antennas are $\left[ 82,345,248,151,53\right] $ degrees, the obtained phase responses after quantization in different antennas will be $\left[ 90,0,270,180,90\right]$.
The differences between the actual and the quantized phase responses may lead to serious phase mismatch in transmission.

We now discuss the rate performance degradation of the SLOS-based MRT precoding due to the phase quantization error.
Then we compare the obtained result to the conventional pilot-aided CSI based MRT.
We derive the relationship between the quantization resolutions of phase shifters and the associated rate performance degradation.
We first denote the equivalent channel of user $k$ after phase quantization as\vspace{-2mm}\footnote{We note that the phase quantization error cannot be modeled as an additive distortion \cite{CLin2017,Ying2015}.}
\begin{equation}
\widehat{\mathbf{h}}_{\mathrm{L},{k}}=\mathbf{\Lambda }_{\Delta k}\mathbf{h}_{\mathrm{L,}{k}},\vspace{-2mm}
\end{equation}%
where the quantization error diagonal matrix is expressed as \vspace{-2mm}\cite{CLin2017,Ying2015}%
\begin{equation}
\mathbf{\Lambda }_{\Delta_k}=\mathrm{diag}\left\{ \Delta \mathbf{\phi}_{k}\right\}.\vspace{-2mm}
\end{equation}%
The phase quantization error of user $k$ is $\Delta \mathbf{\phi}_{k} \in \mathbb{C}^{M\times 1} =\left[
\begin{array}{ccc}
e^{j\overline{{\phi}}_{k,1}},\ldots, e^{j\overline{{\phi}}_{k,M}}%
\end{array}%
\right]^{T}$, where the exact phase quantization error $\overline{{\phi}}_{k,i}$ on the $i$-th phase shifter $( \forall i \in [1, M])$ is uniformly distributed over $\left[-a,+a\right]$, and $a$ is the maximal phase quantization error \cite{Adhikary2013}.
The exact uniformly distributed phase quantization errors on different phase shifters have following property that
\begin{align}
\mathrm{\mathbb{E}}_{\Delta \mathbf{\phi} _{k}}\left[ \mathbf{\Lambda }_{\Delta k}\right]&=\left[\begin{array}{ccc}
\underset{-a}{\overset{a}{\int }}\dfrac{1}{2a}e^{j\Delta \phi _{k,1}}d\Delta \phi _{k,1} & \cdots & {0} \\
\vdots & \ddots  & \vdots \\
{0} & \cdots & \underset{-a}{\overset{a}{%
\int }}\dfrac{1}{2a}e^{j\Delta \phi _{k,M}}d\Delta \phi _{k,M}%
\end{array}\right]\notag\\
&=\frac{\sin
\left( a\right) }{a}\mathbf{I}.\vspace{-0mm}
\end{align}

Then the received signal of user $k$ under limited phase
quantization resolutions mismatch can be expressed as\vspace{-2mm}%
\begin{align}
\widehat{y}_{\mathrm{LMRT}}^{k}
&=\underset{\mathrm{desired}\text{ }\mathrm{signal}}{\underbrace{
\widetilde{\beta }\sqrt{\frac{\upsilon _{k}}{\upsilon _{k}+1}}\mathbf{h}_{%
\mathrm{L,}k}^{H}\mathbf{\Lambda }_{\Delta k}\mathbf{h}_{\mathrm{L,}{k}%
}x_{k}}}+\underset{\mathrm{inter-user}\text{ }\mathrm{interference}}{\underbrace{\widetilde{\beta }\sqrt{\frac{\upsilon _{k}}{\upsilon _{k}+1}}%
\overset{N}{\underset{j=1,j\neq k}{\sum }}\mathbf{h}_{\mathrm{L,}{j}}^{H}%
\mathbf{\Lambda }_{\Delta j}\mathbf{h}_{\mathrm{L,}{k}}x_{j}}}  \notag \\
&+\underset{\mathrm{scattering}\text{ }\mathrm{interference}}{\underbrace{\widetilde{\beta }\sqrt{\frac{1}{\upsilon _{k}+1}}\overset{N}{\underset{%
j=1}{\sum }}\mathbf{h}_{\mathrm{L,}{j}}^{H}\mathbf{\Lambda }_{\Delta j}%
\mathbf{h}_{\mathrm{S,}k}x_{j}}}+z_{k}.\vspace{-0mm}
\end{align}%
We denote the received signal power of user $k$ as\vspace{-2mm}
\begin{equation}
\mathrm{S}_{k}=\dfrac{\upsilon _{k}\widetilde{\beta } ^{2}E_{s}}{\upsilon _{k}+1}\mathbf{h%
}_{\mathrm{L,}k}^{H}\mathrm{\mathbb{E}}_{\Delta \mathbf{\phi} _{k}}\left[\mathbf{\Lambda }_{\Delta k}\right]\mathbf{h}_{\mathrm{L,}{k%
}}\mathbf{h}_{\mathrm{L,}k}^{H}\mathrm{\mathbb{E}}_{\Delta \mathbf{\phi} _{k}}\left[\mathbf{\Lambda }_{\Delta k}^{H}\right]\mathbf{h}_{\mathrm{L,}{k}}.  \label{E_33}\vspace*{-1mm}
\end{equation}%
Then we can approximate the inter-user interference by ignoring small random part as:\vspace{-2mm}
\begin{align}
\mathrm{I}_{\mathrm{LMRT}}^{k} \approx & \dfrac{\widetilde{\beta } ^{2}E_{s}MN}{\upsilon _{k}+1}\notag\\
&+\dfrac{\upsilon _{k}\widetilde{\beta } ^{2}E_{s}}{\upsilon _{k}+1}\overset{N}{\underset%
{j=1,j\neq k}{\sum }}\mathbf{h}_{\mathrm{L,}{k}}^{H}\mathrm{\mathbb{E}}_{\Delta \phi _{j}} \left[\mathbf{%
\Lambda }_{\Delta j}\right] \mathbf{h}_{\mathrm{L,}j}\mathbf{h}_{\mathrm{%
L,}j}^{H}\mathrm{\mathbb{E}}_{\Delta \phi _{j}}\left[\mathbf{\Lambda }_{\Delta j}^{H} \right]\mathbf{h}_{%
\mathrm{L,}k}.\vspace{-2mm}
\end{align}
For different AoAs, the phase quantization errors are different.
We then present the approximated average SINR expression in the high SNR regime as\vspace{-2mm}%
\begin{align}%
\widehat{\mathrm{SINR}}_{\mathrm{LMRT}}^{k}\overset{\mathrm{SNR}\rightarrow \infty }{\approx }&\mathrm{\mathbb{E}}\left[ \frac{\mathrm{S}_{k}}{\mathrm{I}_{\mathrm{LMRT}}^{k}}\right]
\notag\\
&\hspace{-8mm}{\approx }\frac{\dfrac{\upsilon _{k}\beta ^{2}E_{s}}{%
\upsilon _{k}+1}\mathbf{h}_{\mathrm{L,}k}^{H}\mathrm{\mathbb{E}}_{\Delta \phi _{k}}\left[ \mathbf{\Lambda }%
_{\Delta k}\right] \mathbf{h}_{\mathrm{L,}{k}}\mathbf{h}_{\mathrm{L,}%
k}^{H}\mathrm{\mathbb{E}}_{\Delta \phi _{k}}\left[ \mathbf{\Lambda }_{\Delta k}^{H}\right] \mathbf{h}_{\mathrm{%
L,}k}}{\overset{N}{\underset{j=1,j\neq k}{\sum }}\mathbf{h}_{\mathrm{L,}{%
j}}^{H}\mathrm{\mathbb{E}}_{\Delta \phi _{j}}\left[ \mathbf{\Lambda }_{\Delta j}\right] \mathbf{h}_{\mathrm{L,}%
k}\mathbf{h}_{\mathrm{L,}k}^{H}\mathrm{\mathbb{E}}_{\Delta \phi _{j}}\left[ \mathbf{\Lambda }_{\Delta j}^{H}%
\right] \mathbf{h}_{\mathrm{L,}j}+\dfrac{MN}{\upsilon _{k}}}.
\label{E_331}\vspace{-2mm}
\end{align}%
By exploiting the property that\vspace{-1mm}
\begin{align}
\mathrm{\mathbb{E}}_{\Delta \phi _{k}}\left[ \mathbf{\Lambda }_{\Delta k,n}\right] &=\underset{-a}{\overset{a}{%
\int }}\dfrac{1}{2a}e^{j\Delta \phi _{k}}d\Delta \phi _{k}=\frac{\sin
\left( a\right) }{a}
=\mathrm{\mathbb{E}}_{\Delta \phi _{k}}\left[ \mathbf{\Lambda }_{\Delta
k,n}^{H}\right] ,\vspace{-1mm}
\end{align}%
and the SINR expression (\ref{E_331}) can be rewritten as\vspace{-2mm}%
\begin{align}
&\widehat{\mathrm{SINR}}_{\mathrm{LMRT}}^{k}\overset{\mathrm{SNR}%
\rightarrow \infty }{\approx }\notag\\
&\frac{\left[ M\sinc\left( a\right) \right] ^{2}}{\dfrac{MN}{\upsilon _{k}}+\overset{N}{\underset{j=1,j\neq k}{\sum }}\mathbf{h}_{\mathrm{L,}{%
j}}^{H}\mathrm{\mathbb{E}}_{\Delta \phi _{j}} \left[\mathbf{\Lambda }_{\Delta j} \right]\mathbf{h}_{\mathrm{L,}%
k}\mathbf{h}_{\mathrm{L,}k}^{H}\mathrm{\mathbb{E}}_{\Delta \phi _{j}}\left[ \mathbf{\Lambda }_{\Delta j}^{H}\right]\mathbf{h}_{\mathrm{L,}j}} .\vspace{-2mm}
\end{align}%
For a special case of $a=0,$ the sum of sinc function is $\overset{M}{\underset{n=1}{\sum }}\dfrac{\sin \left( a\right) }{a}=M.$
In the considered system, as we assume that all users are separated by hundreds of wavelength and the AoAs angular separation condition in (\ref{Eq_60}) is satisfied, the inter-user interferences are from the side lobes.
For $M\rightarrow \infty,$ $N\rightarrow \infty $, and $\dfrac{N}{M}$ is fixed, we have the following asymptotic SINR expression for the SLOS-based MRT in the high SNR regime:\vspace{-1mm}
\begin{equation}
\widehat{\mathrm{SINR}}_{\mathrm{SLOS-MRT}}^{k}\overset{M,N\rightarrow \infty
}{\underset{\mathrm{SNR}\rightarrow \infty }{\approx }}\dfrac{\left[ \sinc^{2}%
\left( a\right) \right]}{\dfrac{1}{\upsilon _{k}}\dfrac{N}{M}+\dfrac{1}{%
4}\left( \dfrac{N}{M}\right) ^{2}}.  \label{Eq_95}\vspace{-1mm}
\end{equation}%
For $M\rightarrow \infty ,$ $N\rightarrow \infty $, and $\dfrac{N}{M}$ is fixed, we have the following asymptotic SINR expression for the PAC-based MRT in the high SNR regime:\vspace{-0mm}%
\begin{equation}
\widehat{\mathrm{SINR}}_{\mathrm{PAC-MRT}}^{k}\overset{M,N\rightarrow \infty }%
{\underset{\mathrm{SNR}\rightarrow \infty }{\approx }}\frac{\left[ \sinc^{2}%
\left( a\right) \right]}{\dfrac{1}{4}\left( \dfrac{\upsilon _{k}}{%
\upsilon _{k}+1}\right) ^{2}\left( \dfrac{N}{M}\right) ^{2}+\dfrac{N}{M}%
\dfrac{\left( 2\upsilon _{k}+1\right) }{\left( \upsilon _{k}+1\right) ^{2}}+%
\dfrac{N}{M}\delta ^{2}}.  \label{Eq_96}\vspace{-0mm}
\end{equation}%
From Equations (\ref{Eq_95}) and (\ref{Eq_96}), we observe that the SINR expressions are affected by the range of phase quantization error $a$.
With the fixed quantization error, the performance degradation may not vanish, even if more antennas are adopted for transmission.

By comparing Equations (\ref{Eq_95}) to (\ref{Eq_96}), the SLOS-based MRT still achieves a better rate performance than that of the PAC-based MRT, despite the presence of phase quantization errors.

Accordingly, the considered mmWave massive MIMO system requires the support of high phase quantization resolutions RF hardware in mmWave channels.
\begin{rema}
In order to maintain the system performance (no more than a $3$-dB of SINR degradation compared to ideal hardware), the quantization resolutions of phase shifters in each RF chain should satisfy:\vspace{-0mm}
\begin{equation}
\left( \overset{M}{\underset{n=1}{\sum }}\dfrac{\sin \left( a\right) }{a}\right) ^{2}\geqslant \dfrac{M^{2}}{2}.
\end{equation}
\end{rema}\vspace{-0mm}
\begin{figure}[t]
\centering
\includegraphics[width=4.5in,]{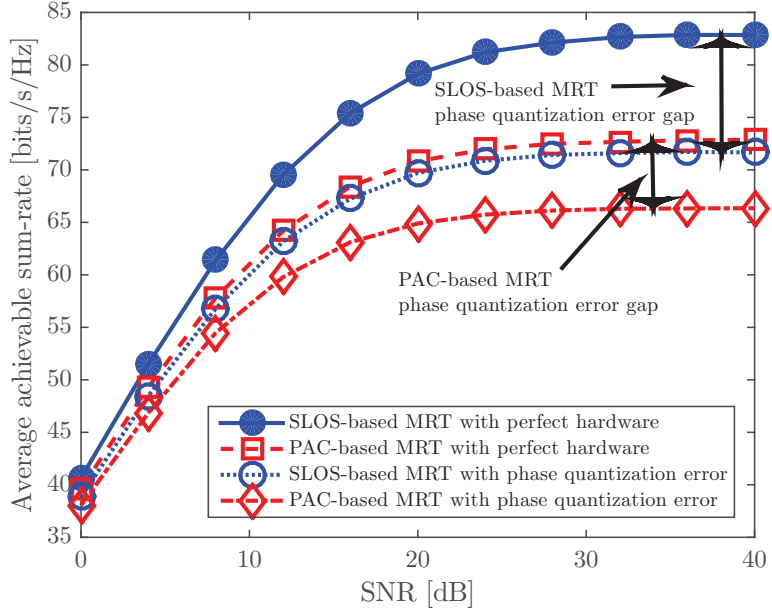}
\caption{Average achievable sum-rate [bits/s/Hz] under phase
quantization error versus SNR [dB], for $N=10$, $M=100$, the normalized $\mathrm{MSE}%
=0.02$ and Rician K-factor of $50$.}\label{Fig008}
\end{figure}
Based on Equations (\ref{Eq_95}) and (\ref{E_61}), we can obtain the rate performance degradation of the SLOS-based MRT under different quantization resolutions in the following simulation.
In Figures \ref{Fig008} and \ref{Fig009}, we illustrate the impact of phase quantization error on system rate performance.
We assume that the resolutions of the phase quantization per RF chain is $2$ bits. In Figure \ref{Fig008}, we show the performance degradation for the SLOS-based MRT and the PAC-based MRT for $M=100,$ $N=10$.
In the high SNR regime, the system rate performance degradation due to the limited phase quantization resolutions is significant.
However, the SLOS-based MRT can still achieve a better sum rate, which verifies the correctness of our derived results in Equations (\ref{Eq_95}) and (\ref{Eq_96}).

In Figure \ref{Fig009}, we demonstrate the derived results for different system loads. In particular, we observe that the impact of phase quantization errors will not vanish with the increase of the number of antennas.
The rate performance degradation caused by phase quantization errors is $2.4$ bits/s/Hz per user, which is close to the predicted value ($2.98$ bits/s/Hz per user) via Equation (\ref{Eq_95}).
It indicates that the hardware requirement of mmWave massive MIMO system is very stringent and should be taken into account in system design.

\begin{figure}[t]
\centering
\includegraphics[width=4.5in,]{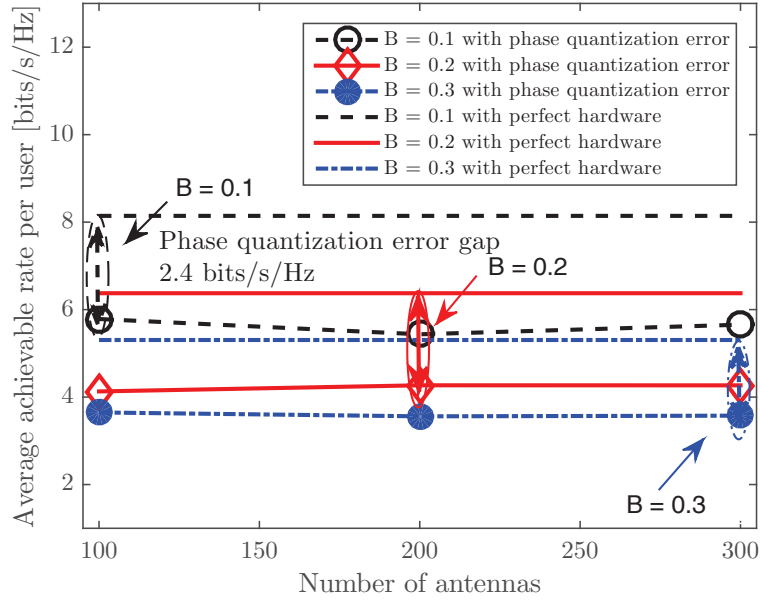}
\caption{The impact of $2$-bit
phase quantization resolutions on average achievable rate per user of SLOS-based MRT versus the number of antennas with the normalized $\mathrm{MSE}=0.02$ and Rician K-factor of $50$.}\label{Fig009}
\end{figure}

\section{Summary}
In this chapter, we first proposed a frequency tone-based AoAs channel estimation algorithm in mmWave massive MIMO systems and we analyzed the performance of the associated SLOS-based MRT and the SLPS-based ZF precoder.
We also derived the achievable rates of the PAC-based MRT and the PAC-based ZF precoding under the impact of pilot contamination.
In particular, we compared the achievable rates of the SLOS-based MRT and the SLPS-based ZF precoding to the achievable rates of the PAC-based MRT and the PAC-based ZF precoding, respectively.
We showed that, for the same MSE of channel estimation and large numbers of antennas and users, the achievable rates of the SLPS-based and the PAC-based ZF precoding are identical.
Besides, our proposed SLOS-based MRT precoder can achieve a higher rate than that of the traditional PAC-based MRT precoder.
Furthermore, we analyzed the impact of phase quantization error on system rate performance, revealing that such impact cannot be mitigated in mmWave massive MIMO systems by increasing the number of antennas.



\chapter{Hybrid mmWave MIMO Systems: {Multi-user Precoding and Channel Estimation}
}\label{C3:chapter3}

%
\section{Introduction}
{\color{black}In this chapter, we consider a TDD MU hybrid mmWave system.
In the previous chapter, we proposed a frequency tone-based channel estimation
algorithm for the estimation of strongest AoAs.
Besides, we proved that our proposed SLOS-based MRT precoder can achieve a higher rate performance than that of the traditional PAC-based MRT precoder.
Based on these prior information, we propose and detail a novel non-feedback non-iterative channel estimation algorithm which is applicable to both non-sparse and sparse mmWave channels.

The proposed three-step channel estimation algorithm is an extension of the tone-based AoA channel estimation algorithm proposed for fully digital mmWave MIMO in Chapter $2$.
Also, we analyze the achievable rate performance of the mmWave system using digital ZF precoding based on the estimated channel information.}
Furthermore, we analyze the performance degradation under some practical hardware imperfections, such as random phase errors, RF transceiver beamforming errors, and channel estimation errors.

Our main contributions are summarized as follows:

\begin{itemize}

\item We propose a three-step MU channel estimation scheme for mmWave channels.
In the first two steps, we estimate the strongest AoAs at both the BS and the users sides instead of estimating the combination of multiple AoAs.
The estimated strongest AoAs will be exploited for the design of analog beamforming matrices at the BS and users.
In the third step, all the users transmit orthogonal pilot symbols to the BS along the beamforming paths of the strongest AoA directions to facilitate the equivalent channel estimation, which will be exploited to design the digital ZF precoder at the BS for the downlink transmission.
Our proposed hybrid scheme can suppress the downlink MU interference effectively via its analog beamforming and digital precoder.
Firstly, the proposed analog beamforming allow signal transmission and reception along the strongest AoA direction, which reduces the interference outside the strongest AoA directions and utilizes the transmission power more efficiently.
Secondly, the digital ZF precoder can suppress the MU interference within the strongest AoA directions.

\item We analyze the achievable rate performance of the proposed scheme based on the estimated equivalent channel CSI, analog beamforming matrices, and digital ZF precoding.
While assuming the equivalent CSI is perfectly known at the BS, we derive a tight performance upper bound on the achievable rate of our proposed scheme.
Also, we quantify the performance gap between the proposed hybrid scheme and the fully digital system in terms of achievable rate per user.
It is interesting to note that the performance gap is determined by the ratio between the power of the strongest AoA component and the power of the scattering component, Rician K-factor $\upsilon$.
The performance gap of the average achievable rate per user between the hybrid system and the fully digital system is only $\left\vert\log_2 \left( \frac{\upsilon }{\upsilon +1}\right)\right\vert$ bits/s/Hz in the large numbers of antennas regime.

\item We further analyze the system performance degradation and derive the closed-form approximation of achievable rate under various types of errors, i.e., random phase errors, transceiver analog beamforming errors, and equivalent channel estimation errors, in the high receiver signal-to-noise ratio (SNR) and the large numbers of antennas regimes.
Interestingly, our results confirm that the impact of phase errors and transceiver beamforming errors will not cause a performance ceiling in terms of achievable rate.
Besides, the performance gap in terms of the achievable rate between the system under phase errors and transceiver beamforming errors and the system with perfect hardware is approximated.

\end{itemize}

The rest of the chapter is organized as follows. Section 3.2 describes the system
model considered throughout the chapter. In Section 3.3, we detail the proposed
channel estimation algorithm for hybrid systems. In Section 3.4, we investigate the downlink transmission rate performance based on the equivalent CSI estimated in Section 3.3. Besides, we further have rate performance analysis under the impact of hardware impairments in Section 3.5. Numerical results and related discussions are presented in Section 3.6. Finally, Section 3.7 summarizes the chapter.
\vspace*{-0mm}
\section{System Model}
\begin{figure}[h]\vspace{-0mm}
\centering
\includegraphics[width=4.5in]{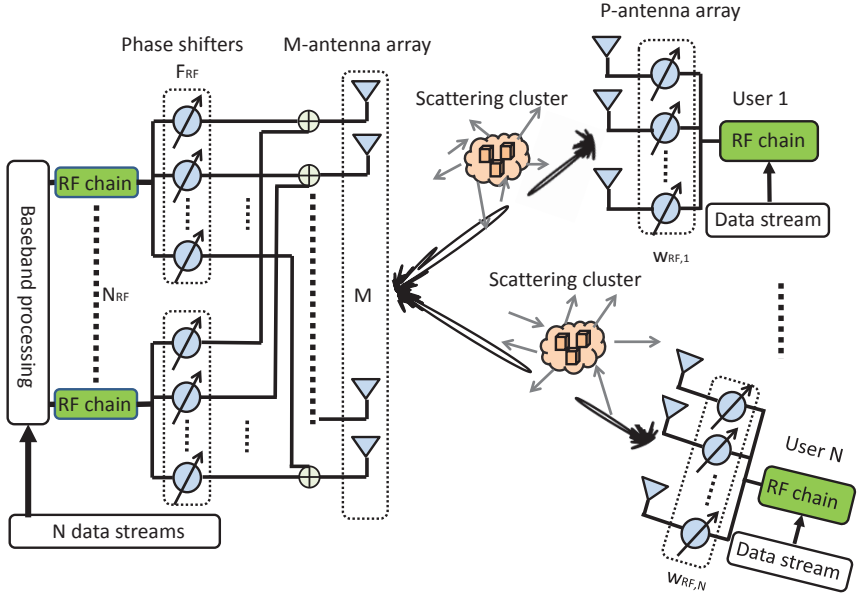}
\vspace{-0mm}
\caption{A mmWave massive MIMO communication system with a hybrid system of
transceivers.}
\label{fig:hybrid}\vspace{-0mm}
\end{figure}
We consider a MU hybrid mmWave system which consists of one BS and $N$ users in a single cell, as shown in Figure \ref{fig:hybrid}.
{Generally, there are two kind of hybrid structures which are widely adopted by researchers \cite{AZhang2015,Alkhateeb2015}: the full access hybrid architecture and the subarray hybrid architecture. The full access hybrid architecture, where each RF chain is connected to all the antennas, can provide a higher array gain and a narrower beam width than that of the subarray hybrid architecture, where each RF chain is connected to a part of the antennas. In this chapter, we adopt the full access hybrid architecture since it offers higher flexibility in the design of channel estimation algorithm.}
We assume that the BS is equipped with $M\geq 1$ antennas and $N_{\mathrm{RF}}$ RF chains to serve the $N$ users.
{Besides, each user is equipped with $P$ antennas and a single RF chain. We also assume that $M\geqslant N_{\mathrm{RF}}\geqslant N$.}
In the following sections, we set $N = N_{\mathrm{RF}}$ to simplify the analysis\footnote{We note that our proposed channel estimation scheme, precoding scheme, and analysis can be generalized to the case of $N_{\mathrm{RF}}\geq N$,  at the expense of a more involved notation.}.

\begin{figure}[t]
\centering\vspace*{-0mm}
\includegraphics[width=4.5in]{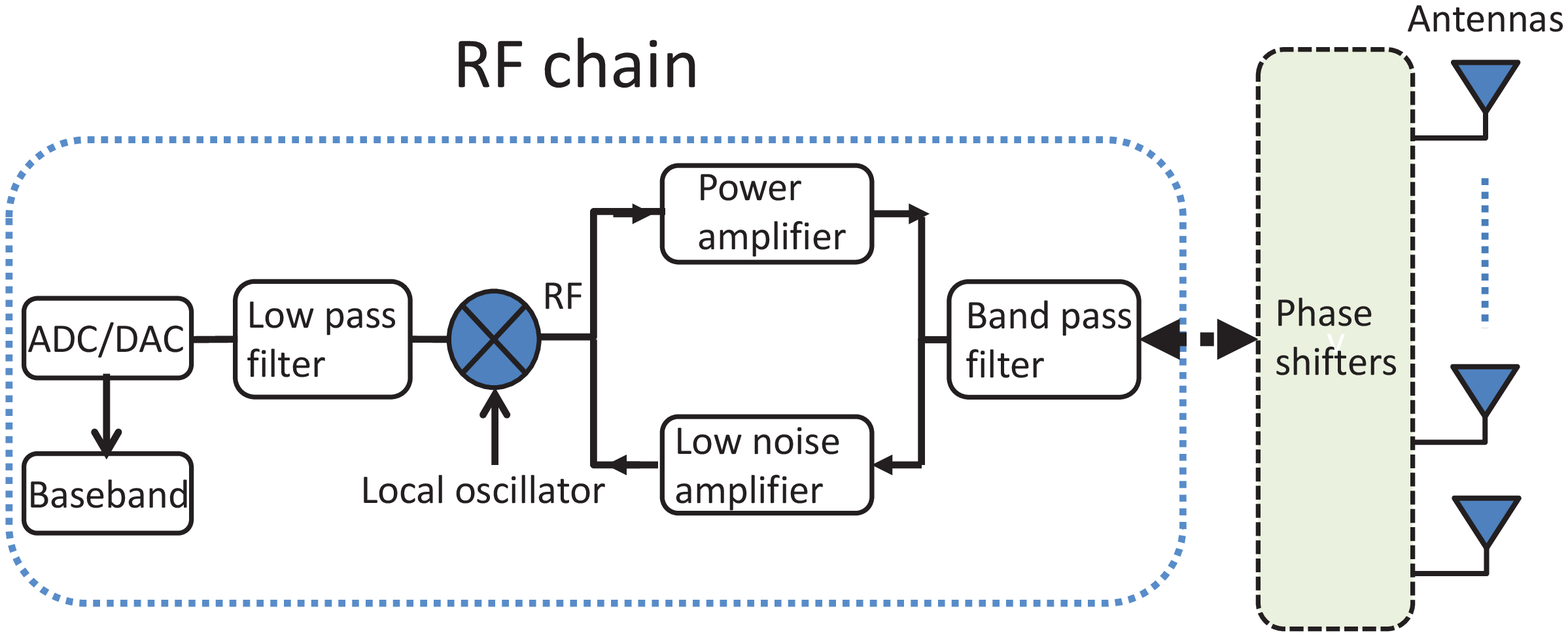}
\vspace{-0mm}
\caption{A block diagram of an RF chain for an antenna array.}
\label{fig:RF_chain2}\vspace{-0mm}
\end{figure}

Each RF chain at the BS can access to all the antennas by using $M$ phase shifters, as shown in Figure \ref{fig:RF_chain2}.
At each BS, the number of phase shifters is $M\times N_{\mathrm{RF}}$.
Due to significant propagation attenuation at mmWave frequency, the system is dedicated to cover a small area, e.g. cell radius is $\sim 150$ m.
We assume that the users and the BS are fully synchronized and TDD is adopted to facilitate uplink and downlink communications \cite{Marzetta2010}.

In previous work \cite{Alkhateeb2015}, mmWave channels were assumed to have sparse propagation paths between the BS and the users.
{Yet, in recent field tests, especially in the urban microcell environments, both a strong LOS component and non-negligible scattering components may exist in mmWave propagation channels \cite{Buzzi2016b,Rappaport2015,Hur2016}.
Therefore, for the urban short-distance propagation environment, mmWave channels are more suitable to be modeled by non-sparse Rician fading and with a large Rician K-factor \cite{Hur2016,Rappaport2015,Al-Daher2012}.
On the other hand, for the suburban long-distance propagation environment, mmWave channels can be modeled by a sparse channel model. The reason is that scattering components will vanish during the long-distance propagation because of the high reflection loss and large propagation path loss.}
{In addition, the blockage of LOS component is critical for mmWave systems and widely considered in previous works \cite{Bai2015b,Andrews2016}. In \cite{Bai2015b}, the authors proved that the optimal cell size to achieve the maximum SINR scales with the average size of the area that is LOS to a user. Therefore, with a properly designed radius of cells, one should expect the existence of at least one LOS component from the BSs to any user. When the LOS component from a certain BS to a user is blocked, the user can still exploit other existing LOS components from other BSs for channel estimation and data transmission.}

Let $\mathbf{H}_{k}\in\mathbb{C}^{M\times P}$ be the uplink channel matrix between user $k$ and the BS in the cell.
We assume that $\mathbf{H}_{k}$ is a slow time-varying block Rician fading channel, i.e., the channel is constant in a block but varies slowly from one block to another.
Then, in this chapter, we assume that the channel matrix $\mathbf{H}_k$ can be decomposed into a deterministic LOS channel matrix $\mathbf{H}_{\mathrm{L},k}\in\mathbb{C}^{M\times P}$ and a scattered channel matrix $\mathbf{H}_{\mathrm{S,}k}\in\mathbb{C}^{M\times P}$ \cite{Buzzi2016b}, i.e., \vspace*{-1mm}
\begin{equation}\label{eqn:LOS_channel31}
\vspace*{-0mm}
\mathbf{H}_{k}=\underset{\mathrm{LOS}\text{\ }\mathrm{component}}{\underbrace{\mathbf{H}_{\mathrm{L,}k}\mathbf{G}_{\mathrm{L,}k}}}+\underset{\mathrm{Scattering}\text{\ }\mathrm{component}}{\underbrace{\mathbf{H}_{\mathrm{S,}k}\mathbf{G}_{\mathrm{S,}k}}},\vspace*{-1mm}
\end{equation}\vspace*{-0mm}%
where $\mathbf{G}_{\mathrm{L,}k}\in\mathbb{C}^{P\times P}$ and $\mathbf{G}_{\mathrm{S},k}\in\mathbb{C}^{P\times P}$ are diagonal matrices with entries\vspace*{-1mm}
\begin{equation}
\vspace*{-0mm}
\mathbf{G}_{\mathrm{L,}k}=\mathrm{diag}\left\{ \sqrt{\frac{\upsilon _{k}}{\upsilon _{k}+1}}\right\} \text{ and } \mathbf{G}_{\mathrm{S,}k}=\mathrm{diag}\left\{ \sqrt{\frac{1}{\upsilon _{k}+1}}\right\},\label{Eq32}\vspace*{-1mm}
\end{equation}%
respectively, and $\upsilon _{k}>0$ is the Rician K-factor of user $k$.
{{Besides, Equations (\ref{eqn:LOS_channel31}) and (\ref{Eq32}) are the generalization of mmWave channel models, which capture both the scattered and non-scattered components.}
In general, we can adopt different array structures, e.g. ULA and uniform panel array (UPA) for both the BS and the users.
Here, we adopt the ULA as it is commonly implemented in practice \cite{Alkhateeb2015}.
We assume that all the users are separated by hundreds of wavelengths or more \cite{Marzetta2010}.
Thus, we can express the deterministic LOS channel matrix $\mathbf{H}_{\mathrm{L},k}$ of user $k$ as \cite{book:wireless_comm}\vspace*{-1mm}
\begin{equation}
\mathbf{H}_{\mathrm{L,}k}=\mathbf{h}_{\mathrm{L,}k}^{\mathrm{BS}}\mathbf{h}_{%
\mathrm{L,}k}^{H},\vspace*{-1mm}
\end{equation}\vspace*{-0mm}%
where $\mathbf{h}_{\mathrm{L},k}^{\mathrm{BS}}$ $\in\mathbb{C}^{M\times 1}$ and $\mathbf{h}_{\mathrm{L,}k}$ $\in\mathbb{C}^{P\times 1}$ are the antenna array response vectors of the BS and user $k$ respectively.

In particular, $\mathbf{h}_{\mathrm{L,}k}^{\mathrm{BS}}$ and $\mathbf{h}_{\mathrm{L,}k}$ can be expressed as \cite{book:wireless_comm,Trees2002}\vspace*{-0mm}
\begin{align}
\vspace*{-0mm}
\mathbf{h}_{\mathrm{L},k}^{\mathrm{BS}}& =\left[
\begin{array}{ccc}
1, & \ldots
, & \text{ }e^{-j2\pi \left( M-1\right) \tfrac{d}{\lambda }\cos \left(
\theta _{k}\right) }%
\end{array}%
\right] ^{T} \text{and} \\
\mathbf{h}_{\mathrm{L},k}& =\left[
\begin{array}{ccc}
1, & \ldots ,
& \text{ }e^{-j2\pi \left( M-1\right) \tfrac{d}{\lambda }\cos \left( \phi
_{k}\right) }%
\end{array}%
\right] ^{T},\vspace*{-0mm}
\end{align}\vspace*{-0mm}
respectively, where $d$ is the distance between the neighboring antennas and $\lambda $ is the wavelength of the carrier frequency.
Variables $\theta _{k}\in \left[ 0,+\pi \right]$ and $\phi _{k}\in \left[ 0,+\pi \right] $ are the angles of incidence of the LOS path at antenna arrays of the BS and user $k$, respectively.
For convenience, we set $d=\dfrac{\lambda }{2}$ for the rest of the chapter which is an assumption commonly adopted in the literature \cite{Trees2002,book:wireless_comm}.
Without loss of generality, we assume that the scattering component $\mathbf{H}_{\mathrm{S,}k}$ consists $N_{\mathrm{cl}}$ clusters and each cluster contributes $N_{\mathrm{l},i}$ propagation paths \cite{Buzzi2016b}, which can be expressed as\vspace*{-1mm}
\begin{equation}
\vspace*{-0mm}
\mathbf{H}_{\mathrm{S,}k}=\sqrt{\tfrac{1}{{\sum }_{i=1}^{N_{\mathrm{cl}}}{N_{\mathrm{l},i}}}}\overset{N_{\mathrm{cl}}}{\underset{i=1}{\sum }}\overset{N_{\mathrm{l},i}}{\underset{l=1}{\sum }}{\alpha _{i,l}}\mathbf{h}_{i,l}^{\mathrm{BS}}\mathbf{h}_{k,i,l}^{H}= \left[  \begin{array}{ccccc} \mathbf{h}_{\mathrm{S},1}, &\ldots,& \mathbf{h}_{\mathrm{S},k}, &\ldots,& \mathbf{h}_{\mathrm{S},P} \end{array}\right],\vspace*{-1mm}
\end{equation}\vspace*{-0mm}%
where $\mathbf{h}_{i,l}^{\mathrm{BS}}\in\mathbb{C}^{M\times 1}$ and $\mathbf{h}_{k,i,l}\in\mathbb{C}^{P\times 1}$ are the antenna array response vectors of the BS and user $k$ associated to the $\left( i,l\right) $-th propagation path, respectively.
Here $\alpha _{i,l}\sim \mathcal{CN}\left( 0,1\right) $ represents the path attenuation of the $\left(i,l\right)$-th propagation path and $\mathbf{h}_{\mathrm{S},k}\in\mathbb{C}^{M\times 1}$ is the $k$-th column vector of $\mathbf{H}_{\mathrm{S},k}$.
With the increasing number of clusters, the path attenuation coefficients and the AoAs between the users and the BS become randomly distributed \cite{Buzzi2016b,Hur2016}.
Therefore, we model the entries of scattering component $\mathbf{H}_{\mathrm{S,}k}$ in a general manner as an i.i.d. random variable\footnote{To facilitate the study of the downlink hybrid precoding, we assume that perfect long-term power control is performed to compensate for path loss and shadowing at the desired users and equal power allocation among different data streams of the users\cite{Alkhateeb2015,Ni2016,Yang2013}. Thus, the entries of scattering component $\mathbf{H}_{\mathrm{S,}k}$ are modeled by i.i.d. random variables. } $\mathcal{CN}\left( 0,1\right)$.
\vspace*{-0mm}
\section{Proposed Channel Estimation for Hybrid System}

In this section, we propose and detail our mmWave channel estimation for hybrid mmWave systems.
In practice, the hybrid system imposes a fundamental challenge for mmWave channel estimation.
Unfortunately, the conventional pilot-aided channel estimation algorithm for fully digital systems, e.g. \cite{Kokshoorn2016,Alkhateeb2015}, is not applicable to the considered hybrid mmWave system.
The reasons are that the number of RF chains is much lower than the number of antennas equipped at the BS and the transceiver beamforming matrix cannot be acquired.

To address this important issue, we propose a novel channel estimation algorithm, which contains three steps as shown in Figure \ref{fig:CEI} and Algorithm \ref{a1}.
In the first and second steps, we introduce unique unmodulated frequency tones to estimate the strongest AoAs at the BS and user sides. The unique frequency tones and linear search algorithm are inspired by signal processing in monopulse passive electronically scanned array (PESA) radar and sonar systems \cite{book:wireless_comm}.
{\color{black}The unmodulated frequency tone-based AoA beam sweeping search signal processing procedures are different while adopting different hardware architectures, e.g. hybrid architecture and fully digital architecture.}
These estimated strongest AoAs will be exploited to develop analog transmit and receive beamforming matrices at the BS and users.
In the third step, the users transmit orthogonal pilot symbols to the BS along the beamforming paths in order to estimate the equivalent channel via the strongest AoA directions.
Then, the estimated channel will be used for the design of BS digital baseband precoder for the downlink transmissions by exploiting the reciprocity between the uplink and downlink channels.
\begin{figure}[t]
\centering\vspace{-0mm}
\includegraphics[width=4.5in]{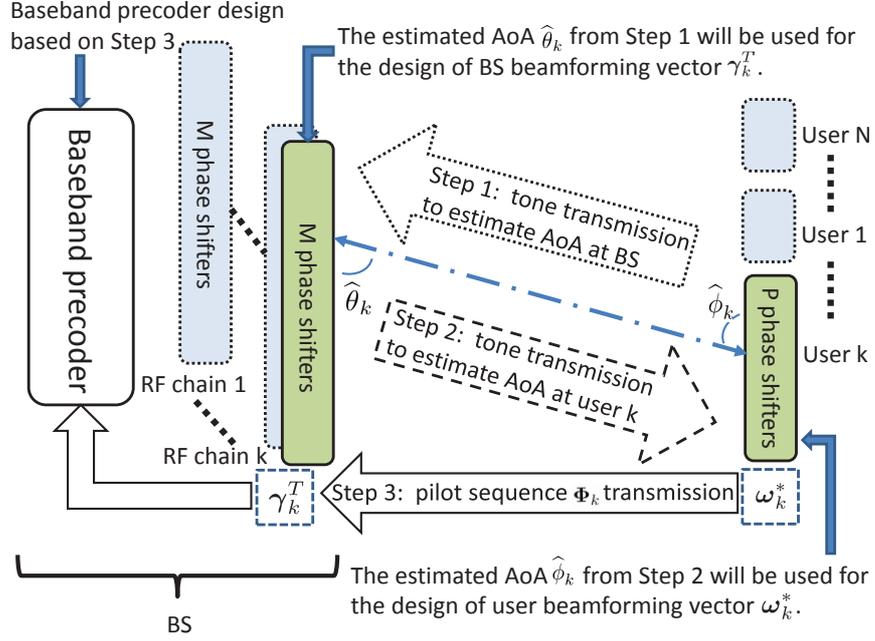}\vspace{-0mm}
\caption{An illustration of the proposed channel estimation algorithm for hybrid mmWave systems.}
\label{fig:CEI}\vspace*{-0mm}
\end{figure}
\begin{algorithm}
\caption{Channel Estimation Algorithm for Hybrid Systems}
\label{a1}
\begin{algorithmic} [1]
\REQUIRE Multiple single-carrier frequency tone signals $[f_1, \text{}\ldots, f_N]$, pilot sequences matrix $\mathbf{\Psi }$,
         the detection matrices $[\bm{\gamma}_1, \text{}\ldots, \bm{\gamma}_N] $ of AoA at the BS,
         and the detection matrices $[\mathbf{\Omega}_1, \text{}\ldots, \mathbf{\Omega}_N]$ of AoA at the users\\\vspace*{+2mm}
{{ STEP 1: \underline{Estimate the AoA at the BS and}}}\\
{{ \quad \quad \quad \quad \underline{design BS analog beamforming matrix}}}\vspace*{+1mm}
\STATE  The estimation of AoA at the BS: all the users transmit their unique frequency tones by using only one omni-directional antenna
\STATE  The BS calculates $r_{k,i}^{\mathrm{BS}}, \text{\ } i\in \{1,\ldots,J\}$, as shown in (\ref{AoA_D}) to estimate the uplink AoA of user $k$ and its corresponding beamforming vector:
{$\widetilde{\bm{\gamma}}_{k}=\underset{\forall \gamma_{k,i}, \text{\ }i\in \{1,\ldots,J\}}{\arg \max \left\vert r_{k,i}^{\mathrm{BS}}\right\vert }$}
\STATE The BS analog beamforming matrix is \\
$\mathbf{F}_{\mathrm{RF}}=\left[\begin{array}{ccc}\widetilde{\bm{\gamma}}_{1},\ldots, \widetilde{\bm{\gamma}}_{N}\end{array}\right]$\\\vspace*{+2mm}
{{ STEP 2: \underline{Estimate the AoA at the users and}}}\\
{{\quad \quad \quad \quad \underline{design user analog beamforming matrix}}}\vspace*{+1mm}
\STATE  The estimation of AoA at the users: the BS transmits frequency tones back to all the users using $\mathbf{F}_{\mathrm{RF}}$ as a transmit beamforming matrix
\STATE  Calculate $r _{k,i}^{\mathrm{UE}},\text{\ } i\in \{1,\ldots,J\}$, as shown in (\ref{AoA_D2}) to estimate the downlink AoA of user $k$ and its corresponding beamforming vector: $\widetilde{\bm{\omega }}_{k}^{\ast }=\underset{\forall \bm{\omega }_{k,i}, \text{\ }i\in \{1,\ldots,J\}}{\arg \max \left\vert \gamma
_{k,i}^{\mathrm{UE}}\right\vert }$
\STATE The users analog beamforming matrix is \\
$\mathbf{Q}_{\mathrm{RF}}=\left[
\begin{array}{ccc}
\widetilde{\bm{\omega }}_{1}^{\ast}, \ldots, \widetilde{\mathbf{\omega}}_{N}^{\ast }\end{array}\right] $\\ \vspace*{+2mm}
{{STEP 3: \underline{Estimate equivalent channel and}}}\\
{{ \quad \quad \quad \quad \underline{design digital ZF precoder}}}\vspace*{+1mm}
\STATE All the users transmit orthogonal pilot sequences by using $\mathbf{Q}_{\mathrm{RF}}$ as beamforming matrix and the BS uses $\mathbf{F}^{T}_{\mathrm{RF}}$ as beamforming matrix to receive pilot sequences
\STATE The BS obtains and calculates $\widehat{\mathbf{H}}_{\mathrm{eq}}^{T}$ as shown in (\ref{EHC_1}) $\widehat{\mathbf{H}}_{\mathrm{eq}}^{T}=\mathbf{\Psi }^{H}\left[\begin{array}{ccccc}\mathbf{s}_{1},\text{}\ldots ,\mathbf{s}_{N}\end{array}\right]$\\
\STATE The BS sets the baseband digital ZF precoder as \\ $\overline{\mathbf{W}}_{\mathrm{eq}}=\widehat{\mathbf{H}}_{\mathrm{eq}}^{\ast }(\widehat{\mathbf{H}}_{\mathrm{eq}}^{T}\widehat{\mathbf{H}}_{\mathrm{eq}}^{\ast })^{-1}$
\end{algorithmic}
\end{algorithm}

\subsection{Details of Proposed Channel Estimation}

\quad\emph{Step $1$}, \underline{Line 1 in Algorithm \ref{a1}:} Firstly, all the users transmit unique frequency tones to the desired BS in the uplink simultaneously.
For user $k$, an unique unmodulated frequency tone, $x_{k}=\cos \left( 2\pi f_{k}t\right), k\in \{1,\cdots ,N\}$, is transmitted from one of the omni-directional antennas in the antenna array to the BS.
Here, $f_{k}$ is the single carrier frequency and $t$ stands for time and $f_{k}\neq f_{j},  \forall k\neq j$.
For the AoA estimation, if the condition $\dfrac{f_{k}-f_{j}}{f_\mathrm{c}} < 10^{-4}, \text{ }\forall k\neq j$, is satisfied, the AoA estimation difference by using different tones is generally negligible \cite{Trees2002}, where $f_\mathrm{c}$ is the system carrier frequency.
The pass-band received signal of user $k$ at the BS, $\mathbf{y}_{k}^{\mathrm{BS}}$, is given by\vspace{-2mm}
\begin{equation}
\vspace*{-1mm}
\mathbf{y}_{k}^{\mathrm{BS}}=\left( \sqrt{\dfrac{\upsilon _{k}}{%
\upsilon _{k}+1}}\mathbf{h}_{\mathrm{L},k}^{\mathrm{BS}}+\sqrt{\dfrac{1}{%
\upsilon _{k}+1}}\mathbf{h}_{\mathrm{S},k}\right) x_{k}+\mathbf{z}_{\mathrm{%
BS}}, \vspace{-2mm}
\end{equation}\vspace*{-0mm}%
where $\mathbf{z}_{\mathrm{BS}}$ denotes the thermal noise at the antenna array of the BS, $\mathbf{z}_{\mathrm{BS}}\sim\mathcal{CN}\left( \mathbf{0},{\sigma_{\mathrm{BS}}^{2}}\mathbf{I} \right)$, and ${\sigma} _{\mathrm{BS}}^{2}$ is the noise variance at each antenna of the BS.
To facilitate the estimation of AoA, we perform a linear search in the angular domain ranged from $0^{\circ}$ to $180^{\circ}$ with an angle search step size of $\dfrac{180}{J}$.
{Therefore, the AoA detection matrix $\mathbf{\Gamma }_{k}\in\mathbb{C}^{M\times J}$, $\mathbf{\Gamma }_{k}= \left[ \begin{array}{ccc} \bm{\gamma }_{k,1},\ldots,\bm{\gamma }_{k,J} \end{array}\right]$, contains $J$ column vectors. In general, the typical value of the minimum search steps $J$ depends on the number of antennas $M$ used for the AoA search. In particular, $J\approx\frac{2M}{1.782}$ \cite{Trees2002}.}
The $i$-th vector $\bm{\gamma }_{k,i}\in\mathbb{C}^{M\times 1}, i\in \{1,\cdots,J\}$, stands for a potential AoA of user $k$ at the BS and is given by\vspace{-3mm}
\begin{equation}
\vspace*{-0mm}
\bm{\gamma}_{k,i}=
\frac{1}{\sqrt{M}}\left[
\begin{array}{ccc}
1, & \ldots , & \text{ }e^{j2\pi \left( M-1\right) \tfrac{d}{\lambda }\cos
\left( \widehat{\theta }_{i}\right) }
\end{array}
\right] ^{T},\vspace{-2mm}
\end{equation}\vspace*{-0mm}%
where $\widehat{\theta }_{i}= \left(i-1\right)\frac{180}{J}, i\in \{1,\cdots,J\}$, is the assumed AoA and $\bm{\gamma }_{k,i}^{H}\bm{\gamma }_{k,i}=1$.
For the AoA estimation of user $k$, $\mathbf{\Gamma }_{k}$ is implemented in the $M$ phase shifters connected by the $k$-th RF chain.
The local oscillator (LO) of the $k$-th RF chain at the BS generates the same carrier frequency $f_{k}$ to down convert the received signals to the baseband, as shown in Figure \ref{fig:RF_chain2}.
After the down-conversion, the signals will be filtered by a low-pass filter which can remove other frequency tones.
The equivalent received signal at the BS from user $k$ at the $i$-th potential AoA is given by\vspace{-1mm}
\begin{equation}
\vspace*{-0mm}
r_{k,i}^{\mathrm{BS}}= \sqrt{\dfrac{\upsilon _{k}}{\upsilon
_{k}+1}}\bm{\gamma }_{k,i}^{T}\mathbf{h}_{\mathrm{L},k}^{\mathrm{BS}}+\sqrt{\dfrac{1}{\upsilon _{k}+1}}\bm{\gamma }_{k,i}^{T}\mathbf{h}_{\mathrm{S},k} +\bm{\gamma }_{k,i}^{T}\mathbf{z}_{\mathrm{BS}\vspace*{-0mm}
}.  \label{AoA_D}\vspace{-2mm}
\end{equation}\vspace*{-0mm}
The potential AoA, which leads to the maximum value among the $J$ observation directions, i.e.,\vspace{-1mm}
\begin{equation}
\vspace*{-0mm}
\widetilde{\bm{\gamma}}_{k}=\underset{\forall \gamma_{k,i}, \text{\ }i\in \{1,\cdots,J\}}{\arg \max \left\vert r_{k,i}^{\mathrm{BS}}\right\vert },  \label{RFBE}\vspace{-1mm}
\end{equation}\vspace*{-0mm}%
is considered as the strongest AoA of user $k$.
{In addition, the strongest AoA estimation shown in Equations (\ref{AoA_D}) and (\ref{RFBE}) can be performed by using either a series of analog comparators in analog domain or digital buffer in digital domain.}
Besides, vector $\widetilde{\bm{\gamma}}_{k}$ corresponding to the AoA with the maximum value in (\ref{RFBE}) will be exploited for the design of the analog beamforming vector of user $k$ at the BS.
As a result, we can also estimate all other users' uplink AoAs at the BS from their corresponding transmitted signals simultaneously.
For notational simplicity, we denote $\mathbf{F}_{\mathrm{RF}}=\left[\begin{array}{ccc}\widetilde{\bm{\gamma}}_{1},\ldots, \widetilde{\bm{\gamma}}_{N}\end{array}\right] \in\mathbb{C}^{M\times N}$ as the BS analog beamforming matrix.

\quad\emph{Step $2$}, \underline{Line 4 in Algorithm \ref{a1}:} The BS sends unique frequency tones to all the users exploiting analog beamforming matrix\footnote{This procedure can be done simultaneously in all the RF chains for all the users.} $\mathbf{F}_{\mathrm{RF}}$ obtained in Step $1$.
This facilitates the downlink AoAs estimation at the users and this AoA information will be used to design analog beamforming vectors to be adopted at the users.
The received signal $\mathbf{y}_{k}^{\mathrm{UE}}$ at user $k$ can be expressed as\vspace{-3mm}
\begin{equation}
\vspace*{-0.0mm}
\mathbf{y}_{k}^{\mathrm{UE}}=\left[ \mathbf{G}_{\mathrm{L,}k}\mathbf{h}_{\mathrm{L,}k}^{\ast }\left(\mathbf{h}_{\mathrm{L},k}^{\mathrm{BS}}\right)^{T}+\mathbf{G}_{\mathrm{S,}k}\mathbf{H}_{\mathrm{S,}k}^{T}\right]\widetilde{\mathbf{\gamma }}_{k}x_{k} +\mathbf{z}_{\mathrm{MS}},\label{RFBF}\vspace{-3mm}
\end{equation}\vspace*{-0mm}%
where $\mathbf{z}_{\mathrm{MS}}$ denotes the thermal noise at the antenna array of the users, $\mathbf{z}_{\mathrm{MS}}\sim \mathcal{CN}\left( \mathbf{0},{\sigma_{\mathrm{MS}}^{2}}\mathbf{I}\right),$ and ${\sigma} _{\mathrm{MS}}^{2}$ is the noise variance for all the users.

The AoA detection matrix for user $k$, $\bm{\omega }_{k}\in\mathbb{C}^{P\times J}$, which also contains $J$ estimation column vectors, is implemented at phase shifters of user $k$.
The $i$-th column vector of matrix $\bm{\omega }_{k}$ for user $k$, $\bm{\omega }_{k,i}\in\mathbb{C}^{P\times 1}, i\in \{1,\cdots,J\}$, is given by\vspace{-2mm}
\begin{equation}
\vspace*{-0mm}
\bm{\omega }_{k,i}=\frac{1}{\sqrt{P}}\left[
\begin{array}{ccc}
1,  & \ldots , & e^{j2\pi \left( P-1\right) \tfrac{d}{\lambda }\cos \left(\widehat{\phi }_{i}\right) }%
\end{array}
\right] ^{T},\vspace{-2mm}
\end{equation}\vspace*{-0mm}%
where $\widehat{\phi }_{i}= \left(i-1\right)\frac{180}{J}, i\in \{1,\cdots,J\}$, is the $i$-th potential AoA of user $k$ and $\bm{\omega }_{k,i}^{H}\bm{\omega }_{k,i}=1$.
With similar  procedures as shown in Step $1$, the equivalent received signal from the BS at user $k$ of the $i$-th potential AoA is given by\vspace{-3mm}
\begin{equation}
\vspace*{-0.0mm}
r _{k,i}^{\mathrm{UE}}= \bm{\omega }_{k,i}^{H}\sqrt{\dfrac{\upsilon _{k}}{\upsilon
_{k}+1}}\mathbf{h}_{\mathrm{L,}k}^{\ast }\left(\mathbf{h}_{\mathrm{L},k}^{\mathrm{BS}}\right)^{T}\widetilde{\mathbf{\gamma }}_{k}
+\mathbf{\omega}_{k,i}^{H}\sqrt{\dfrac{1}{\upsilon_{k}+1}}\mathbf{H}_{\mathrm{S,}k}^{T}\widetilde{\mathbf{\gamma }}_{k}+\bm{\omega }_{k,i}^{H}\mathbf{z}_{\mathrm{MS}}. \label{AoA_D2}\vspace{-3mm}
\end{equation}\vspace*{-0.0mm}%
Similarly, we search for the maximum value among $J$ observation directions and design the analog beamforming vector based on the estimated AoA of user $k$.
The beamforming vector for user $k$ is given by\vspace{-2mm}
\begin{equation}
\vspace*{-0mm}
\widetilde{\bm{\omega }}_{k}^{\ast}=\underset{\forall \bm{\omega }_{k,i}, \text{\ } i\in \{1,\cdots,J\}}{\arg \max \left\vert r
_{k,i}^{\mathrm{UE}}\right\vert }\vspace{-2mm}
\end{equation}\vspace*{-0.0mm}and we denote $\mathbf{Q}_{\mathrm{RF}}=\left[
\begin{array}{ccc}
\widetilde{\bm{\omega }}_{1}^{\ast}, \ldots, \widetilde{\bm{\omega }}_{N}^{\ast }\end{array}\right] \in\mathbb{C}^{P\times N}$ as the users analog beamforming matrix.

\quad\emph{Step $3$}, \underline{Line 7 in Algorithm \ref{a1}:} The BS and users analog beamforming matrices based on estimated uplink AoAs and downlink AoAs are designed via Step $1$ and Step $2$, respectively.
After that, all the users transmit orthogonal pilot sequences to the BS via user beamforming vectors $\widetilde{\bm{\omega }}_{k}^{\ast}$.
In the meanwhile, the BS receives pilot sequences via the BS analog beamforming matrix $\mathbf{F}_{\mathrm{RF}}^{T}$.
With the analog beamforming matrices, we have the equivalent channel between the BS and the users along the strongest AoA paths\footnote{The equivalent channel consists of the BS analog beamforming matrix, the mmWave channel, and the users analog beamforming matrix.}.

We denote the pilot sequences of user $k$ in the cell as $\mathbf{\Phi }_{k}=\left[ \vartheta _{k}\left( 1\right)
,....,\vartheta _{k}\left( N\right) \right]^{T}$, $\mathbf{\Phi }_{k}\in\mathbb{C}^{N\times 1}$, stands for $N$ symbols transmitted across time.
The pilot symbols used for the equivalent channel estimation are transmitted in sequence from symbol $\vartheta _{k}\left( 1\right)$ to symbol $\vartheta _{k}\left(N\right)$.
The pilot symbols for all the $N$ users form a matrix, $\mathbf{\Psi \in\mathbb{C}}^{N\times N}\mathbf{,}$ where $\mathbf{\Phi }_{k}$ is a column vector of
matrix $\mathbf{\Psi }$ given by
$\mathbf{\Psi }=\ \sqrt{E_{\mathrm{P}}}\left[
\begin{array}{ccc}
\mathbf{\Phi }_{1}, & \ldots, & \mathbf{\Phi }_{N}%
\end{array}\right]$, $\mathbf{\Phi }_{i}^{H}\mathbf{\Phi }_{j}=0$, $\forall i\neq j$, $i,\text{ }j\in \left\{ 1,\ldots, N\right\}$,
where $E_{\mathrm{P}}$ represents the transmitted pilot symbol energy.
Note that $\mathbf{\Psi }^{H}\mathbf{\Psi }=E_{\mathrm{P}}\mathbf{I}_{N}$.
Meanwhile, the BS analog beamforming matrix $\mathbf{F}_{\mathrm{RF}}$ is utilized to receive pilot sequences at all the RF chains.
As the length of the pilot sequences is equal to the number of users, we obtain an $N \times N$ observation matrix from all the RF chains at the BS.
In particular, the received signal at the $k$-th RF chain at the BS is $\mathbf{s}_{k}^{T}\in\mathbb{C}^{1\times N}$, which is given by\vspace{-0mm}
\begin{equation}
\vspace*{-0mm}
\mathbf{s}_{k}^{T}  =\widetilde{\bm{\gamma}}_{k}^{T}\overset{N}{\underset{i=1}%
{\sum }}\mathbf{H}_{i}\widetilde{\bm{\omega }}_{i}^{\ast }\sqrt{E_{%
\mathrm{P}}}\mathbf{\Phi }_{i}^{T}+\widetilde{\bm{\gamma}}_{k}^{T}\mathbf{Z},\label{RSIRFC}\vspace{-0mm}
\end{equation}\vspace*{-0mm}%
where $\mathbf{Z}\in\mathbb{C}^{M\times N}$ denotes the additive white Gaussian noise matrix at the BS and the entries of $\mathbf{Z}$ are modeled by i.i.d. random variable with distribution $\mathcal{CN}\left( 0,\sigma _{\mathrm{BS}}^{2}\right)$.

After $\left[\begin{array}{ccc}\mathbf{s}_{1}, \ldots, \mathbf{s}_{N}\end{array}\right]$ is obtained, we then adopt the LS method for our equivalent channel estimation.
We note here, the LS method is widely used in practice since it does not require any prior channel information.
Subsequently, with the help of orthogonal pilot sequences, we can construct an equivalent uplink channel matrix $\widehat{\mathbf{H}}_{\mathrm{eq}}\in\mathbb{C}^{N\times N}$ formed by the proposed scheme via the LS estimation method.
Then, by exploiting the channel reciprocity, the equivalent downlink channel of the hybrid system $\widehat{\mathbf{H}}_{\mathrm{eq}}^{T}$ can be expressed as:\vspace{-0mm}
\begin{align}
\widehat{\mathbf{H}}_{\mathrm{eq}}^{T}&=\mathbf{\Psi }^{H}\left[\begin{array}{ccc}\mathbf{s}_{1} & \ldots & \mathbf{s}_{N}\end{array}\right]
=\left[\begin{array}{c}
\widehat{\mathbf{h}}_{\mathrm{eq,}1}^{T} \\
\vdots \\
\widehat{\mathbf{h}}_{\mathrm{eq,}N}^{T}
\end{array}\right]
=\underset{\mathbf{H}_{\mathrm{eq}}^{T}}{\underbrace{\left[
\begin{array}{c}
\widetilde{\bm{\omega }}_{1}^{H}\mathbf{H}_{1}^{T}\mathbf{F}_{\mathrm{RF}%
} \\
\vdots \\
\widetilde{\mathbf{\omega}}_{N}^{H}\mathbf{H}_{N}^{T}\mathbf{F}_{\mathrm{RF}}\end{array}\right]}}+\underset{\mathrm{Effictive}\text{\ }\mathrm{noise}}{\underbrace{\frac{1}{\sqrt{E_{\mathrm{P}}}}\left[
\begin{array}{c}
\mathbf{\Phi}_{1}^{H}\mathbf{Z}^{T}\mathbf{F}_{\mathrm{RF}} \\
\vdots \\
\mathbf{\Phi} _{N}^{H}\mathbf{Z}^{T}\mathbf{F}_{\mathrm{RF}}
\end{array}\right] }},  \label{EHC_1}\vspace{-0mm}
\end{align}\vspace*{-0mm}%
where $\widehat{\mathbf{h}}_{\mathrm{eq,}k}$ is the $k$-th column vector of matrix $\widehat{\mathbf{H}}_{\mathrm{eq}}$.
From Equation (\ref{EHC_1}), we observe that the proposed channel estimation algorithm can obtain all users' equivalent CSI simultaneously.

\vspace*{-5mm}
\subsection{Performance Analysis of Proposed Channel Estimation}
In the high SNR regime, the effective noise component is negligible and Equation (\ref{EHC_1}) can be simplified\footnote{The performance degradation due to the high SNR assumption will be verified by analysis and simulation in the following sections.} as\vspace{-0mm}
\begin{equation}
\vspace*{-0mm}
\mathbf{H}_{\mathrm{eq}}^{T}=\left[
\begin{array}{c}
\widetilde{\bm{\omega }}_{1}^{H}\mathbf{H}_{1}^{T}\mathbf{F}_{\mathrm{RF}%
} \\
\vdots \\
\widetilde{\bm{\omega }}_{N}^{H}\mathbf{H}_{N}^{T}\mathbf{F}_{\mathrm{RF}%
}%
\end{array}%
\right] =\left[
\begin{array}{ccc}
\widetilde{\bm{\omega }}_{1}^{H} & \cdots & \mathbf{0} \\
\vdots & \ddots  & \vdots \\
\mathbf{0} & \cdots & \widetilde{\bm{\omega }}_{N}^{H}%
\end{array}\right]
\left[
\begin{array}{c}
\mathbf{H}_{1}^{T} \\
\vdots \\
\mathbf{H}_{N}^{T}
\end{array}
\right]
\underset{\mathbf{F}_{\mathrm{RF}}}{\underbrace{\left[
\begin{array}{ccc}
\widetilde{\bm{\gamma}}_{1} & \ldots & \widetilde{\bm{\gamma}}_{N}\end{array}\right] }}.  \label{EHC_2}\vspace*{-0mm}
\end{equation}\vspace*{-0mm}%
From Equation (\ref{EHC_2}), we can see that without the impact of noise, we can perfectly estimate the equivalent channels, which consists of the BS analog beamforming matrix $\mathbf{F}_{\mathrm{RF}}$, the users analog beamforming matrix $\mathbf{Q}_{\mathrm{RF}}^{T}$, and mmWave channels between the BS and all the users.

Now, we analyze the performance of the proposed channel estimation.
With the normalization factor $\frac{1}{\sqrt{MP}}$, the normalized mean squared error (NMSE) of equivalent channel estimation $\widehat{\mathbf{h}}_{\mathrm{eq,}k}^{T}$ is given by\vspace{-1mm}
\begin{align}
\vspace*{-0mm}
&\mathrm{NMSE}_{\mathrm{eq,}k}\notag\\
& =\frac{1}{N}\mathrm{tr}\left\{
\mathrm{\mathbb{E}}_{\mathbf{h}_{\mathrm{S},k}}\left[ \left( \dfrac{1}{\sqrt{MP}}\widehat{\mathbf{h}}_{\mathrm{eq,%
}k}^{T}-\dfrac{1}{\sqrt{MP}}\mathbf{h}_{\mathrm{eq},k}^{T}\right) ^{H} \left(\dfrac{1}{\sqrt{MP}}\widehat{\mathbf{h}}_{\mathrm{eq,}k}^{T}-\dfrac{1}{\sqrt{MP}}\mathbf{h}_{\mathrm{eq},k}^{T}\right) \right] \right\}  \notag \\
& =\frac{\sigma _{\mathrm{BS}}^{2}\mathrm{tr}\left[ \mathbf{F}_{\mathrm{RF}%
}^{H}\mathbf{F}_{\mathrm{RF}}\right] }{E_{\mathrm{P}}NMP}=%
\frac{\sigma _{\mathrm{BS}}^{2}}{E_{\mathrm{P}}MP}.  \label{MSE_NN}\vspace{-1mm}
\end{align}\vspace*{-0mm}%
From (\ref{MSE_NN}), we observe that the NMSE of the equivalent channel of user $k$ decreases with an increasing transmitted pilot symbol power $E_{\mathrm{P}}$ as well as the numbers of antennas equipped at the BS and at each user.
As the numbers of antennas $M\ $and $P$ approach infinity, the impact of noise will vanish asymptotically.
In contrast, the channel estimation errors caused by noise in conventional fully digital massive MIMO systems cannot be mitigated by increasing the number of antennas equipped at the BS and the users \cite{Marzetta2010,Jose2011}.
Therefore, the proposed channel estimation for the hybrid system outperforms the conventional pilot-aided channel estimation for the fully digital system in terms of noise mitigation.
The NMSE analysis result will be verified via simulation.
Here, we would point out that the proposed channel estimation scheme can exploit large array gains offered by the antenna arrays via using the BS analog beamforming matrix $\mathbf{F}_{\mathrm{RF}}$ and the users analog beamforming matrix $\mathbf{Q}_{\mathrm{RF}}$ to enhance the receive SNR of pilot symbols \cite{book:wireless_comm}.
Therefore, it is expected that the performance of the proposed channel estimation scheme improves with increasing the numbers of antennas equipped at the BS and each user.
\begin{figure}[t]
\centering
\includegraphics[width = 4.5in]{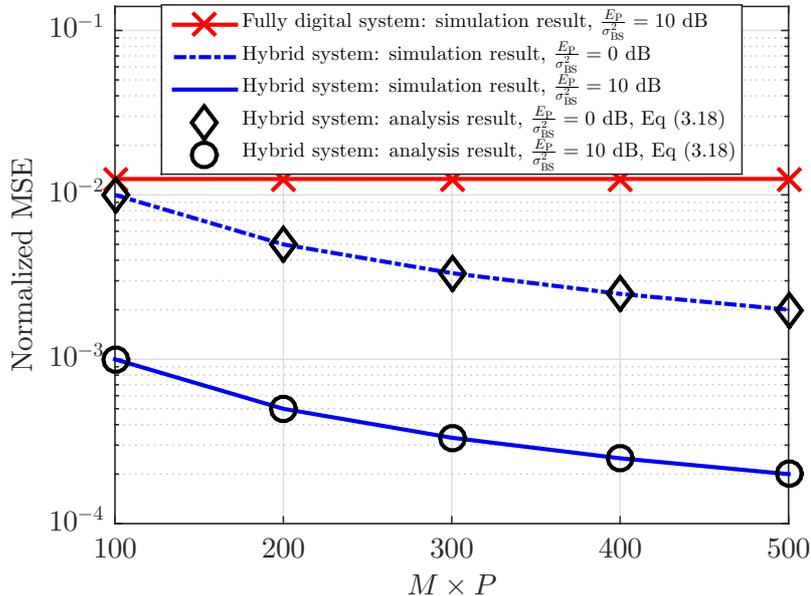}\vspace{-0mm}
\caption{The NMSE performance comparison between the proposed
pilot-aided channel estimation algorithm for the hybrid system and the conventional
pilot-aided LS for the fully digital system.}
\label{fig:antiPC}\vspace{-0mm}
\end{figure}
For a multi-cell scenario, the received pilot sequences at the RF chains of the desired BS will be affected by the reused pilot sequences from neighboring cells \cite{Marzetta2010,Jose2011,Wu2016}. The threat of pilot contamination attack will be detailed and discussed in the next chapter.

Figure \ref{fig:antiPC} shows the NMSE of the equivalent channel estimation of user $k$ versus the total numbers of antennas equipped at the BS and the user.
In Figure \ref{fig:antiPC}, we see that the simulation results match with analytical results derived in Equation (\ref{MSE_NN}).
For the SNR of the transmitted pilot at $\dfrac{E_{\mathrm{P}}}{\sigma _{\mathrm{BS}}^{2}}=10$ dB, it can be seen that the NMSE of the proposed algorithm decreases with an increasing $M\times P$.
On the contrary, we have the observation that the numbers of antennas equipped at the BS and users have no significant impact on the conventional pilot-aided LS channel estimation for the fully digital system.
Furthermore, Figure \ref{fig:antiPC} also shows that a high transmitted pilot energy is helpful to lower the NMSE of the proposed algorithm, since the impact of noise is reduced proportional to $\dfrac{1}{E_{\mathrm{P}}MP}$ as shown in Equation (\ref{MSE_NN}).
It is interesting to note that, to meet a certain required NMSE of channel estimation, we can either increase the number of antennas equipped at the BS or the number of antennas at each user. This indicates that increasing the number of antennas equipped at the BS can always improve the system performance, despite the possibly limited numbers of antennas equipped at the users.
\vspace*{-2mm}
\section{ZF Precoding and Performance Analysis}
In this section, we illustrate and analyze the achievable rate performance per user of the considered hybrid mmWave system under digital ZF downlink transmission.
The digital ZF downlink precoding is based on the estimated equivalent channel $\widehat{\mathbf{H}}_{\mathrm{eq}}$, {\color{black}which is a concatenation of the BS analog beamforming matrix} $\mathbf{F}_{\mathrm{RF}}$ and the users analog beamforming matrix $\mathbf{Q}_{\mathrm{RF}}$.
We derive a closed-form upper bound of achievable rate per user of the ZF precoding in the considered hybrid system.
Also, we compare the system achievable rate upper bound obtained by the fully digital system exploiting the ZF precoding for a large number of antennas.
The achievable rate performance gap between the considered hybrid mmWave system and the fully digital system is characterized, which is verified via analysis and simulation results.
\vspace*{-5mm}
\subsection{ZF Precoding}

Now, we utilize the estimated equivalent channel for downlink digital ZF precoding.
To study the best achievable rate performance, we first assume that the equivalent channel is estimated in the high SNR regime.
In this case, the equivalent channel is considered as perfectly estimated as the proposed channel estimation is only affected by noise as shown in Equation (\ref{MSE_NN}).
Therefore, the baseband digital ZF precoder $\overline{\mathbf{W}}_{\mathrm{eq}}\in\mathbb{C}^{N\times N}$ based on $\mathbf{H}_{\mathrm{eq}}$ is given by\vspace*{-2mm}
\begin{equation}
\vspace*{-0mm}
\overline{\mathbf{W}}_{\mathrm{eq}}=\mathbf{H}_{\mathrm{eq}}^{\ast }(\mathbf{H}_{\mathrm{eq}}^{T}\mathbf{H}_{\mathrm{eq}}^{\ast})^{-1}=\left[\begin{array}{ccc}\overline{\mathbf{w}}_{\mathrm{eq,}1},\ldots ,\overline{\mathbf{w}}_{\mathrm{eq,}N}
\end{array}%
\right] ,  \label{P31}\vspace*{-2mm}
\end{equation}%
where $\overline{\mathbf{w}}_{\mathrm{eq,}k}\in\mathbb{C}^{N \times 1}$ is the $k$-th column of ZF precoder for user $k$.
As each user is equipped with only one RF chain, one superimposed signal is received at each user at each time instant with hybrid transceivers.
The received signal at user $k$ after beamforming can be expressed as:\vspace*{-2mm}%
\begin{equation}
\vspace*{-0mm}
y_{\mathrm{ZF}}^{k}=\underset{\mathrm{desired}\text{ }\mathrm{signal}}{%
\underbrace{\widetilde{\bm{\omega }}_{k}^{H}\mathbf{H}_{k}^{T}\mathbf{F}%
_{\mathrm{RF}}\overline{\beta }\overline{\mathbf{w}}_{\mathrm{eq,}k}x_{k}}} +\underset{\mathrm{interference}}{\underbrace{\widetilde{\bm{\omega }}_{k}^{H}\mathbf{H}_{k}^{T}\overset{N}{\underset{j=1,j\neq k}{\sum }}\mathbf{F}%
_{\mathrm{RF}}\overline{\beta }\overline{\mathbf{w}}_{\mathrm{eq,}j}x_{j}}}+%
\underset{\mathrm{noise}}{\underbrace{\widetilde{\bm{\omega }}_{k}^{H}%
\mathbf{z}_{\mathrm{MS},k}}},  \label{P2}\vspace*{-2mm}
\end{equation}\vspace*{-0mm}%
where $x_{k}\in\mathbb{C}^{1\times 1}$ is the transmitted symbol energy from the BS to user $k$, $\mathrm{\mathbb{E}}\left[ \left\vert x_{k}^{2}\right\vert \right]=E_{s}$, $E_{s}$ is the average transmitted power for each user, $\overline{\beta }=\sqrt{\tfrac{1}{\mathrm{tr}(\overline{\mathbf{W}}%
_{\mathrm{eq}}\overline{\mathbf{W}}_{\mathrm{eq}}^{H})}}$ is the
transmission power normalization factor, and the effective
noise part $\mathbf{z}_{\mathrm{MS,}%
k}\sim \mathcal{CN}\left( \mathbf{0},{\sigma_{\mathrm{MS}}^{2}}\mathbf{I}\right) $.
Due to the fact that the MU interference within the AoA directions can be suppressed by the digital ZF precoder, thus\vspace*{-2mm}%
\begin{align}
\vspace*{-0mm}
&\mathbf{h}_{\mathrm{eq},i}^{T}\overline{\mathbf{w}}_{\mathrm{eq,}j}=0,\text{
}\forall i\neq j, \text{\ } \mbox{and} \text{\ }\widetilde{\bm{\omega }}_{k}^{H}\mathbf{H}_{k}^{T}\overset{N}{\underset{j=1,j\neq k}{\sum }}\mathbf{F}%
_{\mathrm{RF}}\left( \overline{\mathbf{w}}_{\mathrm{eq,}j}\right) x_{j}=0.
\label{P3}\vspace*{-2mm}
\end{align}\vspace*{-0mm}%
Then we express the SINR of user $k$ as \vspace*{-2mm}
\begin{equation}
\mathrm{SINR}_{\mathrm{ZF}}^{k}=\frac{\overline{\beta }^{2}E_{s}}{\sigma _{%
\mathrm{MS}}^{2}}.  \label{Eq_1520}\vspace*{-2mm}
\end{equation}\vspace{-0mm}%
In the sequal, we study the performance of the considered hybrid mmWave systems.
For simplicity, we assume that channels of all the users have the same Rician K-factor, i.e., $\upsilon _{k}=\upsilon,\forall k$.

\vspace*{-0mm}
\subsection{Performance Upper Bound of ZF Precoding}
Now, exploiting the SINR expression in Equation (\ref{Eq_1520}), we summarize the upper bound of achievable rate per user of the digital ZF precoding with the proposed channel estimation algorithm in the following theorem.\vspace*{-2mm}
\begin{theo}\label{thm:Theo_31}
The achievable rate per user of the proposed ZF precoding is bounded by\vspace*{-3mm}
\begin{align}
R_{\mathrm{HB}} &\leqslant R_{\mathrm{HB}}^{\mathrm{upper}}\notag\\
&=  \log _{2}\left\{ 1+\left[ \left( \dfrac{\upsilon}{\upsilon +1}\right)MP  \| \mathbf{F}_{\mathrm{RF}}^{H}\mathbf{F}_{\mathrm{RF}} \|_{\mathrm{F}}^{2} +\left( \dfrac{1}{\upsilon+1}\right) N^{2}\right] \frac{1}{N^{2}} \dfrac{E_{s}}{\sigma _{\mathrm{MS}}^{2}}\right\}.  \label{Theo_1}
\end{align}
\end{theo}\vspace*{-2mm}
\emph{\quad Proof: } Please refer to Appendix C.\QEDA

From Equation (\ref{Theo_1}), we see that the upper bound of achievable rate per user of the proposed ZF precoding depends on the Rician K-factor, $ \upsilon$.
Also, we can further observe that the upper bound of the achievable rate per user also depends on the BS analog beamforming matrix $\mathbf{F}_{\mathrm{RF}}$ designed in Step $2$ of the proposed channel estimation algorithm.
We note that since the analog beamforming only allows the BS to transmit each user's signal via its strongest AoA direction, the proposed scheme can utilize the transmission power more effectively.
In addition, the interference outside the strongest AoA directions is reduced as less transmission power will leak to undesired users.
On other hand, with an increasing number of antennas at the BS, the communication channels are more likely to be orthogonal to each other.
Therefore, it is interesting to evaluate the asymptotic upper bound $R_{\mathrm{HB}}^{\mathrm{upper}}$ for the case of a large number of antennas. We note that, even if the number of antennas equipped at the BS is sufficiently large, the required number of RF chains is still only required to equal to the number of users in the hybrid mmWave system structures. 
\vspace*{-2mm}
\begin{coro}\label{Coro_1}
In the large numbers of antennas regime, i.e., $M\rightarrow\infty $, such that $\mathbf{F}_{\mathrm{RF}}^{H}\mathbf{F}_{\mathrm{RF}}\overset{a.s.}{\rightarrow }$ $\mathbf{I}_{N},$ the asymptotic achievable rate per user of the hybrid system is bounded above by\vspace*{-2mm}
\begin{equation}
\hspace*{-0mm}
R_{\mathrm{HB}}^{\mathrm{upper}}\underset{M\rightarrow \infty }{\overset{a.s.}{\rightarrow }}\log _{2}\left\{ 1+\left[
\frac{MP}{N}\frac{\upsilon }{\upsilon +1} +\dfrac{1}{\upsilon +1}\right] \dfrac{E_{s}}{\sigma _{\mathrm{MS}}^{2}}\right\}.\label{HSUB_LA}
\end{equation}
\end{coro}\vspace*{-2mm}%
\emph{\quad Proof: } The result follows by substituting $\mathbf{F}_{\mathrm{RF}}^{H}\mathbf{F}_{\mathrm{RF}}\underset{M\rightarrow \infty }{\overset{a.s.}{\rightarrow }}\mathbf{I}_{N}$ into (\ref{Theo_1}).\QEDA
\vspace*{+0mm}%

From Equation (\ref{HSUB_LA}), we have an intuitive observation that the asymptotic performance of the proposed precoding is mainly determined by the numbers of equipped antennas and RF chains.
\vspace*{-0mm}
\subsection{Comparison with Fully Digital Systems}

In this section, we derive the achievable rate performance of a fully digital mmWave system in the large numbers of antennas regime.
The obtained analytical results in this section will be used as a reference for comparing to the proposed hybrid system.
To this end, for the fully digital mmWave system, we assume that each user is equipped with one RF chain and $P$ antennas.
The $P$ antenna array equipped at each user can provide $10\log(P)$ dB array gain.
We note that, the number of antennas equipped at the BS is $M$ and the number of RF chains equipped at the BS is equal to the number of antennas.
The channel matrix from the BS to user $k$ is given by\vspace*{-1mm}
\begin{equation}
\vspace*{-0mm}
\mathbf{H}_{k}^{T}=\mathbf{h}_{k}^{\ast }\mathbf{h}_{\mathrm{BS,}k}^{T}.\vspace*{-1mm}
\end{equation}\vspace*{-0mm}%
We assume that the CSI is perfectly known to the users and the BS.
The BS with the fully digital system is adopted to illustrate the maximal performance gap in terms of achievable rate between the fully digital system and the considered hybrid system.
The CSI at the BS for the downlink information transmission is given by\vspace*{-1mm}
\begin{equation}
\vspace*{-0mm}
\mathbf{H}_{\mathrm{FD}}^{T}=\left[
\begin{array}{ccccc}
{\mathbf{h}}_{\mathrm{BS,}1}^{T} & ... & {\mathbf{h}}_{%
\mathrm{BS,}k}^{T} & ... & {\mathbf{h}}_{\mathrm{BS,}N}^{T}%
\end{array}%
\right] .\vspace*{-1mm}
\end{equation}\vspace*{-0.0mm}%
The ZF precoder for the equivalent channel $\mathbf{H}_{\mathrm{FD}}^{T}$ is
denoted as\vspace*{-3mm}
\begin{equation}
\mathbf{W}_{\mathrm{FD}}=\mathbf{H}_{\mathrm{FD}}^{\ast }\left( \mathbf{H}_{%
\mathrm{FD}}^{T}\mathbf{H}_{\mathrm{FD}}^{\ast }\right) ^{-1}.\vspace*{-3mm}
\end{equation}\vspace*{-0mm}%
Therefore, the achievable rate per user of the fully digital system is bounded by\vspace*{-3mm}
\begin{align}
R_{\mathrm{FD}}&=\log _{2}\left[ 1+\dfrac{P}{\mathrm{tr}\left[ \mathbf{W}_{\mathrm{FD}}\mathbf{W}_{\mathrm{FD}}^{H}\right] }\dfrac{E_{s}}{\sigma _{\mathrm{MS}}^{2}}\right] \notag \\
& \overset{(c)}{\leqslant }\log _{2}\left[ 1+\frac{P }{N^{2}}\mathrm{tr}\left[ \mathbf{H}_{\mathrm{FD}}^{H}\mathbf{H}_{\mathrm{FD}}\right] \dfrac{E_{s}}{\sigma_{\mathrm{MS}} ^{2}}\right] =R_{\mathrm{FD}}^{\mathrm{upper}},  \label{FDUB}\vspace*{-3mm}
\end{align}
where $(c)$ follows (\ref{Proof_3}) in Appendix C.%
\vspace*{-1mm}
\begin{coro}
In the large numbers of antennas regime, the asymptotic achievable rate per user of the fully digital system is bounded by\vspace*{-2mm}
\begin{equation}
\vspace*{-0mm}
R_{\mathrm{FD}}\leqslant R_{\mathrm{FD}}^{\mathrm{upper}}\underset{%
M\rightarrow \infty }{\overset{a.s.}{\rightarrow }}\log _{2}\left[ 1+\frac{MP%
}{N}\frac{E_{s}}{\sigma _{\mathrm{MS}}^{2}}\right] .  \label{FDUB_LA}\vspace*{-0mm}
\end{equation}
\end{coro}\vspace*{-2mm}\vspace*{-0mm}
\emph{\quad Proof: } The result follows by substituting $\frac{1}{M}\mathbf{H}_{\mathrm{FD}}^{H}%
\mathbf{H}_{\mathrm{FD}}\underset{M\rightarrow \infty }{\overset{a.s.}{%
\rightarrow }}\mathbf{I}_{N}$ into (\ref{FDUB}).\QEDA
\vspace{+0mm}

Based on Equations (\ref{HSUB_LA}) and (\ref{FDUB_LA}), we further quantify the achievable rate performance gap between the considered hybrid system and the fully digital system in the large numbers of antennas regime.\vspace{-1mm}%
\begin{coro}
In the large numbers of antennas regime, the gap between the achievable rate upper bounds for the hybrid system and the fully digital system can be expressed as:\vspace*{-2mm}
\begin{equation}
\Delta_{\mathrm{GAP}}=R_{\mathrm{HB}}^{\mathrm{upper}}-R_{\mathrm{FD}}^{\mathrm{upper}}\underset{M\rightarrow \infty }{\overset{\mathrm{(S)}}{\approx }}\log _{2}\left\{ \frac{\upsilon}{\upsilon+1}\right\} \leqslant 0,\label{GAP_LA}
\end{equation}%
where $\mathrm{(S)}$ {}stands for $\mathrm{SNR\rightarrow \infty .}$ \label{coro3}
\end{coro}
\vspace*{-2mm}
\emph{\quad Proof: }The result follows by substituting (\ref{HSUB_LA}) and (\ref{FDUB_LA}) into (\ref{GAP_LA}).\QEDA
\vspace*{+0mm}%

In the large numbers of antennas regime, based on Equations (\ref{HSUB_LA}) and (\ref{FDUB_LA}), it is interesting to observe that with an increasing Rician K-factor $\upsilon $, the performance upper bounds of the two considered systems will coincide.
Intuitively, as the Rician K-factor increases, the LOS component becomes the dominant element of the communication channel, as shown in Equation \eqref{eqn:LOS_channel31}.
Therefore, the BS analog beamforming matrix based on the estimated strongest AoA will allocate a smaller portion of the transmitted signal energy to the scattering component.
At the same time, the interference caused by other users is suppressed by the digital baseband ZF precoding, $\overline{\mathbf{W}}_{\mathrm{eq}}$.

In Figure \ref{fig:HBvsFD}, we present a comparison between the achievable rate per user of the hybrid system and the fully digital system for $M=100,$ $N=10$, and a Rician K-factor of $\upsilon_{k} =2,\forall k$.
Firstly, our simulation results verify the tightness of derived upper bounds in Equations (\ref{HSUB_LA}) and (\ref{FDUB_LA}).
It can be observed from Figure \ref{fig:HBvsFD} that even for a small value of Rician K-factor, e.g. $\upsilon =2$, our proposed channel estimation scheme with ZF precoding can achieve a considerable high sum rate performance due to its interference suppression capability.

\begin{figure}[t]
\centering
{\includegraphics[width=4.5in,]{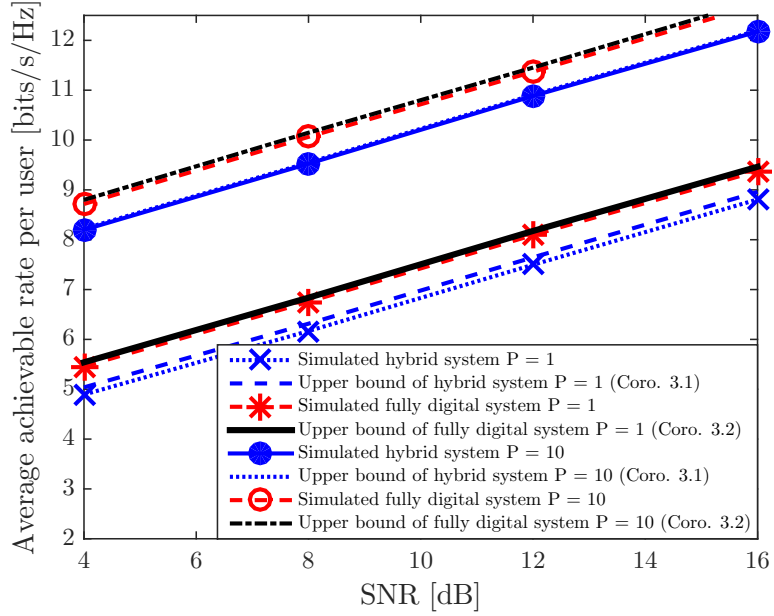}}
\caption{Average achievable rate per user [bits/s/Hz] versus SNR for the hybrid and the fully digital systems with $M=100$, $N=10$ and $\upsilon =2$.}
\label{fig:HBvsFD}\vspace{-0mm}
\end{figure}
\begin{figure}[t]
\centering
\includegraphics[width=4.5in,]{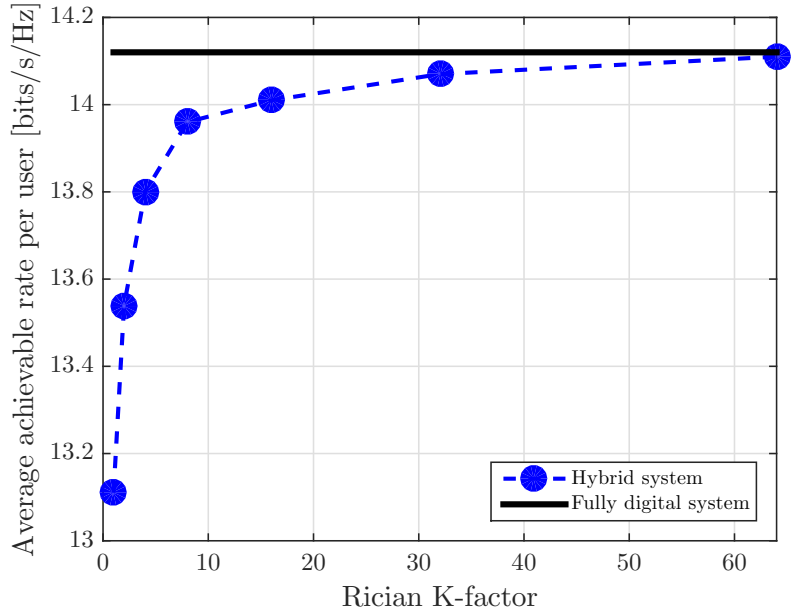}
\caption{Average achievable rate [bits/s/Hz] versus Rician K-factor for different systems with $\mathrm{SNR}=20$ dB, $M=100$, $P=16$, and $N=4$.}
\label{fig:HBFDvsRK}\vspace{-0mm}
\end{figure}
In Figure \ref{fig:HBFDvsRK}, the achievable rate performance gap between the fully digital system and the hybrid system decreases with the increasing Rician K-factor, which is predicted by Equation (\ref{GAP_LA}).
In particular, with a sufficiently large Rician K-factor, the achievable rates of these two systems will coincide.

\vspace*{-2mm}
\section{Performance Analysis with Hardware Impairments}

In the last section, we study the proposed mmWave hybrid system in ideal hardware and ideal estimation conditions.
{In practice, hardware components may have various types of impairments that may degrade the achievable rate performance\footnote{{The authors of \cite{Bjornson2015bb} proved that for a fully digital massive MIMO system, the additive distortion caused by hardware impairments create finite ceilings on the channel
estimation accuracy and on the uplink/downlink capacity, which are irrespective of the SNR and the
number of base station antennas.
In addition, work \cite{Ying2015} concluded that the impact of phase error on hybrid beamforming is a further reduction on the number of effective antenna per user.
In fact, the achievable rate degradation caused by phase errors can be compensated by simply employing more transmit antennas.}}, e.g. phase errors in phase shifters induced by thermal noise, transceiver RF beamforming errors caused by AoA estimation errors, and channel estimation errors affected by independent additive distortion noises \cite{Bjornson2015bb,Zhang2016,Ying2015}.}
In this section, we analyze the rate performance degradation under hardware impairments.
\vspace*{-2mm}
\subsection{Transceiver Beamforming Errors and Random Phase Errors}

Here, we first discuss the scenario that the equivalent channels are estimated in the high pilot transmit power regime, i.e., $\dfrac{E_{\mathrm{P}}}{\sigma _{\mathrm{BS}}^{2}}\rightarrow \infty $, where errors caused by thermal noise are negligible.
As a result, we focus on the effect of the phase errors in phase shifters and transceiver analog beamforming errors on the achievable rate performance.

Firstly, we start from quantifying the impact of transceiver beamforming errors, which is caused by AoA estimation errors.
As the number of antennas equipped at the BS is sufficiently large, the beamwidth of the antenna array is narrow.
Therefore, even a small AoA estimation error may cause significant impacts on the system performance.
Here, the BS analog beamforming error matrix of user $k$, $\mathbf{\Delta  }_{k}\in\mathbb{C}^{M\times M}$, which is caused by the AoA estimation error, can be expressed as\vspace*{-2mm} %
\begin{equation}
\vspace*{-0mm}
\mathbf{\Delta }_{k}=\frac{1}{\sqrt{M}}\left[
\begin{array}{ccc}
1 &  &  \\
& \ddots &  \\
&  & e^{j2\pi \left( M-1\right) \tfrac{d}{\lambda }\cos \left( \Delta \theta
_{\mathrm{BS,}k}\right) }%
\end{array}%
\right] ,\vspace*{-2mm}
\end{equation}\vspace*{-0mm}%
where the AoA estimation error $\Delta \theta _{\mathrm{BS,}k}$ at the BS for user $k$ is modeled by i.i.d. random variable with the distribution $\mathcal{CN}\left( 0, \varrho _{\mathrm{BS}}^{2}\right)$.
In addition, the user beamforming error matrix $\mathbf{\Theta }_{k}\in\mathbb{C}^{P\times P}$, which is caused by AoA estimation error $\Delta \theta _{\mathrm{MS,}k}$, can also be similarly formulated.
Now, we assume that the hybrid system is only affected by AoA estimation errors.
Therefore, the received pilot symbols at the BS during the third step of channel estimation is expressed as:\vspace*{-2mm}
\begin{equation}
\vspace*{-0mm}
\left(\widehat{\mathbf{s}}_{k}^{\mathrm{BE}}\right)^{T}=\widetilde{\bm{\gamma}}_{k}^{T}\mathbf{\Delta }%
_{k}\overset{N}{\underset{i=1}{\sum }}\mathbf{H%
}_{i}\mathbf{\Theta }_{i}\widetilde{\bm{\omega }}_{%
i}^{\ast }\sqrt{E_{\mathrm{P}}}\mathbf{\Phi }_{i}^{T}+\underset{\mathrm{effective}\text{ }\mathrm{noise}}{\underbrace{\widehat{\mathbf{z}}_{\mathrm{eq}}}},\vspace*{-2mm}
\end{equation}\vspace*{-0mm}%
where the entries of effective noise part $\widehat{\mathbf{z}}_{\mathrm{eq}}=\mathbf{f}_{k}^{T}\mathbf{\Delta }_{k}\mathbf{\Xi }_{k}\mathbf{Z}$ can still be modeled by i.i.d. random variable with distribution $\mathcal{CN}\left( 0,\sigma _{\mathrm{BS}}^{2}\right)$.
As we assumed that $\dfrac{E_{\mathrm{P}}}{\sigma _{\mathrm{BS}}^{2}}\rightarrow \infty $, the estimated equivalent channel under transceiver analog beamforming errors, $\mathbf{H}_{\mathrm{eq}}^{\mathrm{BE}}$, is given by\vspace*{-2mm}%
\begin{equation}
\left( \mathbf{H}_{\mathrm{eq}}^{\mathrm{BE}}\right) ^{T}= \left[
\begin{array}{ccc}
\widetilde{\bm{\omega }}_{1}^{H}\mathbf{\Theta }_{1}&\cdots&\mathbf{0%
} \\
\vdots & \ddots & \vdots \\
\mathbf{0}&\cdots&\widetilde{\bm{\omega }}_{N}^{H}\mathbf{\Theta }%
_{N}%
\end{array}%
\right] \left[
\begin{array}{c}
\mathbf{H}_{1}^{T} \\
\vdots \\
\mathbf{H}_{N}^{T}%
\end{array}%
\right] \left[
\begin{array}{ccc}
\mathbf{\Delta }_{1}\widetilde{\bm{\gamma}}_{1}\ldots\mathbf{\Delta }%
_{N}\widetilde{\bm{\gamma}}_{N}%
\end{array}%
\right].\vspace*{-2mm}
\end{equation}\vspace*{-0mm}%
Following the similar signal processing procedures as in Equations (\ref{P31})--(\ref{P3}), the average achievable rate per user under the BS and users analog beamforming errors can be expressed as\vspace*{+1mm}%
\begin{equation}
\vspace*{-0.0mm}
\widehat{R}_{\mathrm{HB}}^{\mathrm{BE}}=\mathrm{\mathbb{E}}_{\mathbf{H}_{\mathrm{S}},\mathrm{\Delta \theta }_{\mathrm{MS}}\mathrm{,\Delta \theta }_{%
\mathrm{BS}}}\left\{ \log _{2}\left[ 1+\frac{(\widehat{\beta}^{\mathrm{BE}})^{2}E_{s}}{%
\sigma _{\mathrm{MS}}^{2}}\right] \right\} , \label{HBBFE}\vspace*{-0mm}
\end{equation}\vspace*{+1mm}%
where $\widehat{\beta}^{\mathrm{BE}}=\sqrt{\frac{1}{\mathrm{tr}[(\widehat{\mathbf{W}}_{\mathrm{eq}}^{\mathrm{BE}})(\widehat{\mathbf{W}}_{\mathrm{eq}}^{\mathrm{BE}})^{H}]}}$ is the transmission power normalization factor and $\widehat{\mathbf{W}}_{\mathrm{eq}}^{\mathrm{BE}}$ is the downlink ZF precoder based on $\mathbf{H}_{\mathrm{eq}}^{\mathrm{BE}}$.
From Equation (\ref{HBBFE}), it is interesting to observe that the achievable rate performance is not bounded by above.
This is due to the fact that the impact of transceiver analog beamforming errors on different RF chains can be estimated by the pilot matrix $\mathbf{\Psi}$, treated as parts of the channel, and compensated by the digital ZF transmission.
Therefore, it is expected that our proposed scheme is robust against transceiver analog beamforming errors.

Observing Equations (\ref{HBBFE}) and (\ref{Eq_1520}), we see that the performance degradation caused by AoA estimation errors can be quantified by comparing power normalization factors $\widehat{\beta}^{\mathrm{BE}}$ to $\overline{\beta}$.
In particular, the hybrid system with hardware impairments incurs a power loss in the received SINR at the users.
However, it is difficult to provide an explicit mathematical expression to quantify the loss.
Therefore, we borrow the ``power loss'' concept from the literature of array signal processing.
Specifically, the power loss due to AoA estimation errors is related to the HPBW \cite{Trees2002} and the HPBW can be approximated (p. 48, \cite{Trees2002}) %
\begin{equation}
\vspace*{-0.0mm}
\mathrm{HPBW} \approx \dfrac{1.782}{M},\vspace*{-0mm}
\end{equation}\vspace*{-0mm}%
where $M$ is the number of antennas equipped in the array.
Hence, we can approximate the impact of AoA estimation errors by introducing a power loss coefficient $\xi \in \left( 0,1\right]$, which is determined by the beam pattern of the array and the variance of the distribution of AoA estimation errors.
For example, if the variance of AoA estimation errors is assumed no larger than half of the HPBW of the antenna array\footnote{In general, this assumption is valid for practical communication systems with a sufficiently large number of antennas.}, i.e., $\varrho _{\mathrm{BS}}^{2}\leqslant \dfrac{1.782}{2M}$, the power loss coefficient is given by $\xi \approx 0.5$ according to the half power loss principle \cite{Trees2002}. 
With an increasing AoA estimation error variance $\varrho _{\mathrm{BS}}^{2}$, $\xi $ decreases significantly\footnote{The calculation of $\xi $ for different numbers of antennas and AoA errors is well studied for linear arrays and interested readers may refer to \cite{Trees2002} for a detailed discussion.}. 
With the help of $\xi$, we now can express the approximation of $\widehat{R}_{\mathrm{HB}}^{\mathrm{BE}}$, the average achievable rate per user under transceiver beamforming errors, in the large numbers of antennas regime as\vspace*{-0mm}
\vspace*{-0mm}%
\begin{equation}
\widehat{R}_{\mathrm{HB}}^{\mathrm{BE}}\overset{M\rightarrow \infty }{%
{\approx }}\log _{2}\left\{ 1+\left[
\dfrac{\upsilon }{\upsilon +1}\frac{MP}{N}\xi +\dfrac{1}{\upsilon +1}\right] \dfrac{E_{s}}{\sigma _{\mathrm{MS}}^{2}}\right\}.\label{HPBW_D}\vspace*{-0mm}
\end{equation}\vspace*{-0mm}%
From (\ref{HPBW_D}), it is interesting to note that, the beamforming errors matrices $\mathbf{\Theta }_{k}$ and $\mathbf{\Delta }_{k}$ only lead to a certain power loss, which depends on $\xi$. Besides, this loss can be compensated at the expense of a higher transmit power, i.e., increase $E_{\mathrm{S}}$.

Now, we discuss the impact of phase errors, which are caused by the AWGN and the limited quantization resolution of the phase shifters.
At the users sides, phase errors of user $k$ are modeled by \cite{Ying2015}\vspace*{-0mm}
\begin{equation}
\vspace*{-0mm}
\mathbf{\Lambda }_{k}=\mathrm{diag}\left\{ e^{j\Delta \phi
_{k,p}}\right\}\in \mathbb{C}^{P\times P} ,\text{ }p\in \left\{ 1,\ldots ,P\right\},\vspace*{-0mm}
\end{equation}\vspace*{-0mm}%
where phase errors $\Delta \phi _{k,p}, \forall p$, are uniformly distributed over $\left[ -a,\text{ }a\right] \ $and $a>0$ is the maximal phase error of user $k$.
Similarly, for the phase shifters connected with the $k$-th RF chain in the BS, the associated phase errors are given by\vspace*{-1mm}
\begin{equation}
\vspace*{-0mm}
\mathbf{\Xi }_{k}=\mathrm{diag}\left\{ e^{j\Delta \psi
_{k,m}}\right\}\in\mathbb{C}^{M\times M} ,\text{ }m\in \left\{ 1,\ldots ,M\right\} ,\vspace*{-1mm}
\end{equation}\vspace*{-0mm}%
where the errors $\Delta\psi _{k,m}, \forall m$, are uniformly distributed over $\left[-b,\ b\right]$ and $b>0$ is the maximal phase error of the BS.
The property of $\mathbf{\Lambda }_{k}$ and $\mathbf{\Xi }_{k}$ can be expressed as\vspace*{-3mm}
\begin{align}
&\mathrm{\mathbb{E}}\left[ \mathbf{\Lambda }_{k}\right] =\underset{-a}{%
\overset{a}{\int }}\,\dfrac{1}{2a}e^{j\Delta \phi _{k,p}}\,d\Delta \phi _{k,p}=%
\frac{\sin \left( a\right) }{a} \mathbf{I}_{\mathrm{P}}\text{ and}\notag \\
&\mathrm{\mathbb{E}}\left[
\mathbf{\Xi }_{k}\right] =\underset{-b}{\overset{b}{\int }}%
\dfrac{1}{2b}e^{-j\Delta \psi _{k,m}}\,d\Delta \psi _{k,m}=\frac{\sin \left(
b\right)}{b}\mathbf{I}_{\mathrm{M}}, \label{Prop}\vspace*{-3mm}
\end{align}\vspace*{-0mm}%
respectively. Then, the received pilot symbols $\widehat{\mathbf{s}}_{k}^{T}$ used for the equivalent channel estimation at the $k$-th RF chain, which is under the impact of phase errors and transceiver analog beamforming errors, can be expressed as\vspace*{-0mm}%
\begin{equation}
\widehat{\mathbf{s}}_{k}^{T}=\widetilde{\bm{\gamma}}_{k}^{T}\mathbf{\Delta }%
_{k}\mathbf{\Xi }_{k}\overset{N}{\underset{i=1}{\sum }}\mathbf{H%
}_{i}\mathbf{\Lambda }_{i}\mathbf{\Theta }_{i}\widetilde{\bm{\omega }}_{%
i}^{\ast }\sqrt{E_{\mathrm{P}}}\mathbf{\Phi }_{i}^{T}+\underset{\mathrm{effective}\text{ }\mathrm{noise}}{\underbrace{\widehat{\mathbf{z}}_{\mathrm{eq}}}}.\vspace*{-0mm}
\end{equation}\vspace*{-0.0mm}%
Similarly, we can express the average achievable rate per user as\vspace*{-0mm}%
\begin{equation}
\vspace*{-0.0mm}
\widehat{R}_{\mathrm{HB}}=\mathrm{\mathbb{E}}_{\mathbf{H}_{\mathrm{S}}\mathrm{,\Delta
\phi ,\Delta \psi ,\Delta \theta }_{\mathrm{MS}}\mathrm{,\Delta \theta }_{%
\mathrm{BS}}}\left\{ \log _{2}\left[ 1+\frac{\widehat{\beta }^{2}E_{s}}{%
\sigma _{\mathrm{MS}}^{2}}\right] \right\} ,\vspace*{-0mm}
\end{equation}\vspace*{-0.00mm}%
where $\widehat{\beta}$ is the transmission power normalization factor under phase errors and transceiver beamforming errors.
Based on Equations (\ref{HSUB_LA}), (\ref{HPBW_D}), and (\ref{Prop}), the approximation of the average achievable rate per user $\widehat{R}_{\mathrm{HB}}$ in the large numbers of antennas regime is given by\vspace*{-0mm}%
\begin{equation}
\widehat{R}_{\mathrm{HB}}\overset{M\rightarrow \infty }{\approx }\log
_{2}\left\{ 1+ \left[  \dfrac{\upsilon}{%
\upsilon+1}\frac{MP}{N}\widehat{\xi} +\dfrac{1}{\upsilon+1}%
 \right] \dfrac{E_{s}}{\sigma _{\mathrm{MS}}^{2}}\right\},
\label{QHI}\vspace*{-0mm}
\end{equation}\vspace*{-0.0mm}%
where $\widehat{\xi}=\left( \frac{\sin \left( a\right) }{a}\right) ^{2}\left( \frac{\sin \left( b\right) }{b}\right) ^{2}\xi$ is the equivalent power loss coefficient.
From Equation (\ref{QHI}), we note that the joint impact of random phase errors and transceiver beamforming errors on the average achievable rate cause performance degradation compared to the case of perfect hardware. Besides, using extra transmission power can compensate the performance degradation.

\vspace*{-5mm}
\subsection{Hardware Impairment and  Imperfect
Channel Estimation}
In this section, we further study the system performance by jointly taking account the impact of hardware impairments and equivalent channel estimation errors caused by noise.
The estimated equivalent channel $\widetilde{\mathbf{H}}_{\mathrm{eq}}^{T}$ can be expressed as\vspace*{-2mm}
\begin{equation}
\vspace*{-0.0mm}
\widetilde{\mathbf{H}}_{\mathrm{eq}}^{T}=\widehat{\mathbf{H}}_{\mathrm{eq}%
}^{T}+\Delta \widehat{\mathbf{H}}_{\mathrm{eq}}^{T}\mathbf{,}\vspace*{-2mm}
\end{equation}\vspace*{-0mm}%
where $\widehat{\mathbf{H}}_{\mathrm{eq}}$ is the equivalent channel under random phase errors and transceiver beamforming errors and the entries of the normalized channel estimation error $\frac{1}{\sqrt{MP}}\Delta \widehat{\mathbf{H}}_{\mathrm{eq}}^{T}$ are modeled by i.i.d. random variables with distribution $\mathcal{CN}\left( 0,\delta ^{2}\right)$, $\delta ^{2}=\mathrm{MSE}_{\mathrm{eq}}$, which is given by Equation (\ref{MSE_NN}).
The ZF precoding matrix for the considered hybrid system based on the imperfect CSI is given by\vspace*{-2mm}%
\begin{align}
\vspace*{-0mm}
\widetilde{\mathbf{W}}_{\mathrm{eq}}=\widetilde{\mathbf{H}}_{\mathrm{eq}%
}^{\ast }\left[ \widetilde{\mathbf{H}}_{\mathrm{eq}}^{T}\widetilde{\mathbf{H}%
}_{\mathrm{eq}}^{\ast }\right] ^{-1}=\widehat{\mathbf{W}}_{\mathrm{eq}%
}+\Delta \widehat{\mathbf{W}}_{\mathrm{eq}}
=\left[
\begin{array}{ccc}
\widetilde{\mathbf{w}}_{\mathrm{eq,}1} &\cdots& \widetilde{\mathbf{w}}_{%
\mathrm{eq,}N}%
\end{array}%
\right] ,\vspace*{-2mm}
\end{align}\vspace*{-0mm}%
where $\widetilde{\mathbf{w}}_{\mathrm{eq,}k}\in\mathbb{C}^{N\times 1}$ is the $k$-th column vector of $\widetilde{\mathbf{W}}_{\mathrm{eq}}$ and $\widehat{\mathbf{W}}_{\mathrm{eq}}$ is the precoder based on $\widehat{\mathbf{H}}_{\mathrm{eq}}$. 
The received signal at user $k$ under imperfect CSI is given by\vspace*{-2mm}%
\begin{equation}
\vspace*{-1mm}
\widetilde{y}_{\mathrm{ZF}}^{k}=\underset{\mathrm{desired}\text{ }\mathrm{%
signal}}{\underbrace{\widetilde{\beta }x_{k}}}+\underset{\mathrm{intra-cell}%
\text{ }\mathrm{interference}}{\underbrace{\widetilde{\beta }\widehat{%
{\bm{\omega }}}_{k}^{H}\mathbf{H}_{k}^{T}\widehat{\mathbf{F}}_{%
\mathrm{RF}}\Delta \widehat{\mathbf{W}}_{\mathrm{eq}}\mathbf{x}}}+z_{k},\vspace*{-0mm}
\end{equation}\vspace*{-0mm}%
where $\widetilde{\beta }=\sqrt{\tfrac{1}{\mathrm{tr}\left( \widetilde{\mathbf{W}}_{\mathrm{eq}}\widetilde{\mathbf{W}}_{\mathrm{eq}}^{H}\right) }}$ is the power normalization factor. $\Delta \widehat{\mathbf{w}}_{\mathrm{eq,}j}\in\mathbb{C}^{M\times 1}$ denotes the $j$-th column vector of the ZF precoder error matrix $\Delta\widehat{\mathbf{W}}_{\mathrm{eq}}\mathbf{=}\widetilde{\mathbf{W}}_{\mathrm{eq}}-\widehat{\mathbf{W}}_{\mathrm{eq}}$ and $\mathbf{x}=[x_{1},\,x_{2},\ldots ,$ $x_{N}]^{T}$ denotes the transmitted signal for all users.

We then express the SINR of user $k$ as \vspace*{-1mm}
\begin{equation}
\hspace*{-0mm}
\widetilde{\mathrm{SINR}}_{\mathrm{ZF}}^{k}=\frac{E_{s}{\widetilde{\beta }^{2}}%
}{{\widetilde{\beta }^{2}}E_{s}\widehat{\mathbf{h}}_{\mathrm{eq,}k}^{T}\mathrm{%
E}_{\mathrm{\Delta}\widehat{\mathrm{\mathbf{H}}}_{\mathrm{eq}%
}}\left[ \Delta \widehat{\mathbf{W}}_{\mathrm{eq}}\Delta \widehat{\mathbf{W}}%
_{\mathrm{eq}}^{H}\right] \widehat{\mathbf{h}}_{\mathrm{eq,}k}^{\ast
}+{\sigma_{\mathrm{MS}}^{2}}}.  \label{SINR_with_error1}\vspace*{-1mm}
\end{equation}\vspace*{-0mm}%
Now we summarize the achievable rate per user in the high SNR regime in the following theorem.
\begin{theo}\label{thm:Theo_32}
As the receive $\mathrm{SNR=}\dfrac{E_{s}}{\sigma _{\mathrm{MS}}^{2}}$ approaches infinity, the approximated achievable rate of user $k$ of ZF precoding under imperfect hybrid CSI is given by\vspace*{-1mm}
\begin{align}
\widetilde{R}_{\mathrm{ZF},k}&\approx\log _{2} \left\{ 1+\left[ \left(
\sqrt{1+\delta ^{2}}-1\right) ^{2}-2\sqrt{1+\delta ^{2}}\left( \sqrt{1+\delta ^{2}}-1\right) \delta ^{2}N\eta _{kk}\right. \right. \notag\\
& \left. \left. +\left( \sqrt{1+\delta ^{2}}\right) \left( 2-\sqrt{1+\delta
^{2}}\right) \delta ^{2}\mathrm{tr}\left( \mathbf{K}^{-1}\right) \right]
^{-1}\right\} ,  \label{Theo_IP_RL}
\end{align}%
where $\mathbf{K}=\widehat{\mathbf{H}}_{\mathrm{eq}}^{T}\widehat{\mathbf{H}}%
_{\mathrm{eq}}^{\ast }$, $\eta _{kk}$ represents the $k$-th diagonal element
of $\mathbf{K}^{-1}$, and $\delta ^{2}$ is the NMSE of channel estimation given by Equation (\ref{MSE_NN}).
\end{theo}\vspace*{-1mm}
\emph{\quad Proof: }Please refer to Appendix D.\QEDA
\begin{coro}\vspace{-0.0mm}%

In the large numbers of antennas regime, i.e., $M\rightarrow \infty$, %
the asymptotic average achievable rate per user of the hybrid system is approximated by\vspace*{-2mm}
\begin{align}
\hspace*{-0mm}
\widetilde{R}_{\mathrm{ZF}} \approx &\log _{2}\left\{ 1+\left[ \left(
\sqrt{1+\delta ^{2}}-1\right) ^{2}\right. \right.  \notag \\
&-2\sqrt{1+\delta ^{2}}\left( \sqrt{%
1+\delta ^{2}}-1\right)\frac{\delta ^{2}N}{\widehat{\xi} MP}\frac{%
\upsilon +1}{\upsilon } \notag \\
& \left. \left. +\left( \sqrt{1+\delta ^{2}}\right) \left( 2-\sqrt{1+\delta
^{2}}\right)\frac{\delta ^{2}N}{\widehat{\xi} MP}\frac{\upsilon +1}{%
\upsilon }\right] ^{-1}\right\} .  \label{Coro_RZFK1}\vspace*{-3mm}
\end{align}
\end{coro}\vspace*{-2mm}
\emph{\quad Proof: }The result follows by substituting $\mathbf{K}\underset{M\rightarrow \infty }{\overset{a.s.}{\rightarrow }}\widehat{\xi} MP \frac{\upsilon}{%
\upsilon+1}\mathbf{I}_N$ into (\ref{Theo_IP_RL}).\QEDA
\vspace{+1mm}%

Based on (\ref{QHI}) and (\ref{Coro_RZFK1}), the additional achievable rate performance degradation, which is caused jointly by random phase errors and transceiver beamforming errors, is further summarized in the following corollary.
\begin{coro}\vspace{-0.0mm}%
In the large numbers of antennas regime, i.e., $M\rightarrow \infty$, the approximated achievable rate per user performance gap between the system with ideal hardware and the system under phase errors and transceiver beamforming errors is given by\vspace*{-0mm}
\begin{equation}
\vspace*{-0mm}
\Delta \mathrm{Gap}\approx \log _{2}\left[ \frac{1}{\widehat{\xi} }\right].  \label{Coro_GAPforHI}
\end{equation}
\end{coro}\vspace{-0.0mm}
\emph{\quad Proof: }The result comes after some mathematical manipulation on (\ref{QHI}) and (\ref{Coro_RZFK1}).\QEDA

\vspace*{-0mm}
\section{Simulation and Discussion}

In this section, we present further numerical results to validate our analysis.
We consider a small single-cell hybrid mmWave system.
\begin{figure}[t]
\centering
\includegraphics[width=4.5in]{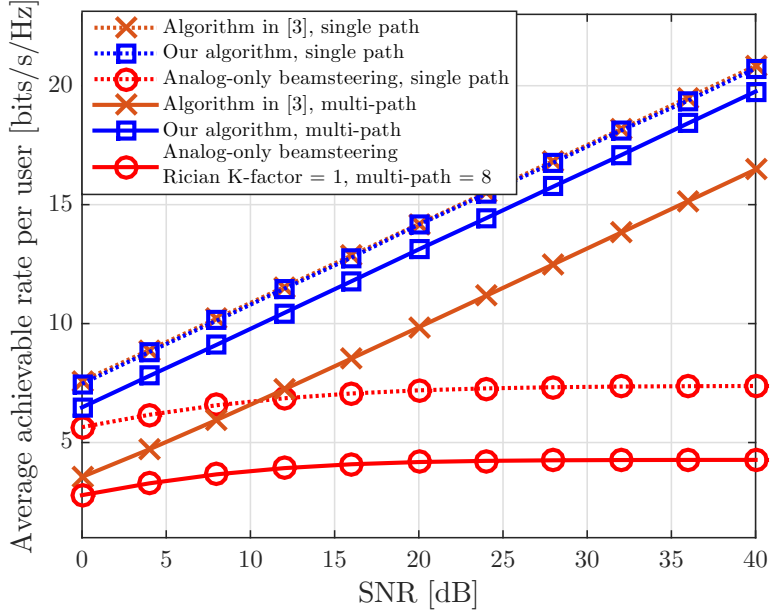}\vspace{-0mm}
\caption{The average achievable rate per user [bits/s/Hz] versus SNR for algorithm proposed in [3] and our proposed algorithm.}
\label{fig:Comp_HB_ALH_ours}\vspace{-0mm}
\end{figure}
In Figure \ref{fig:Comp_HB_ALH_ours}, we compare the achievable rates using the proposed algorithm and the algorithm proposed by \cite{Alkhateeb2015} for sparse and non-sparse mmWave channels.
We assume perfect channel estimation with $M=100,$ $N=4,$ and $P=16$. For non-sparse mmWave channels, $\upsilon_{k} =1,\forall k$.
Firstly, we illustrate the effectiveness of the proposed channel estimation algorithm over non-sparse mmWave channels.
For sparse single-path channels, the achievable rate of the proposed algorithm matches with the algorithm proposed in \cite{Alkhateeb2015}.
For non-sparse mmWave channels, e.g. with the number of multi-paths $N_{\mathrm{l}}=8$, we observe that the proposed algorithm achieves a better system performance than that of the algorithm proposed in \cite{Alkhateeb2015}.
The reasons are two-fold. Firstly, the proposed algorithm can effectively reduce the MU interference via the proposed analog beamformers and digital ZF precoder.
In particular, the analog beamformers adopted at the BS and the users allow transmission and receiving along the strongest AoA directions, respectively.
Therefore, the interference outside the strongest AoA directions is reduced.
Secondly, the digital ZF precoder designed based on the equivalent channels can remove the MU interference within the strongest AoAs.
We note that since the analog beamforming only allows the transmission along the strongest AoA directions, the proposed scheme can utilize the transmission power more efficiently.
In contrast, the algorithm proposed in \cite{Alkhateeb2015}, which aims to maximize the desired signal energy, does not suppress the MU interference as effective as our proposed algorithm.
Furthermore, Figure \ref{fig:Comp_HB_ALH_ours} also illustrates that a significant achievable rate gain can be brought by the proposed channel estimation and the digital ZF precoding over a simple analog-only beamforming steering scheme.
\begin{figure}[t]
\centering\vspace{-0mm}
\includegraphics[width=4.5in,]{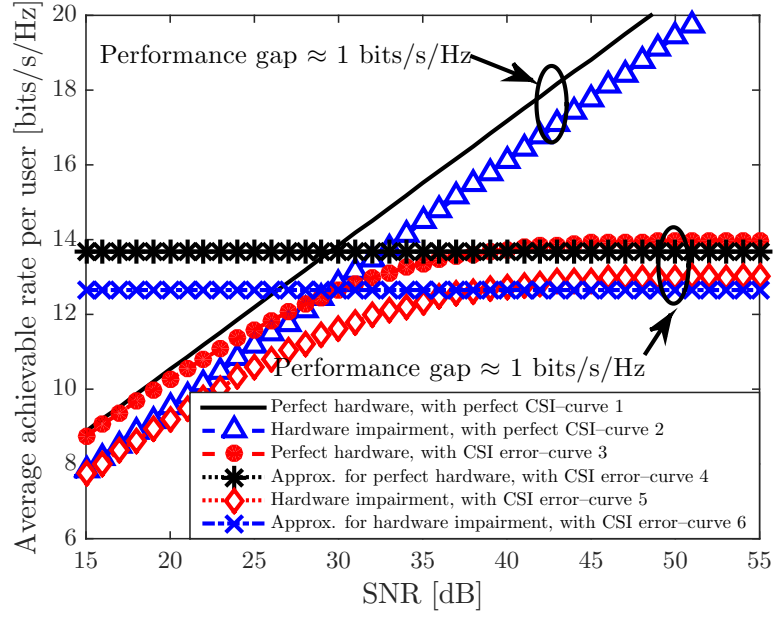}\vspace{-0mm}
\caption{The average achievable rate per user [bits/s/Hz] versus SNR for the hybrid system.}
\label{fig:HWI_CSIerror}\vspace{-0mm}
\end{figure}

In Figure \ref{fig:HWI_CSIerror}, we illustrate two sets of comparisons to validate our derived results in Equations (\ref{QHI}) and (\ref{Coro_RZFK1}), and to show the impact of hardware impairments on system performance.
In the simulations, we set the Rician K-factor as $\upsilon_{k} =2, \forall k$, the number of antennas equipped at the BS is $M=100$, the number of antennas equipped at each user is $P=8$, and the number of users is $N=8$.
We assume that the maximum phase error value of phase shifters is $\sigma _{\Delta }=3$ degrees and the variance of AoA estimation errors at the BS side is $\varrho _{\mathrm{BS}}^{2}= \dfrac{1.782}{2M}$.
In this case, the equivalent coefficient is $\widehat{\xi} \approx 0.5$, which predicts that there is a $3$-dB loss in SNR caused by AoA estimation errors and phase errors.
First, we observe that even with transceiver beamforming errors and random phase errors, the achievable rate of the proposed scheme (curve $2$) scales with increasing SNR and is unbounded by above.
The small gap between the achievable rate per user with perfect hardware (curve $1$) and the achievable rate per user with hardware impairment (curve $2$) confirms the robustness of the proposed scheme against random phase errors as well as transceiver beamforming errors, which is predicted by Equation (\ref{QHI}).
In addition, the $3$ dB of extra power consumption caused by hardware impairments is also verified via the comparison between curves $1$ and $2$.
Figure \ref{fig:HWI_CSIerror} also illustrates a comparison between the proposed scheme with imperfect equivalent CSI under perfect hardware (curve
$3$) and the proposed scheme with imperfect CSI under random phase errors and transceiver beamforming errors (curve $5$), which validates the correctness of Theorem \ref{thm:Theo_32}.
These two simulation curves assume an identical normalized CSI error variance $\delta^{2}$.
In particular, we set the NMSE of CSI as $\delta^{2}=0.005$, which is achieved by the proposed channel estimation scheme as shown in Figure \ref{fig:antiPC}.
In the high receive SNR and large numbers of antennas regimes, curve $3$ and curve $5$ can be asymptotically approximated by curve $4$ and curve $6$, respectively.
Interestingly, the performance gaps between curve $3$ and curve $5$ as well as curve $1$ and curve $2$ are approximately $\Delta \mathrm{Gap}\approx \log _{2}\left( \dfrac{1}{\widehat{\xi} }\right)= 1$ bits/s/Hz, which are accurately characterized by Equation (\ref{Coro_GAPforHI}).

\begin{figure}[t]
\centering\vspace{-0mm}
\includegraphics[width=4.5in]{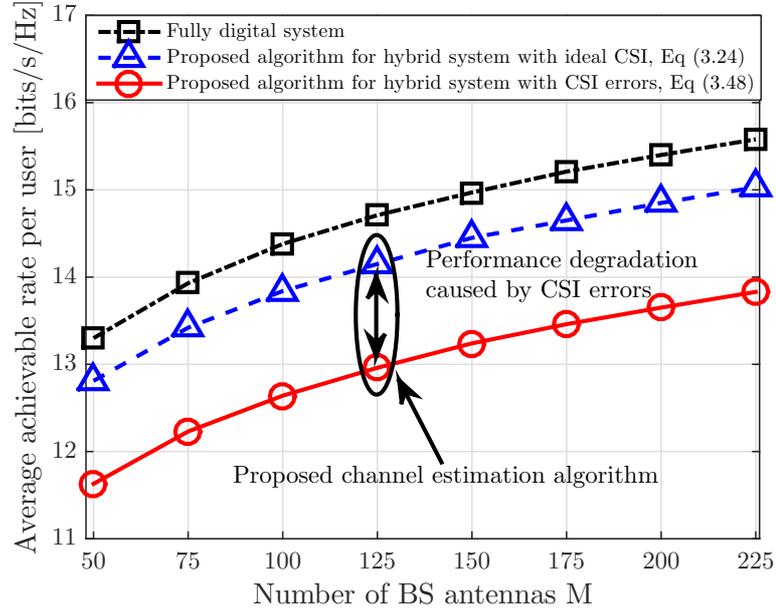}
\caption{Average achievable rate [bits/s/Hz] versus different number of BS
antennas for different structures with $P=8$, the number of users $N=8$, the
normalized CSI $\mathrm{NMSE}=0.005$, $\mathrm{SNR}=30$ dB, and the Rician
K-factor of $2$.}
\label{Fig004}
\end{figure}
Now, we evaluate the achievable rate versus the number of antennas in Figures \ref{Fig004} and \ref{Fig005}.
\begin{figure}[t]
\centering\vspace{-0mm}
\includegraphics[width=4.5in]{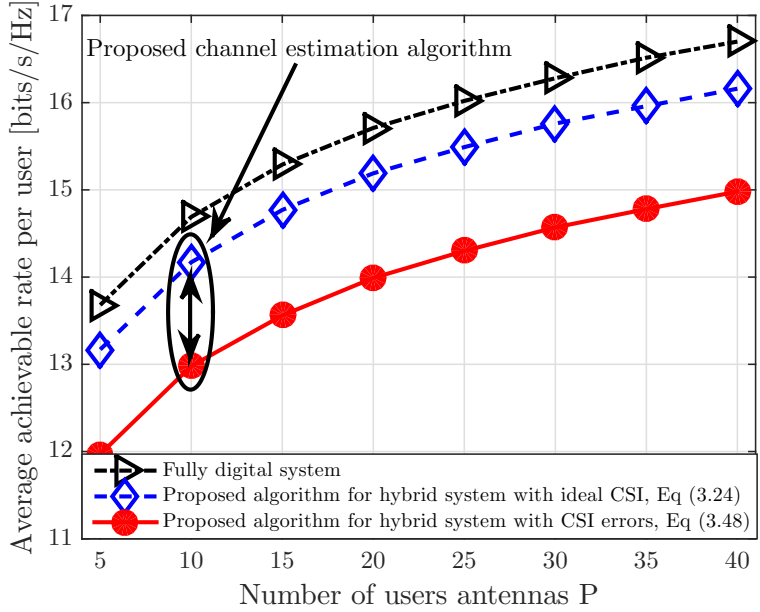}
\caption{ Average achievable rate [bits/s/Hz] versus different number of users antennas for different structures with $M=100$, the number of users $N=8$, the
normalized CSI $\mathrm{NMSE}=0.005$, $\mathrm{SNR}=30$ dB, and the Rician
K-factor of $2$.}
\label{Fig005}
\end{figure}
The setup in Figure \ref{Fig004} is considered at receive $\mathrm{SNR}=30$ dB and the number of antennas equipped at each user is $P=8$, with different numbers of BS antennas.
Figure \ref{Fig004} shows that the achievable rate performance of the proposed digital ZF precoding increases with increasing numbers of antennas at the BS, despite the existence of CSI estimation errors.
This is mainly due to the fact that equipping more antennas at the BS and the users can lead to higher array gains.
It can be observed that the achievable rate of the proposed system scales with the number of BS antennas with a similar slope as the fully digital system, which shows the effectiveness of the proposed scheme for the hybrid system in exploiting the spatial degrees of freedom.
In Figure \ref{Fig005}, the SNR setup and $M=100$ are the same as Figure \ref{Fig004} but with different numbers of users antennas.
Similar phenomena are observed as in Figure \ref{Fig004}.
With the same value of $MP$ and identical NMSE $\delta^{2}$, it is shown in Figure \ref{Fig004} and Figure \ref{Fig005} that the achievable rate performance degradations (the gap between the dash lines and the solid lines) caused by CSI errors are identical.

\section{Summary}

In this chapter, we proposed a low-complexity mmWave channel estimation algorithm exploiting the strongest AoA for the MU hybrid mmWave systems, which is applicable for both sparse and non-sparse mmWave channel environments.
The NMSE performance of the proposed channel estimation was analyzed and verified via numerical simulation.
The achievable rate performance of designed analog beamforming and digital ZF precoding based on the proposed channel estimation scheme was derived and compared to that of the fully digital system.
The analytical and simulation results indicated that the proposed scheme can approach the rate performance achieved by the fully digital system with sufficiently large Rician K-factors.
By taking into consideration of the effects of random phase errors, transceiver beamforming errors, and CSI errors in the rate performance analysis, closed-form approximation of achievable rate in the high SNR regime was derived and verified via simulation.
Our results showed that the proposed scheme is robust against random phase errors and transceiver beamforming errors.

\chapter{Multi-cell Hybrid mmWave MIMO Systems: {Pilot Contamination and Interference Mitigation}}\label{C4:chapter4}

\nomenclature{$a$}{The number of angels per unit area}%
\nomenclature{$N$}{The number of angels per needle point}%
\nomenclature{$A$}{The area of the needle point}%

\ifpdf
    \graphicspath{{1_introduction/figures/PNG/}{1_introduction/figures/PDF/}{1_introduction/figures/}}
\else
    \graphicspath{{1_introduction/figures/EPS/}{1_introduction/figures/}}
\fi
\section{Introduction}
In this chapter, we consider a multi-cell MU hybrid mmWave system.
In the previous chapter, we propose and detail a novel non-feedback non-iterative channel estimation algorithm which is applicable to both non-sparse and sparse mmWave channels.
{\color{black}In particular, we apply the non-feedback TDD-based mmWave channel estimation algorithm proposed in Chapter $3$ and \cite{Zhao2017} to the considered multi-cell scenario and study the corresponding performance of channel estimation under the impact of pilot contamination.}
In addition, we analyze the downlink achievable rate performance of the hybrid mmWave system by {\color{black}taking into account} the CSI errors caused by pilot contamination and inter-cell interferences originated from the BSs in neighboring cells.
{\color{black}
Note that, for pilot contamination mitigation, this chapter does not require any information of covariance matrices of pilot-sharing users in neighbouring cells as required by the multi-cell MMSE-based precoding algorithm proposed in \cite{Bjornson2017a}.
In addition, the algorithm adopted in this chapter can work well while mean values of cross-cell channels from pilot-sharing users in neighbouring cells to the desired BS are non-zero.}
Our main contributions are summarized as follows:

\begin{itemize}

\item We study the impact of pilot contamination on the uplink mmWave channel estimation due to the reuse of orthogonal pilot symbols among different cells.
Our results reveal that in the phase of channel estimation, the receive analog beamforming matrix adopted at the desired BS forms a spatial filter which blocks the signal reception of the undesired pilot symbols from neighboring cells.
In particular, with an increasing number of antennas equipped at each RF chain, the main lobe beamwidth of the spatial filter, which aligns to the strongest AoA direction, becomes narrower and the amplitude of sidelobes becomes lower.
Thus, the impact of pilot contamination caused by the users outside strongest AoA directions can be mitigated.
We mathematically prove that the normalized MSE performance of the channel estimation algorithm improves proportionally with the increasing number of antennas equipped at each RF chain, which is different from previous results in \cite{Marzetta2010,Jose2011}.

\item We adopt ZF precoding for the downlink transmission based on the estimated CSI.
Taking into account the impact of pilot contamination and the inter-cell interference from the neighboring BSs, we analyze and derive the closed-form approximation of the average achievable rate performance in the large number of antennas regimes.
In addition, we show that the achievable rate scales with $M$ in the multi-cell scenario, where $M$ is the number of antennas equipped at each RF chain.
We derive a scaling law of the achievable rate per users, $\log_{2}{M}$.
In contrast, the average achievable rate of fully digital systems adopting the conventional LS channel estimation algorithm suffers from the severe impact of pilot contamination.
All the derived analytical results are verified via simulations.


\end{itemize}

The rest of the chapter is organized as follows. Section 4.2 describes the system
model considered throughout the chapter. In Section 4.3, we detail the uplink channel estimation under the pilot contamination. In Section 4.4, we investigate the downlink transmission rate performance under pilot contamination. Numerical results and related discussions are presented in Section 4.5. Finally, Section 4.6 summarizes the chapter.
\vspace{-2mm}
\section{System Model}
In this chapter, a multi-cell MU hybrid subarray mmWave system is considered. The system consists of $L$ neighboring cells and there are one BS and $N$ users in each cell, cf. Figure $4.1$.
The BS in each cell is equipped with $N_{\mathrm{RF}}$ RF chains serving the $N$ users simultaneously.
We assume that each RF chain equipped at the BS can access to a  ULA with $M$ antennas by using $M$ phase shifters.
Besides, each user is equipped with one RF chain and a $P$-antenna array.
In addition, we focus on $M\geqslant N_{\mathrm{RF}}$, which exploits a large antenna array gain with limited number of RF chains.

{\color{black}
Work \cite{Sohrabi2016} proved that when the number of RF chains equipped at the BS is more than twice of  the number of users, the hybrid systems can realize the rate performance of the fully digital systems.
In general, for the case of $N_{\mathrm{RF}} = N$, the hybrid system can provide enough degrees of freedom to serve $N$ users simultaneously.
If the BS can equip with more RF chains, more degrees of freedom can be exploited for the design of analog beamformers and digital precoders.
In addition, the extra RF chains adopted by the BS can be used for user scheduling design and multipath diversity exploitation.
Thus, for the case that $N_{\mathrm{RF}}>N$, hybrid systems can achieve higher rate performance than that of $N_{\mathrm{RF}}=N$.
In addition, authors of \cite{Zhang2005,Zhang2014aa,Bogale2016} employed RF chains with sets of digitally controlled phase paired phase-shifters and switches to achieve the same rate performance as that of the fully digital system.
Furthermore, with the support of an advanced hybrid architecture \cite{Zhang2005,Zhang2014aa,Bogale2016}, hybrid systems can achieve higher rate performance than that of adopting a simple ULA antenna array case.

However, how to exploit extra RF chains as well as advanced hybrid architectures to further improve the rate performance of hybrid systems is beyond the scope of this chapter.
We may consider the case of $N_{\mathrm{RF}} > N$ and the case of adopting advanced hybrid architectures for channel estimation and downlink transmission precoding in our future works.
To simplify the analysis in the following sections, without loss of generality, we set $N_{\mathrm{RF}} = N$ and each cell has the same number of RF chains equipped at the BS.}

According to the widely adopted setting for multi-cell TDD in uplink channel estimation and downlink data transmission, we assume that the users and the BSs in all cells are fully synchronized in time \cite{Marzetta2010,Jose2011}.
We denote $\mathbf{H}_{k}\in\mathbb{C}^{M\times P}$ as the uplink channel matrix between the desired BS and user $k$ in the desired cell.
Besides, $\mathbf{H}_{k}$ is a narrowband slow time-varying block fading channel.
Recent field tests show that both strongest AoA components and non-negligible scattering components may exist in mmWave propagation channels \cite{Rappaport2015,Hur2016}, especially in the urban areas.
In this chapter, without loss of generality, $\mathbf{H}_{k}$, consists of a strongest AoA component $\mathbf{H}_{\mathrm{SAoA},k}\in\mathbb{C}^{M\times P}$ and $N_{\mathrm{cl}}$ scattering clusters/paths $\mathbf{H}_{\mathrm{S,}k,i}\in\mathbb{C}^{M\times P}$, $i \in \{1,\cdots,N_{\mathrm{cl}}\}$ \cite{Saleh1987,Raghavan2016a,Raghavan2017a,Hur2014,Dai2016}, which can be expressed as\vspace*{-0mm}
\vspace{-1mm}
\begin{align}
\vspace*{-0mm}
\mathbf{H}_{k}=\sqrt{\frac{\varpi_{{k}}}{|\alpha_{k,0}|^2+{\frac{1}{N_{\mathrm{cl}}}}\overset{N_{\mathrm{cl}}}{\underset{i=1}{\sum }}|\alpha_{k,i}|^2}}\left[\underset{\mathrm{Strongest\text{\ }AoA}}{\underbrace{\alpha_{k,0}\mathbf{H}_{\mathrm{SAoA,}k}}}+\underset{\mathrm{Scattering\text{\ }components}}{\underbrace{\sqrt{\tfrac{1}{{N_{\mathrm{cl}}}}} \overset{N_{\mathrm{cl}}}{\underset{i=1}{\sum }}{\alpha _{k,i}}\mathbf{H}_{\mathrm{S,}k,i}}}\right],\label{Eq:2}\vspace{-1mm}
\end{align}\vspace*{-0mm}%
where $\varpi_{{k}}$ accounts for the corresponding large-scale path loss, $\alpha _{k,0}$ is the complex gain corresponding to the strongest AoA component and $\alpha_{k,i}\sim\mathcal{CN}\left(0,1\right)$, $i \in \{1,\cdots,N_{\mathrm{cl}}\}$ represents the complex gain corresponding to the $i$-th cluster.
We can also assume that $\{\alpha_{k,i}\}$ are in non-increasing order, i.e., $|\alpha_{k,0}|\geq |\alpha_{k,1}|\geq \cdots \geq |\alpha_{k,N_{\mathrm{cl}}}|$.
We note here, the strongest AoAs can be either from the LOS components or the NLOS components.
\begin{figure}[t]\vspace{-0mm}
\centering
\includegraphics[width=4.5in]{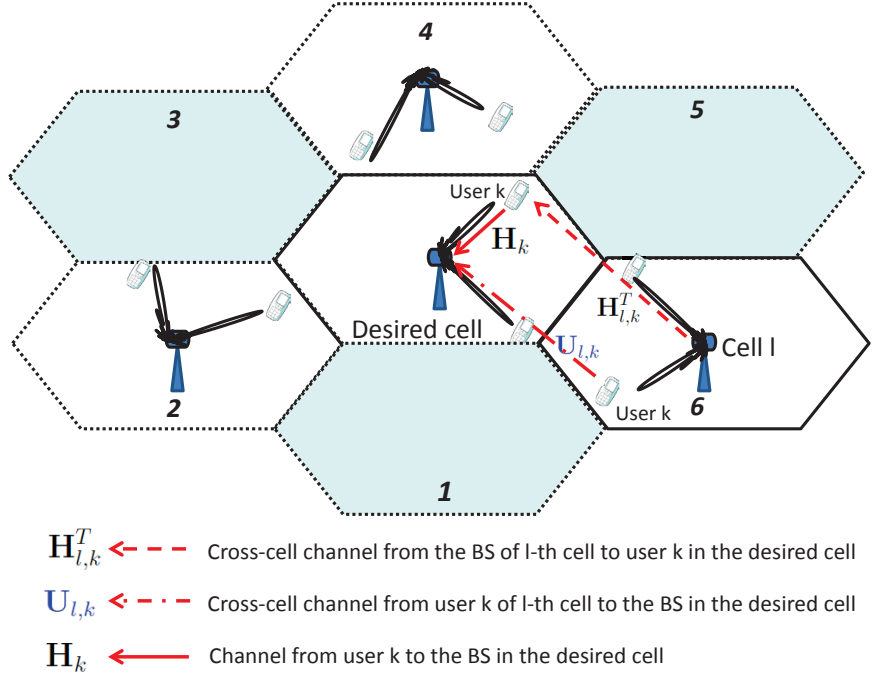}
\vspace{-0mm}
\centering
\caption{A multi-cell MU mmWave cellular system with $L=6$ neighboring cells.}
\label{fig:MCM}\vspace{-0mm}
\end{figure}
Then, the strongest AoA component of user $k$ in the desired cell, $\mathbf{H}_{\mathrm{SAoA},k}$, can be expressed as \cite{book:wireless_comm}\vspace*{-0.0mm}
\begin{equation}
\mathbf{H}_{\mathrm{SAoA,}k}=\mathbf{h}_{\mathrm{SAoA,}k}^{\mathrm{BS}}\mathbf{h}_{%
\mathrm{SAoA,}k}^{H},\label{Eq:3}
\end{equation}\vspace*{-0.0mm}%
where $\mathbf{h}_{\mathrm{SAoA},k}^{\mathrm{BS}}$ $\in\mathbb{C}^{M\times 1}$ and $\mathbf{h}_{\mathrm{SAoA,}k}$ $\in\mathbb{C}^{P\times 1}$ are the antenna array response vectors of the BS and user $k$, respectively.
In particular, $\mathbf{h}_{\mathrm{SAoA,}k}^{\mathrm{BS}}$ and $\mathbf{h}_{\mathrm{SAoA,}k}$ can be expressed as \cite{book:wireless_comm}\vspace*{-0mm}
\begin{align}
\mathbf{h}_{\mathrm{SAoA},k}^{\mathrm{BS}}& =\left[
\begin{array}{cccc}
1,  e^{-j2\pi \tfrac{d}{\lambda }\cos \left( \theta _{k}\right) },  \ldots
,e^{-j2\pi \left( M-1\right) \tfrac{d}{\lambda }\cos \left(
\theta _{k}\right) }%
\end{array}\label{Eq:4}
\right] ^{T}\text{ and} \\
\mathbf{h}_{\mathrm{SAoA},k}& =\left[
\begin{array}{cccc}
1,  e^{-j2\pi \tfrac{d}{\lambda }\cos \left( \phi _{k}\right) },  \ldots ,
e^{-j2\pi \left( P-1\right) \tfrac{d}{\lambda }\cos \left( \phi
_{k}\right) }%
\end{array}%
\right] ^{T}, \label{Eq:5}
\end{align}\vspace*{-0mm}%
respectively, where $d$ is the distance between the neighboring antennas at the BS and users and $\lambda $ is the wavelength of the carrier frequency.
Variables $\theta _{k}\in \left[ 0,\text{\ }\pi \right]$ and $\phi _{k}\in \left[ 0,\text{\ }\pi \right] $ are the angle of incidence of the strongest path at antenna arrays of the desired BS and user $k$, respectively.
\color{black}As commonly adopted in the literature \cite{book:wireless_comm}, we set $d=\frac{\lambda }{2}$ for convenience.
Similarly, the scattering components of user $k$ in the desired cell, $\mathbf{H}_{\mathrm{S,}k}$, can be expressed as
\begin{equation}\label{eqn:NLOS_channel} \mathbf{H}_{\mathrm{S,}k}=\sqrt{\tfrac{1}{{N_{\mathrm{cl}}}}}\overset{N_{\mathrm{cl}}}{\underset{i=1}{\sum }}{\alpha _{k,i}}\mathbf{H}_{\mathrm{S,}k,i}=\sqrt{\tfrac{1}{{N_{\mathrm{cl}}}}}\overset{N_{\mathrm{cl}}}{\underset{i=1}{\sum }}{\alpha _{k,i}}\mathbf{h}_{k,i}^{\mathrm{BS}}\mathbf{h}_{k,i}^{H},
\end{equation}
where $\mathbf{h}_{k,i}^{\mathrm{BS}}\in\mathbb{C}^{M\times 1}$ and $\mathbf{h}_{k,i}\in\mathbb{C}^{P\times 1}$ are the antenna array response vectors of the BS and user $k$ associated to the $i$-th propagation path, respectively.
By introducing the power ratio of the strongest cluster power over the other scattered paths of user $k$\footnote{{\color{black}According to field test results summarized in Table I of \cite{Hur2014} in the case of LOS and NLOS environments, the power ratio of the strongest cluster power over the other scattered paths' power, $\varsigma _{k}$, is larger than $1$.}}, $\varsigma_{k}=\frac{|\alpha_{k,0}|^2}{{\frac{1}{N_{\mathrm{cl}}}}\overset{N_{\mathrm{cl}}}{\underset{i=1}{\sum }}|\alpha_{k,i}|^2}$, we can rewrite Equation (\ref{Eq:2}) as\vspace{-2mm}
\begin{equation}
\mathbf{H}_{k}= \sqrt{\varpi_{{k}}}\left[ \underset{\mathrm{Strongest\text{\ }AoA}}{\underbrace{\sqrt{\frac{\varsigma _{k}}{\varsigma _{k}+1}}\mathbf{h}_{\mathrm{SAoA},k}^{\mathrm{BS}}\mathbf{h}_{\mathrm{SAoA},k}^{H} }}+\underset{\mathrm{Scattering\text{\ }components}}{\underbrace{ \sqrt{\frac{1}{\varsigma _{k}+1}}\sqrt{\tfrac{1}{{N_{\mathrm{cl}}}}} \overset{N_{\mathrm{cl}}}{\underset{i=1}{\sum }}{\alpha _{k,i}}\mathbf{h}_{k,i}^{\mathrm{BS}}\mathbf{h}_{k,i}^{H}}}\right].\vspace{-2mm}
\end{equation}}%
With the increasing number of clusters, the path attenuation coefficients and the AoAs between the users and the BS behave randomly \cite{Hur2016}.
Let {\color{black}$\mathbf{U}_{l,k}\in\mathbb{C}^{M\times P}$} be the inter-cell mmWave uplink channel between user $k$ in the $l$-th neighboring cell and the desired BS, $l\in \left\{ 1,\ldots, L\right\}$, cf. Figure \ref{fig:MCM}.
Let ${\mathbf{H}}^{T}_{l,k}\in\mathbb{C}^{P\times M}$ be the inter-cell downlink mmWave channel between the BS of the $l$-th neighboring cell and user $k$ in the desired cell, cf. Figure \ref{fig:MCM}.
Since the inter-cell distance is shortened in small-cell systems, the inter-cell channels usually contain the strongest AoA components.
{\color{black}
Thus, the inter-cell uplink and downlink mmWave channels can be expressed as
\begin{align}
\mathbf{U}_{l,k}&=\sqrt{\overline{{\varpi}}_{l,{k}}}\underset{\mathrm{Strongest}\text{\ }\mathrm{AoA}}{\underbrace{\sqrt{\frac{\varsigma _{l,k}}{\varsigma _{l,k}+1}}{\mathbf{U}}_{\mathrm{SAoA,}l,k}}}+\sqrt{\overline{{\varpi}}_{l,{k}}}\underset{\mathrm{Scattering}\text{\ }\mathrm{component}}{\underbrace{\sqrt{\frac{1}{\varsigma _{l,k}+1}}{\mathbf{U}}_{\mathrm{S,}l,k}}}\text{ and} \\
{\mathbf{H}}^{T}_{l,k}&=\sqrt{\widetilde{\varpi}_{l,{k}}}\underset{\mathrm{Strongest}\text{\ }\mathrm{AoA}}{\underbrace{\sqrt{\frac{\widetilde{\varsigma} _{l,k}}{\widetilde{\varsigma} _{l,k}+1}}\mathbf{H}^{T}_{\mathrm{SAoA,}l,k}}}+\sqrt{\widetilde{\varpi}_{l,{k}}}\underset{\mathrm{Scattering}\text{\ }\mathrm{component}}{\underbrace{\sqrt{\frac{1}{\widetilde{\varsigma} _{l,k}+1}}\mathbf{H}^{T}_{\mathrm{S,}l,k}}}\text{\ \ \ \ \ \ \ ,}
\end{align}
respectively, where $\overline{\varpi}_{l,{k}}$ and $\widetilde{\varpi}_{l,{k}}$ are the corresponding large-scale path loss coefficients.
We note that $\varsigma _{l,k}$ and $\widetilde{\varsigma}_{l,k}$ are the power ratios of the strongest cluster power over the other scattered paths' power for the uplink and downlink inter-cell channels, respectively.
The strongest AoA components ${\mathbf{U}}_{\mathrm{SAoA,}l,k}={\mathbf{u}}_{\mathrm{SAoA,}l,k}^{\mathrm{BS}}{\mathbf{u}}_{%
\mathrm{SAoA,}l,k}^{H}$, and $\mathbf{H}^{T}_{\mathrm{SAoA,}l,k}=\mathbf{h}_{%
\mathrm{SAoA,}l,k}^{\ast}\left(\mathbf{h}_{\mathrm{SAoA,}l,k}^{\mathrm{BS}}\right)^{T}$ follow similar assumptions as in Equations $(\ref{Eq:3})-(\ref{Eq:5})$.
Besides, the scattering components ${\mathbf{U}}_{\mathrm{S,}l,k}$ and $\mathbf{H}^{T}_{\mathrm{S,}l,k}$ follow similar assumption as in Equation $(\ref{eqn:NLOS_channel})$.
}
All these mentioned inter-cell propagation path loss coefficients are related to the propagation distance and modeled as in \cite{Hur2016,Akdeniz2014,Rappaport2015}.
According to recent field measurements, e.g. \cite{Hur2016,Akdeniz2014,Rappaport2015,Al-Daher2012,Eldeen2010}, the typical values of the power ratios of the strongest cluster power over the other scattered paths' power $\varsigma _{l,k}$ and ${\widetilde{\varsigma}} _{l,k}$ for {\color{black}$\mathbf{U}_{l,k}$} and ${\mathbf{H}}^{T}_{l,k}$ are in $(0, \text{\ }5]$, respectively.

\vspace{-0mm}
\section{Multi-cell Uplink Channel Estimation Performance Analysis}

\begin{figure}[t]
\centering\vspace*{-0mm}
\includegraphics[width=4.0in]{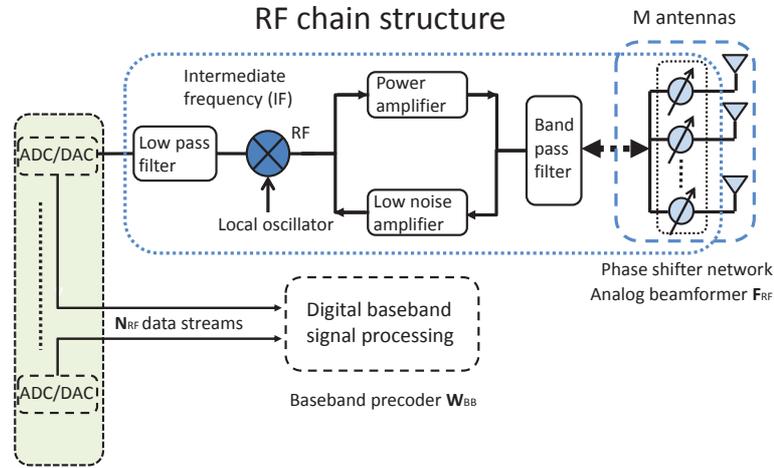}
\vspace{-0mm}
\caption{A block diagram of an RF chain of a subarray antenna structure.}
\label{fig:RF_chain}\vspace{-0mm}
\end{figure}

In this section, we adopt the algorithm proposed in \cite{Zhao2017} for the estimation of an equivalent mmWave channel which comprise the physical mmWave channels and analog beamforming matrices adopted at the desired transceivers.
The proposed algorithm is suitable for both the conventional fully digital systems and the emerging hybrid systems with fully access and subarray implementation structures.
For the sake of presentation, we provide a summary of the algorithm proposed in \cite{Zhao2017} in the following.
For illustration, we adopt the subarray structure as an example, as shown in Figure \ref{fig:RF_chain}.
Specifically, each RF chain can access to the $M$ antennas via a phase shifter network.

There are three steps in the algorithm proposed in \cite{Zhao2017}.
In first and second step, strongest AoAs between the desired BS and the users are estimated.
Then, the desired BS and the users design their analog beamforming matrices by aligning the beamforming direction to the estimated strongest AoAs.
In the third step, orthogonal pilot symbols are transmitted from the users to the BS by using the pre-designed analog beamforming matrices.
Then, by exploiting the channel reciprocity, the equivalent downlink channel can be estimated and {\color{black}adopted} at the BS as the input for the digital baseband precoder.

Note that, orthogonal pilot symbols are used for the estimation of the equivalent channels in the third step.
For the multi-cell scenario, pilot symbols are reused among different cells which results in pilot contamination and cause a severe impact on the equivalent channel estimation performance.
We note that only single-cell scenario was considered in \cite{Zhao2017} and it is unclear if the channel estimation algorithm provides robustness against pilot contamination.
Furthermore, the performance analysis studied in \cite{Zhao2017} does not take into account the impact of potential out-of-cell interference on the performance of channel estimation.
In the following sections, we investigate the impact of pilot contamination on the mmWave channel estimation performance for small-cell scenarios.

\vspace{-2mm}
\subsection{Channel Estimation and Pilot Contamination Analysis}

Basically, step one and step two of the proposed algorithm in \cite{Zhao2017} provide the analog beamforming matrices at the desired BS and the desired users to facilitate the estimation of equivalent channel.
In particular, the analog beamforming matrices pair the desired BS and the users and align the directions of data stream transmission.
Due to the inter-cell large-scale propagation path loss, the impact of multi-cell interference on the design of the analog beamforming matrices is usually negligible\footnote{For strongest AoAs estimation at the desired BS in a multi-cell scenario, the received power of the reused pilot symbols transmitted from the users in neighboring cells is smaller than that of the desired pilot symbols transmitted from the users in the desired cell. In addition, strongest AoAs estimation may not rely on pilot symbols \cite{Zhao2017b}.}.
Besides, simulation results show that the pilot contamination does not affect strongest AoAs estimation result\footnote{The simulation result of the strongest AoAs estimation is omitted here due to the space limitation. However, the impact of pilot contamination on the strongest AoAs estimation will be captured in the final simulation.}.
Therefore, to facilitate the performance analysis of the multi-cell equivalent channel estimation, we assume that strongest AoAs among the users and the BS are perfectly estimated and the desired signals always fall in the main lobe.

{\color{black}Based on the strongest AoAs, which are perfectly estimated at the users and the BS}, the analog receive beamforming vector of user $k$ adopted at the desired BS is given by\vspace{-2mm}
\begin{equation}
\widehat{\bm{{\color{black}\nu}}}^{T}_{k}\in\mathbb{C}^{1\times M} =
\frac{1}{\sqrt{M}}\left[
\begin{array}{ccc}
1,  \ldots , e^{j2\pi \left( M-1\right) \tfrac{d}{\lambda }\cos
\left(\theta_{k}\right)}
\end{array}\right]\vspace{-2mm}
\end{equation}
and the analog transmit beamformer of user $i$ in the desired cell is given by\vspace{-2mm}
\begin{equation}
\widehat{\bm{\omega}}^{\ast}_{i} \in\mathbb{C}^{P\times 1}=
\frac{1}{\sqrt{P}}\left[
\begin{array}{ccc}
1, \ldots , e^{j2\pi \left( P-1\right) \tfrac{d}{\lambda }\cos
\left(\phi_{i}\right) }
\end{array}\right] ^{H}.\vspace{-2mm}
\end{equation}
In addition, we denote the analog beamforming matrix at the desired BS as\vspace{-2mm}
\begin{equation}
\mathbf{F}_{\mathrm{RF}}\in\mathbb{C}^{M\times N}=\left[\begin{array}{ccc}\widehat{\bm{{\color{black}\nu}}}_{1},\ldots, \widehat{\bm{{\color{black}\nu}}}_{N}\end{array}\right].\vspace{-2mm}
\end{equation}
Let $\mathbf{\Phi }_{k}\in\mathbb{C}^{N\times 1}$ denote the pilot symbols of user $k$ in the desired cell.
The pilot symbols for all the $N$ users in the desired cell form a matrix, $\mathbf{\Psi \in\mathbb{C}}^{N\times N}\mathbf{,}$ where $\mathbf{\Phi }_{k}$ is a column vector of
matrix $\mathbf{\Psi }$ given by
$\mathbf{\Psi }=\ \sqrt{E_{\mathrm{P}}}\left[
\begin{array}{ccc}
\mathbf{\Phi }_{1},\ldots,\mathbf{\Phi }_{N}%
\end{array}\right]$, $\mathbf{\Phi }_{i}^{H}\mathbf{\Phi }_{j}=0$, $ i\neq j$, $i,\text{ }j\in \left\{ 1,\ldots, N\right\}$,
where $E_{\mathrm{P}}$ represents the transmitted pilot symbol energy.
In the equivalent channel estimation, user $k$ transmits the pilot symbols $\mathbf{\Phi }_{k}$ via transmit beamformer $\widehat{\bm{\omega}}^{\ast}_{i} $ and the desired BS receives the pilot symbols utilizing the analog beamforming matrix $\mathbf{F}_{\mathrm{RF}}$.
However, the reuse of pilot symbols in neighboring cells affects the performance of equivalent channel estimation.
The received signal of the $k$-th RF chain at the desired BS in the uplink is given by\vspace{-2mm}
\begin{equation}
\widehat{\mathbf{s}}_{k}^{T}  =\widehat{\bm{{\color{black}\nu}}}_{k}^{T}\overset{N}{\underset{i=1}{\sum }}\mathbf{H}_{i}\widehat{\bm{\omega }}_{i}^{\ast }\sqrt{E_{%
\mathrm{P}}}\mathbf{\Phi }_{i}^{T}+\underset{\mathrm{Pilot\text{ \ }contamination}}{\underbrace{\widehat{\bm{{\color{black}\nu}}}_{k}^{T}\overset{L}{\underset{l=1}{\sum }}\overset{N}{\underset{i=1}{\sum }}\left({\color{black}\mathbf{U}_{l,i}}\widehat{\bm{\omega}}^{\ast}_{l,i}\sqrt{E_{\mathrm{P}}}\mathbf{\Phi }_{i}^{T}\right)}}+\widehat{\bm{{\color{black}\nu}}}_{k}^{T}\mathbf{Z},\vspace{-2mm}
\end{equation}
where $\widehat{\bm{\omega}}_{l,i}\in\mathbb{C}^{P\times 1}=
\frac{1}{\sqrt{P}}\left[
\begin{array}{ccc}
1,  \ldots , e^{j2\pi \left( P-1\right) \tfrac{d}{\lambda }\cos
\left(\phi_{l,k}\right) }
\end{array}\right] ^{T}$ is the analog beamforming vector of user $i$ in the $l$-th cell, the entries of noise matrix, $\mathbf{Z}$, are modeled by i.i.d. random variables with distribution $\mathcal{CN}\left( 0,\sigma _{\mathrm{BS}}^{2}\right)$.
To facilitate the investigation of channel estimation and downlink transmission, we assume that long-term power control is performed to compensate the different strongest AoAs' path loss among different desired users in the desired cell.
As a result, it can be considered that large-scale propagation path losses of different users in the desired cell are identical.
Thus, we can express the estimated equivalent downlink channel $\widehat{\mathbf{H}}_{\mathrm{eq}}^{T}\in\mathbb{C}^{N\times N}$ at the desired BS under the impact of pilot contamination as\vspace{-2mm}
\begin{align}
&\widehat{\mathbf{H}}_{\mathrm{eq}}^{T}= \mathbf{B}\left({\mathbf{H}}_{\mathrm{eq}}^{T} + \Delta \widehat{\mathbf{H}}_{\mathrm{eq}}^{T}\right)=  \mathbf{B}\left(\frac{\mathbf{\Psi }^{H}}{E_\mathrm{P}}\left[\begin{array}{ccc}\widehat{\mathbf{s}}_{1} & \ldots & \widehat{\mathbf{s}}_{N}\end{array}\right]\right)\notag\\
=&\mathbf{B}\left(\underset{\mathbf{H}_{\mathrm{eq}}^{T}}{\underbrace{\left[
\begin{array}{c}
\widehat{\bm{\omega }}_{1}^{H}\mathbf{H}_{1}^{T}\mathbf{F}_{\mathrm{RF}%
} \\
\vdots \\
\widehat{\bm{\omega}}_{N}^{H}\mathbf{H}_{N}^{T}\mathbf{F}_{\mathrm{RF}}\end{array}\right]}}+\underset{{\color{black}\mathrm{Effective}}\text{\ }\mathrm{noise}}{\underbrace{\frac{1}{\sqrt{E_{\mathrm{P}}}}\left[
\begin{array}{c}
\mathbf{\Phi}_{1}^{H}\mathbf{Z}^{T}\mathbf{F}_{\mathrm{RF}} \\
\vdots \\
\mathbf{\Phi} _{N}^{H}\mathbf{Z}^{T}\mathbf{F}_{\mathrm{RF}}
\end{array}\right] }} +
\underset{\mathrm{Pilot}\text{\ }\mathrm{contamination}}{\underbrace{\left[
\begin{array}{c}
\overset{L}{\underset{l=1}{\sum }}\left(\widehat{\bm{\omega}}^{H}_{l,1}{\color{black}\mathbf{U}_{l,1}^{T}}\right)\mathbf{F}_{\mathrm{RF}} \\
\vdots \\
\overset{L}{\underset{l=1}{\sum }}\left(\widehat{\bm{\omega}}^{H}_{l,N}{\color{black}\mathbf{U}_{l,N}^{T}}\right)\mathbf{F}_{\mathrm{RF}}
\end{array}\right] }}\right), \label{Eq_eq}\vspace{-4mm}
\end{align}
where $\Delta \widehat{\mathbf{H}}_{\mathrm{eq}}^{T}$ is the equivalent channel estimation error caused by pilot contamination and noise, and the path loss compensation matrix $\mathbf{B}\in\mathbb{C}^{N\times N}$ is given by
\begin{equation}
\mathbf{B}=\left[
\begin{array}{ccc}
\frac{1}{\sqrt{\varpi_{{1}}}} & \cdots &  {0} \\
\vdots & \ddots  & \vdots \\
{0} & \cdots & \frac{1}{\sqrt{\varpi_{{N}}}}%
\end{array}\right].
\end{equation}
In the following, {\color{black}for notational simplicity}, we denote $\widehat{\rho}_{l,k}= \sqrt{\frac{\widehat{\varpi}_{l,{k}}}{\varpi_{{k}}}}$ as the inter-cell propagation path loss coefficients.
Now, to evaluate the impact of pilot contamination, we introduce a theorem which reveals the normalized MSE performance of equivalent channel estimation.
\begin{theo}\label{thm:Theo_1}
The normalized MSE of the equivalent channel estimation with respect to the $k$-th RF chain under the impacts of pilot contamination and noise can be approximated as
\begin{align}
\mathrm{NMSE}_{\mathrm{eq,}k} =&\frac{1}{N}
\mathbb{E}_{{\color{black}\mathrm{\mathbf{U}_{l,i}}}}\left[ \left( \dfrac{1}{\sqrt{MP}}\Delta\widehat{\mathbf{h}}_{\mathrm{eq},k}^{T}\right)\left(\dfrac{1}{\sqrt{MP}}
\Delta\widehat{\mathbf{h}}_{\mathrm{eq},k}^{\ast}\right) \right]   \notag \\
\approx & \frac{1}{NMP}\left[\overset{L}{\underset{l=1}{\sum }}{\frac{\widehat{\rho} _{l,k}^{2}{\color{black}\varsigma}_{l,k}}{{\color{black}\varsigma}_{l,k}+1}}N\right]+\frac{1}{MP}\overset{L}{\underset{l=1}{\sum }}{\frac{\widehat{\rho}_{l,k}^{2}}{{\color{black}\varsigma}_{l,k}+1}}+\frac{\sigma _{\mathrm{BS}}^{2}\mathrm{tr}\left[ \mathbf{F}_{\mathrm{RF}%
}^{H}\mathbf{F}_{\mathrm{RF}}\right] }{{\varpi_{{k}}}E_{\mathrm{P}}NMP}\notag \\
= & \underset{\mathrm{Multi-cell\text{ \ }pilot\text{ \ }contamination}}{\underbrace{\frac{1}{MP}\overset{L}{\underset{l=1}{\sum }}\left( \widehat{\rho} _{l,k}^{2} \right)}}+\underset{\mathrm{Noise}}{\underbrace{\frac{\sigma_{\mathrm{BS}}^{2}}{{\varpi_{{k}}}E_{\mathrm{P}}MP}}}\text{\ } . \label{PC}
\end{align}%
In particular, when the number of antennas equipped at the desired BS and the users are sufficiently large, we have\vspace{-2mm}
\begin{equation}
\underset{M,P\rightarrow\infty}{\lim}\mathrm{NMSE}_{\mathrm{eq,}k}\approx  0 \text{\ \ \ \ }  .\vspace{-2mm}
\end{equation}
\end{theo}
\emph{\quad Proof: } Please refer to Appendix A.\QEDA

In Equation (\ref{PC}), the impact of the multi-cell pilot contamination term on the normalized MSE performance is inversely proportional to {\color{black}the number of antennas} $M$ and $P$.
In addition, the noise term decreases with the increasing transmit pilot symbol energy and the number of antennas, $M$ and $P$.
{\color{black}
It is important to note that, the impact of noise on channel estimation will vanish in the high SNR regime, e.g. $E_{p}\gg 1 $.
However, the impact of pilot contamination on the MSE performance cannot be mitigated by simply increasing the transmit pilot symbol energy $E_{p}$.
  }

It is known that the conventional massive MIMO pilot-aided LS channel estimation performance under the impact of pilot contamination cannot be improved by increasing the number of antennas equipped at the BS \cite{Marzetta2010,Jose2011}.
Interestingly, the result of Theorem $\ref{thm:Theo_1}$ unveils that the impacts of pilot contamination and noise on the equivalent channel estimation will vanish  asymptotically with the increasing number of antennas equipped at each RF chain, $M$ and $P$.
Actually, the numbers of antennas $M$ and $P$ have an identical effect on the normalized MSE performance.

This is because the direction of analog beamforming matrices adopted at the desired BS and the desired users align with the strongest AoA path.
Hence, the analog beamforming matrices adopted at the desired BS and the users form a pair of spatial filters which block the pilot signals from undesired users to the desired BS via non-strongest paths.
In addition, transmitting the pilot signals from the desired users via the analog beamforming matrix can reduce the potential energy leakage to other undesired cells, which further reduces the impact of pilot contamination.


In order to help the readers to understand Theorem \ref{thm:Theo_1} from the physical point of view, we provide a detail discussion in this subsection on the result of Equation (\ref{PC}) via an illustrative example of a single-antenna user in the high SNR regime.

In the single-antenna user scenario, we focus on the normalized estimated equivalent channel of user $k$, $\widehat{\mathbf{h}}_{\mathrm{eq,}k}^{T}\in\mathbb{C}^{1\times N}$, in the high SNR regime which is given by\vspace{-3mm}
\begin{equation}
\frac{1}{{\sqrt{M}}}\widehat{\mathbf{h}}_{\mathrm{eq,}k}^{T}
=\frac{1}{{\sqrt{M}}}\left[\mathbf{h}_{k}^{T}\underset{\mathrm{\mathbf{F}_{RF}}}{\underbrace{\left[
\begin{array}{ccc}
\widehat{\bm{{\color{black}\nu}}}_{1} & \ldots & \widehat{\bm{{\color{black}\nu}}}_{N}\end{array}\right]}} +\underset{\mathrm{Pilot\text{ \ }contamination}}{\underbrace{\overset{L}{\underset{l=1}{\sum }}{\color{black}\mathbf{u}_{l,k}^{T}}\left[
\begin{array}{ccc}
\widehat{\bm{{\color{black}\nu}}}_{1} & \ldots & \widehat{\bm{{\color{black}\nu}}}_{N}\end{array}\right] }}\right]. \label{Eq_ILS}\vspace{-3mm}
\end{equation}%
From Equation (\ref{Eq_ILS}), it is shown that the equivalent channel $\widehat{\mathbf{h}}_{\mathrm{eq,}k}^{T}$ is formed by projecting the actual mmWave channel $\mathbf{h}_{k}$ on the analog beamforming matrix $\mathbf{F}_{\mathrm{RF}}$.
Then, $\eta_{k,i}\in\mathbb{C}^{1\times 1}$, the $i$-th entry of the estimated equivalent channel vector $\widehat{\mathbf{h}}_{\mathrm{eq,}k}^{T}$, representing the projection of channel of user $k$ in the desired cell on the $i$-th strongest AoA direction, is given by\vspace{-3mm}
\begin{align}
\hspace{-0mm}
\eta_{k,i}= & \frac{1}{{\sqrt{M}}}\mathbf{h}_{k}^{T}\widehat{\bm{{\color{black}\nu}}}_{i} + \frac{1}{\sqrt{{M}}}\overset{L}{\underset{l=1}{\sum }}{\color{black}\mathbf{u}_{l,k}^{T}}\widehat{\bm{{\color{black}\nu}}}_{i} \notag\\
= & \underset{\mathrm{Strongest\text{ \ }AoA}}{\underbrace{\sqrt{\frac{{\color{black}\varsigma}_{k}}{{\color{black}\varsigma}_{k}+1}}\frac{\sin\left[ M \pi\frac{d}{\lambda}\left( \cos \left(\theta_{k}\right)-\cos\left(\theta_{i}\right)   \right)\right]}{{M}\sin\left[\pi\frac{d}{\lambda}\left( \cos \left(\theta_{k}\right)-\cos\left(\theta_{i}\right) \right)\right]}}}+ \underset{\mathrm{Scattering\text{ \ }component}}{\underbrace{ \sqrt{\frac{1}{{\color{black}\varsigma}_{k}+1}}\frac{1}{\sqrt{{M}}}\mathbf{h}_{\mathrm{S},k}^{T}\widehat{\bm{{\color{black}\nu}}}_{i}}}    \notag\\
+ & \underset{\mathrm{Pilot\text{ \ }contamination}}{\underbrace{\overset{L}{\underset{l=1}{\sum }}\sqrt{\frac{{\color{black}\varsigma}_{l,k}}{{\color{black}\varsigma}_{l,k}+1}}\frac{\sin\left[ M \pi\frac{d}{\lambda}\left( \cos \left({\theta}_{l,k}\right)-\cos\left(\theta_{i}\right)   \right)\right]}{{M}\sin\left[\pi\frac{d}{\lambda}\left( \cos \left({\theta}_{l,k}\right)-\cos\left(\theta_{i}\right) \right)\right]}+\overset{L}{\underset{l=1}{\sum }}\sqrt{\frac{1}{{\color{black}\varsigma}_{l,k}+1}}\frac{1}{\sqrt{{M}}}{\color{black}\mathbf{u}_{\mathrm{S},l,k}^{T}}\widehat{\bm{{\color{black}\nu}}}_{i}}},\label{Eq_VC}
\vspace{-3mm}\end{align}
where ${\theta}_{l,k}$ is the AoA from user $k$ in the $l$-th cell to the desired BS, $\mathbf{h}_{\mathrm{S},k}^{T}$ is the scattering component of the channel of user $k$ in the desired cell, and ${\color{black}\mathbf{u}_{\mathrm{S},l,k}^{T}}$ is the scattering component of the channel from user $k$ in the $l$-th cell to the desired BS.
For the projection of channel of user $k$ on the $k$-th RF chain's analog beamforming vector $\widehat{\bm{{\color{black}\nu}}}_{k}$, we have $\widehat{\bm{{\color{black}\nu}}}_{k}^{T}=\left(\mathbf{h}_{\mathrm{SAoA},k}^{\mathrm{BS}}\right)^{H}$. 
Thus, the $k$-th entry of the vector $\widehat{\mathbf{h}}_{\mathrm{eq,}k}^{T}$ can be expressed as\vspace{-2mm}
\begin{align}
\eta_{k,k} = \left[\underset{\mathrm{Strongest\text{ \ }AoA}}{\underbrace{\sqrt{\frac{{\color{black}\varsigma}_{k}}{{\color{black}\varsigma}_{k}+1}}}} + \underset{\mathrm{Scattering\text{ \ }component}}{\underbrace{\varepsilon_{k}\frac{1}{{\sqrt{M}}}\sqrt{\frac{1}{{\color{black}\varsigma}_{k}+1}}}} + \underset{\mathrm{Pilot\text{ \ }contamination}}{\underbrace{\frac{1}{{\sqrt{M}}}\mu_{k}}}\right], \label{Eq_SC}\vspace{-2mm}
\end{align}
where $\varepsilon_{k}= \mathbf{h}_{\mathrm{S},k}^{T}\widehat{\bm{{\color{black}\nu}}}_{k}$ denotes the projection of the scattering component $\sqrt{\frac{1}{{\color{black}\varsigma}_{k}+1}}\mathbf{h}_{\mathrm{S},k}$ on $\widehat{\bm{{\color{black}\nu}}}_{k}$ and $\mu_{k}=\overset{L}{\underset{l=1}{\sum }}{\color{black}\mathbf{u}_{l,k}^{T}}\widehat{\bm{{\color{black}\nu}}}_{k}$ denotes the projection of inter-cell channels $\overset{L}{\underset{l=1}{\sum }}{\color{black}\mathbf{u}_{l,k}^{T}}$ on $\widehat{\bm{{\color{black}\nu}}}_{k}$.

\begin{figure}[H]
\centering\vspace*{-0mm}
\subfigure[Channel estimation algorithm explanation via a vector space model.]
{\label{fig:HEQ_ILLU1}
\includegraphics[width=3.3in,]{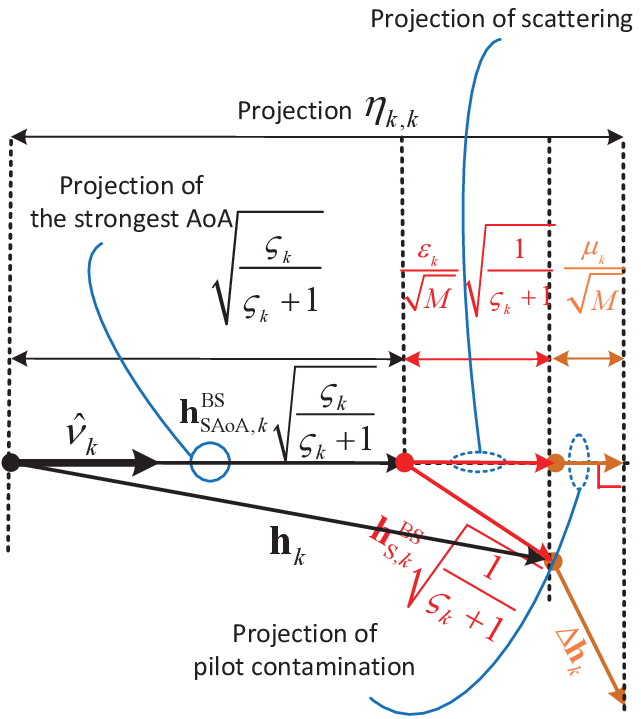}}
\subfigure[Antenna array beam pattern response.]
{\label{fig:HEQ_ILLU2}
\includegraphics[width=4.5in,]{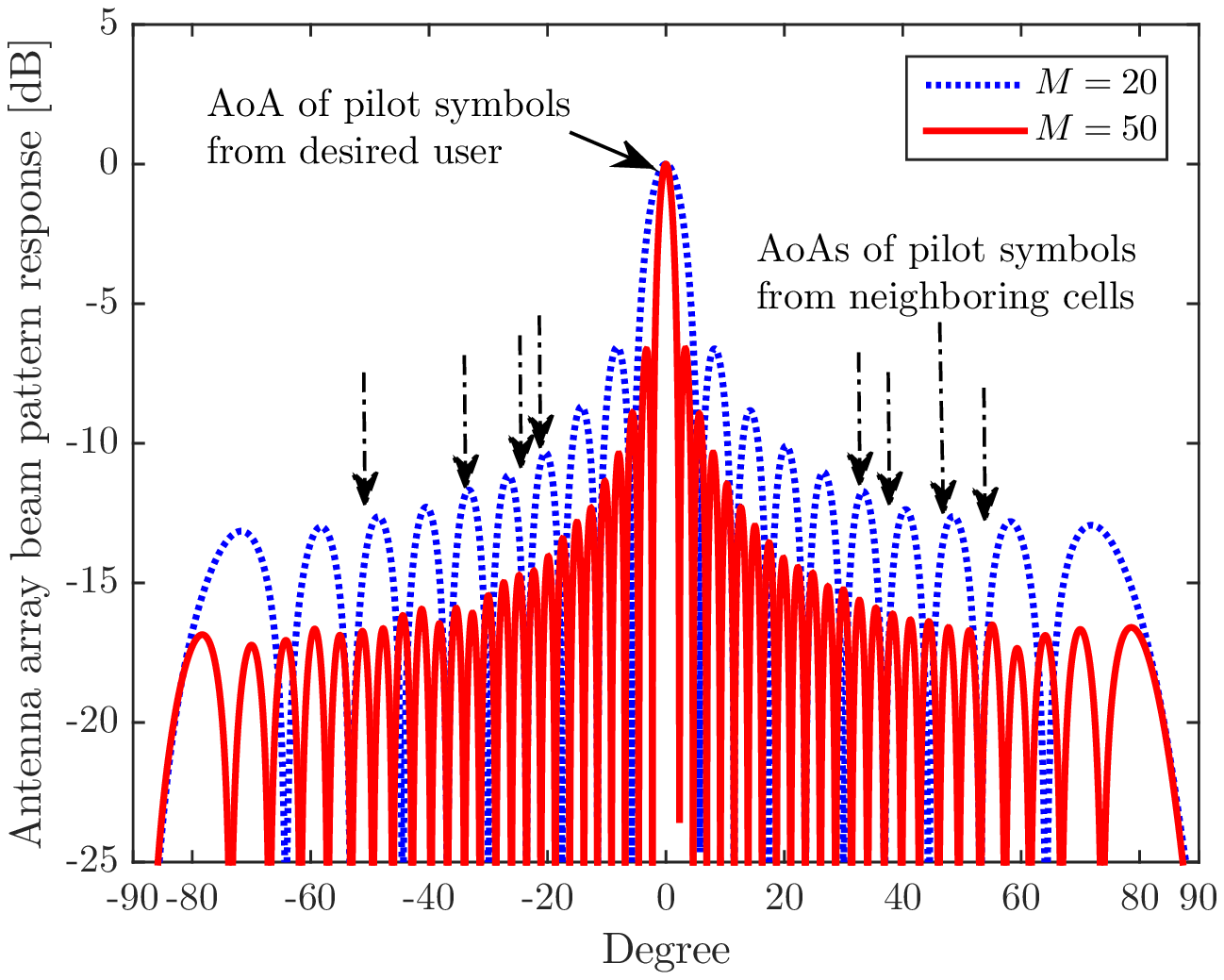}}
\caption{(a) The illustration of equivalent channel estimation under the impact of pilot contamination. (b)  The illustration of sidelobe suppression for different numbers of antennas $M$.}
\label{fig:HBFD}\vspace{-0mm}
\end{figure}

The computation of Equation (\ref{Eq_SC}) can be illustrated graphically via the concept of vector space with projection as shown in Figure \ref{fig:HEQ_ILLU1}.
The first term of Equation (\ref{Eq_SC}), $\sqrt{\frac{{\color{black}\varsigma}_{k}}{{\color{black}\varsigma}_{k}+1}}$, which is the projection of the strongest AoA component on the analog beamforming direction, is a constant and is independent of $M$.
For the second term and the third term of Equation (\ref{Eq_SC}), $\varepsilon_{k}$ and $\mu_{k}$, which consist of multiple paths with various AoAs, cannot enjoy the array gain $M$.
Actually, $\frac{\varepsilon_{k}}{{\sqrt{M}}}\sqrt{\frac{1}{{\color{black}\varsigma}_{k}+1}}$ and $\frac{\mu_{k}}{{\sqrt{M}}}$, these two terms will vanish asymptotically with the increasing number of antennas $M$.

{\color{black}
We note that for the conventional pilot-aided channel estimation algorithms, e.g. LS-based algorithms, they estimate the channels from all the directions.
Thus, BSs adopting these algorithms receiving reused pilot symbols from the undesired users and cannot be distinguished from the desired pilot symbols.
It is known as pilot contamination. }
However, adopting analog beamforming matrices for receiving pilot symbols at the desired BS via the strongest AoA directions forms a spatial filter, which blocks the undesired pilot symbols from neighboring cells via different AoA paths.
Furthermore, the ``blocking capability" improves with the increasing number of antennas equipped at the BS.
Specifically, the beamwidth of main lobe becomes narrower and the magnitude of sidelobes is lower, which is illustrated in Figure \ref{fig:HEQ_ILLU2}.
In fact, this is an important feature for mitigating the impact of pilot contamination.

Therefore, with the equivalent channel estimation proposed in \cite{Zhao2017}, the impact of pilot contamination vanishes asymptotically with an increasing number of antennas $M$ equipped at the desired BS.



\vspace{-3mm}
\section{{\color{black}Multi-cell Downlink Achievable Rate}}
In the last section, we mathematically prove that the impact of pilot contamination on the equivalent channel estimation can be effectively mitigated.
Analytical and simulation results show that CSI errors caused by pilot contamination vanish asymptotically with the increasing numbers of antennas equipped at each RF chain of the BS and the users, $M$ and $P$.

Now, we aim to study the downlink performance of multi-cell hybrid mmWave networks, which takes into account the impact of inter-cell interference caused by the neighboring BSs as well as the intra-cell interference caused by channel estimation errors.
In this section, we illustrate and derive the closed-form approximations of achievable rate per user under the impact of intra-cell and the inter-cell interference of the considered hybrid system.
Our results reveal that the intra-cell interference as well as the inter-cell interference can be mitigated effectively by simply increasing the number of antennas equipped at the desired BS.
\vspace{-2mm}
\subsection{Downlink Achievable Rate Performance}
We now detail the received information signal at user $k$ in the desired cell.
Due to the channel reciprocity, we utilize the estimated equivalent channel $\widehat{\mathbf{H}}_{\mathrm{eq}}^{T}$ in Equation ({\ref{Eq_eq}}) for the digital baseband precoder design during the downlink transmission.
To suppress the inter-user interference among different users, we adopt a ZF precoder for the downlink transmission.
In addition, equal power allocation is used among different data streams of the users \cite{Alkhateeb2015,Ni2016,Yang2013}.
The desired baseband digital ZF precoder $\overline{\mathbf{W}}_{\mathrm{eq}}\in\mathbb{C}^{N\times N}$ is given by\vspace*{-2mm}
\begin{equation}
\overline{\mathbf{W}}_{\mathrm{eq}}={\mathbf{H}}_{\mathrm{eq}}^{\ast }({\mathbf{H}}_{\mathrm{eq}}^{T}{\mathbf{H}}_{\mathrm{eq}}^{\ast})^{-1}=\left[\begin{array}{ccc}\overline{\mathbf{w}}_{\mathrm{eq,}1},\ldots ,\overline{\mathbf{w}}_{\mathrm{eq,}N}
\end{array}%
\right].  \label{P1}\vspace*{-2.0mm}
\end{equation}%
The received signal at user $k$ in the desired cell under the intra-cell and inter-cell interference is given by
\begin{align}
\vspace*{-2mm}
y_{\mathrm{ZF}}^{k}=&\underset{\mathrm{Desired}\text{ }\mathrm{signal}}{%
\underbrace{{\widehat{\bm{\omega }}_{k}^{H}}\mathbf{H}_{k}^{T}\mathbf{F}%
_{\mathrm{RF}}{\overline{\beta }}\overline{\mathbf{w}}_{\mathrm{eq,}k}x_{k}}}+%
\underset{\mathrm{Intra-cell\text{ \ }interference}}{\underbrace{{\widehat{\bm{\omega }}_{k}^{H}}\mathbf{H}_{k}^{T}\mathbf{F}_{\mathrm{RF}}\overset{N}{\underset{j=1}{\sum }}{\overline{\beta }}\Delta\overline{\mathbf{w}}_{\mathrm{eq,}j}x_{j}}}\notag\\
+&\underset{\mathrm{Inter-cell\text{ \ }interference}}{\underbrace{{\widehat{\bm{\omega }}_{k}^{H}}\overset{L}{\underset{l=1}{\sum }}\mathbf{H}_{l,k}^{T}\mathbf{F}_{\mathrm{RF},l}\widehat{\beta_{l}}\widehat{\mathbf{W}}_{\mathrm{eq},l}\mathbf{x}_{l}}}%
+\underset{\mathrm{Noise}}{\underbrace{\widehat{\bm{\omega }}_{k}^{H}%
\mathbf{z}_{\mathrm{MS},k}}},  \label{P2}\vspace*{-2mm}
\end{align}
where $x_{k}\in\mathbb{C}^{1\times 1}$ is the transmitted symbol from the desired BS to the desired user $k$,
$\mathbb{E}\left[ \left\vert x_{k}^{2}\right\vert \right]=E_{s}$ is the average transmitted symbol energy for each user,
$\overline{\beta}=\sqrt{\tfrac{1}{\mathrm{tr}(\overline{\mathbf{W}}_{\mathrm{eq}}\overline{\mathbf{W}}_{\mathrm{eq}}^{H})}}$ is the transmission power normalization factor of the desired BS,
$\Delta \overline{\mathbf{w}}_{\mathrm{eq,}j}\in\mathbb{C}^{N\times 1}$ denotes the $j$-th column vector of the ZF precoder error matrix $\Delta\overline{\mathbf{W}}_{\mathrm{eq}}=\widehat{\mathbf{W}}_{\mathrm{eq}}-\overline{\mathbf{W}}_{\mathrm{eq}}=\left[\begin{array}{ccc}\Delta\overline{\mathbf{w}}_{\mathrm{eq,}1},\ldots ,\Delta\overline{\mathbf{w}}_{\mathrm{eq,}N}
\end{array}%
\right]$,
$\widehat{\mathbf{W}}_{\mathrm{eq}}=\widehat{\mathbf{H}}_{\mathrm{eq}}^{\ast }(\widehat{\mathbf{H}}_{\mathrm{eq}}^{T}\widehat{\mathbf{H}}_{\mathrm{eq}}^{\ast})^{-1}$ is the designed ZF precoder based on the estimated equivalent channel, and the effective noise $\mathbf{z}_{\mathrm{MS,} k}\sim \mathcal{CN}\left( \mathbf{0},{\sigma_{\mathrm{MS}}^{2}}\mathbf{I}\right)$.
In addition, $\widehat{\mathbf{W}}_{\mathrm{eq},l}=\widehat{\mathbf{H}}_{\mathrm{eq},l}^{\ast }(\widehat{\mathbf{H}}_{\mathrm{eq},l}^{T}\widehat{\mathbf{H}}_{\mathrm{eq},l}^{\ast})^{-1}$ is the digital precoder of the BS in cell $l$ based on the estimated equivalent channel $\widehat{\mathbf{H}}_{\mathrm{eq},l}$, $\mathbf{F}_{\mathrm{RF},l}$ is the analog beamforming matrix of the BS in cell $l$,
$\mathbf{x}_{l}=[x_{l,1},\,x_{l,2},\ldots ,$ $x_{l,N}]^{T}$ denotes the transmitted signal for all the users in cell $l$,
and $\widehat{\beta_{l}}=\sqrt{\tfrac{1}{\mathrm{tr}\left(\widehat{\mathbf{W}}_{\mathrm{eq},l}\widehat{\mathbf{W}}_{\mathrm{eq},l}^{H}\right) }}$ is the power normalization factor of the BS in cell $l$.
Compared to the system model in \cite{Zhao2017}, the intra-cell interference, which is caused by CSI errors due to pilot contamination effects, is captured and investigated.
In addition, the model considered here also captures the inter-cell interference caused by the simultaneous transmission from the neighboring BSs.
We then express the average receive SINR expression of user $k$ in the desired cell as \vspace{-1mm}
\begin{align}
&\mathbb{E}_{\mathrm{\Delta}\widehat{\mathrm{\mathbf{H}}}^{T}_{\mathrm{eq}},\mathbf{H}_{l,k}}\left[\widehat{\mathrm{SINR}}_{\mathrm{ZF}}^{k}\right] \approx \frac{}{}\notag \\
&\frac{\overline{\beta }^{2}{{\varpi_{{k}}}}E_{s}%
}{\underset{\mathrm{Intra-cell\text{ \ }interference\text{ \ }}\Upsilon_{k}}{\underbrace{\overline{\beta }^{2}{{\varpi_{{k}}}}E_{s}\widehat{\mathbf{h}}_{\mathrm{eq,}k}^{T}\mathbb{%
E}_{\mathrm{\Delta}\widehat{\mathrm{\mathbf{H}}}^{T}_{\mathrm{eq}%
}}\left[ \Delta \overline{\mathbf{W}}_{\mathrm{eq}}\Delta \overline{\mathbf{W}}%
_{\mathrm{eq}}^{H}\right] \widehat{\mathbf{h}}_{\mathrm{eq,}k}^{\ast
}}}+ \underset{\mathrm{Inter-cell\text{ \ }interference\text{ \ } }\Omega_{k}}{\underbrace{\mathbb{%
E}_{\mathbf{H}_{l,k}}\left[\left|\widehat{\bm{\omega }}_{k}^{H}\overset{L}{\underset{l=1}{\sum }}\mathbf{H}_{l,k}^{T}\mathbf{F}_{\mathrm{RF},l}\widehat{\beta_{l}}\widehat{\mathbf{W}}_{\mathrm{eq},l}\mathbf{x}_{l}\right|^{2}\right]}} + \sigma _{\mathrm{MS}}^{2}}.  \label{SINR_with_error}\vspace{-1mm}
\end{align}
\subsection{Intra-cell and Inter-cell Interference Performance Analysis}

In this section, we aim at deriving a closed-form expression of the average achievable rate per user.
To this end, we first focus on the intra-cell interference term $\Upsilon_{k}$ shown in Equation (\ref{SINR_with_error}).
{Note that the intra-cell interference $\Upsilon_{k}$ is caused by CSI errors, which is due to the impact of pilot contamination on channel estimation.}
Then we detail and analyze the inter-cell interference $\Omega_{k}$.
The expression of intra-cell interference, $\Upsilon_{k}$, can be summarized in the following lemma.

\begin{lemm}\vspace{-0.0mm}%
In the large number of antennas regimes, the asymptotic intra-cell interference under the impact of pilot contamination can be approximated by\vspace*{-0mm}
\begin{align}
\hspace*{-0mm}
\Upsilon_{k} &={{\overline{\beta }^{2}{{\varpi_{{k}}}}E_{s}\widehat{\mathbf{h}}_{\mathrm{eq,}k}^{T}\mathbb{%
E}_{\mathrm{\Delta}\widehat{\mathrm{\mathbf{H}}}^{T}_{\mathrm{eq}%
}}\left[ \Delta \overline{\mathbf{W}}_{\mathrm{eq}}\Delta \overline{\mathbf{W}}%
_{\mathrm{eq}}^{H}\right] \widehat{\mathbf{h}}_{\mathrm{eq,}k}^{\ast
}}} \notag \\
&\hspace*{-3mm}\overset{{M}\rightarrow \infty} \approx  \overline{\beta }^{2}{{\varpi_{{k}}}}E_{s}\left[\left(
\sqrt{1+\overset{L}{\underset{l=1}{\sum }}\frac{{\widehat{\rho} _{l,k}^{2}}}{MP}}-1\right) ^{2}
+\left( {1+\overset{L}{\underset{l=1}{\sum }}\frac{{\widehat{\rho} _{l,k}^{2}}}{MP}}\right)\overset{L}{\underset{l=1}{\sum }}\frac{{\widehat{\rho} _{l,k}^{2}}N}{MP}\frac{{\color{black}\varsigma}_{k} +1}{{\color{black}\varsigma}_{k} }\right]. \label{Coro_RZFK}\vspace*{-0mm}
\end{align}
\end{lemm}\vspace{-0.0mm}
\emph{\quad Proof: }The proof follows a similar approach as in \cite{Zhao2017}. Due to the page limitation, we omit the proof here.\QEDA

Now, we introduce the following lemma which simplifies the expression of the intra-cell interference.
\begin{lemm}\vspace{-0.0mm}%
In the large numbers of antennas regime, the asymptotic inter-cell interference, which is caused by the BSs in neighboring cells, can be approximated as\vspace{-1mm}
\begin{align}
\Omega_{k} = {{{{\mathbb{E}_{\mathbf{H}_{l,k}}\left[\left|\widehat{\bm{\omega }}_{k}^{H}\overset{L}{\underset{l=1}{\sum }}\mathbf{H}_{l,k}^{T}\mathbf{F}_{\mathrm{RF},l}\widehat{\beta_{l}}\widehat{\mathbf{W}}_{\mathrm{eq},l}\mathbf{x}_{l}\right|^{2}\right]}} } }
\overset{{M}\rightarrow \infty} \approx  E_{s}\left[\overset{L}{\underset{l=1}{\sum}}\left({\widetilde{\varpi}_{l,{k}}}
\right) \right].\label{Eq_CORO}\vspace{-1mm}
\end{align}
\end{lemm}
\emph{\quad Proof: }Please refer to Appendix B.\QEDA

From Equation (\ref{Eq_CORO}), the inter-cell interference is mainly contributed by the downlink transmission of neighboring BSs.
Besides, the power normalization factor $\overline\beta$ can be approximated as \cite{Zhao2017}\vspace{-1mm}
\begin{equation}
\overline\beta \approx \sqrt{\frac{{\color{black}\varsigma}_{k}}{{\color{black}\varsigma}_{k} + 1}\frac{MP}{N}}. \label{Eq_22}\vspace{-1mm}
\end{equation}
Then, we summarize the average achievable rate of user $k$ in the large number of antennas and high SNR regime in the following theorem.
\begin{theo}\label{thm:Theo_2}
For a large number of antennas regime, the average achievable rate of user $k$ adopting the ZF precoding under the intra-cell and the inter-cell interference is approximated by
\begin{align}
\widetilde{R}_{\mathrm{ZF}}^k 
&\approx \log_{2}\left\{1+\left[\left(
\sqrt{1+\frac{1}{MP}{\overset{L}{\underset{l=1}{\sum }}\left( {\widehat{\rho} _{l,k}^{2}} \right)}}-1\right) ^{2}+  \frac{N({\color{black}\varsigma}_{k} + 1)}{{{{\color{black}\varsigma}_{k}} MP}}\overset{L}{\underset{l=1}{\sum}}\left(\zeta_{l,k}^{2}\right)\right.\right.\notag\\
&\hspace{+8mm}\left.\left.+\left[ {1+\frac{1}{MP}{\overset{L}{\underset{l=1}{\sum }}\left( \widehat{\rho} _{l,k}^{2} \right)}}\right]\frac{N\left({\color{black}\varsigma}_{k} +1\right)}{{{\color{black}\varsigma}_{k} }MP}{\overset{L}{\underset{l=1}{\sum }}\left( \widehat{\rho} _{l,k}^{2}\right)}+ \frac{\sigma _{\mathrm{MS}}^{2}}{\overline{\beta }^{2}{{\varpi_{{k}}}}E_{s}} \right]^{-1}\right\}, \label{SINR_final}
\end{align}
where ${\zeta_{l,k}}=\sqrt{\dfrac{\widetilde{\varpi}_{l,{k}}}{{\varpi_{{k}}}}}$.
\end{theo}
\emph{\quad Proof: }The result follows by substituting Equations (\ref{Coro_RZFK}), (\ref{Eq_CORO}), and (\ref{Eq_22}) into (\ref{SINR_with_error}).\QEDA

In the high SNR and large antenna regimes, the average achievable rate under the impact of intra-cell interference caused by CSI errors and the inter-cell interference caused by neighboring BSs can be mitigated by simply increasing the number of antennas $M$. Thus, the average achievable rate per user can increase unboundedly with an increasing antenna $\log_{2}{M}$.

For comparison, we also derive the achievable rate of user $k$ in a single-cell hybrid system.
The performance of this system, $R_{\mathrm{HB}}^{\mathrm{upper}_k}$, serves as a performance upper bound of the considered multi-cell system as there is no out-of-cell interference.
The expression of $R_{\mathrm{HB}}^{\mathrm{upper}_k}$ is given by (Equation (24) of Corollary $1$ in \cite{Zhao2017})
\begin{equation}
R_{\mathrm{HB}}^{\mathrm{upper}_k}\underset{M\rightarrow \infty }{\overset{a.s.}{\rightarrow }}\log _{2}\left\{ 1+\left[
\frac{MP}{N}\frac{{\color{black}\varsigma}_{k} }{{\color{black}\varsigma}_{k} +1} +\dfrac{1}{{\color{black}\varsigma}_{k} +1}\right] \dfrac{E_{s}}{\sigma _{\mathrm{MS}}^{2}}\right\}.
\end{equation}
\begin{coro}
Now, we derive the scaling law of the average achievable rate of user $k$ in the high SNR regime which is given by
\begin{align}
\underset{{M}\rightarrow\infty}{\lim}\frac{\widetilde{R}_{\mathrm{ZF}}^k}
{{\log_{2}{M}}}=\underset{\mathrm{M}\rightarrow\infty}{\lim}\frac{R_{\mathrm{HB}}^{\mathrm{upper}_k}}{{\log_{2}{M}}}=1.\label{Final_Conclusion}
\end{align}\label{Coro3}
\end{coro}\vspace*{-8mm}
\emph{\quad Proof: }The result comes after some straightforward mathematical manipulation. Due to the page limitation, we omit the proof here.\QEDA

It is found that the system data rate scales with $\log_{2}{M}$.
In other words, pilot contamination is not a fundamental problem for massive MIMO with hybrid mmWave systems, when the CSI estimation algorithm proposed in \cite{Zhao2017} is adopted.
{In contrast, the average achievable rate performance of multi-cell MU massive MIMO networks adopting a conventional LS-based channel estimation algorithm is limited by the effects of pilot contamination, e.g. \cite{Rusek2013}.}
{\color{black}Note that the algorithm proposed in \cite{Zhao2017} does not require any information of covariance matrices of pilot-sharing users in neighbouring cells as required by the multi-cell MMSE-based precoding algorithm proposed in \cite{Bjornson2017a}.
In addition, the algorithm proposed in \cite{Zhao2017} can work well while mean values of cross-cell channels from pilot-sharing users in neighbouring cells to the desired BS are non-zero.}
%
\begin{rema}
{\color{black}Hybrid mmWave systems are the special case of fully digital systems, the algorithm proposed in \cite{Zhao2017} can be extended to the case of fully digital systems.
Under such circumstance, the derived analysis can be applied to the latter systems.}
\end{rema}

\vspace{-0mm}

\section{Discussions and Simulations}

In this section, we will discuss and verify the derived results via simulations.
In the following simulations unless specified otherwise, simulation settings are listed as follows.
The transmit antenna gain of the BS antennas is assumed as $14$ dBi.
The maximum BS transmit power is set as $46$ dBm.
We set the number of neighboring cells as $L=6$ and the number of users per cell $N=10$.
For the adopted simulation parameters, the large-scale path loss coefficients\footnote{The commonly adopted empirical propagation path loss model is considered as a function of distance and carrier frequency, which is given by $\varpi_{{k}}=10^{-\left(\tfrac{10\alpha \log_{10}{d_{k}}+\varrho \log_{10}\left( 4\pi\frac{1}{\lambda}\right)}{10}\right)}\label{Eq_PL}$, where $d_{k}$ is the distance between the desired BS and user $k$ in the desired cell, $\alpha $ and $\varrho $ are the least square fits of floating intercept and slope over the measured distance \cite{Akdeniz2014}.}, $\alpha $ and $\varrho $, may have different values for different scenarios.
We have $\alpha = 1.9$ and $\varrho =20$ for corresponding large-scale path loss calculation \cite{Akdeniz2014}.
In addition, the receiving thermal noise power is $\kappa= -90$ dBm\footnote{ Thermal noise power is determined by the signal bandwidth $B_{W}$ and the noise spectral density level $N_{0}=  B_{\mathrm{W}}e_{\mathrm{B}}T_{k}$, where $B_{\mathrm{W}}=250$ MHz, $T_{k}=300$ K, and $e_{\mathrm{B}}=1.38\times10^{-23}$ J/K is the Boltzmann constant \cite{Akdeniz2014}.}.
The channel estimation errors of a user caused by pilot contamination is given by $\xi^{2}=\overset{L}{\underset{l=1}{\sum }}\left( \widehat{\rho} _{l,k}^{2} \right)$, $ k\left\{ 1,\ldots, N\right\}$.
Besides, we define the average achievable rate per user as $\widetilde{R}_{\mathrm{AVE}}=\dfrac{1}{N}{\overset{N}{\underset{k=1}{\sum }}\widetilde{R}_{\mathrm{ZF}}^k}$.


\begin{figure}[t]
\centering{\vspace{-0mm}}
\includegraphics[width=4.5in]{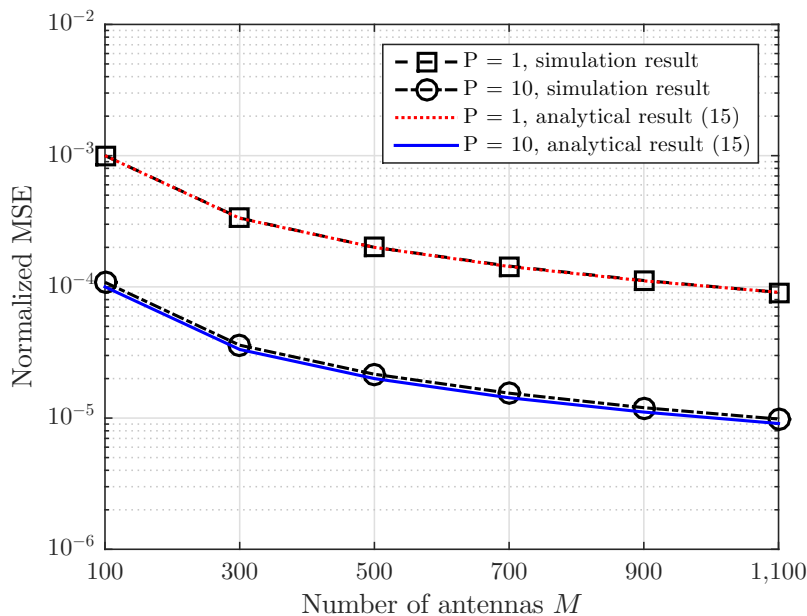}
\caption{The illustration of multi-cell normalized MSE performance under the impact of pilot contamination versus the number of antennas $M$ in the high SNR regime, i.e., maximum transmit power $46$ [dBm], for $N_{\mathrm{RF}}=N=10$, $\varsigma _{l,k}=5$, $k\in\{1,\cdots,N\}$ and channel estimation error $\xi^{2} = 0.01$.}{\vspace{-0mm}}
\label{fig:CE_PC}
\end{figure}{\vspace{-0mm}}
{\color{black}
To verify the correctness of the analytical results derived in Equation (\ref{PC}), here, we provide some simulation results in Figure \ref{fig:CE_PC}, which illustrates the normalized MSE performance of the equivalent channel estimation versus the number of antennas equipped at the desired BS, $M$, under the impact of pilot contamination.
In the simulation, we take into account the inter-cell propagation path loss and AoA estimation errors in estimating the strongest AoA paths.
For illustration, channel estimation errors caused by pilot contamination is set as $\xi^{2}=0.01$.
Due to the existence of the inter-cell strongest AoA components, this setting can be considered as the worst case scenario of pilot contamination as the impact of inter-cell interference is magnified.
In Figure \ref{fig:CE_PC}, we can observe that with an increasing number of antennas $M$, the normalized MSE decreases monotonically.
In addition, an increasing number of antennas equipped at the users $P$ can also improve the normalized MSE.
Besides, the simulation results match the analytical results derived in Equation (\ref{PC}).
Thus, the impact of pilot contamination on the channel estimation of multi-cell hybrid mmWave systems vanishes asymptotically, for a sufficiently large number of antennas equipped at the desired BS $M$.}

\begin{figure}[t]
\centering{\vspace{-0mm}}
\includegraphics[width=4.5in]{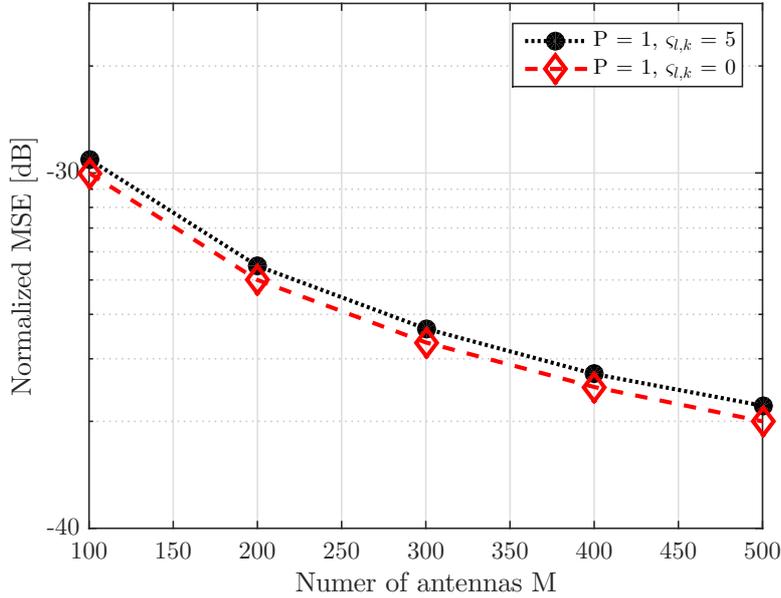}
\caption{The illustration of multi-cell normalized MSE performance under the impact of pilot contamination versus the number of antennas $M$ in the high SNR regime with different $\varsigma _{l,k}$ settings.}{\vspace{-0mm}}
\label{fig:ADD}
\end{figure}{\vspace{-0mm}}

{\color{black}In Chapter $4$, to simplify the performance analysis, we assume the same average large scale fading coefficient for LOS components and scattering components. Thus, the change of cross-cell LOS to NLOS power ratio $\varsigma _{l,k}$ does not change the MSE performance of the channel estimation, which is demonstrated in Equation (\ref{PC}) of Theorem \ref{thm:Theo_1}.
However, in practice, large scale propagation path loss coefficients for cross-cell LOS and NLOS components are supposed to be different. In Figure \ref{fig:ADD}, we set a simulation with different large scale propagation path loss coefficients for cross-cell LOS and NLOS components (the large scale path loss coefficient of NLOS components is larger than that of LOS components). The simulation results reveal that with increasing the cross-cell LOS to NLOS power ratio $\varsigma _{l,k}$, the normalized MSE performance decreases.}

\begin{figure}[t]
\centering{\vspace{-0mm}}
\includegraphics[width=4.5in]{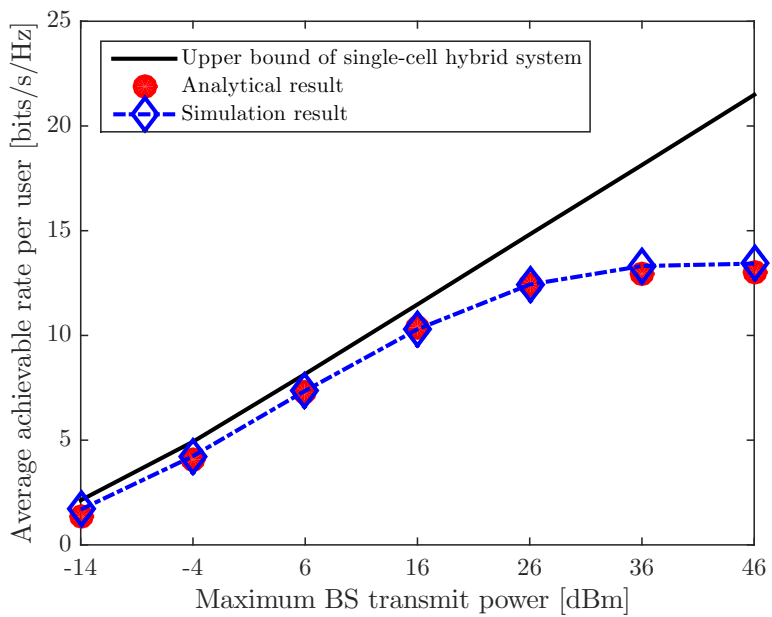}
\caption{The average achievable rate per user [bits/s/Hz] under the impact of pilot contamination and inter-cell downlink transmission interference versus maximum transmit power at the desired BS [dBm] with $M=200$. }{\vspace{-0mm}}
\label{fig:3}
\end{figure}

In Figure \ref{fig:3}, we verify the tightness of the derived approximation in Equation (\ref{SINR_final}) of Theorem \ref{thm:Theo_2}.
Figure \ref{fig:3} illustrates the average achievable rate per user under the intra-cell and inter-cell interference versus SNR.
Besides, we illustrate the corresponding upper performance which is based on a single-cell hybrid mmWave system.
In Figure \ref{fig:3}, the number of antennas equipped at the users is set to $P=10$, ${\color{black}\varsigma}_{k}= 4$, ${\color{black}\varsigma}_{l,k} = 2$, $ k \in \left\{ 1,\ldots, N\right\}$, the channel estimation error caused by pilot contamination is set as $\xi^{2}=0.01$.
We can observe that the derived approximation is tight for a wide range of SNR.
It also can be observed that, with a finite number of antennas, there is still a ceiling of the achievable rate performance in the high SNR regime.

\begin{figure}[H]
\centering\vspace*{-0mm}
\subfigure[]
{\label{fig:20}
\includegraphics[width=4.5in]{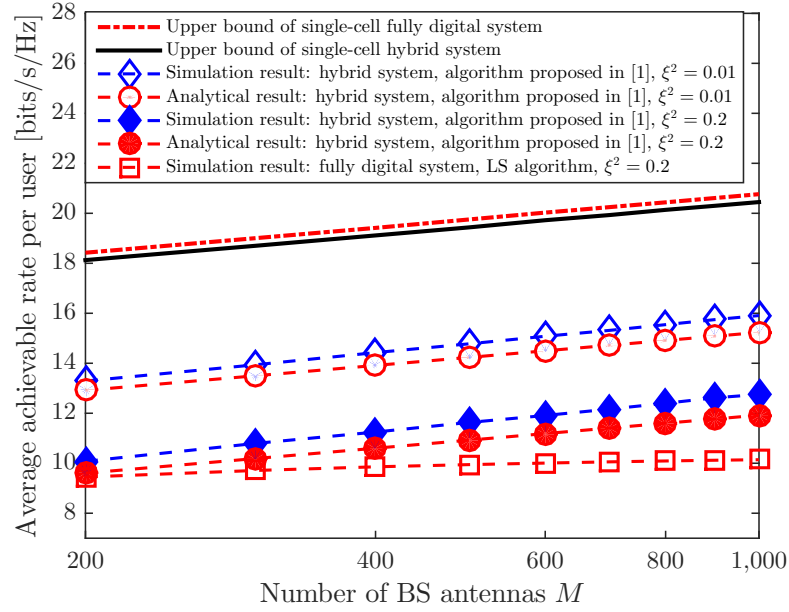}}
\subfigure[]
{\label{fig:21}
\includegraphics[width=4.5in]{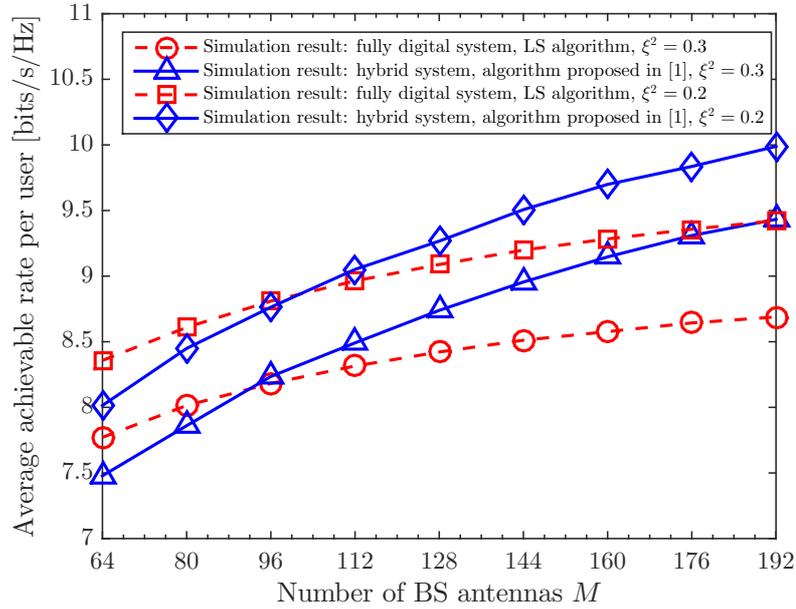}}
\caption{The achievable rate per user [bits/s/Hz] under the impact of pilot contamination effects and inter-cell interference versus the number of antennas $M$ in the high SNR regime (maximum BS transmit power is $46$ [dBm]).}
\label{fig:2}\vspace{-0mm}
\end{figure}

The setup in Figure \ref{fig:2} is considered at ${\color{black}\varsigma}_{k}= 4$, ${\color{black}\varsigma}_{l,k} = 2$, $ k \in \left\{ 1,\ldots, N\right\}$, and the number of antennas equipped at each user is $P=10$.
In Figure \ref{fig:2}, we first illustrate the upper bounds of the average achievable rate per user of the single-cell hybrid mmWave system and the single-cell fully digital system as the system benchmarks.
Without any intra-cell interference and inter-cell interference, the corresponding upper performance of the fully digital system as well as the hybrid mmWave system increases linearly with the increasing number of BS antennas $M$.
Then, we compare the achievable rate per user of the fully digital system using a conventional LS-based CSI estimation algorithm to that of the hybrid system adopting CSI estimation algorithm proposed in \cite{Zhao2017} in the high SNR regime.
For the fully digital system, we assume that the receive CSI at the users' side is perfectly known.
Thus, the $P$ antenna array equipped at each user can provide $10\log_{10}{P}$ dB array gain.

In addition, we assume that channel estimation errors of a user at the desired BS due to the impact of pilot contamination effects for both the fully digital system and the hybrid system are set as $\xi^{2}=0.2$.
As expected, in the large number of antennas and high SNR regimes, the average achievable rate of the hybrid system increases unboundedly and logarithmically with the increasing number of antennas ${M}$, despite the existence of pilot contamination as well as inter-cell interference.
The average achievable rate of the hybrid system scales with the number of BS antennas in the order of $\log_{2}{M}$, which is the same as that of the upper performance of the single-cell hybrid system, and verifies the correctness of the derived scaling law (Equation (\ref{Final_Conclusion}) in Corollary \ref{Coro3}).

However, the average achievable rate of the fully digital system adopting the conventional LS-based CSI estimation algorithm under the impact of pilot contamination is saturated with the increasing number of antennas at the desired BS.
Thus, pilot contamination is the performance bottleneck of massive MIMO systems when the conventional LS-based channel estimation algorithm is adopted \cite{Rusek2013}.
In contrast, when the channel estimation algorithm proposed in \cite{Zhao2017} is applied to the multi-cell MU mmWave hybrid system, the impacts of CSI errors caused by pilot contamination and the inter-cell interference caused by neighboring BSs on the achievable rate performance will vanish asymptotically for a sufficiently large number of antennas equipped at the desired BS.
{{\color{black}
For a practical number of antennas setup, $M \in [64,\text{\ }192]$, the average rate performance achieved by the algorithm proposed in $[1]$ can also increase with an increasing number of antennas $M$ and outperforms that of a fully digital system, which is illustrated in Figure \ref{fig:21}.}
Meanwhile, the number of RF chains required in the hybrid system is significantly lower than that of fully digital system.
In addition, our simulation results in Figure \ref{fig:20} also verify the tightness of the analytical approximation derived in Theorem \ref{thm:Theo_2}.
{\vspace{-0mm}}
\section{Summary}
In this chapter, we investigated the normalized MSE performance of the channel estimation proposed in \cite{Zhao2017} for multi-cell hybrid mmWave systems.
The derived closed-form approximation of the normalized channel estimation MSE performance revealed that the channel estimation error caused by the impact of pilot contamination and noise would vanish asymptotically with the increasing number of antennas.
Furthermore, based on the estimated CSI, we adopted ZF precoding for the downlink transmission and derived the closed-form approximation of the average achievable rate per user in the multi-cell scenario.
The analytical and simulation results showed that the intra-cell interference caused by pilot contamination as well as the inter-cell interference incurred by neighboring BSs can be mitigated effectively with an increasing antenna array gain.
It is an excellent feature for multi-cell hybrid mmWave systems with small-cell radius for improving the network spectral efficiency.

\chapter{Conclusions and Future Works}\label{C5:chapter5}

%
\ifpdf
    \graphicspath{{1_introduction/figures/PNG/}{1_introduction/figures/PDF/}{1_introduction/figures/}}
\else
    \graphicspath{{1_introduction/figures/EPS/}{1_introduction/figures/}}
\fi

We conclude this thesis by summarizing our contributions and discussing some potential future works.

\section{Conclusions}

This thesis mainly focused on channel estimation, analog beamformer and digital precoder design, and performance analysis that address key challenges in mmWave systems, by considering, the hardware constraints, the hardware impairments, non-sparse mmWave channels, the channel acquisition overhead, the analog beamformer/digital precoder design complexity, and the cross-cell interference caused by users and BSs in neighbouring cells.

First, we proposed a tone-based AoA estimation channel estimation algorithm for mmWave channels to reduce the channel acquisition overhead significantly.
Then, we proposed efficient analog beamformer/digital precoder technologies for MU hybrid mmWave MIMO systems.
Finally, we applied this solution to multi-small-cell scenarios.
We proved that this channel estimation and analog beamformer/digital precoder design scheme could effectively mitigate cross-cell interference during uplink and downlink phases.

In Chapter $2$, for the channel estimation problem of the massive MIMO system in mmWave frequencies over Rician fading channels, we developed a tone-based AoA estimation algorithm leveraging the angular resolution.
Based on estimated CSIs, we proposed an MU SLOS-based MRT strategy to achieve higher rate performance than that of the conventional PAC-based MRT strategy.
The rate performance gap between the SLOS-based MRT and the PAC-based MRT was analytically evaluated and verified via simulation.
Besides, the impact of phase quantization errors on the system rate performance had also been analyzed.

In Chapter $3$, we proposed a low complexity channel estimation algorithm for MU TDD hybrid mmWave MIMO systems during the uplink phase, which was applicable to both non-sparse
and sparse mmWave channel environments.
The proposed channel estimation solution leverages the characteristics of the mmWave channels (the strongest AoAs), the orthogonal pilot symbols, and the hybrid structures.
Exploiting the estimated strongest AoAs, analog beamforming matrices could be designed.
Then, leveraging the designed analog beamforming matrices and orthogonal pilot symbols, the BS could estimate the equivalent mmWave channels for the design of digital precoding matrices.
The performance of the proposed channel estimation algorithm was analyzed.
Analytical and simulation results indicated that the downlink achievable rate based on the proposed estimated CSI could obtain a considerable achievable rate of fully digital systems.
Furthermore, the robustness of the proposed channel estimation and MU downlink precoding scheme against hardware impairments was illustrated via analytical and simulation results.

In Chapter $4$, we investigated the system rate performance of a multi-small-cell MU hybrid mmWave MIMO network.
We applied the channel estimation algorithm and the analog beamforming/digital precoding design scheme proposed in Chapter $3$ to multi-small-cell scenarios.
Different than the conventional massive MIMO system with the conventional LS channel estimation algorithm, our analytical and simulation results showed that the channel estimation error incurred by the impact of pilot contamination and noise vanished asymptotically with an increasing number of antennas equipped at the desired BS.
The results unveiled that even facing severe intra-cell interference and inter-cell interference, the average achievable rate performance does not saturate with increasing the number of antennas equipped at the desired BS.
It was an excellent feature for multi-cell hybrid mmWave systems with small-cell radius
for improving the network spectral efficiency.
The obtained analytical and simulation results could be generalized and extended for sub-$6$ GHz massive MIMO systems in Rician fading channels.
In other words, pilot contamination was no longer the fundamental problem for massive MIMO systems in Rician fading channels.

\section{Future Works}

There are several possible directions for future research.

\subsubsection{Hybrid architectures with extra RF chains:}
The hybrid architecture adopted in Chapter $3$ considered that the number of RF chains is equal to the number of users.
This assumption is adopted to simplify the notation and the derivation.
In general, for the case of $N_{\mathrm{RF}} = N$, the hybrid system can provide enough degrees of freedom to serve $N$ users simultaneously.
However, in the work \cite{Sohrabi2016}, the authors proved that when the number of RF chains equipped at the BS is more than twice of the number of users, the hybrid systems can realize the rate performance of the fully digital systems.
If the BS can equip with more RF chains, more degrees of freedom can be exploited for the design of analog and digital precoders.
In addition, the extra RF chains adopted by the BS can be used for user scheduling design, anti-blocking design, and multipath diversity exploitation.
Thus, for the case that $N_{\mathrm{RF}}>N$, hybrid systems can achieve a higher rate than that of $N_{\mathrm{RF}}=N$.
However, how to exploit the extra RF chains to further improve the rate performance and the channel estimation performance of hybrid systems is not considered in this thesis.
The case of $N_{\mathrm{RF}} > N$ should be considered in the future works.
{\color{black}In addition, authors of \cite{Zhang2005,Zhang2014aa,Bogale2016} employed RF chains with sets of digitally controlled phase paired phase-shifters and switches to achieve the same rate performance as that of the fully digital system.
Furthermore, with the support of an advanced hybrid architecture \cite{Zhang2005,Zhang2014aa,Bogale2016}, hybrid systems can achieve higher rate performance than that of adopting a simple ULA antenna array case.
However, how to exploit extra RF chains as well as advanced hybrid architectures to further improve the rate performance of hybrid systems is still an unexplored/interesting research topic and is beyond the scope of this thesis.
We may consider the case of $N_{\mathrm{RF}} > N$ and the case of adopting advanced hybrid architectures, e.g. phase shifts with digital controlled phase and amplitude, for channel estimation and downlink transmission precoding in our future works.}


\subsubsection{Distributed hybrid mmWave MIMO network to recapture the geographic degree of freedom:}
The multi-small-cell hybrid mmWave MIMO network adopted in Chapter $4$ pointed out that the inter-cell and intra-cell interference during the downlink transmission can be suppressed by simply increasing the number of antennas equipped at each BS.
These features are useful for small cell scenarios and the dense deployment of BSs.
Thus, cooperating multiple BSs across a large geographic area with small cell radius can facilitate the reuse of the same piece of spectrum and further improve the network spectral efficiency \cite{Miao2014}.
Besides, with the dense deployment of BSs, beamforming transmission technology offers robustness against blockages via inter-BS beam switching.
In addition, the desired user may have multiple paths to multiple BSs.
If users are equipped with multiple RF chains, the SNR maximization principle \cite{Alkhateeb2015} for the design of analog beamforming matrices does not exploit the full channel capacity.
Thus, it would be interesting to develop analog beamforming matrices and channel estimation solutions jointly to strike a balance among the channel capacity, the robustness of mmWave systems, the inter-BS handover, and the energy efficiency.
{\color{black}
\subsubsection{Extension to the wideband frequency selective :}
In general, wideband mmWave channels are assumed as frequency selective. Since the scalable OFDM structure has been accepted as the standard of 5G NR and mmWave systems are attractive due to their wide bandwidth (in practice, mmWave are mostly operated at wideband frequencies to support higher data streams), developing efficient mmWave channel estimation for frequency selective channels is of great importance.

Current works in the literature assume that the analog beamforming matrices are frequency flat and are generated using a fully connected architecture with digitally controlled quantized phase shifters. However, analog beamforming matrices are frequency selective and have been discussed in the literature about optimal array signal processing.
To extend the proposed tone-based algorithm, which estimates mmWave channels by sending narrow bandwidth pulses, to frequency selective wideband mmWave channels, there are several possible directions for future research:
1)	It is necessary to discuss the spatial beam pattern distortion of the analog beamforming in frequency selective mmWave channels.
2)	If the digitally controlled quantized amplifier can be introduced to the hybrid architecture (combined with digitally controlled quantized phase shifters), can advanced array signal processing algorithms be introduced for the frequency selective compensation/equalization?}

\appendix
\onehalfspacing
\fancyhead[CE]{\leftmark}
\fancyhead[CO]{\leftmark}
\chapter{Proof of Theorem 2.1}\label{appendix_2a}

The SINR expression of SLPS-based ZF is given by
\begin{align}
\mathrm{SINR}^{k}_{\mathrm{ZF}}=\frac{{\widehat{\beta }^{2}E_{s}}}{{%
{\widehat{\beta }^{2}E_{s}}{\mathbf{h}_{k}^{T}\mathrm{\mathbb{E}}_{\Delta\mathbf{ H}%
}\left[ \Delta \mathbf{W}\Delta \mathbf{W}^{H}\right] \mathbf{h}_{k}^{\ast }}%
}+{\sigma ^{2}}}.
\end{align}%
To express $\mathrm{\mathbb{E}}_{\Delta\mathbf{ H}}[\Delta\mathbf{ W}\Delta \mathbf{W}^{H}]$, first we present the expression of $\Delta\mathbf{W}$ as
\begin{align}
\Delta\mathbf{ W}& =\widehat{\mathbf{W}}\mathbf{-W}\notag\\
&=\sqrt{1+\delta ^{2}}(\mathbf{H}+\Delta \mathbf{H})^{\ast }\left[ (\mathbf{H}+\Delta \mathbf{H})^{T}(\mathbf{H}+\Delta \mathbf{H})^{\ast }\right] ^{-1}-\mathbf{H}^{\ast }\left(\mathbf{H}^{T}\mathbf{H}^{\ast }\right) ^{-1},  \label{Eq_222}\\
&(\mathbf{H}^{T}\mathbf{H}^{\ast})=\mathbf{K}=(\mathbf{H}^{H}\mathbf{H}),\label{Eq_223}\\
&\mathbf{D}=(\mathbf{H}^{H}\Delta\mathbf{ H}+ \Delta\mathbf{ H}^{H}\mathbf{H}+\Delta\mathbf{ H}^{H}\Delta\mathbf{ H}),
\end{align}
and $\mathbf{H}^{\ast }\left( \mathbf{H}^{H}\mathbf{H}\right) ^{-1}=\mathbf{W=\mathbf{H}}^{\ast }\mathbf{K}^{-1}.$ Then, exploiting the matrix inversion approximation \cite{book:MCB}
\begin{equation}
\mathbf{(K+D)}^{-1}\approx\left[ \mathbf{K}^{-1}-\mathbf{K}^{-1}\left(
\mathbf{I+DK}^{-1}\right) ^{-1}\mathbf{DK}^{-1}\right],
\end{equation}%
we re-express (\ref{Eq_222}) as
\begin{align}
\Delta \mathbf{W}=\sqrt{1+\delta ^{2}}\left(\mathbf{H}^{\ast }+\mathbf{\Delta H}^{\ast }\right)\left[ \mathbf{K}^{-1}-\mathbf{K}^{-1}\left( \mathbf{I+DK}^{-1}\right)^{-1}\mathbf{DK}^{-1}\right]- \mathbf{H}^{\ast }\left(\mathbf{H}^{H}\mathbf{H}\right)^{-1}.\vspace{-0mm}
\end{align}%
Finally, we could express the closed-form approximation of
\begin{align}
\mathrm{\mathbb{E}}_{\Delta\mathbf{ H}}=[\Delta\mathbf{ W}\Delta \mathbf{ W}^{H}]&=\overset{\left(a \right)}\approx (\sqrt{1+\delta ^{2}}-1)^{2}\mathbf{WW}^{H}\notag\\
&+2(\sqrt{1+\delta ^{2}}-1-\delta ^{2})\delta ^{2}\left[ M\mathbf{WK}^{-1}\left( \mathbf{%
I+DK}^{-1}\right) ^{-1}\mathbf{W}^{H}\right]   \notag \\
&+2(\sqrt{1+\delta ^{2}}-1-\delta ^{2})\delta ^{2}\left[ \mathrm{tr}%
(\mathbf{K}^{-1})\mathbf{W\left( \mathbf{I+DK}^{-1}\right) ^{-1}W}^{H}\right]
\notag \\
&+(\sqrt{1+\delta ^{2}})^{2}\delta ^{2}\mathrm{tr}\left[ \mathbf{K}%
^{-1}\left( \mathbf{I+DK}^{-1}\right) ^{-1}\right] \mathbf{WW}^{H}  \notag
\\
&\overset{\left(b \right)}\approx(\sqrt{1+\delta ^{2}}-1)^{2}\mathbf{WW}^{H}\notag\\
&+\frac{2\sqrt{1+\delta ^{2}}(1-\sqrt{1+\delta ^{2}})\delta ^{2}\mathrm{tr}%
\left( \mathbf{K}^{-1}\right) }{1+\delta ^{2}\mathrm{tr}\left( \mathbf{K}%
^{-1}\right) }\mathbf{WW}^{H}  \notag \\
&+\frac{2\sqrt{1+\delta
^{2}}(1-\sqrt{1+\delta ^{2}})\delta ^{2}M}{1+\delta ^{2}\mathrm{tr}(\mathbf{K%
}^{-1})}\mathbf{WK}^{-1}\mathbf{W}^{H}\notag\\
&+(\sqrt{1+\delta ^{2}})^{2}\delta ^{2}\frac{%
\mathrm{tr}(\mathbf{K}^{-1})}{1+\delta ^{2}\mathrm{tr}(\mathbf{K}^{-1})}%
\mathbf{WW}^{H}.  \label{Eq_301}
\end{align}%
In (a), we omit some negligibly small random scattering correlation parts. Also, due to the reason that $(\mathbf{I+DK})^{-1}$ is symmetric and small, we can obtain the approximation $(\mathbf{I+DK})^{-1}\approx\frac{1}{1+\delta ^{2}\mathrm{tr}(\mathbf{K}^{-1})}\mathbf{I}$ in (b).
We substitute (\ref{Eq_301}) into (\ref{Eq_151}), then the result comes immediately after some straight forward mathematical manipulation.

\chapter{Proof of Corollary 2.1}\label{appendix_2b}
Now we prove Corollary \ref{Coro_21}. The desired signal part $\Phi _{k}$ of the
received signal at user $k$ is given by%
\begin{align}
\Phi _{k}=\beta ^{2}E_{s}\mathbf{h}_{k}^{H}\mathbf{h}_{k}\mathbf{h}_{k}^{H}%
\mathbf{h}_{k}=M^{2}\beta ^{2}E_{s}.
\end{align}%
Then we express the interference part $\Omega _{k}$ as%
\begin{align}
\Omega _{k}& =\beta ^{2}\mathrm{\mathbb{E}}_{\mathbf{h}_{\mathrm{S}},\Delta\mathbf{h}} \left[ \left( \overset{N}{\underset{j=1,j\neq k}{\sum }}\mathbf{h}%
_{j}^{H}x_{j}+\overset{N}{\underset{j=1}{\sum }}\Delta \mathbf{h}%
_{j}^{H}x_{j}\right) \mathbf{h}_{k}\mathbf{h}_{k}^{H}\left( \overset{N}{%
\underset{j=1,j\neq k}{\sum }}x_{j}^{H}\mathbf{h}_{j}+\overset{N}{\underset{%
j=1}{\sum }}x_{j}^{H}\Delta \mathbf{h}_{j}\right) \right]  \notag \\
& =\beta ^{2}E_{s}\underset{{Q }}{\underbrace{\mathrm{\mathbb{E}}_{\mathbf{h}%
_{\mathrm{S}}}\left[ \overset{N}{\underset{j=1,j\neq k}{\sum }}\left(
\mathbf{h}_{j}^{H}\mathbf{h}_{k}\mathbf{h}_{k}^{H}\mathbf{h}_{j}\right) %
\right] }}+\beta ^{2}E_{s}\underset{{U }}{\underbrace{\mathrm{\mathbb{E}}_{%
\mathbf{h}_{\mathrm{S}}}\left[ \overset{N}{\underset{j=1}{\sum }}E_{\mathrm{%
\Delta h}}\left( \Delta \mathbf{h}_{j}^{H}\mathbf{h}_{k}\mathbf{h}%
_{k}^{H}\Delta \mathbf{h}_{j}\right) \right] }},
\end{align}%
where ${Q}$ is given by
\begin{align}
{Q}&=\left( \dfrac{\upsilon _{k}}{\upsilon _{k}+1}\right) ^{2}\overset{N}{%
\underset{j=1,j\neq k}{\sum }}\mathbf{h}_{\mathrm{L,}j}^{H}\mathbf{h}_{%
\mathrm{L,}{k}}\mathbf{h}_{\mathrm{L,}{k}}^{H}\mathbf{h}_{\mathrm{L,}%
j}+\dfrac{1}{\upsilon _{k}+1}\overset{N}{\underset{j=1,j\neq k}{\sum }}\mathrm{\mathbb{E}}_{%
\mathbf{h}_{\mathrm{ds}}}\left[ \mathbf{h}_{j}^{H}\mathbf{h}_{\mathrm{S,}{k}}%
\mathbf{h}_{\mathrm{S,}{k}}^{H}\mathbf{h}_{j}\right]  \notag \\
&+\dfrac{\upsilon _{k}}{\left( \upsilon _{k}+1\right) ^{2}}\overset{N}{%
\underset{j=1,j\neq k}{\sum }}\mathrm{\mathbb{E}}_{\mathbf{h}_{\mathrm{S}}}\left[ \mathbf{h}_{%
\mathrm{S,}j}^{H}\mathbf{h}_{\mathrm{L,}{k}}\mathbf{h}_{\mathrm{L,}{k}%
}^{H}\mathbf{h}_{\mathrm{S,}j}\right]  +\dfrac{1}{\left( \upsilon
_{k}+1\right) ^{2}}\overset{N}{\underset{j=1,j\neq k}{\sum }}\mathrm{\mathbb{E}}_{\mathbf{h}_{%
\mathrm{S}}}\left[ \mathbf{h}_{\mathrm{S,}j}^{H}\mathbf{h}_{\mathrm{S,}k}%
\mathbf{h}_{\mathrm{S,}k}^{H}\mathbf{h}_{\mathrm{S,}j}\right]  \notag \\
& \overset{\left(c \right)}%
{\approx } \left( \dfrac{\upsilon _{k}}{\upsilon _{k}+1}\right) ^{2}\overset{N%
}{\underset{j=1,j\neq k}{\sum }}\mathbf{h}_{\mathrm{L,}j}^{H}\mathbf{h}_{%
\mathrm{L,}{k}}\mathbf{h}_{\mathrm{L,}{k}}^{H}\mathbf{h}_{\mathrm{L,}%
j} +\dfrac{(N-1)M\left( 2\upsilon _{k}+1\right) }{\left( \upsilon
_{k}+1\right) ^{2}}
\end{align}%
and ${U}$ is given by
\begin{align}
{U} &=\overset{N}{\underset{j=1}{\sum }}\mathrm{\mathbb{E}}_{\Delta\mathbf{ h}}\left[
\left( \Delta \mathbf{h}_{j}^{H}\mathbf{h}_{k}\mathbf{h}_{k}^{H}\Delta
\mathbf{h}_{j}\right) \right]\notag\\& =\mathrm{tr}\left( \mathbf{h}_{k}^{H}\mathbf{h}%
_{k}\right) N\delta ^{2}=MN\delta ^{2}.
\end{align}%
In (c), we omit the negligibly small random scattering correlation part.
Then, the expression in Equation (\ref{E_513}) comes immediately after some straight forward mathematical manipulation.

\chapter{Proof of Theorem 3.1}\label{appendix_3a}

Now we prove Theorem \ref{thm:Theo_31}.
\begin{equation}
R_{\mathrm{HB}} =\mathrm{\mathbb{E}}_{%
\mathbf{H}_{\mathrm{S}}}\left\{ \log _{2}\left[ 1+\left[ \mathrm{tr}\left[ (%
\mathbf{H}_{\mathrm{eq}}^{T}\mathbf{H}_{\mathrm{eq}}^{\ast })^{-1}\right] %
\right] ^{-1}\dfrac{E_{s}}{\sigma _{\mathrm{MS}}^{2}}\right] \right\} .
\label{Proof_1}
\end{equation}\vspace*{-0.0mm}%
First, we introduce some preliminaries.

Since $\mathbf{H}_{\mathrm{eq}}^{T}\mathbf{H}_{\mathrm{eq}}^{\ast}$ is a positive definite Hermitian matrix, by eigenvalue decomposition, it can be decomposed as $\mathbf{H}_{\mathrm{eq}}^{T}\mathbf{H}_{\mathrm{eq}}^{\ast }=\mathbf{U\Lambda V}^{H}$, $\mathbf{\Lambda}\in\mathbb{C}^{N\times N}$ is the positive diagonal eigenvalue matrix, while $\mathbf{V}\in\mathbb{C}^{N\times N}$ and $\mathbf{U}\in\mathbb{C}^{N\times N}$ are unitary matrixes, $\mathbf{U=V}^{H}$.

The trace of the eigenvalues of $\mathbf{H}_{\mathrm{eq}}^{T}\mathbf{H}_{\mathrm{eq}}^{\ast}$ equals to the trace of matrix $\mathbf{\Lambda }$.

Then we can rewrite the power normalization factor in (\ref{Proof_1}) as\vspace*{-0.0mm}%
\begin{align}
\vspace*{-0mm}
\dfrac{N}{\mathrm{tr}\left[ (\mathbf{H}_{\mathrm{eq}}^{T}\mathbf{H}_{\mathrm{eq}}^{\ast })^{-1}\right] }&=N\left[ \mathrm{tr}\left[ \mathbf{U\Lambda U}^{H}\right] ^{-1}\right]^{-1}\label{Proof_2}\\
&=N\left[ \mathrm{tr}\left( \mathbf{\Lambda }^{-1}\right) \right] ^{-1}=\left[ \overset{N}{\underset{i=1}{\sum }}\frac{1}{N}\lambda _{i}^{-1}\right]^{-1}, \vspace*{-1mm} \notag
\end{align}\vspace{-0.0mm}%
In addition, $f(x)=x^{-1}$, $x > 0$,  is a strictly decreasing convex function and exploiting the convexity, we have the following results \cite{book:infotheory}\vspace{-0mm}%
\begin{equation}
\vspace{-0.0mm}
\left[ \overset{N}{\underset{i=1}{\sum }}\frac{1}{N}\lambda _{i}^{-1}\right]
^{-1}\leqslant \overset{N}{\underset{i=1}{\sum }}\frac{1}{N}\left[ \left(
\lambda _{i}^{-1}\right) ^{-1}\right] =\overset{N}{\underset{i=1}{\sum }}%
\frac{1}{N}\lambda _{i}.  \label{JIE}\vspace{-0.0mm}
\end{equation}\vspace{-0.0mm}%
Therefore, based on (\ref{Proof_2}) and (\ref{JIE}), we have the following inequality: \vspace{-0mm}%
\begin{align}
\vspace{-0.0mm}
\dfrac{1}{\mathrm{tr}\left[ \left( \mathbf{H}_{\mathrm{eq}}^{T}\mathbf{H}_{%
\mathrm{eq}}^{\ast }\right) ^{-1}\right] }\leqslant& \overset{N}{\underset{i=1%
}{\sum }}\frac{1}{N^{2}}\lambda _{i}=\frac{1}{N^{2}}\overset{N}{\underset{%
i=1}{\sum }}\lambda _{i}
=\frac{1}{N^{2}}\mathrm{tr}\left[ \mathbf{H}_{\mathrm{eq}}^{T}\mathbf{H}_{\mathrm{eq}}^{\ast }\right] .  \label{Proof_3} \vspace{-0.0mm}
\end{align}\vspace{-0.0mm}%

Based on (\ref{Proof_3}), Equation (\ref{Proof_1}) can be rewritten as\vspace{-1.0mm}%
\begin{align}
R_{\mathrm{HB}}\overset{(a)}{\leqslant }&\mathrm{\mathbb{E}}_{\mathbf{H}_{\mathrm{S}%
}}\left\{ \log _{2}\left[ 1+\frac{1}{N^{2}}\mathrm{tr}\left[ \mathbf{H}_{%
\mathrm{eq}}^{T}\mathbf{H}_{\mathrm{eq}}^{\ast }\right] \dfrac{E_{s}}{\sigma
_{\mathrm{MS}}^{2}}\right] \right\} \notag\\
\overset{(b)}{\leqslant }&\log _{2}\left\{
1+\frac{1}{N^{2}}\mathrm{\mathbb{E}}_{\mathbf{H}_{\mathrm{S}}}\left[ \mathrm{tr}%
\left( \mathbf{H}_{\mathrm{eq}}^{T}\mathbf{H}_{\mathrm{eq}}^{\ast }\right) %
\right] \dfrac{E_{s}}{\sigma _{\mathrm{MS}}^{2}}\right\}  \notag \\
=&\log _{2}\left\{ 1+\frac{1}{N^{2}}\dfrac{E_{s}}{\sigma _{\mathrm{MS}}^{2}}%
\left[ \left( \dfrac{\upsilon}{\upsilon+1}\right) MP\| \mathbf{F}_{\mathrm{RF}}^{H}\mathbf{F}_{\mathrm{RF}} \|_{\mathrm{F}}^{2} \right]  +\left( \dfrac{1}{%
\upsilon+1}\right)\dfrac{E_{s}}{\sigma _{\mathrm{MS}}^{2}} \right\}.
\label{Proof_HBUB}
\end{align}\vspace{-0.0mm}%
In $(a),$ we follow (\ref{Proof_3}) and in $(b),$ we adopt the Jensen's inequality. This completes the proof.

\chapter{Proof of Theorem 3.2}\label{appendix_3b}

Now we prove Theorem \ref{thm:Theo_32}.

The receive SINR of user $k$ is given by \vspace*{-0.0mm}%
\begin{equation}
\vspace*{-0mm}
\widetilde{\mathrm{SINR}}_{\mathrm{ZF}}^{k}=\frac{\widetilde{\beta }^{2}E_{s}%
}{\underset{\mathrm{Interference}\text{\ }\mathrm{term}\text{\ }\mathrm{due}\text{\ }\mathrm{to}\text{\ }\mathrm{errors}}{\underbrace{\widetilde{\beta }^{2}E_{s}\widehat{\mathbf{h}}_{%
\mathrm{eq,}k}^{T}\mathrm{\mathbb{E}}_{{\Delta}\widehat{\mathbf{H}}_{\mathrm{eq}}}\left[ \Delta \widehat{\mathbf{W}}_{\mathrm{eq}%
}\Delta \widehat{\mathbf{W}}_{\mathrm{eq}}^{H}\right] \widehat{\mathbf{h}}_{%
\mathrm{eq,}k}^{\ast }}}+\sigma _{\mathrm{MS}}^{2}}.\vspace*{-0mm}
\end{equation}\vspace{-0.0mm}%
To express $\mathrm{\mathbb{E}}_{{\Delta}\widehat{\mathbf{H}%
}_{\mathrm{eq}}}\left[ \Delta \widehat{\mathbf{W}}_{\mathrm{eq}}\Delta
\widehat{\mathbf{W}}_{\mathrm{eq}}^{H}\right]$, first we present the
normalized expression of\textbf{\ }$\Delta \widehat{\mathbf{W}}_{\mathrm{eq}%
} =$\vspace*{-0.0mm}
\begin{align}
\sqrt{1+\delta ^{2}}&(\widehat{\mathbf{H}}_{\mathrm{eq}}^{\ast}+\Delta \widehat{%
\mathbf{H}}_{\mathrm{eq}}^{\ast})\left[ (\widehat{\mathbf{H}}_{\mathrm{eq}%
}+\Delta \widehat{\mathbf{H}}_{\mathrm{eq}})^{T}(\widehat{\mathbf{H}}_{%
\mathrm{eq}}^{\ast}+\Delta \widehat{\mathbf{H}}_{\mathrm{eq}}^{\ast})\right] ^{-1}-\widehat{\mathbf{H}}_{\mathrm{eq}}^{\ast }\left( \widehat{\mathbf{H}}_{%
\mathrm{eq}}^{T}\widehat{\mathbf{H}}_{\mathrm{eq}}\right)^{-1},
\label{Eq_2222}\vspace*{-0.0mm}
\end{align}%
where $\mathbf{K}=(\widehat{\mathbf{H}}_{\mathrm{eq}}^{T}\widehat{\mathbf{H}}%
_{\mathrm{eq}}^{\ast })$, \vspace*{+1mm}$\mathbf{D}=(\Delta \widehat{\mathbf{H}}_{\mathrm{%
eq}}^{T}\widehat{\mathbf{H}}_{\mathrm{eq}}^{\ast }+\widehat{\mathbf{H}}_{%
\mathrm{eq}}^{T}\Delta \widehat{\mathbf{H}}_{\mathrm{eq}}^{\ast }+\Delta
\widehat{\mathbf{H}}_{\mathrm{eq}}^{T}\Delta \widehat{\mathbf{H}}_{\mathrm{eq%
}}^{\ast }),$\vspace*{+1mm} and $\widehat{\mathbf{W}}_{\mathrm{eq}}=\widehat{%
\mathbf{H}}_{\mathrm{eq}}^{\ast }\mathbf{K}^{-1}.$

The matrix inversion
approximation is given by\vspace*{-0.0mm}
\begin{equation}
\vspace*{-0mm}
\mathbf{(K+D)}^{-1}\approx \left[ \mathbf{K}^{-1}-\mathbf{K}^{-1}\mathbf{DK}%
^{-1}\right] .\vspace*{-0mm}
\end{equation}\vspace*{-0.0mm}%
In this case, we re-express (\ref{Eq_2222}) as\vspace*{-0.0mm}
\begin{align}
\vspace*{-0mm}
\Delta \widehat{\mathbf{W}}_{\mathrm{eq}}\approx &\sqrt{1+\delta ^{2}}\left(
\widehat{\mathbf{H}}_{\mathrm{eq}}^{\ast }+\Delta \widehat{\mathbf{H}}_{%
\mathrm{eq}}^{\ast }\right) \left( \mathbf{K}^{-1}-\mathbf{K}^{-1}\mathbf{DK}^{-1}\right)-\widehat{\mathbf{H}}_{\mathrm{eq}}^{\ast }\mathbf{K}^{-1}.\vspace*{-0mm}
\end{align}\vspace*{-0mm}%
Finally, we have
\begin{align}
&E_{\mathrm{\Delta}\widehat{\mathrm{\mathbf{H}}}_{\mathrm{eq}%
}}\left[ \Delta \widehat{\mathbf{W}}_{\mathrm{eq}}\Delta \widehat{\mathbf{W}}%
_{\mathrm{eq}}^{H}\right]\notag\\
& \overset{\left( b\right) }{\approx } (\sqrt{1+\delta ^{2}}-1)^{2}\left[
\widehat{\mathbf{W}}_{\mathrm{eq}}\widehat{\mathbf{W}}_{\mathrm{eq}}^{H}%
\right] +2\sqrt{1+\delta ^{2}}(1-\sqrt{1+\delta ^{2}})\delta ^{2}N\left[ \widehat{\mathbf{W}}_{\mathrm{eq}}\mathbf{K}^{-1}\widehat{\mathbf{%
W}}_{\mathrm{eq}}^{H}\right]  \notag \\
& +(2\sqrt{1+\delta ^{2}}-\sqrt{1+\delta ^{2}})\delta ^{2}\left[ \mathrm{tr}(%
\mathbf{K}^{-1})\widehat{\mathbf{W}}_{\mathrm{eq}}\widehat{\mathbf{W}}_{%
\mathrm{eq}}^{H}\right]  \label{Eq_30121} \\
&\overset{(d)}{\approx } (\sqrt{1+\delta ^{2}}-1)^{2}\left[ \widehat{\mathbf{W}}_{%
\mathrm{eq}}\widehat{\mathbf{W}}_{\mathrm{eq}}^{H}\right]+\frac{2N}{\xi MP}\sqrt{1+\delta ^{2}}(1-\sqrt{1+\delta ^{2}})\delta ^{2}\widehat{%
\mathbf{W}}_{\mathrm{eq}}\mathbf{G}_{\mathrm{L}}^{-1}\widehat{\mathbf{W}}_{%
\mathrm{eq}}^{H}  \notag \\
& +(2\sqrt{1+\delta ^{2}}-\sqrt{1+\delta ^{2}})\delta ^{2}\left[ \frac{N}{\xi MP}\widehat{\mathbf{W}}_{\mathrm{eq}}\mathbf{G}_{\mathrm{%
L}}^{-1}\widehat{\mathbf{W}}_{\mathrm{eq}}^{H}\right].  \label{Eq_3011}\vspace{-1.0mm}
\end{align}\vspace{-0.0mm}%
In (b), we omit some negligibly small parts which neither dominate the performance nor scale with $M$.
In (d), while the number of antennas $M\rightarrow \infty $, $\mathbf{K}=\widehat{\mathbf{H}}_{\mathrm{eq}}^{T}\widehat{\mathbf{H}}_{\mathrm{eq}}^{\ast }\underset{M\rightarrow \infty }{\overset{a.s.}{\approx }} \xi MP\mathbf{G}_{\mathrm{L}}$ holds, where $\xi \in \left( 0,1\right]$.
We substitute (\ref{Eq_30121}) into (\ref{SINR_with_error1}) and substitute (\ref{Eq_3011}) into (\ref{SINR_with_error1}), the Equations (\ref{Theo_IP_RL}) and (\ref{Coro_RZFK1}) come immediately after some straight forward mathematical manipulation.\vspace*{-0mm}

\chapter{Proof of Theorem 4.1}\label{appendix_4a}
The normalized MSE performance of equivalent channel estimation under the impact of pilot contamination is given by
\begin{align}
\hspace{-0mm}\mathrm{NMSE}_{\mathrm{eq,}k} &=\frac{1}{N}
\mathbb{E}_{{\color{black}\mathbf{U}}_{l,i}}\left[ \left( \dfrac{1}{\sqrt{MP}}\Delta\widehat{\mathbf{h}}_{\mathrm{eq},k}^{T}\right)\left(\dfrac{1}{\sqrt{MP}}
\Delta\widehat{\mathbf{h}}_{\mathrm{eq},k}^{\ast}\right) \right]   \notag \\
&\overset{(a)}{\approx}\frac{1}{{\varpi_{{k}}}NMP}\underset{\mathrm{Pilot}\text{ }\mathrm{contamination}}{\underbrace{\mathbb{E}_{{\color{black}\mathbf{U}}_{l,i}}\left[\overset{L}{\underset{l=1}{\sum }}\left(\widehat{\bm{\omega}}^{H}_{l,k}{\color{black}\mathbf{U}}_{l,k}^{T}\mathbf{F}_{\mathrm{RF}}\right)\right]\left[\overset{L}{\underset{l=1}{\sum }}\left(\mathbf{F}_{\mathrm{RF}}^{H}{\color{black}\mathbf{U}}_{l,k}^{\ast}\widehat{\bm{\omega}}_{l,k}\right)\right]}}\notag\\
&+\underset{\mathrm{Effective}\text{ }\mathrm{noise}}{\underbrace{\frac{\sigma _{\mathrm{BS}}^{2}\mathrm{tr}\left[ \mathbf{F}_{\mathrm{RF}%
}^{H}\mathbf{F}_{\mathrm{RF}}\right] }{{\varpi_{ {k}}}E_{\mathrm{P}}NMP}}}.\label{Eq_AP1}
\end{align}
In (a), we omit some small part.
Due to the small-cell radius, the inter-cell uplink propagation channels contain the strongest AoA components from the users in the neighboring cells to the desired BS.
Thus, the part associated with pilot contamination can be expressed as
\begin{align}
&\frac{1}{{\varpi_{ {k}}}}\mathbb{E}_{{\color{black}\mathbf{U}}_{l,k}}\left[\overset{L}{\underset{l=1}{\sum }}\left(\widehat{\bm{\omega}}^{H}_{l,k}{\color{black}\mathbf{U}}_{l,k}^{T}\mathbf{F}_{\mathrm{RF}}\mathbf{F}_{\mathrm{RF}}^{H}{\color{black}\mathbf{U}}_{l,k}^{\ast}\widehat{\bm{\omega}}_{l,k}\right)\right] \notag\\
=& \mathbb{E}_{{\color{black}\mathbf{U}}_{l,k}}\left[\overset{L}{\underset{l=1}{\sum }}\widehat{\bm{\omega}}^{H}_{l,k}\left(\sqrt{\frac{{\color{black}\varsigma}_{l,k}}{{\color{black}\varsigma}_{l,k}+1}}\widehat{\rho} _{l,k}{\color{black}\mathbf{U}}_{\mathrm{SAoA},l,k}^{T}+\sqrt{\frac{1}{{\color{black}\varsigma}_{l,k}+1}}\widehat{\rho} _{l,k}{\color{black}\mathbf{U}}_{\mathrm{S},l,k}^{T}\right)\mathbf{F}_{\mathrm{RF}}\right.\notag \\
& \times\left.\mathbf{F}_{\mathrm{RF}}^{H}\left(\sqrt{\frac{{\color{black}\varsigma}_{l,k}}{{\color{black}\varsigma}_{l,k}+1}}\widehat{\rho} _{l,k}{\color{black}\mathbf{U}}_{\mathrm{SAoA},l,k}^{\ast}+\sqrt{\frac{1}{{\color{black}\varsigma}_{l,k}+1}}\widehat{\rho} _{l,k}{\color{black}\mathbf{U}}_{\mathrm{S},l,k}^{\ast}\right)\widehat{\bm{\omega}}_{l,k}\right] \notag\\
\approx &{{\mathbb{E}_{{\color{black}\mathbf{U}}_{\mathrm{SAoA},l,k}}\underset{\mathrm{Inter-cell}\text{ }\mathrm{interference}\text{ }\mathrm{caused}\text{ }\mathrm{by}\text{ }\mathrm{strongest}\text{ }\mathrm{AoA}\text{ }\mathrm{components}}{\underbrace{\left[\overset{L}{\underset{l=1}{\sum }}\left({\frac{\widehat{\rho} _{l,k}^{2}{\color{black}\varsigma}_{l,k}}{{\color{black}\varsigma}_{l,k}+1}}\widehat{\bm{\omega}}^{H}_{l,k} {\color{black}\mathbf{U}}_{\mathrm{SAoA},l,k}^{T}\mathbf{F}_{\mathrm{RF}}\mathbf{F}_{\mathrm{RF}}^{H} {\color{black}\mathbf{U}}_{\mathrm{SAoA},l,k}^{\ast}\widehat{\bm{\omega}}_{l,k}\right)\right]}}}}\notag\\
&+
{\mathbb{E}_{{\color{black}\mathbf{U}}_{\mathrm{S},l,k}}\underset{\mathrm{Inter-cell}\text{ }\mathrm{interference}\text{ }\mathrm{caused}\text{ }\mathrm{by}\text{ }\mathrm{scattering}\text{ }\mathrm{components}}{\underbrace{\left[\overset{L}{\underset{l=1}{\sum }}\left({\frac{\widehat{\rho} _{l,k}^{2}}{{\color{black}\varsigma}_{l,k}+1}}\widehat{\bm{\omega}}^{H}_{l,k}{\color{black}\mathbf{U}}_{\mathrm{S},l,k}^{T}\mathbf{F}_{\mathrm{RF}}\mathbf{F}_{\mathrm{RF}}^{H}{\color{black}\mathbf{U}}_{\mathrm{S},l,k}^{\ast}\widehat{\bm{\omega}}_{l,k}\right)\right]}}},
\end{align}
where $\widehat{\rho}_{l,k}= \sqrt{\dfrac{\widehat{\varpi}_{l,{k}}}{\varpi_{{k}}}}$.
Then, the inter-cell interference caused by scattering component can be further approximated as
\begin{align}
&{{\left[\overset{L}{\underset{l=1}{\sum }}{\frac{\widehat{\rho} _{l,k}^{2}}{{\color{black}\varsigma}_{l,k}+1}}\widehat{\bm{\omega}}^{H}_{l,k}\mathbb{E}_{{\color{black}\mathbf{U}}_{\mathrm{S},l,k}}\left( {\color{black}\mathbf{U}}_{\mathrm{S},l,k}^{T}\mathbf{F}_{\mathrm{RF}}\mathbf{F}_{\mathrm{RF}}^{H} {\color{black}\mathbf{U}}_{\mathrm{S},l,k}^{\ast}\right)\widehat{\bm{\omega}}_{l,k}\right]}}\notag\\
\approx &\left[\overset{L}{\underset{l=1}{\sum }}{\frac{\widehat{\rho} _{l,k}^{2}}{{\color{black}\varsigma}_{l,k}+1}}\widehat{\bm{\omega}}^{H}_{l,k}{\mathrm{tr}}\left(\mathbf{F}_{\mathrm{RF}}\mathbf{F}_{\mathrm{RF}}^{H}\right)\mathbf{I}_{\mathrm{P}}\widehat{\bm{\omega}}_{l,k}\right]
=\overset{L}{\underset{l=1}{\sum }}{\frac{\widehat{\rho} _{l,k}^{2}}{{\color{black}\varsigma}_{l,k}+1}}N. \label{Eq_AP2}
\end{align}
Now, we would like to approximate the inter-cell interference caused by the multi-cell strongest AoA components
\begin{align}
&\mathbb{E}_{{\color{black}\mathbf{U}}_{\mathrm{SAoA},l,k}}\left[\overset{L}{\underset{l=1}{\sum }}\left({\frac{\widehat{\rho} _{l,k}^{2}{\color{black}\varsigma}_{l,k}}{{\color{black}\varsigma}_{l,k}+1}}\widehat{\bm{\omega}}^{H}_{l,k}{\color{black}\mathbf{U}}_{\mathrm{SAoA},l,k}^{T}\mathbf{F}_{\mathrm{RF}}\mathbf{F}_{\mathrm{RF}}^{H} {\color{black}\mathbf{U}}_{\mathrm{SAoA},l,k}^{\ast}\widehat{\bm{\omega}}_{l,k}\right)\right] \notag \\
=&\mathbb{E}_{{\color{black}\mathbf{U}}_{\mathrm{SAoA},l,k}}\left[\overset{L}{\underset{l=1}{\sum }}\left[{\frac{\widehat{\rho} _{l,k}^{2}{\color{black}\varsigma}_{l,k}}{{\color{black}\varsigma}_{l,k}+1}}\left({\color{black}\mathbf{u}}_{\mathrm{SAoA,}l,k}^{\mathrm{BS}}\right)^{T}\mathbf{F}_{\mathrm{RF}}
\mathbf{F}_{\mathrm{RF}}^{H}\left({\color{black}\mathbf{u}}_{\mathrm{SAoA,}l,k}^{\mathrm{BS}}\right)^{\ast}{\color{black}\mathbf{u}}_{%
\mathrm{SAoA,}l,k}^{T}\widehat{\bm{\omega}}_{l,k}\widehat{\bm{\omega}}^{H}_{l,k}{\color{black}\mathbf{u}}_{%
\mathrm{SAoA,}l,k}^{\ast}\right]\right]. \label{Eq_apdix1}
\end{align}%
In Equation (\ref{Eq_apdix1}), we have
\begin{align}
\widehat{\bm{\omega }}_{l,k}^{H}&=\frac{1}{\sqrt{P}}\left[
\begin{array}{cccc}
1, & e^{j2\pi \tfrac{d}{\lambda }\cos \left( {\phi}_{l,k}\right) }, & \ldots , & e^{j2\pi \left( P-1\right) \tfrac{d}{\lambda }\cos \left({\phi}_{l,k}\right) }%
\end{array}%
\right] ^{\ast} \text{\ and}\\
{\color{black}\mathbf{u}}_{\mathrm{SAoA},l,k}^{\ast} &=\left[\begin{array}{cccc} 1, & e^{-j2\pi \tfrac{d}{\lambda }\cos \left( \Delta\phi_{l,k}\right) }, & \ldots , & \text{ }e^{-j2\pi \left( P-1\right) \tfrac{d}{\lambda }\cos \left( \Delta\phi_{l,k}\right) }\end{array}\right] ^{H},
\end{align}%
where variables $\phi _{l,k}\in \left[0, \text{\ }\pi \right]$ is the angle of incidence of the strongest AoA path at antenna arrays of user $k$ in cell $l$, and $\Delta\phi _{l,k}\in \left[0, \text{\ }\pi \right] $ is the angle of incidence of the inter-cell strongest AoA path at antenna arrays from user $k$ of cell $l$ to the desired BS.
By defining the array gain function $G_{\mathrm{act},P}{\left[x\right]}$, cf. \cite{book:wireless_comm}, as
\begin{equation}
G_{\mathrm{act},P}{\left[x \right]} = \frac{\left\{\sin\left[ P \pi\frac{d}{\lambda}\left( x \right)\right]\right\}^{2}}{{P}\left\{\sin\left[\pi\frac{d}{\lambda}\left( x \right)\right]\right\}^{2}},
\end{equation}where $\dfrac{d}{\lambda}=\dfrac{1}{2}$.
Then, we have:
\begin{align}
&{\color{black}\mathbf{u}}_{%
\mathrm{SAoA,}l,k}^{T}\widehat{\bm{\omega}}_{l,k}\widehat{\bm{\omega }}_{l,k}^{H}{\color{black}\mathbf{u}}_{\mathrm{SAoA},l,k}^{\ast}=G_{\mathrm{act},P}{\left[ \cos\left(\phi_{l,k}\right) - \cos\left(\Delta\phi_{l,k}\right) \right]}\notag\\
= &\frac{{P}\left\{\sinc\left[ P \frac{\pi }{2}\left( \cos \left(\phi_{l,k}\right)-\cos\left(\Delta\phi_{l,k}\right)   \right)\right]\right\}^{2}}{\left\{\sinc\left[\frac{\pi }{2}\left( \cos \left(\phi_{l,k}\right)-\cos\left(\Delta\phi_{l,k}\right)   \right)\right]\right\}^{2}}.\label{Eq_AP3}
\end{align}
Similarly, we can have following preliminaries, i.e.,
\begin{align}
\left({\color{black}\mathbf{u}}_{\mathrm{SAoA},l,k}^{\mathrm{BS}}\right)^{T} &=\left[\begin{array}{cccc}1, & e^{-j2\pi \tfrac{d}{\lambda }\cos \left( \theta _{l,k}\right) }, & \ldots, & \text{ }e^{-j2\pi \left( M-1\right) \tfrac{d}{\lambda }\cos \left(\theta _{l,k}\right) }\end{array}\right] \text{\ and}\notag\\
\widehat{\bm{{\color{black}\nu}}}_{i}&=
\frac{1}{\sqrt{M}}\left[
\begin{array}{cccc}
1, & e^{j2\pi \tfrac{d}{\lambda }\cos \left( {\theta }_{i}\right) }, & \ldots , & \text{ }e^{j2\pi \left( M-1\right) \tfrac{d}{\lambda }\cos
\left( {\theta }_{i}\right) }
\end{array}%
\right]^{T}.
\end{align}
Based on expressions mentioned above, we rewrite $\left({\color{black}\mathbf{u}}_{\mathrm{SAoA,}l,k}^{\mathrm{BS}}\right)^{T}\mathbf{F}_{\mathrm{RF}}
\mathbf{F}_{\mathrm{RF}}^{H}\left({\color{black}\mathbf{u}}_{\mathrm{SAoA,}l,k}^{\mathrm{BS}}\right)^{\ast}$
as
\begin{align}
&\left[\begin{array}{ccc}\left({\color{black}\mathbf{u}}_{\mathrm{SAoA,}l,k}^{\mathrm{BS}}\right)^{T}\widehat{\bm{{\color{black}\nu}}}_{1},\ldots, \left({\color{black}\mathbf{u}}_{\mathrm{SAoA,}l,k}^{\mathrm{BS}}\right)^{T}\widehat{\bm{{\color{black}\nu}}}_{N}\end{array}\right]\notag\\
&\hspace{+20mm}\times\left[\begin{array}{ccc}\widehat{\bm{{\color{black}\nu}}}^{H}_{1}\left({\color{black}\mathbf{u}}_{\mathrm{SAoA,}l,k}^{\mathrm{BS}}\right)^{\ast},\ldots, \widehat{\bm{{\color{black}\nu}}}^{H}_{N}\left({\color{black}\mathbf{u}}_{\mathrm{SAoA,}l,k}^{\mathrm{BS}}\right)^{\ast}\end{array}\right]^{T}\notag\\
=&\overset{N}{\underset{i=1}{\sum }}\frac{\left\{\sin{\left[ \frac{\pi }{2}M\left( \cos \left( \theta _{l,k}\right) - \cos \left( {\theta }_{i}\right) \right)\right]}\right\}^{2}}{M\left\{\sin{\left[ \frac{\pi }{2}\cos \left( \theta _{l,k}\right) - \cos \left( {\theta }_{i}\right) \right]}\right\}^{2}}
=\overset{N}{\underset{i=1}{\sum }}G_{\mathrm{act},M}{\left[ \cos \left( \theta _{l,k}\right)  - \cos \left( {\theta }_{i}\right) \right]}. \label{Eq_AP4}
\end{align}%
Then, Equation (\ref{Eq_apdix1}) can be re-expressed as
\begin{align}
&\mathbb{E}_{{\color{black}\mathbf{U}}_{\mathrm{SAoA},l,k}}\left[\overset{L}{\underset{l=1}{\sum }}\left[{\frac{\widehat{\rho} _{l,k}^{2}{\color{black}\varsigma}_{l,k}}{{\color{black}\varsigma}_{l,k}+1}}\left({\color{black}\mathbf{u}}_{\mathrm{SAoA,}l,k}^{\mathrm{BS}}\right)^{T}\mathbf{F}_{\mathrm{RF}}
\mathbf{F}_{\mathrm{RF}}^{H}\left({\color{black}\mathbf{u}}_{\mathrm{SAoA,}l,k}^{\mathrm{BS}}\right)^{\ast}{\color{black}\mathbf{u}}_{%
\mathrm{SAoA,}l,k}^{T}\widehat{\bm{\omega}}_{l,k}\widehat{\bm{\omega}}^{H}_{l,k}{\color{black}\mathbf{u}}_{%
\mathrm{SAoA,}l,k}^{\ast}\right]\right]\notag\\
= & \mathbb{E}_{ \phi_{l,k}, \phi_{l,k},\theta _{l,k}, \theta_{i} }\left\{\overset{L}{\underset{l=1}{\sum }}\left[{\frac{\widehat{\rho} _{l,k}^{2}{\color{black}\varsigma}_{l,k}}{{\color{black}\varsigma}_{l,k}+1}}G_{\mathrm{act},P}{\left[ \cos\left(\phi_{l,k}\right) - \cos\left(\Delta\phi_{l,k}\right) \right]}\right.\right.\notag\\
&\hspace{+50mm}\left.\left.\times\overset{N}{\underset{i=1}{\sum }}G_{\mathrm{act},M}{\left[ \cos \left( \theta _{l,k}\right)  - \cos \left( {\theta }_{i}\right) \right]} \right]\right\},  \label{ffff}
\end{align}
where $\cos\left(\phi_{l,k}\right)$, $\cos\left(\Delta\phi_{l,k}\right)$, $\cos \left( \theta _{l,k}\right)$, and $\cos \left( {\theta }_{i}\right)$ are uniformly distributed over $[-1, \text{\ } 1]$.
Besides, they are independent with each other.
Exploiting the periodic property of function $e^{j2\pi x}$, the linear antenna array gain $G_{\mathrm{act},P}{\left[ \cos\left(\phi_{l,k}\right) - \cos\left(\Delta\phi_{l,k}\right) \right]}$ is equal in distribution to $G_{\mathrm{act},P}{\left[\mu_{l,k}\right]}$ and $G_{\mathrm{act},M}{\left[ \cos \left( \theta _{l,k}\right)  - \cos \left( {\theta }_{i}\right) \right]}$ is equal in distribution to $G_{\mathrm{act},M}{\left[  \epsilon_{l,k,i} \right]}$, where $\mu_{l,k}$, $ k \in \left\{ 1,\ldots,N \right\}$, and $\epsilon_{l,k,i}$, $i \in \left\{ 1,\ldots,N \right\}$ are uniformly distributed over $[-1, \text{\ } 1]$ (Lemma $1$ of \cite{Yu2017}).
In addition, $\mu_{l,k}$ and $\epsilon_{l,k,i}$ is independent with each other.
Thus, we re-express Equation (\ref{ffff}) as
\begin{align}
&\overset{L}{\underset{l=1}{\sum }}{\frac{\widehat{\rho} _{l,k}^{2}{\color{black}\varsigma}_{l,k}}{{\color{black}\varsigma}_{l,k}+1}}\mathbb{E}_{\phi_{l,k}, \phi_{l,k} }\left[G_{\mathrm{act},P}{\left[ \cos\left(\phi_{l,k}\right) - \cos\left(\Delta\phi_{l,k}\right) \right]}\right]\notag\\
&\hspace{+20mm}\times \overset{N}{\underset{i=1}{\sum }}\mathbb{E}_{\theta _{l,k},\theta_{i} }\left[G_{\mathrm{act},M}{\left[ \cos \left( \theta _{l,k}\right)  - \cos \left( {\theta }_{i}\right) \right]} \right]\notag\\
= & \overset{L}{\underset{l=1}{\sum }}{\frac{\widehat{\rho} _{l,k}^{2}{\color{black}\varsigma}_{l,k}}{{\color{black}\varsigma}_{l,k}+1}}\mathbb{E}_{\mu_{l,k}}\left[G_{\mathrm{act},P}{\left[\mu_{l,k}\right]}\right]\mathbb{E}_{{\epsilon}_{l,k,i}}\left(G_{\mathrm{act},M}{\left[  \epsilon_{l,k,i} \right]}\right)    \notag\\
= &\overset{L}{\underset{l=1}{\sum }}{\frac{\widehat{\rho} _{l,k}^{2}{\color{black}\varsigma}_{l,k}}{{\color{black}\varsigma}_{l,k}+1}}\mathbb{E}_{\mu_{l,k}}\left(\frac{\left[\sinc{\left( \frac{\pi }{2}P\mu_{l,k}\right)}\right]^{2}P}{\left[\sinc{\left( \frac{\pi }{2}\mu_{l,k}\right)}\right]^{2}}\right)\mathbb{E}_{{\epsilon}_{l,k,i}}\left(\overset{N}{\underset{i=1}{\sum }}\frac{\left[\sinc{\left( \frac{\pi }{2}M\epsilon_{l,k,i}\right)}\right]^{2}M}{\left[\sinc{\left( \frac{\pi }{2}\epsilon_{l,k,i}\right)}\right]^{2}}\right)\notag \\
\overset{(b)}{\geqslant} & \overset{L}{\underset{l=1}{\sum }}{\frac{\widehat{\rho} _{l,k}^{2}{\color{black}\varsigma}_{l,k}}{{\color{black}\varsigma}_{l,k}+1}}\mathbb{E}_{\mu_{l,k}}\left({\left[\sinc{\left( \frac{\pi }{2}P\mu_{l,k}\right)}\right]^{2}P}\right)\mathbb{E}_{{\epsilon}_{l,k,i}}\left(\overset{N}{\underset{i=1}{\sum }}{\left[\sinc{\left( \frac{\pi }{2}M\epsilon_{l,k,i}\right)}\right]^{2}M}\right) \notag \\
\overset{(c)}{\approx} & \overset{L}{\underset{l=1}{\sum }}{\frac{\widehat{\rho} _{l,k}^{2}{\color{black}\varsigma}_{l,k}}{{\color{black}\varsigma}_{l,k}+1}}N. \label{Eq_AP5}\vspace{-2mm}
\end{align}
%
In (b), we exploit the fact that
\begin{equation}
[\sinc(x)]^{2} = \left(\frac{\sin{x}}{x}\right)^2\leqslant 1.
\end{equation}
In (c), we explore the law of integration of sinc function for the number of antennas $M$ is sufficiently large, i.e.,
\begin{align}
&\mathbb{E}_{{\epsilon}_{l,k,i}}\left[{\left[\sinc{\left( \dfrac{\pi }{2}M\epsilon_{l,k,i}\right)}\right]^{2}}M\right]= M\int^{1}_{-1}\frac{1}{2}{\left[\sinc{\left(\dfrac{\pi}{2}M\epsilon_{l,k,i}\right)}\right]^{2}}d\epsilon_{l,k,i}\notag\\
=&\frac{M}{M\pi}\int^{1}_{-1}{\left[\sinc{\left(\dfrac{\pi}{2}M\epsilon_{l,k,i}\right)}\right]^{2}}d\dfrac{\pi}{2}M\epsilon_{l,k,i}
\overset{M\rightarrow\infty}\approx\frac{1}{\pi}\int^{\infty}_{-\infty}{\left[\sinc{\left(\chi\right)}\right]^{2}}d\chi = 1,
\end{align}
where $\chi=\dfrac{\pi}{2}M\epsilon_{l,k,i}$.
We substitute Equation (\ref{Eq_AP2}) and (\ref{Eq_AP5}) into (\ref{Eq_AP1}), the expression in (\ref{PC}) comes immediately after some straightforward mathematical manipulation.

\chapter{Proof of Lemma 4.2}\label{appendix_4b}
The inter-cell downlink transmission interference is given by
\begin{align}
& \Omega_{k} \geqslant \notag\\
& E_{s} \underset{\mathrm{Inter-cell}\text{ }\mathrm{interfernce}\text{ }\mathrm{caused}\text{ }\mathrm{by}\text{ }\mathrm{strongest}\text{ }\mathrm{AoAs}}{\underbrace{\mathbb{E}_{\mathbf{H}_{\mathrm{SAoA},l,k}}\left[\overset{L}{\underset{l=1}{\sum }}{\widetilde{\varpi}_{l,{k}}}\left(\widehat{\beta_{l}}\right)^{2}\left(
{\frac{{\color{black}\varsigma}_{l,k}}{{\color{black}\varsigma}_{l,k}+1}}\right)\widehat{\bm{\omega }}_{k}^{H}\mathbf{H}_{\mathrm{SAoA},l,k}^{T}\mathbf{F}_{\mathrm{RF},l}\widehat{\mathbf{W}}_{\mathrm{eq},l}\widehat{\mathbf{W}}_{\mathrm{eq},l}^{H}\mathbf{F}_{\mathrm{RF},l}^{H}
\mathbf{H}_{\mathrm{SAoA},l,k}^{\ast}\widehat{\bm{\omega }}_{k} \right]}}\notag \\
&+E_{s} \underset{\mathrm{Inter-cell}\text{ }\mathrm{interference}\text{ }\mathrm{caused}\text{ }\mathrm{by}\text{ }\mathrm{scattering}\text{ }\mathrm{component}}{\underbrace{\mathbb{E}_{\mathbf{H}_{\mathrm{S},l,k}}\left[\overset{L}{\underset{l=1}{\sum }}{\widetilde{\varpi}_{l,{k}}}\left(\widehat{\beta_{l}}\right)^{2}\left(
{\frac{1}{{\color{black}\varsigma}_{l,k}+1}} \right)\widehat{\bm{\omega }}_{k}^{H}\mathbf{H}_{\mathrm{S},l,k}^{T}\mathbf{F}_{\mathrm{RF},l}\widehat{\mathbf{W}}_{\mathrm{eq},l}\widehat{\mathbf{W}}_{\mathrm{eq},l}^{H}\mathbf{F}_{\mathrm{RF},l}^{H}
\mathbf{H}_{\mathrm{S},l,k}^{\ast}\widehat{\bm{\omega }}_{k} \right]}}.\label{ff0}
\end{align}
Following similar approaches shown in Equations (\ref{Eq_apdix1})$-$(\ref{Eq_AP4}), we can re-express the inter-cell interference caused by the strongest AoA components as
\begin{align}
&\mathbb{E}_{\mathbf{H}_{\mathrm{SAoA},l,k}}\left\{\overset{L}{\underset{l=1}{\sum }}{\widetilde{\varpi}_{l,{k}}}\left( \widehat{\beta_{l}}\right)^{2}\left(
{\frac{{\color{black}\varsigma}_{l,k}}{{\color{black}\varsigma}_{l,k}+1}}\right)\right.\notag\\
&\hspace{+20mm}\left.\times\mathrm{tr}\left[\widetilde{\bm{\omega }}_{k}^{H}\mathbf{H}_{\mathrm{SAoA},l,k}^{T}\mathbf{F}_{\mathrm{RF},l}\widehat{\mathbf{W}}_{\mathrm{eq},l}\widehat{\mathbf{W}}_{\mathrm{eq},l}^{H}\mathbf{F}_{\mathrm{RF},l}^{H}
\mathbf{H}_{\mathrm{SAoA},l,k}^{\ast}\widetilde{\bm{\omega }}_{k} \right]\right\} \notag \\
= & \overset{L}{\underset{l=1}{\sum }}\left\{{\widetilde{\varpi}_{l,{k}}}\left(\widehat{\beta_{l}}\right)^{2}\left(
{\frac{{\color{black}\varsigma}_{l,k}}{{\color{black}\varsigma}_{l,k}+1}}\right)\mathbb{E}_{\widehat{\eta}_{l,k}}\left[\tfrac{\sinc^{2}\left[P\frac{\pi}{2}\left(\widehat{\eta}_{l,k}\right)\right]{P}}{\sinc^{2}\left[\frac{\pi}{2}\left(\widehat{\eta}_{l,k}\right)\right]}\right] \right.\notag\\
& \hspace{+10mm}\left.\times\mathrm{tr}\left[\widehat{\mathbf{W}}_{\mathrm{eq},l}
\widehat{\mathbf{W}}_{\mathrm{eq},l}^{H}\cdot\mathbb{E}_{\tau_{l,k,i}}\left[\mathbf{F}_{\mathrm{RF},l}^{H}\left(\mathbf{h}_{\mathrm{SAoA},l,k}^{\mathrm{BS}}\right)^{\ast}\left(\mathbf{h}_{\mathrm{SAoA},l,k}^{\mathrm{BS}}\right)^{T}\mathbf{F}_{\mathrm{RF},l} \right]\right]^{T}\right\}, \label{LOS_formular}
\end{align}
where $ i \in \left\{ 1,\ldots,N \right\}$, $\widehat{\eta}_{l,k}$ and $\tau_{l,k,i}$ are also independent uniformly distributed in $[-1, \text{\ } 1]$. In addition, variables $\Delta\kappa_{l,k}\in \left[ 0,\text{\ }\pi \right]$ is the angle of incidence of the inter-cell strongest AoA path at antenna arrays of user $k$ in the desired cell from the BS in cell $l$, and $\psi _{l,k}\in \left[ 0,\text{\ }\pi \right] $ is the angle of incidence of the inter-cell strongest AoA path at antenna arrays of the BS of cell $l$ for user $k$.
Further, in the large number of antennas regime, $\mathbb{E}_{\tau_{l,k,i}}\left[{\mathbf{F}_{\mathrm{RF},l}^{H}\left(\mathbf{h}_{\mathrm{SAoA},l,k}^{\mathrm{BS}}\right)^{\ast}\left(\mathbf{h}_{\mathrm{SAoA},l,k}^{\mathrm{BS}}\right)^{T}\mathbf{F}_{\mathrm{RF},l}}\right], \text{\ } \ i \in \left\{ 1,\ldots,N \right\}$, can be approximated as
\begin{align}
\hspace{-3mm}\left[
\begin{array}{ccc}
\hspace{-10mm}\mathbb{E}_{\tau_{l,k,1}}\left[\frac{\sinc^{2}\left[M\frac{\pi}{2}\left(\tau_{l,k,1}\right)\right]{M}}{\sinc^{2}\left[\frac{\pi}{2}\left(\tau_{l,k,1}\right)\right]}\right] & \hspace{-40mm}\cdots & \hspace{-20mm}\mathbb{E}_{\tau_{l,k,1},\tau_{l,k,N}}\left[\frac{\sin\left[ M \frac{\pi}{2}\left( \tau_{l,k,1} \right)\right]}{\sqrt{M}\sin\left[\frac{\pi}{2}\left(\tau_{l,k,1}\right)\right]}\frac{\sin\left[ M \frac{\pi}{2}\left( \tau_{l,k,N} \right)\right]}{\sqrt{M}\sin\left[\frac{\pi}{2}\left(\tau_{l,k,N}\right)\right]}\right]  \\
\vdots & \hspace{-20mm}\ddots  & \vdots \\
\mathbb{E}_{\tau_{l,k,1},\tau_{l,k,N}}\left[\frac{\sin\left[ M \frac{\pi}{2}\left( \tau_{l,k,1} \right)\right]}{\sqrt{M}\sin\left[\frac{\pi}{2}\left(\tau_{l,k,1}\right)\right]}\frac{\sin\left[ M \frac{\pi}{2}\left( \tau_{l,k,N} \right)\right]}{\sqrt{M}\sin\left[\frac{\pi}{2}\left(\tau_{l,k,N}\right)\right]}\right]  & \cdots & \hspace{-10mm} \mathbb{E}_{\tau_{l,k,N}}\left[\frac{\sinc^{2}\left[M\frac{\pi}{2}\left(\tau_{l,k,N}\right)\right]{M}}{\sinc^{2}\left[\frac{\pi}{2}\left(\tau_{l,k,N}\right)\right]}\right]
\end{array}\right].\label{Eq_AP6}
\end{align}
%
In the large number of antennas regime, $\mathbb{E}_{\tau_{l,k,i}}\left[\frac{\sinc^{2}\left[M\frac{\pi}{2}\left(\tau_{l,k,i}\right)\right]{M}}{\sinc^{2}\left[\frac{\pi}{2}\left(\tau_{l,k,i}\right)\right]}\right]$, $ i \in \left\{ 1,\ldots, N\right\}$, can be asymptotically approximated as
\begin{equation}
\mathbb{E}_{\tau_{l,k,i}}\left[\frac{\sinc^{2}\left[M\frac{\pi}{2}\left(\tau_{l,k,i}\right)\right]{M}}{\sinc^{2}\left[\frac{\pi}{2}\left(\tau_{l,k,i}\right)\right]}\right]\approx1.
\end{equation}
As variables $\tau_{l,k,i}$ and $\tau_{l,k,j}$ are independent with each other, here we first derive $\mathbb{E}_{\tau_{l,k,i}}\left[\frac{\sin\left[ M \frac{\pi}{2}\left( \tau_{l,k,i}\right)\right]}{\sqrt{M}\sin\left[\frac{\pi}{2}\left(\tau_{l,k,i}\right)\right]}\right]$.
We rewrite $\frac{\sin\left[ M \frac{\pi}{2}\left( \tau_{l,k,i} \right)\right]}{\sqrt{M}\sin\left[\frac{\pi}{2}\left(\tau_{l,k,i}\right)\right]}$ as
\begin{equation}
\frac{\sin\left[ M \frac{\pi}{2}\left( \tau_{l,k,i} \right)\right]}{\sqrt{M}\sin\left[\frac{\pi}{2}\left(\tau_{l,k,i}\right)\right]}
=\sqrt{M}\frac{\frac{\sin\left[ M \frac{\pi}{2}\left( \tau_{l,k,i} \right)\right]}{M\frac{\pi}{2}\left(\tau_{l,k,i}\right)}}{\frac{\sin\left[\frac{\pi}{2}\left(\tau_{l,k,i}\right)\right]}{\frac{\pi}{2}\left(\tau_{l,k,i}\right)}}
=\sqrt{M}\frac{\sinc\left[ M\frac{\pi}{2}\left( \tau_{l,k,i} \right) \right]}{\sinc\left[ \frac{\pi}{2}\left( \tau_{l,k,i} \right) \right]}.\label{0}
\end{equation}
Since the value of $ \tau_{l,k,i} $ is uniformly distributed over $[-1,\text{\ }1]$, we can obtain the following result:
\begin{equation}
\sinc\left({\frac{\pi}{2}}\right)\leq\sinc\left[ \frac{\pi}{2}\left( \tau_{l,k,i} \right) \right]\leq1. \label{1}
\end{equation}
Based on Equation (\ref{1}) and Equation (\ref{0}), we have the following inequality
\begin{equation}
\sqrt{M}\sinc\left[ M\frac{\pi}{2}\left( \tau_{l,k,i} \right) \right]\leq\sqrt{M}\frac{\sinc\left[ M\frac{\pi}{2}\left( \tau_{l,k,i} \right) \right]}{\sinc\left[ \frac{\pi}{2}\left( \tau_{l,k,i} \right) \right]}<2\sqrt{M}\sinc\left[ M\frac{\pi}{2}\left( \tau_{l,k,i} \right) \right]. \label{2}
\end{equation}
Now, we can obtain the following result by exploring the law of integration of sinc function for a sufficiently large number of antennas $M$:
\begin{align}
&\mathrm{E}_{\tau_{l,k,i}}\left[\sqrt{M}\sinc\left[ M\frac{\pi}{2}\left( \tau_{l,k,i} \right) \right]\right]\notag\\
=&\sqrt{M}\int^{1}_{-1}\frac{1}{2}{\sinc{\left[\dfrac{\pi}{2}M\left(\tau_{l,k,i}\right)\right]}}d\tau_{l,k,i}
=\frac{\sqrt{M}}{M\pi}\int^{1}_{-1}{\sinc{\left[\dfrac{\pi}{2}M\left(\tau_{l,k,i}\right)\right]}}d\dfrac{\pi}{2}M\tau_{l,k,i}\notag\\
\approx & \frac{1}{\sqrt{M}\pi}\int^{\infty}_{-\infty}{\sinc{\left(\chi_{2}\right)}}d\chi_{2} \overset{{M}\rightarrow\infty}\approx 0\notag\\
=&\mathrm{E}_{\tau_{l,k,i}}\left[2\sqrt{M}\sinc\left[ M\frac{\pi}{2}\left( \tau_{l,k,i} \right) \right]\right],
\end{align}
where $\chi_{2}=\dfrac{\pi}{2}M\tau_{l,k,i}$.

According to the squeeze theorem \cite{book:sohrab2014}, we have
\begin{equation}
\mathrm{E}_{\tau_{l,k,i}}\left[\frac{\sin\left[ M \frac{\pi}{2}\left( \tau_{l,k,i}\right)\right]}{\sqrt{M}\sin\left[\frac{\pi}{2}\left(\tau_{l,k,i}\right)\right]}\right]\overset{{M}\rightarrow\infty}\approx0. \label{Eq_XX}
\end{equation}
Thus, for $i\neq j$, $i,\text{ }j\in \left\{ 1,\ldots, N\right\}$ and the large number of antennas regime, $\mathbb{E}_{\tau_{l,k,i},\tau_{l,k,j}}\left[\frac{\sin\left[ M \frac{\pi}{2}\left( \tau_{l,k,i} \right)\right]}{\sqrt{M}\sin\left[\frac{\pi}{2}\left(\tau_{l,k,i}\right)\right]}\frac{\sin\left[ M \frac{\pi}{2}\left( \tau_{l,k,j} \right)\right]}{\sqrt{M}\sin\left[\frac{\pi}{2}\left(\tau_{l,k,j}\right)\right]}\right]$ can be approximated as
\begin{equation}
\mathbb{E}_{\tau_{l,k,i}}\left[\frac{\sin\left[ M \frac{\pi}{2}\left( \tau_{l,k,i} \right)\right]}{\sqrt{M}\sin\left[\frac{\pi}{2}\left(\tau_{l,k,i}\right)\right]}\right]\mathbb{E}_{\tau_{l,k,j}}\left[\frac{\sin\left[ M \frac{\pi}{2}\left( \tau_{l,k,j} \right)\right]}{\sqrt{M}\sin\left[\frac{\pi}{2}\left(\tau_{l,k,j}\right)\right]}\right]\approx0.
\end{equation}
In the large number of antennas regime, Equation (\ref{Eq_AP6}) can be asymptotically approximated as
\begin{equation}
\mathbb{E}_{\tau_{l,k,i}}\left[\mathbf{F}_{\mathrm{RF},l}^{H}\left(\mathbf{h}_{\mathrm{SAoA},l,k}^{\mathrm{BS}}\right)^{\ast}\left(\mathbf{h}_{\mathrm{SAoA},l,k}^{\mathrm{BS}}\right)^{T}\mathbf{F}_{\mathrm{RF},l} \right] \approx \mathbf{I}_{\mathrm{ N}}, \text{\ }  i \in \left\{ 1,\ldots,N \right\}.
\end{equation}
Thus, Equation (\ref{LOS_formular}) can be asymptotically approximated as
\begin{align}
&\mathbb{E}_{\mathbf{H}_{\mathrm{SAoA},l,k}}\left[\overset{L}{\underset{l=1}{\sum}}\left({\frac{{\widetilde{\varpi}_{l,{k}}}{\color{black}\varsigma}_{l,k}}{{\color{black}\varsigma}_{l,k}+1}}\right)
\left[\frac{\sinc^{2}\left[P\frac{\pi}{2}\left(\eta_{l,k}\right)\right]{P}}{\sinc^{2}\left[\frac{\pi}{2}\left(\eta_{l,k}\right)\right]}\right]
\widehat{\beta}^{2}\mathrm{tr}{\left(\widehat{\mathbf{W}}_{\mathrm{eq},l}\widehat{\mathbf{W}}_{\mathrm{eq},l}^{H}\right)} \right]\notag \\
&\overset{M\rightarrow\infty}{\approx} \overset{L}{\underset{l=1}{\sum}}\left({\frac{{\widetilde{\varpi}_{l,{k}}}{\color{black}\varsigma}_{l,k}}{{\color{black}\varsigma}_{l,k}+1}}\right)
\mathbb{E}_{\eta_{l,k}}\left[\frac{\sinc^{2}\left[P\frac{\pi}{2}\left(\eta_{l,k}\right)\right]{P}}{\sinc^{2}\left[\frac{\pi}{2}\left(\eta_{l,k}\right)\right]}\right] \notag\\
&{\approx}\overset{L}{\underset{l=1}{\sum}}\left({\frac{{\widetilde{\varpi}_{l,{k}}}{\color{black}\varsigma}_{l,k}}{{\color{black}\varsigma}_{l,k}+1}}\right). \label{ff1}
\end{align}
On the other hand, we re-express the inter-cell interference caused by scattering components in the large number of antennas regime as
\begin{align}
&\overset{L}{\underset{l=1}{\sum }}\widetilde{\bm{\omega }}_{k}^{H}\left(
{\frac{{\widetilde{\varpi}_{l,{k}}}}{{\color{black}\varsigma}_{l,k}+1}} \right){\frac{1}{\mathrm{tr}\left( \widehat{\mathbf{W}}_{\mathrm{eq},l}\widehat{\mathbf{W}}_{\mathrm{eq},l}^{H}\right)}}\mathrm{tr}\left(\mathbf{F}_{\mathrm{RF},l}\widehat{\mathbf{W}}_{\mathrm{eq},l}
\widehat{\mathbf{W}}_{\mathrm{eq},l}^{H}
\mathbf{F}_{\mathrm{RF},l}^{H}\right)\widetilde{\bm{\omega }}_{k}
\notag\\
&\overset{M\rightarrow\infty}{\approx}\overset{L}{\underset{l=1}{\sum}}\left({\frac{{\widetilde{\varpi}_{l,{k}}}}{{\color{black}\varsigma}_{l,k}+1}}\right).\label{ff2}
\end{align}
By substituting Equations (\ref{ff1}) and (\ref{ff2}) into Equation (\ref{ff0}), we have
\begin{equation}
\Omega_{k}\approx E_{s}\left[\overset{L}{\underset{l=1}{\sum}}\left({\frac{{\widetilde{\varpi}_{l,{k}}}}{{\color{black}\varsigma}_{l,k}+1}}\right)+\overset{L}{\underset{l=1}{\sum}}\left({\frac{{\widetilde{\varpi}_{l,{k}}}{\color{black}\varsigma}_{l,k}}{{\color{black}\varsigma}_{l,k}+1}}\right)
\right] = E_{s}\left[\overset{L}{\underset{l=1}{\sum}}\left({\widetilde{\varpi}_{l,{k}}}
\right) \right].
\end{equation}
This completes the proof.

\clearpage{\pagestyle{empty}\cleardoublepage}

\renewcommand{\bibname}{Bibliography}
\addcontentsline{toc}{chapter}{\protect\numberline{}{Bibliography}}
\singlespacing
{\bibliographystyle{IEEEtran}
\bibliography{L_Z}
}

\end{document}